\documentclass[12pt,a4paper]{article}

\usepackage{ifthen} 
\newboolean{pdflatex}
\setboolean{pdflatex}{true} 

\newboolean{articletitles}
\setboolean{articletitles}{true} 

\newboolean{uprightparticles}
\setboolean{uprightparticles}{false} 

\usepackage[top=1in, bottom=1.25in, left=1in, right=1in]{geometry}

\usepackage{microtype}
\usepackage{lineno}  
\usepackage{xspace} 
\usepackage{caption} 

\usepackage{graphicx}  
\usepackage{color}
\usepackage{colortbl}
\graphicspath{{./figs/}} 

\usepackage{amsmath} 
\usepackage{amssymb}
\usepackage{amsfonts}
\usepackage{upgreek} 

\newcommand*\patchAmsMathEnvironmentForLineno[1]{%
\expandafter\let\csname old#1\expandafter\endcsname\csname #1\endcsname
\expandafter\let\csname oldend#1\expandafter\endcsname\csname
end#1\endcsname
 \renewenvironment{#1}%
   {\linenomath\csname old#1\endcsname}%
   {\csname oldend#1\endcsname\endlinenomath}%
}
\newcommand*\patchBothAmsMathEnvironmentsForLineno[1]{%
  \patchAmsMathEnvironmentForLineno{#1}%
  \patchAmsMathEnvironmentForLineno{#1*}%
}
\AtBeginDocument{%
\patchBothAmsMathEnvironmentsForLineno{equation}%
\patchBothAmsMathEnvironmentsForLineno{align}%
\patchBothAmsMathEnvironmentsForLineno{flalign}%
\patchBothAmsMathEnvironmentsForLineno{alignat}%
\patchBothAmsMathEnvironmentsForLineno{gather}%
\patchBothAmsMathEnvironmentsForLineno{multline}%
\patchBothAmsMathEnvironmentsForLineno{eqnarray}%
}

\usepackage{hyperxmp}

\usepackage[pdftex]{hyperref}

\usepackage[colorinlistoftodos,textsize=scriptsize]{todonotes}

\usepackage[bottom,flushmargin,hang,multiple]{footmisc}

\usepackage[all]{hypcap} 

\usepackage{multirow}

\usepackage{xspace} 
\usepackage{upgreek}

\def\lhcb   {\mbox{LHCb}\xspace}

\def\babar  {\mbox{BaBar}\xspace}
\def\belle  {\mbox{Belle}\xspace}

\def\cdf    {\mbox{CDF}\xspace}

\def\MagUp {\mbox{\em Mag\kern -0.05em Up}\xspace}

\ifthenelse{\boolean{uprightparticles}}%
{

 \def\Peta        {\ensuremath{\upeta}\xspace}

 \def\Pmu         {\ensuremath{\upmu}\xspace}                 
 \def\Pnu         {\ensuremath{\upnu}\xspace}                 
                  
 \def\Ppi         {\ensuremath{\uppi}\xspace}                 
                  
 \def\Prho        {\ensuremath{\uprho}\xspace}                 
                  
 \def\Ptau        {\ensuremath{\uptau}\xspace}

 \def\Pchi        {\ensuremath{\upchi}\xspace}                 
 \def\Ppsi        {\ensuremath{\uppsi}\xspace}

 \def\PDelta      {\ensuremath{\Delta}\xspace}                 
 \def\PXi         {\ensuremath{\Xi}\xspace}                 
 \def\PLambda     {\ensuremath{\Lambda}\xspace}                 
 \def\PSigma      {\ensuremath{\Sigma}\xspace}                 
 \def\POmega      {\ensuremath{\Omega}\xspace}                 
 \def\PUpsilon    {\ensuremath{\Upsilon}\xspace}

 \def\PB      {\ensuremath{\mathrm{B}}\xspace}                 
                  
 \def\PD      {\ensuremath{\mathrm{D}}\xspace}

 \def\PJ      {\ensuremath{\mathrm{J}}\xspace}                 
 \def\PK      {\ensuremath{\mathrm{K}}\xspace}

 \def\PW      {\ensuremath{\mathrm{W}}\xspace}

 \def\Pb      {\ensuremath{\mathrm{b}}\xspace}                 
 \def\Pc      {\ensuremath{\mathrm{c}}\xspace}                 
 \def\Pd      {\ensuremath{\mathrm{d}}\xspace}                 
 \def\Pe      {\ensuremath{\mathrm{e}}\xspace}

 \def\Pi      {\ensuremath{\mathrm{i}}\xspace}

 \def\Pp      {\ensuremath{\mathrm{p}}\xspace}

 \def\Ps      {\ensuremath{\mathrm{s}}\xspace}                 
 \def\Pt      {\ensuremath{\mathrm{t}}\xspace}                 
 \def\Pu      {\ensuremath{\mathrm{u}}\xspace}

 \def\thebaroffset{0.0em}
}
{

 \def\Peta        {\ensuremath{\eta}\xspace}

 \def\Pmu         {\ensuremath{\mu}\xspace}                 
 \def\Pnu         {\ensuremath{\nu}\xspace}                 
                  
 \def\Ppi         {\ensuremath{\pi}\xspace}                 
                  
 \def\Prho        {\ensuremath{\rho}\xspace}                 
                  
 \def\Ptau        {\ensuremath{\tau}\xspace}

 \def\Pchi        {\ensuremath{\chi}\xspace}                 
 \def\Ppsi        {\ensuremath{\psi}\xspace}                 
                  
 \mathchardef\PDelta="7101
 \mathchardef\PXi="7104
 \mathchardef\PLambda="7103
 \mathchardef\PSigma="7106
 \mathchardef\POmega="710A
 \mathchardef\PUpsilon="7107
                  
 \def\PB      {\ensuremath{B}\xspace}                 
                  
 \def\PD      {\ensuremath{D}\xspace}

 \def\PJ      {\ensuremath{J}\xspace}                 
 \def\PK      {\ensuremath{K}\xspace}

 \def\PW      {\ensuremath{W}\xspace}

 \def\Pb      {\ensuremath{b}\xspace}                 
 \def\Pc      {\ensuremath{c}\xspace}                 
 \def\Pd      {\ensuremath{d}\xspace}                 
 \def\Pe      {\ensuremath{e}\xspace}

 \def\Pi      {\ensuremath{i}\xspace}

 \def\Pp      {\ensuremath{p}\xspace}

 \def\Ps      {\ensuremath{s}\xspace}                 
 \def\Pt      {\ensuremath{t}\xspace}                 
 \def\Pu      {\ensuremath{u}\xspace}

 \def\thebaroffset{0.18em}
}
\newcommand{\offsetoverline}[2][\thebaroffset]{\kern #1\overline{\kern -#1 #2}}%

\makeatletter
\ifcase \@ptsize \relax
  \newcommand{\miniscule}{\@setfontsize\miniscule{4}{5}}
\or
  \newcommand{\miniscule}{\@setfontsize\miniscule{5}{6}}
\or
  \newcommand{\miniscule}{\@setfontsize\miniscule{5}{6}}
\fi
\makeatother

\DeclareRobustCommand{\optbar}[1]{\shortstack{{\miniscule (\rule[.5ex]{1.25em}{.18mm})}
  \\ [-.7ex] $#1$}}

\def\epem       {{\ensuremath{\Pe^+\Pe^-}}\xspace}

\def\mup        {{\ensuremath{\Pmu^+}}\xspace}
\def\mun        {{\ensuremath{\Pmu^-}}\xspace} 

\def\mumu       {{\ensuremath{\Pmu^+\Pmu^-}}\xspace}

\def\tautau     {{\ensuremath{\Ptau^+\Ptau^-}}\xspace}

\def\ellm       {{\ensuremath{\ell^-}}\xspace}
\def\ellp       {{\ensuremath{\ell^+}}\xspace}

\def\ellell     {\ensuremath{\ell^+ \ell^-}\xspace}

\def\neu        {{\ensuremath{\Pnu}}\xspace}

\def\neum       {{\ensuremath{\neu_\mu}}\xspace}

\def\Wpm    {{\ensuremath{\PW^\pm}}\xspace}

\def\uquark    {{\ensuremath{\Pu}}\xspace}
\def\uquarkbar {{\ensuremath{\overline \uquark}}\xspace}

\def\dquark    {{\ensuremath{\Pd}}\xspace}
\def\dquarkbar {{\ensuremath{\overline \dquark}}\xspace}
\def\ddbar     {{\ensuremath{\dquark\dquarkbar}}\xspace}
\def\squark    {{\ensuremath{\Ps}}\xspace}
\def\squarkbar {{\ensuremath{\overline \squark}}\xspace}
\def\ssbar     {{\ensuremath{\squark\squarkbar}}\xspace}
\def\cquark    {{\ensuremath{\Pc}}\xspace}
\def\cquarkbar {{\ensuremath{\overline \cquark}}\xspace}
\def\ccbar     {{\ensuremath{\cquark\cquarkbar}}\xspace}
\def\bquark    {{\ensuremath{\Pb}}\xspace}
\def\bquarkbar {{\ensuremath{\overline \bquark}}\xspace}
\def\bbbar     {{\ensuremath{\bquark\bquarkbar}}\xspace}
\def\tquark    {{\ensuremath{\Pt}}\xspace}

\def\pion   {{\ensuremath{\Ppi}}\xspace}
\def\piz    {{\ensuremath{\pion^0}}\xspace}
\def\pip    {{\ensuremath{\pion^+}}\xspace}
\def\pim    {{\ensuremath{\pion^-}}\xspace}

\def\pimp   {{\ensuremath{\pion^\mp}}\xspace}

\def\rhomeson {{\ensuremath{\Prho}}\xspace}
\def\rhoz     {{\ensuremath{\rhomeson^0}}\xspace}

\def\kaon    {{\ensuremath{\PK}}\xspace}
\def\Kbar    {{\ensuremath{\offsetoverline{\PK}}}\xspace}

\def\KorKbar {\kern \thebaroffset\optbar{\kern -\thebaroffset \PK}{}\xspace}
\def\Kz      {{\ensuremath{\kaon^0}}\xspace}

\def\Kp      {{\ensuremath{\kaon^+}}\xspace}
\def\Km      {{\ensuremath{\kaon^-}}\xspace}
\def\Kpm     {{\ensuremath{\kaon^\pm}}\xspace}

\def\KS      {{\ensuremath{\kaon^0_{\mathrm{S}}}}\xspace}

\def\Kstarz  {{\ensuremath{\kaon^{*0}}}\xspace}
\def\Kstarzb {{\ensuremath{\Kbar{}^{*0}}}\xspace}
\def\Kstar   {{\ensuremath{\kaon^*}}\xspace}

\def\Kstarp  {{\ensuremath{\kaon^{*+}}}\xspace}
\def\Kstarm  {{\ensuremath{\kaon^{*-}}}\xspace}

\def\Dbar    {{\ensuremath{\offsetoverline{\PD}}}\xspace}
\def\D       {{\ensuremath{\PD}}\xspace}

\def\DorDbar {\kern \thebaroffset\optbar{\kern -\thebaroffset \PD}\xspace}
\def\Dz      {{\ensuremath{\D^0}}\xspace}
\def\Dzb     {{\ensuremath{\Dbar{}^0}}\xspace}
\def\Dp      {{\ensuremath{\D^+}}\xspace}
\def\Dm      {{\ensuremath{\D^-}}\xspace}

\def\DpDm    {\ensuremath{\Dp {\kern -0.16em \Dm}}\xspace}
\def\Dstar   {{\ensuremath{\D^*}}\xspace}

\def\Dstarz  {{\ensuremath{\D^{*0}}}\xspace}
\def\Dstarzb {{\ensuremath{\Dbar{}^{*0}}}\xspace}

\def\Dstarp  {{\ensuremath{\D^{*+}}}\xspace}
\def\Dstarm  {{\ensuremath{\D^{*-}}}\xspace}

\def\Ds      {{\ensuremath{\D^+_\squark}}\xspace}
\def\Dsp     {{\ensuremath{\D^+_\squark}}\xspace}
\def\Dsm     {{\ensuremath{\D^-_\squark}}\xspace}

\def\B       {{\ensuremath{\PB}}\xspace}
\def\Bbar    {{\ensuremath{\offsetoverline{\PB}}}\xspace}

\def\BorBbar {\kern \thebaroffset\optbar{\kern -\thebaroffset \PB}\xspace}
\def\Bz      {{\ensuremath{\B^0}}\xspace}
\def\Bzb     {{\ensuremath{\Bbar{}^0}}\xspace}
\def\Bd      {{\ensuremath{\B^0}}\xspace}

\def\BdorBdbar {\kern \thebaroffset\optbar{\kern -\thebaroffset \Bd}\xspace}
\def\Bu      {{\ensuremath{\B^+}}\xspace}
\def\Bub     {{\ensuremath{\B^-}}\xspace}
\def\Bp      {{\ensuremath{\Bu}}\xspace}
\def\Bm      {{\ensuremath{\Bub}}\xspace}

\def\Bs      {{\ensuremath{\B^0_\squark}}\xspace}
\def\Bsb     {{\ensuremath{\Bbar{}^0_\squark}}\xspace}
\def\BsorBsbar {\kern \thebaroffset\optbar{\kern -\thebaroffset \Bs}\xspace}
\def\Bc      {{\ensuremath{\B_\cquark^+}}\xspace}
\def\Bcp     {{\ensuremath{\B_\cquark^+}}\xspace}

\def\Bds     {{\ensuremath{\B_{(\squark)}^0}}\xspace}

\def\jpsi     {{\ensuremath{{\PJ\mskip -3mu/\mskip -2mu\Ppsi}}}\xspace}
\def\psitwos  {{\ensuremath{\Ppsi{(2S)}}}\xspace}

\def\etac     {{\ensuremath{\Peta_\cquark}}\xspace}

\def\chic     {{\ensuremath{\Pchi_\cquark}}\xspace}

\def\chicone  {{\ensuremath{\Pchi_{\cquark 1}}}\xspace}
\def\chictwo  {{\ensuremath{\Pchi_{\cquark 2}}}\xspace}

\def\Y#1S{\ensuremath{\PUpsilon{(#1S)}}\xspace}

\def\FourS {{\Y4S}}

\def\chibone  {{\ensuremath{\Pchi_{\bquark 1}}}\xspace}
\def\chibtwo  {{\ensuremath{\Pchi_{\bquark 2}}}\xspace}

\def\proton      {{\ensuremath{\Pp}}\xspace}
\def\antiproton  {{\ensuremath{\overline \proton}}\xspace}

\def\Lz          {{\ensuremath{\PLambda}}\xspace}
\def\Lbar        {{\ensuremath{\offsetoverline{\PLambda}}}\xspace}
\def\LorLbar     {\kern \thebaroffset\optbar{\kern -\thebaroffset \PLambda}\xspace}
\def\Lambdares   {{\ensuremath{\PLambda}}\xspace}

\def\Sigmares    {{\ensuremath{\PSigma}}\xspace}

\def\Xires       {{\ensuremath{\PXi}}\xspace}

\def\Omegares    {{\ensuremath{\POmega}}\xspace}

\def\Lc          {{\ensuremath{\Lz^+_\cquark}}\xspace}
\def\Lcbar       {{\ensuremath{\Lbar{}^-_\cquark}}\xspace}
\def\Xic         {{\ensuremath{\Xires_\cquark}}\xspace}

\def\Xicp        {{\ensuremath{\Xires^+_\cquark}}\xspace}

\def\Xiccp       {{\ensuremath{\Xires^+_{\cquark\cquark}}}\xspace}
\def\Xiccpp      {{\ensuremath{\Xires^{++}_{\cquark\cquark}}}\xspace}

\def\Omegacc     {{\ensuremath{\Omegares^+_{\cquark\cquark}}}\xspace}

\def\Lbsimple           {{\ensuremath{\Lz_\bquark}}\xspace}
\def\Lb           {{\ensuremath{\Lz^0_\bquark}}\xspace}

\def\Sigmab       {{\ensuremath{\Sigmares_\bquark}}\xspace}

\def\Xib          {{\ensuremath{\Xires_\bquark}}\xspace}
\def\Xibz         {{\ensuremath{\Xires^0_\bquark}}\xspace}
\def\Xibm         {{\ensuremath{\Xires^-_\bquark}}\xspace}

\def\Omegab       {{\ensuremath{\Omegares^-_\bquark}}\xspace}

\def\BF         {{\ensuremath{\mathcal{B}}}\xspace}

\def\to                 {\ensuremath{\rightarrow}\xspace}

\def\CP                {{\ensuremath{C\!P}}\xspace}

\def\rhobar {{\ensuremath{\overline \rho}}\xspace}
\def\etabar {{\ensuremath{\overline \eta}}\xspace}

\def\Vud  {{\ensuremath{V_{\uquark\dquark}^{\phantom{\ast}}}}\xspace}
\def\Vcd  {{\ensuremath{V_{\cquark\dquark}^{\phantom{\ast}}}}\xspace}
\def\Vtd  {{\ensuremath{V_{\tquark\dquark}^{\phantom{\ast}}}}\xspace}

\def\Vts  {{\ensuremath{V_{\tquark\squark}^{\phantom{\ast}}}}\xspace}
\def\Vub  {{\ensuremath{V_{\uquark\bquark}^{\phantom{\ast}}}}\xspace}
\def\Vcb  {{\ensuremath{V_{\cquark\bquark}^{\phantom{\ast}}}}\xspace}
\def\Vtb  {{\ensuremath{V_{\tquark\bquark}^{\phantom{\ast}}}}\xspace}

\def\Vtss  {{\ensuremath{V_{\tquark\squark}^\ast}}\xspace}
\def\Vubs  {{\ensuremath{V_{\uquark\bquark}^\ast}}\xspace}
\def\Vcbs  {{\ensuremath{V_{\cquark\bquark}^\ast}}\xspace}
\def\Vtbs  {{\ensuremath{V_{\tquark\bquark}^\ast}}\xspace}

\newcommand{\dms}{{\ensuremath{\Delta m_{\squark}}}\xspace}

\newcommand{\ACP}{{\ensuremath{{\mathcal{A}}^{\CP}}}\xspace}

\def\AT#1     {\ensuremath{A_{\mathrm{T}}^{#1}}\xspace}           

\def\C#1      {\ensuremath{\mathcal{C}_{#1}}\xspace}                       
\def\Cp#1     {\ensuremath{\mathcal{C}_{#1}^{'}}\xspace}                    
\def\Ceff#1   {\ensuremath{\mathcal{C}_{#1}^{\mathrm{(eff)}}}\xspace}        
\def\Cpeff#1  {\ensuremath{\mathcal{C}_{#1}^{'\mathrm{(eff)}}}\xspace}       
\def\Ope#1    {\ensuremath{\mathcal{O}_{#1}}\xspace}                       
\def\Opep#1   {\ensuremath{\mathcal{O}_{#1}^{'}}\xspace}                    

\def\xprime     {\ensuremath{x^{\prime}}\xspace}
\def\yprime     {\ensuremath{y^{\prime}}\xspace}
\def\ycp        {\ensuremath{y_{\CP}}\xspace}
\def\agamma     {\ensuremath{A_{\Gamma}}\xspace}

\newcommand{\aunit}[1]{\ensuremath{\text{\,#1}}}       

\newcommand{\tev}{\aunit{Te\kern -0.1em V}\xspace}
\newcommand{\gev}{\aunit{Ge\kern -0.1em V}\xspace}
\newcommand{\mev}{\aunit{Me\kern -0.1em V}\xspace}
\newcommand{\kev}{\aunit{ke\kern -0.1em V}\xspace}
\newcommand{\ev}{\aunit{e\kern -0.1em V}\xspace}
 
\newcommand{\mevc}{\ensuremath{\aunit{Me\kern -0.1em V\!/}c}\xspace}
\newcommand{\gevc}{\ensuremath{\aunit{Ge\kern -0.1em V\!/}c}\xspace}
\newcommand{\kevcc}{\ensuremath{\aunit{ke\kern -0.1em V\!/}c^2}\xspace}
\newcommand{\mevcc}{\ensuremath{\aunit{Me\kern -0.1em V\!/}c^2}\xspace}
\newcommand{\gevcc}{\ensuremath{\aunit{Ge\kern -0.1em V\!/}c^2}\xspace}
\newcommand{\gevgevcccc}{\ensuremath{\gev^2\!/c^4}\xspace} 

\def\mbarn{\aunit{mb}\xspace}

\def\pb {\aunit{pb}\xspace}
\def\invpb {\ensuremath{\pb^{-1}}\xspace}
\def\fb   {\ensuremath{\aunit{fb}}\xspace}
\def\invfb   {\ensuremath{\fb^{-1}}\xspace}
\def\ab   {\ensuremath{\aunit{ab}}\xspace}
\def\invab   {\ensuremath{\ab^{-1}}\xspace}

\def\ps   {\ensuremath{\aunit{ps}}\xspace}
\def\fs   {\aunit{fs}}

\def\mhz  {\ensuremath{\aunit{MHz}}\xspace}
\def\khz  {\ensuremath{\aunit{kHz}}\xspace}

\newcommand{\stat}{\aunit{(stat)}\xspace}
\newcommand{\syst}{\aunit{(syst)}\xspace}

\newcommand{\chisq}{\ensuremath{\chi^2}\xspace}

\def\deriv {\ensuremath{\mathrm{d}}}

\def\gsim{{~\raise.15em\hbox{$>$}\kern-.85em
          \lower.35em\hbox{$\sim$}~}\xspace}
\def\lsim{{~\raise.15em\hbox{$<$}\kern-.85em
          \lower.35em\hbox{$\sim$}~}\xspace}

\def\sqs   {\ensuremath{\protect\sqrt{s}}\xspace}
\def\sqsnn {\ensuremath{\protect\sqrt{s_{\scriptscriptstyle\text{NN}}}}\xspace}
\def\pt         {\ensuremath{p_{\mathrm{T}}}\xspace}

\def\degrees{\ensuremath{^{\circ}}\xspace}

\def\tell1  {TELL1\xspace}
\def\ukl1   {UKL1\xspace}

\newcommand{\eg}{\mbox{\itshape e.g.}\xspace}
\newcommand{\ie}{\mbox{\itshape i.e.}\xspace}

\newcommand{\etc}{\mbox{\itshape etc.}\xspace}

\newcommand{\phz}{\phantom{0}}

\usepackage{cite} 
\usepackage{mciteplus}

\usepackage{graphicx}

\begin{document}

\begin{titlepage}
\noindent
\begin{tabular*}{\linewidth}{lc@{\extracolsep{\fill}}r@{\extracolsep{0pt}}}
 & & 12 December 2022 \\ 
\end{tabular*} 

\vspace*{4cm}
{\huge\bf 
\begin{center}  
Heavy Flavour Physics and CP Violation at LHCb: a Ten-Year Review
\end{center}
}
\vspace{1.0cm}
\begin{center}
Shanzhen Chen$^1$, Yiming Li$^1$, Wenbin Qian$^2$, Zhihong Shen$^3$, Yuehong Xie$^4$,\\
Zhenwei Yang$^3$, Liming Zhang$^5$, Yanxi Zhang$^3$ \\
\vspace{0.5cm}
{
\normalfont\itshape\footnotesize
$ ^1$Institute of High Energy Physics, Chinese Academy of Sciences, Beijing, China\\
$ ^2$University of Chinese Academy of Sciences, Beijing, China\\
$ ^3$State Key Laboratory of Nuclear Physics and Technology {\rm\&} School of Physics, Peking University, Beijing, China\\
$ ^4$Key Laboratory of Quark and Lepton Physics of Ministry of Education {\rm \&} Institute of Particle Physics, Central China Normal University, Wuhan, Hubei, China\\
$ ^5$Department of Engineering Physics {\rm\&} Center for High Energy Physics, Tsinghua University, Beijing, China\\
}
\end{center}
\vspace{\fill}

\begin{abstract}
Heavy flavour physics provides excellent opportunities to indirectly search for new physics at very high energy scales and to  study hadron properties for 
deep understanding of the strong interaction. 
The LHCb experiment has been playing a leading role in the study of heavy flavour physics since the start of the LHC operations about ten years ago, 
and made a range of high-precision measurements  and unexpected discoveries, which may have far-reaching implications on  the field of particle physics.
This review highlights a selection of the most influential physics results on CP violation, rare decays, and heavy flavour production and  spectroscopy obtained by LHCb using the data collected during the first two operation periods of the LHC.
The upgrade plan of LHCb and the physics prospects are also briefly discussed. 

\end{abstract}
\vspace{\fill}
\end{titlepage}


\setcounter{secnumdepth}{3}
\setcounter{tocdepth}{2}
\tableofcontents
\cleardoublepage

\section{Introduction}
\label{sec:introduction}


With the discovery of the Higgs boson at the Large Hadron Collider (LHC) in 2012~\cite{Bruening:782076,ATLAS:2012yve,CMS:2012qbp}, all fundamental particles expected in the Standard Model (SM) of particle physics are found and the SM is finally completed.  The SM has achieved tremendous success in explaining experimental results in high-energy physics.
However, there are still many key questions that are  not answered in the SM, such as the mechanism to generate the matter-antimatter asymmetry in the Universe, the origin of the three generations of fermions and their mixing,  
and the nature of dark matter and dark energy.
It is commonly believed that new physics (NP) beyond the SM should exist at or above the $\kern -0.25em\tev$ energy scale. Flavour physics can provide a unique approach to indirectly probe NP at energy scales far above  \tev via precision study of charge-parity  (CP) violation and rare phenomena, complementary to the direct search for new particles and interactions at the energy frontier. 
Flavour physics also serves as a natural laboratory to test quantum chromodynamics (QCD), the theory of the strong interaction, via measurements of hadron production and spectroscopy.
The Large Hadron Collider beauty (LHCb) experiment has been playing a leading role in the study of heavy-flavour physics since the start of the LHC, and has made a series of discoveries and improvements in CP violation, rare decays, and hadron production and spectroscopy.
This review aims to present a selection of the high-impact physics results on the above subjects from the LHCb experiment, and to briefly discuss the prospects.
Due to the limited space, not all interesting results can be covered here.
For a complete list of all LHCb physics results, please refer to the official LHCb summary~\cite{LHCbpapers}.

\begin{figure}[!b]
 \begin{center}
  \includegraphics[width=0.70\textwidth]{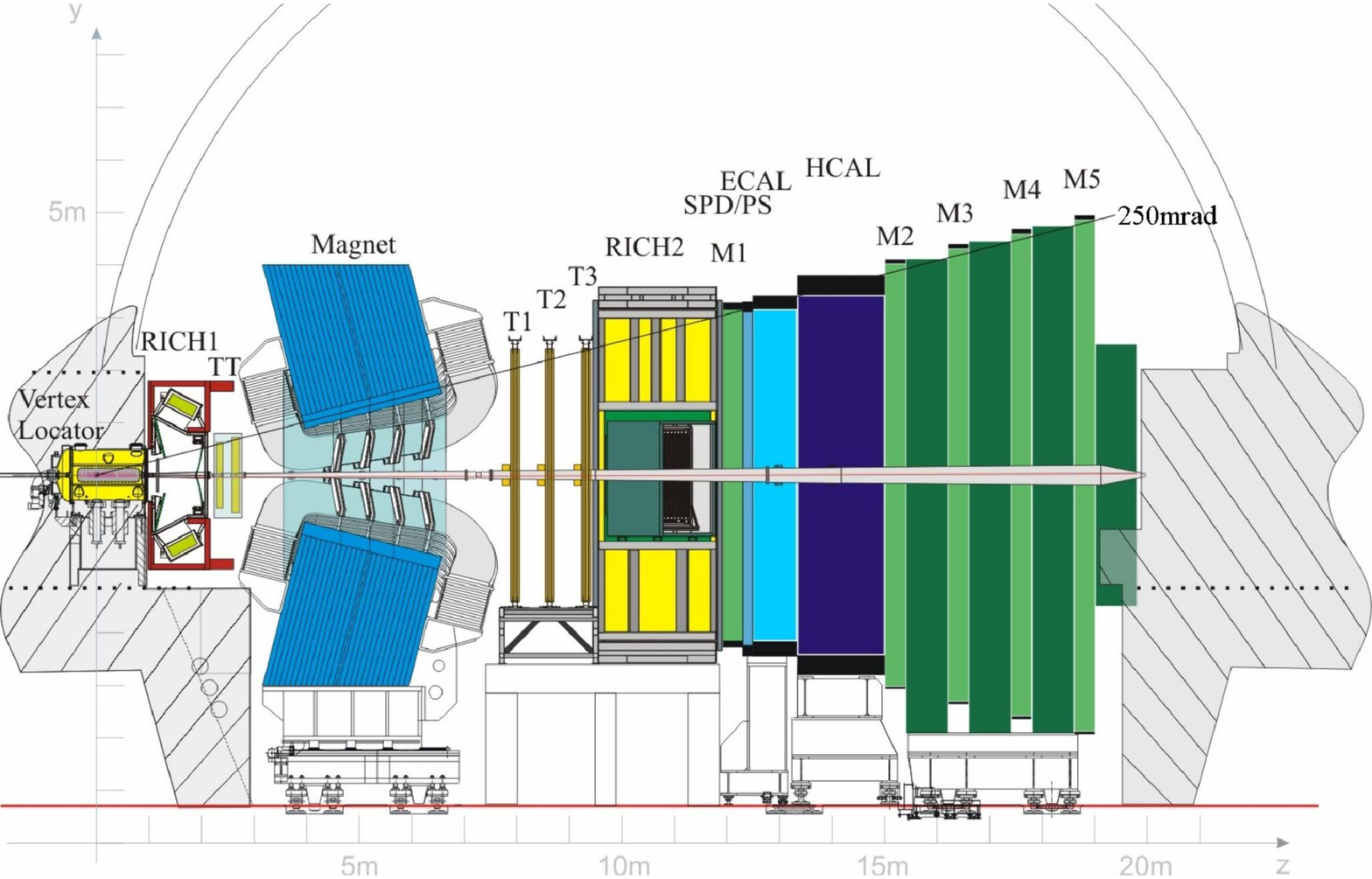}
 \end{center}
 \caption{
Layout of the LHCb detector~\cite{LHCb-DP-2008-001}.
 }
 \label{fig:01:01:LHCbDetector}
\end{figure}

The LHCb detector~\cite{LHCb-DP-2008-001} is optimised for the study of the decays of heavy-flavour hadrons, \ie hadrons containing heavy quarks (\bquark or \cquark quarks, often collectively referred to as $Q$).
In proton-proton ($pp$) collisions at LHC energies, the \bbbar pairs are produced dominantly through the gluon fusion process $gg\to \bbbar$.
Due to the large Lorentz boost along the proton beam in the laboratory frame, the \bquark and \bquarkbar quarks 
generated in a pair are highly correlated in their momentum directions, 
either both in the forward region or both in the backward region in the majority of  cases. 
In order to take advantage of this characteristic of the \bbbar pair at the LHC, 
 the LHCb detector is designed to have a forward geometry
as  shown in Fig.~\ref{fig:01:01:LHCbDetector} to cover the forward region of $pp$ collisions.
The flavour of the  \bquark-hadron 
under study can usually  be tagged by the other \bquark-hadron that is also inside the LHCb acceptance.
This enhances the potential of the LHCb experiment in the study of CP violation and mixing with $\Bz$ and $\Bs$ decays. 
 
Excellent vertex and momentum resolution, particle identification \etc are key ingredients for flavour physics measurements at hadron colliders. 
A silicon vertex locator (VELO) surrounding 
the $pp$
collision region is used to precisely determine  the primary interaction vertices (PVs) and the displaced secondary/tertiary vertices (SVs/TVs) formed by the decay products of heavy-flavour hadrons.
The VELO system offers decay time measurements with a typical resolution better than $50\fs$,  thus allows the LHCb experiment to make precision measurement of hadron lifetimes and resolve the  fast $\Bs$-$\Bsb$ oscillation, which has a period of about 350\fs.
The clear separation of SVs from PVs also allows for substantial suppression of the combinatorial background, which is extremely high in $pp$ collisions. 
In addition to the VELO, the LHCb tracking system includes also four trackers, namely the TT and T1-T3 stations, located  upstream and downstream of the dipole magnet, respectively. 
Together with a magnet that has a bending power of $4\,\mathrm{Tm}$, the tracking system provides precise measurements of the momenta of charged particles. The momentum resolution ($\Delta p/p$) is typically 0.5\% for
low momentum tracks and 1.0\% for track momentum up to $200\gevc$.
The mass resolution for $\bquark$-hadrons can be as good as $8\mevcc$, precise enough to distinguish decays of $\Bz$ and $\Bs$ mesons to the same final state.

There are two ring-imaging Cherenkov detectors (RICH1 and RICH2)  used to identify charged hadrons in the momentum range $p\in(2,100)\gevc$.  
The  RICH detectors are very powerful at suppressing misidentification background for \bquark-hadron  decays to final states containing
charged kaons, pions or protons.
The muon system provides excellent muon identification and is essential for reconstruction of decays with muons in the final states.
The electromagnetic calorimeter is used  for photon and electron reconstruction and identification. Together with the hadronic calorimeter, it also provides information for event trigger.

The trigger system is crucial for the success of the LHCb experiment. 
The hardware trigger at the first level
reduces the data rate from $40\mhz$ down to $1\mhz$, at
which point the flexible software-based trigger takes over to further reduce the rate to around $12\khz$ for offline processing and  analysis. 
The ability to sustain such large rates enables the LHCb experiment to trigger with high efficiency on decay processes across a wide range of final states
and to provide large data samples for study of exclusive rare decay modes as well as for inclusive data mining. 
Of particular importance is the LHCb muon trigger, which allows events containing one muon or two muons to be selected with greater than 95\% efficiency. This is ideal for studying $b$-hadron decays to
$\jpsi$ or $\psitwos$ mesons that further decay to $\mup\mun$ pairs and semileptonic decays into muons.
Meanwhile, there are also highly efficient triggers that are purely based on the multi-body topology of the final state hadrons. These triggers are important for  reconstructing decays of heavy-quark hadrons to final states without muons.

The LHCb experiment  collected a data sample corresponding to an integrated luminosity of  3\invfb in $pp$ collisions  at centre-of-mass energies $\sqs=7$ and $8\tev$ from 2011 to 2012 (Run 1), and another sample of 6\invfb at $\sqs=13\tev$ from 2015 to 2018 (Run 2). 
Results discussed in this review are based on either full Run 1 and Run 2 data samples or a subset.
Since the cross-sections for $c$- and $b$-quark production in  $pp$ collisions at  13\tev  are about twice of the cross-sections at 7 and 8\tev~\cite{LHCb-PAPER-2010-002,LHCb-PAPER-2011-003,LHCb-PAPER-2013-016,LHCb-PAPER-2012-041,LHCb-PAPER-2015-037,LHCb-PAPER-2015-041,LHCb-PAPER-2016-031}, and  the trigger scheme for \mbox{Run 2} has also been improved compared with \mbox{Run 1}, the number of recorded charm and beauty decays available for physics analysis is more than four times higher in the Run 2 data  than in Run 1 data. 
In total, 
more than $10^{11}$ \bquark-hadrons and $10^{12}$ \cquark-hadrons have been produced within the LHCb detector acceptance. The typical trigger and selection efficiencies are of the order of $10^{-3}$ to $10^{-2}$ 
for  decays only to  charged particles 
and  $10^{-4}$ to $10^{-3}$ for decays to final states  involving photons, $\piz$, $\Lz^0$ or $\KS$ particles.
The enormous $b$- and $c$-hadron samples 
 form the basis of 
precision measurements of CP violation, exploration of rare decays, and searches for new hadrons.


This review is structured as follows, with each section
covering a different subject. 
Recent results of heavy-flavour production and spectroscopy are shown in Section~\ref{sec:production}. For heavy-flavour production, recent results of associated production and the studies in heavy-ion collisions are shown; for spectroscopy, the results of conventional hadrons and exotic hadrons are summarised.
Section~\ref{sec:raredecay} discusses rare $B$-hadron decays. The results on purely leptonic $B$-meson decays, semileptonic $b\to\squark\ellp\ellm$ decays, and radiative $b\to\squark\gamma$ decays are shown.\footnote{Charge conjugation is implied throughout unless stated otherwise.} 
Some puzzling results in angular distributions and  lepton flavour universality tests are discussed in details.
Section~\ref{sec:beautyCPV} presents the latest results on CP violation in the beauty sector.
Emphasis is put on the progresses that have been made for precision test of the Cabibbo-Kobayashi-Maskawa (CKM) mechanism, such as significant improvement in the determination of the parameters $\gamma$, $\beta$, $\beta_s$, $\Vub$, $\Vcb$, and $\Delta m_{s/d}$.
Section~\ref{sec:charmCPV} provides recent results of CP violation in the charm sector, including those for charm mixing and the observation of CP violation in $\Dz$ decays. 
The final section provides a picture of the LHCb upgrade, as well as a brief summary of the content in this review. The main goals and modifications to the detector in Upgrade I and Upgrade II are introduced, and the prospects of some key measurements are presented.

\clearpage
\section{Heavy-flavour production and spectroscopy}
\label{sec:production}


Rich information on QCD dynamics can be deciphered from measurements of heavy-flavour production ~\cite{Cacciari:2012ny,Andronic:2015wma,Zhang:2021any,Chen:2021obo,Chen:2021tmf,Wang:2020jjq,Chen:2020ajc,Yang:2019lls,He:2019qqr,Butenschoen:2011yh,Lansberg:2016deg,Zhang:2014ybe,Ma:2012hh,Feng:2018ukp,Wang:2014vsa,Tang:2014tga,Sun:2018yam,Li:2017qmj,Han:2014jya,Han:2014kxa,Zhang:2020atv,Liu:2019iml,Ma:2017xno,Sun:2014gca,Ma:2010jj,Ma:2010yw,Li:2009zu,Ma:2010vd,Zhang:2009ym,Shao:2012iz,Chang:2003cr,Niu:2018tvo,He:2017had,Zhang:2021wjj,Chen:2019ykv,Hu:2021gdg,Jia:2020csg}
and studies of heavy-flavour spectroscopy~\cite{Brambilla:2019esw,Chen:2016qju,Guo:2017jvc,Swanson:2006st,Olsen:2017bmm,Liu:2019zoy,Chen:2016spr,Liu:2013waa,Guo:2019twa,Yuan:2018inv,Liu:2009qhy,Dong:2021bvy,Shi:2018rhk,Ali:2017jda,Esposito:2014rxa,Maiani:2014aja,Peng:2020xrf,Yang:2020atz,Richard:2021nvn,Karliner:2017qhf,Richard:2016eis,Wu:2021tzo}. 
The mass of a heavy quark, which is much larger than the nonperturbative QCD scale $\Lambda_\mathrm{QCD}$,  provides an energy scale that allows for perturbative calculation of heavy-quark production. 
The production of heavy quark pairs, $Q\overline{Q}$,
is predominantly  in the initial stage of the collision, thus can be used to probe properties of the colliding system and the possibly created QCD medium~\cite{Wicks:2005gt,Zhou:2014kka}.
The presence of heavy quark(s) also provides practical benefits for theoretical and experimental studies of spectroscopy.   
Heavy quarks are approximately nonrelativistic in hadrons, which makes it possible to simplify theoretical calculations.
The large mass and the weak decay of heavy-flavour hadrons offer essential features, such as decay products with high transverse momentum, \pt, and vertices displaced from the PVs, which can be exploited to reject the huge QCD background at hadron colliders.

As discussed in Section~\ref{sec:introduction}, 
the excellent performance of the LHCb detector, owing to the dedicated design for heavy-flavour hadrons, enables LHCb to make great achievements in the study of heavy-flavour production and spectroscopy, \eg the observation of pentaquark states and the doubly charmed baryon $\Xires_{cc}^{++}$.
In this section, relevant results from the LHCb experiment are reviewed, with a focus on recent developments.

\subsection{Production} 
A summary of LHCb production measurements for open heavy-flavour hadrons, heavy quarkonia and pairs of heavy-flavour hadrons at LHCb~\cite{LHCb-CONF-2019-004,LHCb-PAPER-2011-003,LHCb-PAPER-2011-013,LHCb-PAPER-2011-019,LHCb-PAPER-2011-030,LHCb-PAPER-2011-034,LHCb-PAPER-2011-036,LHCb-PAPER-2011-043,LHCb-PAPER-2011-045,LHCb-PAPER-2012-003,LHCb-PAPER-2012-028,LHCb-PAPER-2012-039,LHCb-PAPER-2012-041,LHCb-PAPER-2013-004,LHCb-PAPER-2013-028,LHCb-PAPER-2013-052,LHCb-PAPER-2013-066,LHCb-PAPER-2014-004,LHCb-PAPER-2014-015,LHCb-PAPER-2014-029,LHCb-PAPER-2014-031,LHCb-PAPER-2014-040,LHCb-PAPER-2014-050,LHCb-PAPER-2015-016,LHCb-PAPER-2015-032,LHCb-PAPER-2015-037,LHCb-PAPER-2015-041,LHCb-PAPER-2015-045,LHCb-PAPER-2015-046,LHCb-PAPER-2015-058,LHCb-PAPER-2016-042,LHCb-PAPER-2016-057,LHCb-PAPER-2016-064,LHCb-PAPER-2017-014,LHCb-PAPER-2017-015,LHCb-PAPER-2017-037,LHCb-PAPER-2018-002,LHCb-PAPER-2018-021,LHCb-PAPER-2018-035,LHCb-PAPER-2018-047,LHCb-PAPER-2018-048,LHCb-PAPER-2018-049,LHCb-PAPER-2019-024,LHCb-PAPER-2019-033,LHCb-PAPER-2019-035,LHCb-PAPER-2020-010,LHCb-PAPER-2020-023,LHCb-PAPER-2020-046,LHCb-PAPER-2020-048,LHCb-PAPER-2021-026} are listed in Table~\ref{sec:production:tab:productionHF}.
Inclusive hadroproduction of open heavy-flavour hadrons ($H_Q$) factorises into
three components in perturbative QCD (pQCD) calculations: the parton distribution function (PDF) in the two initial projectiles $f_{i,j}$, the parton level cross-section $\sigma_{ij\to Q+X}$ of a heavy-quark $Q$ production, and
the heavy-quark fragmentation function $D_{Q\to H_Q}$~\cite{Pumplin:2002vw}.  Differential cross-section for $H_Q$ production in $A$-$B$ collisions is expressed as
\begin{equation}
    \deriv \sigma_{AB\to H_Q} = \sum_{i,j}(f_i^A\otimes f_j^B )\otimes\deriv\sigma_{ij\to Q+X}\otimes D_{Q\to H_Q},\label{sec:prod:heavyflavour}
\end{equation}
where the indices $i,j$ run over all possible parton species, and at LHC energies heavy-flavour production is dominated by gluons. 
The PDF and fragmentation function include nonperturbative effects, and can be determined from a global fit of available
data~\cite{Pumplin:2002vw,Andronic:2015wma}.
The results of open charm and beauty production at LHCb are consistent with pQCD models, for example the calculation based on
fixed-order plus next-to-leading logs (FONLL)~\cite{Cacciari:1998it}. 
It turns out that the LHCb results on charm and beauty cross-sections have a better precision than theoretical calculations~\cite{LHCb-PAPER-2015-041,LHCb-PAPER-2017-037,LHCb-PAPER-2016-042}, and can be used to reduce the
uncertainties on gluon PDF, in particular in the small Bjorken-$x$ region, $x\lsim 10^{-5}$~\cite{Gauld:2016kpd}.

\begin{table}[!tb]
\caption{LHCb measurements of production cross-sections (or ratios) for various heavy hadrons in $pp$ and $p$Pb collisions at different centre-of-mass energies.}
{\small
    \begin{tabular}{m{0.06\textwidth}|m{0.14\textwidth}|m{0.23\textwidth}|m{0.23\textwidth}|m{0.19\textwidth}}
        \hline
        \multirow{2}{*}{System}  &\multicolumn{4}{c}{$\sqrt{s}_{(NN)}$ ($\kern -0.25em\tev$)}\\
        \cline{2-5}
        & \hspace{0.5cm}$5\,(2.76)$ &  \hspace{1.5cm}$7$ & $\hspace{1.2cm}8\,(8.16)$ & \hspace{1.5cm}$13$\\
        \hline
            $pp$  &        
           \mbox{$\jpsi$\cite{LHCb-PAPER-2012-039}},
           \mbox{$\PUpsilon$\cite{LHCb-PAPER-2013-066}},
           \mbox{$D$\cite{LHCb-PAPER-2016-042}}
            &
           \mbox{$\etac$\cite{LHCb-PAPER-2014-029}},
           \mbox{$\jpsi$\cite{LHCb-PAPER-2011-003}},
           \mbox{$\chi_{c1}(3872)$\cite{LHCb-PAPER-2011-034}},
           \mbox{$\chic/\jpsi$\cite{LHCb-PAPER-2011-030}},
            \mbox{$\chictwo/\chicone$\cite{LHCb-PAPER-2011-019,LHCb-PAPER-2013-028}},
           \mbox{$\psitwos$\cite{LHCb-PAPER-2011-045,LHCb-PAPER-2018-049}},
           \mbox{$\PUpsilon$\cite{LHCb-PAPER-2011-036,LHCb-PAPER-2015-045}},
           \mbox{$\chibtwo/\chibone$\cite{LHCb-PAPER-2014-040,LHCb-PAPER-2014-031}},
           \mbox{$D,\Lc$\cite{LHCb-PAPER-2012-041}},
           \mbox{$B$\cite{LHCb-PAPER-2011-043,LHCb-PAPER-2013-004,LHCb-PAPER-2017-037,LHCb-PAPER-2020-046}},
           \mbox{$\Lb$\cite{LHCb-PAPER-2014-004,LHCb-PAPER-2015-032}},
           \mbox{$\Xib$\cite{LHCb-PAPER-2018-047}},
           \mbox{$\Bc$\cite{LHCb-PAPER-2012-028}},
           \mbox{$\jpsi\jpsi$\cite{LHCb-PAPER-2011-013}},
           \mbox{$\PUpsilon D$\cite{LHCb-PAPER-2015-046}},
           \mbox{$\jpsi D, DD, D\Dzb$\cite{LHCb-PAPER-2012-003}}
            &
           \mbox{$\etac$\cite{LHCb-PAPER-2014-029}},
           \mbox{$\jpsi$\cite{LHCb-PAPER-2015-016}},
           \mbox{$\chibtwo/\chibone$\cite{LHCb-PAPER-2014-040,LHCb-PAPER-2014-031}},
           \mbox{$\chi_{c1}(3872)/\psitwos$\cite{LHCb-PAPER-2020-023}},
           \mbox{$\chi_{c1}(3872)$\cite{LHCb-PAPER-2021-026}},       
           \mbox{$\PUpsilon$\cite{LHCb-PAPER-2015-016,LHCb-PAPER-2015-045}},
           \mbox{$\Bs/\Bd$\cite{LHCb-PAPER-2020-046}},
           \mbox{$\Lb$\cite{LHCb-PAPER-2015-032}},
           \mbox{$\Xib$\cite{LHCb-PAPER-2018-047}},
           \mbox{$\PUpsilon D$\cite{LHCb-PAPER-2015-046}},
           \mbox{$\Bc$\cite{LHCb-PAPER-2014-050}}
            &
           \mbox{$\etac$\cite{LHCb-PAPER-2019-024}},                 
           \mbox{$\jpsi$\cite{LHCb-PAPER-2015-037,LHCb-PAPER-2016-064}},
           \mbox{$\chi_{c1}(3872)$\cite{LHCb-PAPER-2021-026}}, 
           \mbox{$\psitwos$\cite{LHCb-PAPER-2018-049}},
           \mbox{$\PUpsilon$\cite{LHCb-PAPER-2018-002}},
           \mbox{$D$\cite{LHCb-PAPER-2015-041}},
           \mbox{$\Xiccpp$\cite{LHCb-PAPER-2019-035}},
           \mbox{$\Bp$\cite{LHCb-PAPER-2017-037}},
           \mbox{$\Xib$\cite{LHCb-PAPER-2018-047}},
           \mbox{$\Bc$\cite{LHCb-PAPER-2019-033}},
           \mbox{$\Bs/\Bd$\cite{LHCb-PAPER-2020-046}},                 
           \mbox{$\jpsi\jpsi$\cite{LHCb-PAPER-2016-057}}
            \\
        \hline
            $p$Pb &                       
            \mbox{$\jpsi$\cite{LHCb-PAPER-2013-052}},
            \mbox{$\psitwos$\cite{LHCb-PAPER-2015-058}},
            \mbox{$\PUpsilon$\cite{LHCb-PAPER-2014-015}},
            \mbox{$\Dz$\cite{LHCb-PAPER-2017-015}},
            \mbox{$\Lc$\cite{LHCb-PAPER-2018-021}}
            &                 &
           \mbox{$\jpsi$\cite{LHCb-PAPER-2017-014}},
           \mbox{$\PUpsilon$\cite{LHCb-PAPER-2018-035}},
           \mbox{$\Dz$\cite{LHCb-CONF-2019-004}},
           \mbox{$\Bp, \Bz, \Lb$\cite{LHCb-PAPER-2018-048}},
           \mbox{$\chictwo/\chicone$\cite{LHCb-PAPER-2020-048}},
           \mbox{$DD,D\Dzb$\cite{LHCb-PAPER-2020-010}}
            &\\
        \hline
\end{tabular}
\label{sec:production:tab:productionHF}
}
\end{table}

\begin{figure}[tb]
\begin{center}
    \includegraphics[width=0.37\textwidth]{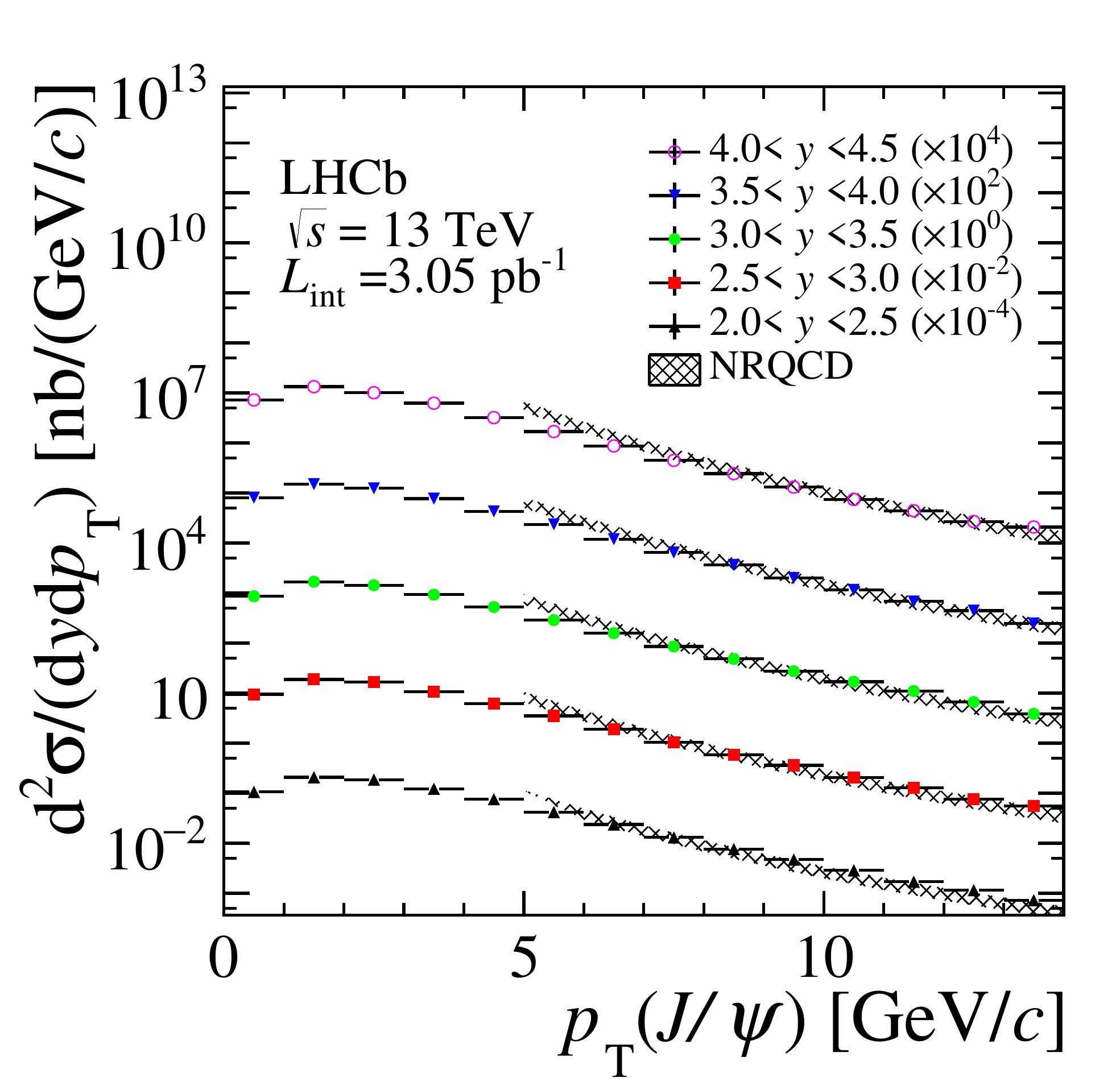}
     \includegraphics[width=0.58\textwidth]{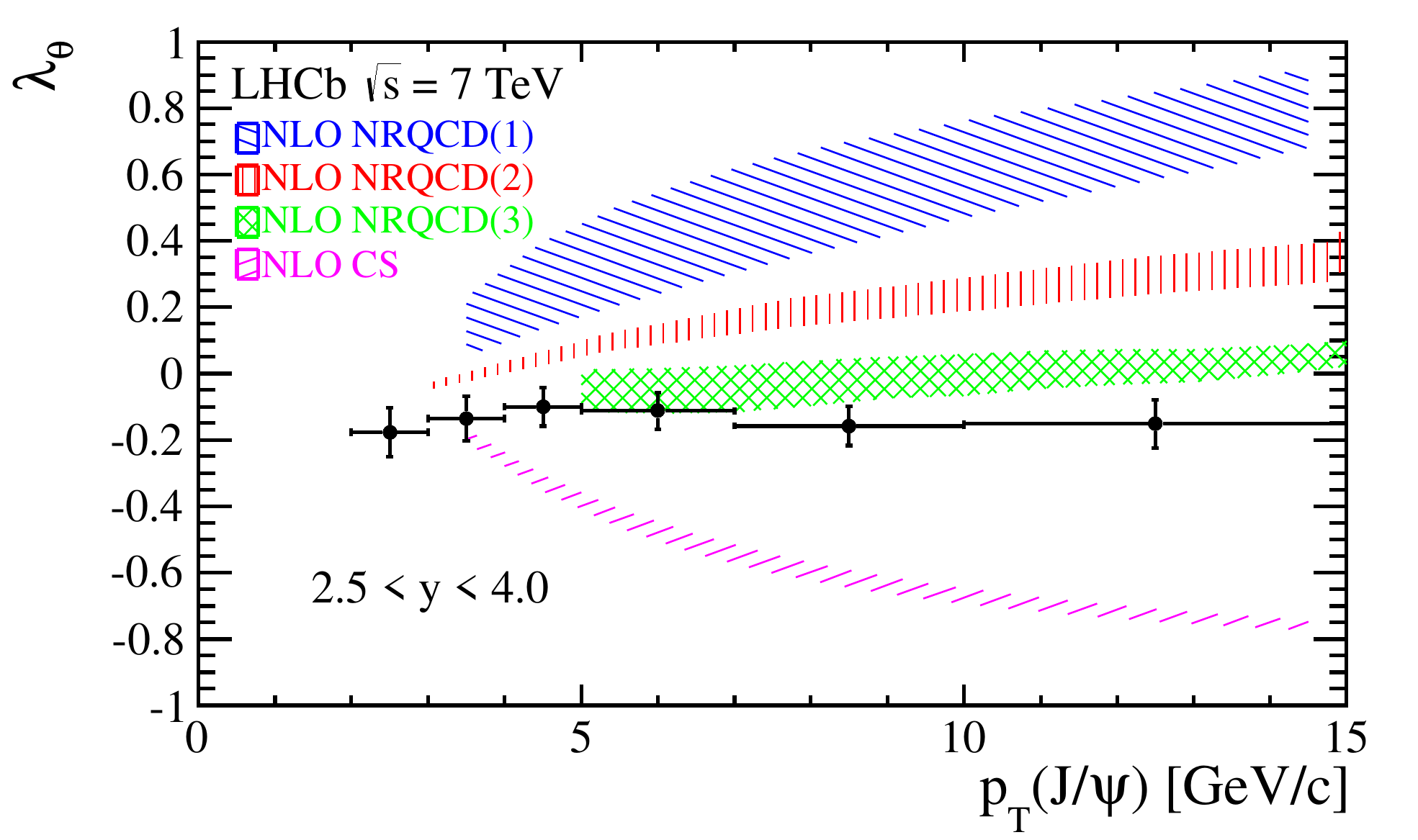}
    \caption{(Left) Differential cross-section of $\jpsi$ in $pp$ collisions at $\sqs=13\tev$, taken from Ref.~\cite{LHCb-PAPER-2015-037}. (Right) $\jpsi$ polarisation in $pp$ collisions at $\sqs=7\tev$, taken from Ref.~\cite{LHCb-PAPER-2013-008}. The measurements are compared with NRQCD~\cite{Chao:2012iv,Butenschoen:2012qh,Gong:2012ug} and colour singlet calculations~\cite{Butenschoen:2012qh}.}
\label{sec:prod:fig:JpsiProduction}
\end{center}
\end{figure}

For quarkonium production,  assumptions have to be put on 
how heavy-quark pairs, $Q\overline{Q}$, produced with various possible colour, spin and parity configurations,  transform into specific colourless
quarkonia~\cite{Ma:2014mri,Zhang:2013igb,Brambilla:2010cs,Brambilla:2004jw}. Cross-section measurements favour calculations using the nonrelativistic QCD (NRQCD)
framework~\cite{Bodwin:1994jh}, as shown on the left of Fig.~\ref{sec:prod:fig:JpsiProduction}, for $\jpsi$ production in $pp$ collisions at $\sqs=13\tev$. The NRQCD framework
introduces long distance matrix elements (LDMEs) as model parameters~\cite{Ma:2016exq} to account for transition probabilities from heavy-quark pairs to quarkonia. LDMEs are
assumed to be independent of quarkonium production environments and kinematics, and are fixed by matching the predicted $\pt$ spectrum to data. 
The polarisation of heavy quarkonia is another observable sensitive to the $Q\overline{Q}$ production mechanism and LDMEs.
Inconsistencies are observed between LHCb data and theoretical calculations on the $\psi$ and $\PUpsilon$
polarisations~\cite{LHCb-PAPER-2017-028,LHCb-PAPER-2013-067,LHCb-PAPER-2013-008}. Only a level of 10\% or smaller polarisation
is observed in the LHCb acceptance, in contrast to a dominant transverse polarisation predicted by NRQCD~\cite{Chao:2012iv,Gong:2012ug,Butenschoen:2012qh}. The measurement of $\jpsi$ polarisation is shown on the right of Fig.~\ref{sec:prod:fig:JpsiProduction}. Even though the discrepancy can be reduced by tuning the LDMEs, a coherent description of production
cross-section and polarisation is still a difficult theoretical problem~\cite{Gong:2012ug,Shao:2014yta,Han:2014jya,Feng:2018ukp,Shao:2014yfa,Butenschoen:2012px,Ma:2018qvc,Shao:2014fca,Shao:2012fs}. If only the $Q\overline{Q}$ state that has the same quantum number as the final quarkonium is considered, the NRQCD framework reduces to the colour singlet model, which underestimates production cross-sections~\cite{Butenschoen:2012qh} and disagrees with data on $\psi$ and $\PUpsilon$ polarisations~\cite{LHCb-PAPER-2013-008,LHCb-PAPER-2013-067,LHCb-PAPER-2017-028}.

\subsubsection{Associated production}
Recent LHCb production measurements focus on associated production of multiple heavy flavours and quantities
probing properties of QCD matter. 
Associated heavy-flavour production provides an approach to study the multiple parton interactions (MPIs).  
MPIs are sensitive to correlations between partons in space, momentum, flavour, colour, spin \etc inside  the colliding projectiles~\cite{Chapon:2020heu,Kom:2011bd,Shao:2019qob,He:2007te,Lansberg:2013qka,Shao:2020kgj}. 
Usually in an MPI process these correlations are assumed to be absent initially such that  each parton-scattering is independent from each other, and then consistency checks are performed to verify this assumption. 
Under this assumption, the cross-section for associated production of $ab$ through a double-parton-scattering (DPS) process is related to the single inclusive production of $a$ and $b$ as~\cite{Chapon:2020heu}
\begin{equation}
\sigma^{ab}=\kappa\frac{\sigma^a\sigma^b}{\sigma_\mathrm{eff}},\label{sec:produ:eq:dps}
\end{equation}
where $\kappa$ is a symmetry factor with $\kappa=1$ if $a\ne
b$ and the effective cross-section $\sigma_\mathrm{eff}$ is assumed to be  universal.  
Heavy-quark fragmentations  in MPIs  are implied to be identical to that in inclusive production defined in Eq.~\ref{sec:produ:eq:dps}.
In particular, the kinematics of $a$ and $b$ is uncorrelated and each of them is similar to that in single particle inclusive production. Besides MPI, the single parton scattering (SPS) is also able to generate associated production, but in SPS the final-state kinematics is correlated. This difference between DPS and SPS is used to identify DPS. Studies of DPS include measurements of the $\sigma_\mathrm{eff}$ parameter and tests of its universality for different states, and investigations of kinematic correlations between $a$ and $b$. One example of correlation variables is the relative azimuthal angle  $\Delta\phi$ between $a$ and $b$ and to infer the correlations between colliding partons. For DPS production $\Delta\phi$ distribution is approximately flat, while in SPS events a concentration at $\Delta\phi\sim0$ or $\pi$ is expected.

Measurements of associated production in $pp$ collisions are made at LHCb for two open charm hadrons~\cite{LHCb-PAPER-2012-003}, a heavy quarkonium plus an open charm~\cite{LHCb-PAPER-2015-046,LHCb-PAPER-2012-003}, and double $\jpsi$ mesons~\cite{LHCb-PAPER-2016-057}.
To subtract the SPS contribution from data, one usually relies on theoretical inputs for cross-sections of SPS or fits to data using templates of correlation variables built for both SPS and DPS production.
Note that theoretical uncertainties are still much larger than
experimental ones for these measurements. 

For some associated production, SPS is estimated or assumed to be negligible, resulting in a sample of approximately pure DPS events. 
In this case the $\sigma_\mathrm{eff}$ parameter measured using Eq.~\ref{sec:produ:eq:dps} is around $15\mbarn$ for $\jpsi D$ and $\PUpsilon D$ production, independent of the $D$ species and 
collision centre-of-mass energies~\cite{LHCb-PAPER-2012-003,LHCb-PAPER-2015-046}. 
The results are similar to those extracted using multi-jet production at Tevatron~\cite{CDF:1997yfa}.  
However, the values obtained using same-sign $DD$ pairs are around $20\mbarn$, 
and that for $\jpsi\jpsi$ pairs is about $7\mbarn$~\cite{LHCb-PAPER-2016-057}. 
The former are consistently higher than the value of $15\mbarn$~\cite{LHCb-PAPER-2012-003}, while the latter is significantly lower.
Higher values of $\sigma_\mathrm{eff}$ for $\jpsi\jpsi$ production are obtained if  a fraction of SPS is subtracted. The SPS fraction is estimated to be between $20\%$ and $40\%$ depending on the choice of control
variables and input templates for SPS and DPS distributions~\cite{LHCb-PAPER-2016-057}.
The smaller $\sigma_\mathrm{eff}$ measurement for $\jpsi\jpsi$ pairs confirms previous
observations by D0, CMS and ATLAS experiments for quarkonium pairs~\cite{ATLAS:2016ydt,D0:2015dyx,Lansberg:2014swa}.  

\begin{figure}[!tb]
\begin{center}
    \includegraphics[width=0.4\textwidth]{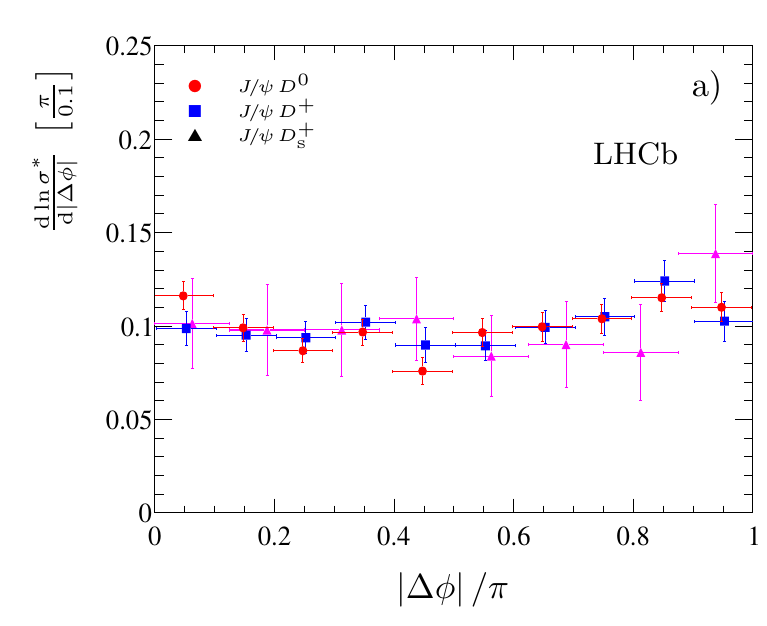}
    \includegraphics[width=0.4\textwidth]{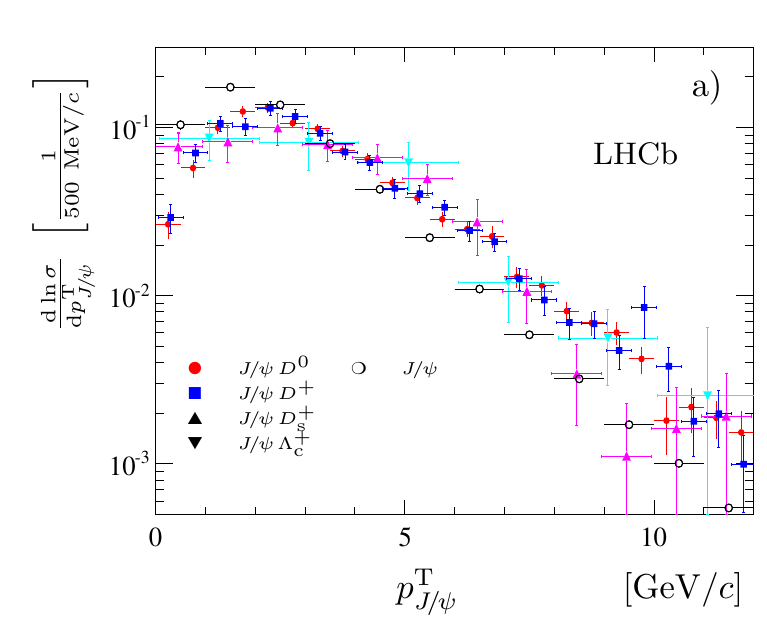}
    \caption{(Left) Distribution of relative azimuthal angle, $\Delta\phi$, between $\jpsi$ and $D$ mesons in pair production, showing flat behaviour; (Right) the $\pt$ distribution of $\jpsi$ mesons in $\jpsi D$ pair production, compared with that in inclusive production (shown in black open circle)~\cite{LHCb-PAPER-2011-003}. Figures are taken from Ref.~\cite{LHCb-PAPER-2012-003}. }
\label{sec:prod:fig:dps}
\end{center}
\end{figure}

For correlation variables, as shown on the left of Fig.~\ref{sec:prod:fig:dps}, the $\Delta\phi$ distributions of $\jpsi D$ events are reasonably flat~\cite{LHCb-PAPER-2012-003}, consistent with the DPS production in which $\jpsi$ and $D$ kinematics is uncorrelated. This observation is a sign
of dominant or pure DPS contribution for $\jpsi D$ samples. For same-sign $DD$ production, the correlation variables also favour DPS dominance~\cite{LHCb-PAPER-2012-003}.

The $\pt$ distribution of each hadron in the pair production is also studied.  For DPS events, it is expected to be similar to that in single inclusive production. For $\PUpsilon D$ samples, both the $\PUpsilon$ and the $D$ meson have a $\pt$ distribution similar  to that in single
inclusive production~\cite{LHCb-PAPER-2015-046}. The same conclusion holds for the \pt distribution of $D$ mesons in $\jpsi D$ events. However, the $\pt$ of $\jpsi$  mesons in $\jpsi D$ events is
significantly harder than that in inclusive production, indicated by the right of Fig.~\ref{sec:prod:fig:dps}. For the same-sign $DD$ production, the $\pt$ distribution of $D$ mesons is also significantly harder than
that in single inclusive $D$ production, but are similar to those in opposite-sign $D\Dzb$ samples~\cite{LHCb-PAPER-2012-003}. This
result is inconsistent with the observation in correlation variables, which hints at a dominant DPS (SPS) contribution in the same-sign (opposite-sign) pair production.  Note that for associated
production of $DD$ pairs, the $\pt$ distribution of $D$ mesons is similar for different $D$ species, indicating that charm hadron fragmentations are not modified, so that the unexpected $\pt$ distribution is not due to the fragmentation process. A detailed theoretical calculation on $\jpsi D$ production was performed recently to understand the problem, but a solid conclusion is not available yet~\cite{Shao:2020kgj}.

\subsubsection{Production in $p$Pb collisions}
\begin{figure}[!tb]
\begin{center}
    \includegraphics[width=0.49\textwidth]{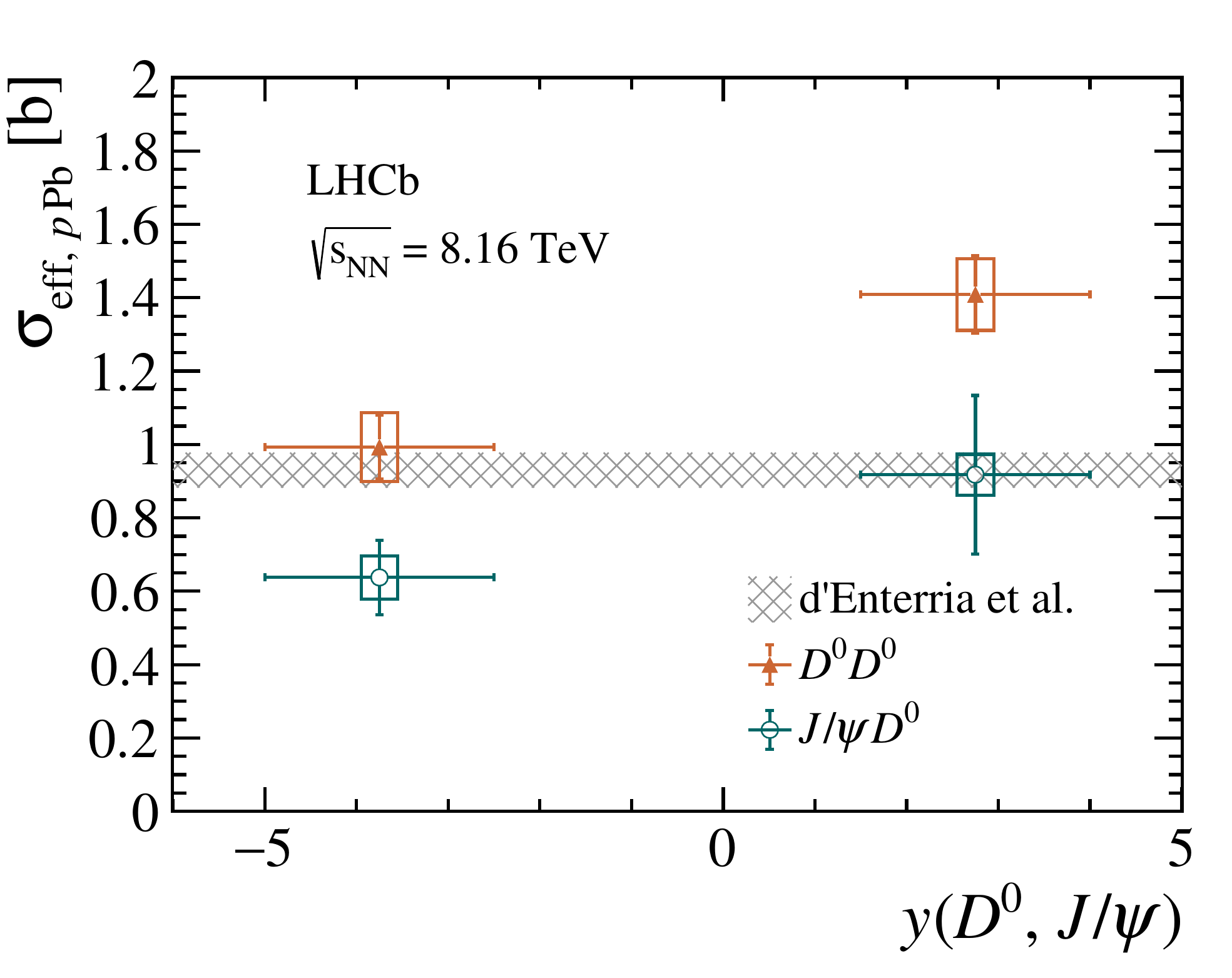}
    \includegraphics[width=0.4\textwidth]{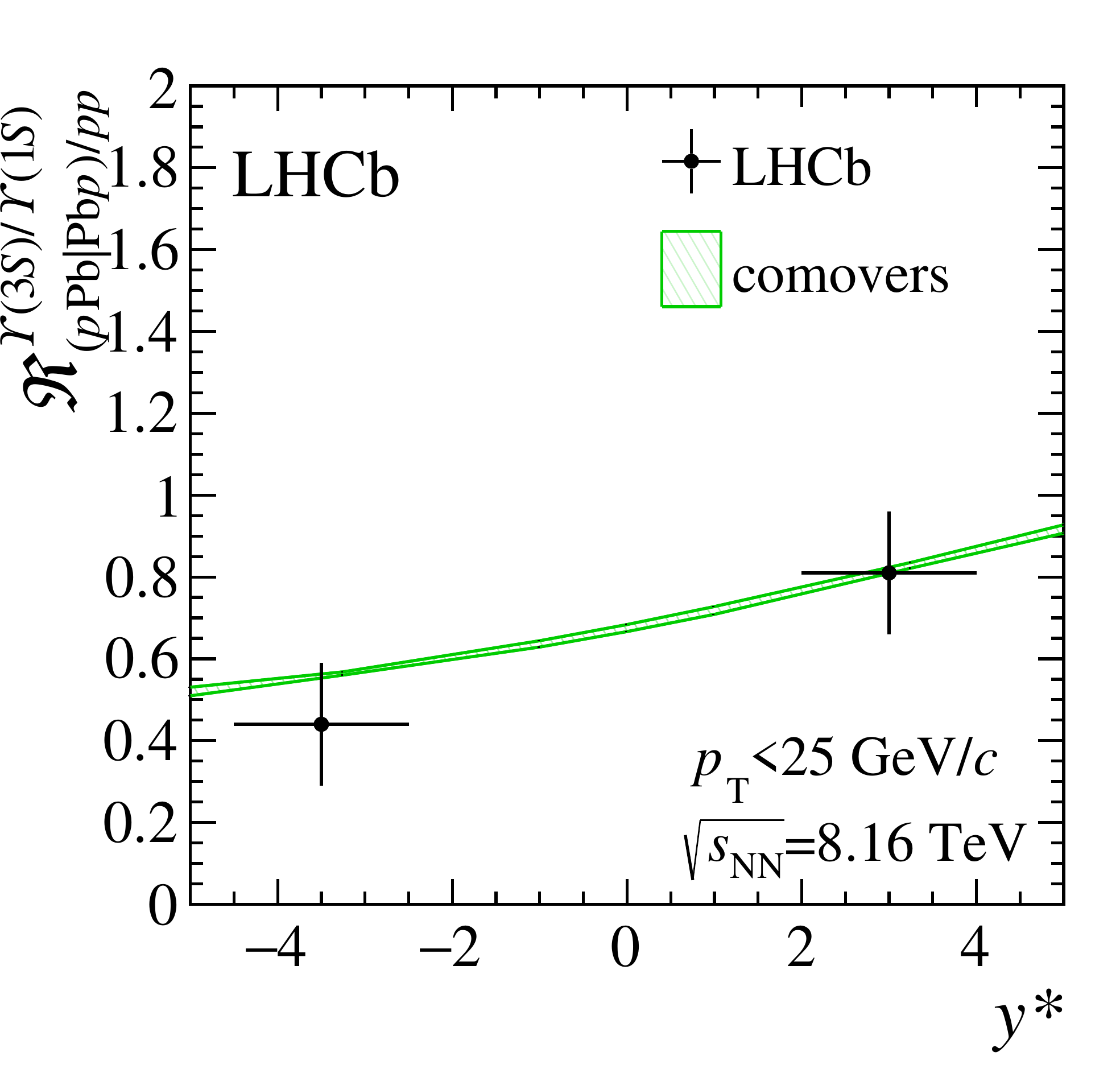}
    \caption{(Left) The parameter $\sigma_\mathrm{eff}$ measured in $p$Pb collisions at $\sqsnn=8.16\tev$ using (green) $\jpsi\Dz$ and (brown) $\Dz\Dz$ production for (positive rapidity) $p$ beam and (negative rapidity) Pb-beam direction, taken from~\cite{LHCb-PAPER-2020-010}. The prediction shown in shaded area is from Ref.~\cite{dEnterria:2012jam}. (Right) The ratio of cross-sections between $\PUpsilon(3S)$ and $\PUpsilon(1S)$ over that in $pp$ data at $\sqsnn=8\tev$, taken from Ref.~\cite{LHCb-PAPER-2018-035}, is compared with the comovers model~\cite{Ferreiro:2018wbd}. The rapidities are defined in the rest frame of two colliding nucleons with respect to proton-beam direction.}
\label{sec:prod:fig:heavyIons}
\end{center}
\end{figure}

Charm pair production is also studied in $p$Pb collisions of a center-of-mass energy per nucleon pair $\sqsnn=8.16\tev$~\cite{LHCb-PAPER-2020-010}. The DPS cross-section in $p$Pb collisions is expected to
scale with three times of the Pb mass number ($A_\mathrm{Pb}=208$)  with respect to that in $pp$ data at the same $\sqsnn$, rather than  a simple scale factor of $A_\mathrm{Pb}$, when nuclear matter effects are not considered~\cite{Baranov:2011ch}. The $A$-scaling is relevant for
SPS production in the absence of nuclear matter effects~\cite{Baranov:2011ch}. In the end DPS production has a factor of three enhancement compared with SPS. The  cross-section ratio between same-sign $DD$ and
opposite-sign $D\Dbar$ signals is measured to be around three times of that in $pp$ collisions~\cite{LHCb-PAPER-2020-010}. The result is in favour of the expected factor-three enhancement. The parameter $\sigma_\mathrm{eff}$ is measured with $\jpsi D$ and same-sign $DD$ production as shown on the left of Fig.~\ref{sec:prod:fig:heavyIons}, for the positive rapidity region, which corresponds to the Pb-beam direction (high Bjorken-$x$ of Pb nucleus), and the negative rapidity region, which corresponds to the $p$-beam direction (low Bjorken-$x$ of Pb nucleus). The measurements show that, similar to the results in $pp$ data, the $\sigma_\mathrm{eff}$ parameter for $\jpsi D$ production
is about 30\% smaller than that for
same-sign $DD$ production. Besides, the results in negative rapidity
hint at smaller values than those in positive rapidity for both $\jpsi D$ and $DD$ pair production. It may be
a sign of universality violation of $\sigma_\mathrm{eff}$. This will be explored with better precision in Run 3 heavy-ion collision data, where ten times more
luminosity is expected to be collected.

Besides the enhancement of DPS production, heavy nuclear collisions have many more new phenomena compared with $pp$ collisions, collectively called nuclear matter
effects. Presence of nuclear matter effects in $p$Pb collisions could modify the PDF, or reduce parton energies or  dissociate heavy quarkonia, which can be probed using heavy-flavour production in $p$Pb data compared with the $A_\mathrm{Pb}$ scaled $pp$ cross-section~\cite{Gavin:1991qk,Armesto:2006ph,Arleo:2012rs}. Measurements of $\Dz$ and
$\Bp$ production in $p$Pb data suggest heavy-quark production in the $p$-beam direction is significantly
suppressed compared with the $A_\mathrm{Pb}$ scaling, by about 30\%, while the production in Pb-beam direction 
approximately scales with $A_\mathrm{Pb}$~\cite{LHCb-PAPER-2017-015,LHCb-PAPER-2018-048}. The results are
consistent with modifications of the gluon PDF in a Pb nucleus compared with that in a free nucleon. The LHCb measurements are found to be able to reduce the
gluon PDF uncertainties in the Pb nucleus by about a factor of three compared with the commonly used nuclear PDF sets~\cite{Kusina:2017gkz}. A new precise measurement of $\Dz$ production in $p$Pb shows that
the magnitude of the $\Dz$ suppression in $p$ beam
direction over that in Pb-beam direction, \ie the forward-backward ratio $R_\mathrm{FB}$, increases significantly at high $\pt(\Dz)$ and 
seems to reach unity at $\pt(\Dz)>8\gevc$~\cite{LHCb-CONF-2019-004}. 
However according to predictions using the nuclear PDF, the $R_\mathrm{FB}$ is
about 70\%, almost independent of $\pt(\Dz)$. The observed trend of $R_\mathrm{FB}$ for $\Dz$ may be caused by the parton energy loss effect which alters heavy-flavour kinematic distribution, whose impact is reduced at high $\pt$~\cite{Arleo:2021bpv}, otherwise the result will require a modification of current knowledge of the nuclear PDF.

Measurement of $\jpsi$ production in
$p$Pb data shows a similar trend of suppression compared with open heavy-flavour hadrons~\cite{LHCb-PAPER-2013-052,LHCb-PAPER-2017-014}, suggesting that they suffer from a common influence by nuclear matter effects. However, the result for $\psitwos$ in $p$Pb data
suggests a stronger suppression compared with $\jpsi$, in particularly in the Pb-beam direction~\cite{LHCb-PAPER-2015-058}. Similarly,
the $\PUpsilon(3S)$ meson is measured to be more suppressed compared with $\PUpsilon(1S)$~\cite{LHCb-PAPER-2018-035}, as shown on the right of Fig.~\ref{sec:prod:fig:heavyIons}. The stronger suppression for excited quarkonium cannot be explained using the nuclear PDF
modification or the parton energy loss effect. The comover model introducing final state interactions between a heavy quarkonium and comoving particles 
is able to explain data~\cite{Ferreiro:2018wbd}. The comovers effect is stronger in events of higher occupancy and for particles with
larger sizes, such that it is more
pronounced in Pb-beam direction and for excited states, reducing their yields more significantly than for the ground state in the $p$-beam direction $A_\mathrm{Pb}$~. The comovers mechanism also exists in $pp$ collisions, and is probed using heavy quarkonium production.
The cross-section ratio between prompt $\chi_{c1}(3872)$ and $\psitwos$ mesons is measured to decrease with the increase of the number of reconstructed tracks in
the vertex detector~\cite{LHCb-PAPER-2020-023}. It suggests that the $\chi_{c1}(3872)$ state has a larger size or a smaller binding energy compared  with the $\psitwos$ meson, and is consistent with a component of $\Dstarz\Dzb+\Dstarzb\Dz$ hadron molecule in the $\chi_{c1}(3872)$ wave function~\cite{Braaten:2020iqw,Esposito:2020ywk}.
The same measurement in $p$Pb data feasible in LHC Run 3 period would be very important to confirm this result.

In the near future, new data will provide enough statistics for associated production of triple heavy flavours and heavy quarkonium pairs beyond $\jpsi\jpsi$, which help to further understand
MPI and heavy quarkonium production mechanism~\cite{Chapon:2020heu}. Concerning studies of nuclear matter effects using heavy-flavour production, a rich program is foreseen in \mbox{Run 3}, including measurements probing the DPS enhancement, the modification of nuclear PDF and heavy-quark
fragmentation and the heavy quarkonium dissociation mechanism.

\subsection{Spectroscopy} 
\begin{figure}[!htpb]
\begin{center}
    \includegraphics[width=0.9\textwidth]{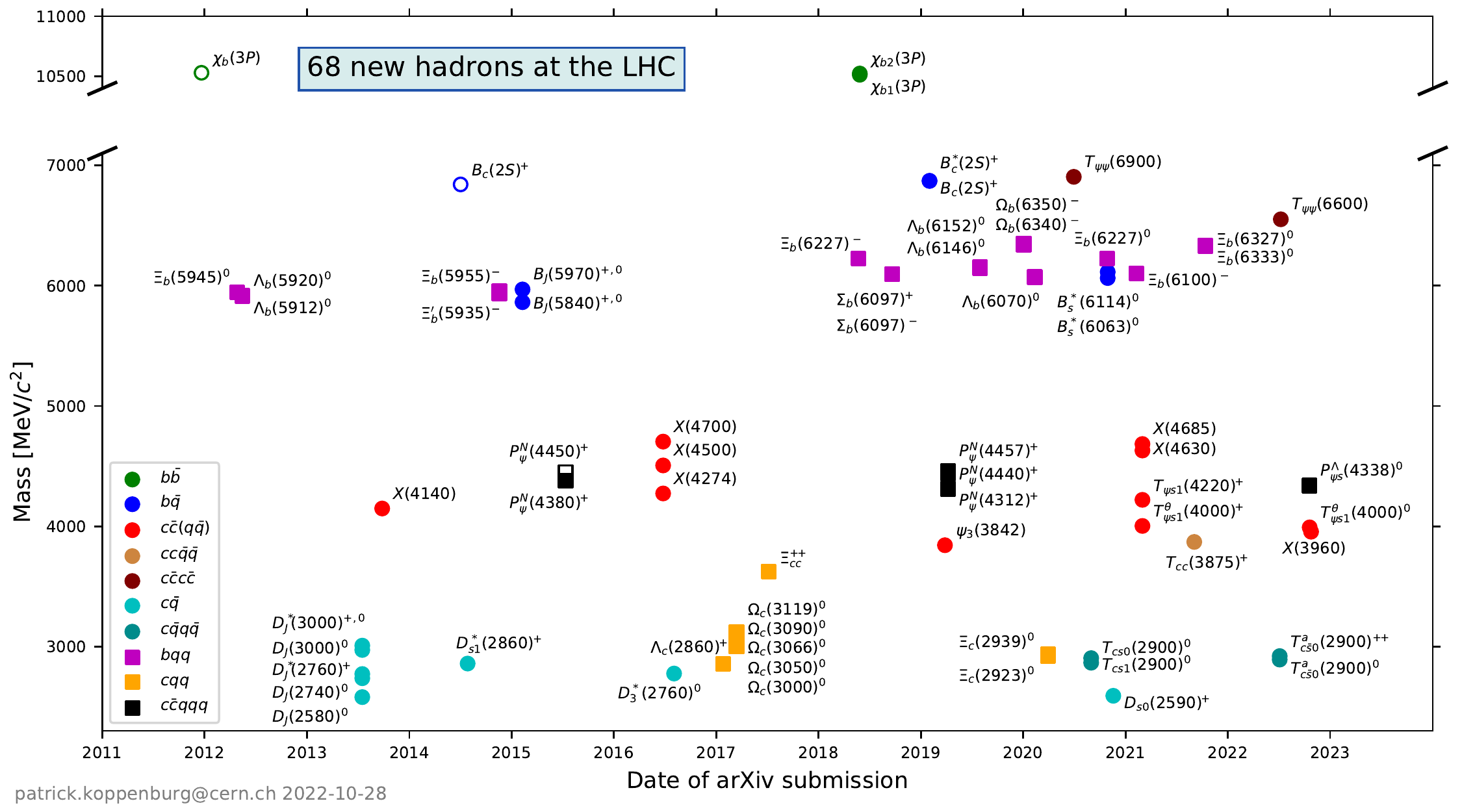}
    \caption{New hadrons that are discovered by the LHC experiments, taken from 
    \cite{lhcHadron}.}
\label{sec:prod:fig:hadrons}
\end{center}
\end{figure}

The strong interaction confines quarks (and/or gluons) to form various colour-singlet hadrons that are accessible
experimentally. This confinement phenomenon is nonperturbative and is not fully understood yet from the current QCD theory. In
analogy to photon spectroscopy in atomic physics, hadron spectroscopy provides a way to understand dynamics of 
QCD at low energies.  Hadrons composed of a quark and an antiquark are called mesons, those of three quarks are called baryons, and those
composed of more than three quarks are usually referred to as exotic hadrons. Existence of exotic hadrons have been predicted since the birth of the quark model and their properties are reexamined by refined theoretical approaches in the past decades~\cite{Gell-Mann:1964ewy,Zweig:570209,Zhu:2003ba,Sun:2011uh,Liu:2009ei,Liu:2008fh,Zhu:2007wz,Chen:2019asm,Liu:2007bf}. The past years witnessed the observations of a plethora of new conventional and
exotic hadrons containing heavy quarks and LHCb is one of the leading players in the field.
Figure~\ref{sec:prod:fig:hadrons} displays the 68 new hadrons that are discovered by the LHC experiments, and most of
them by LHCb.

\subsubsection{Conventional hadrons}

LHCb have filled many gaps in conventional heavy meson and baryon spectra. 
This section will focus on more recent highlights in baryon spectroscopy 
and discovery of doubly heavy hadrons. 
Though not discussed in detail, it should not be ignored that multiple excited beauty~\cite{LHCb-PAPER-2014-067,LHCb-PAPER-2020-026} and charm~\cite{LHCb-PAPER-2013-026} mesons are observed in high energy $pp$ collisions by LHCb.
Moreover, the beauty hadron decay is an ideal place to study excited charm hadron
as $b\to c$ is the dominant transition of the $b$ quark.
Among the many new charm mesons discovered in $B$ decay at LHCb~\cite{LHCb-PAPER-2014-035,LHCb-PAPER-2016-026,LHCb-PAPER-2020-034}, 
an interesting example is a new excited $D_s^+$ meson, $D_s(2590)^+$, consistent with the radial excited state $D_s(2^1S_0)^+$. Its mass and width will help to understand the excitation spectrum of $D_s^+$ mesons which are found to be not fully consistent with the quark model predictions~\cite{Luo:2021dvj,Ni:2021pce,Xie:2021dwe,Yang:2021tvc,Wang:2021orp,Wang:2012wk}.

\begin{figure}[!tb]
\begin{center}
    \includegraphics[width=0.9\textwidth]{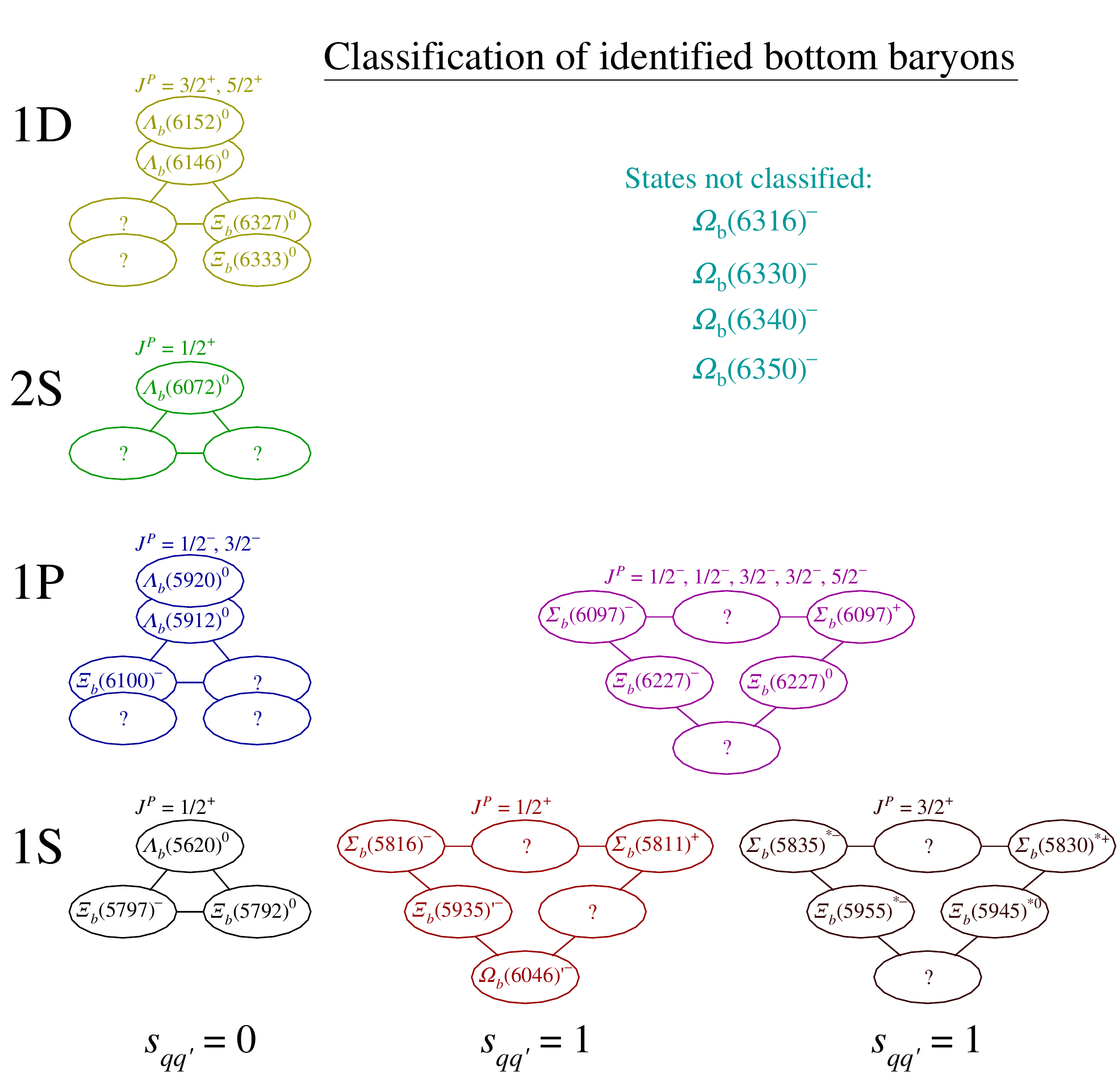}
    \caption{Experimentally identified bottom baryons, grouped according to spin and flavour symmetry of the light quarks as well as the  excitation between the light quark system and $b$ quark. The assignments of quantum numbers are based on measured properties and relevant predictions. Note that for recently discovered states their $J^P$ need confirmation and the  assignments of excited $\Omegares_b$ are not certain, though they likely have $J^P=1^-$. In the figure, 1$P$ excitation of flavour symmetric bottom baryons are gathered into a single multiplet (shown in purple). Only $\lambda$ mode radial and orbital angular momentum excitation are considered. Missing states are marked as question marks in the figure. The notation $s_{qq'}$ indicates the total spin of the light quark system.}
\label{sec:prod:fig:HbB}
\end{center}
\end{figure}

\paragraph{Classification of heavy baryons} Following the heavy-quark symmetry, baryons with a heavy quark $Q$ are organised into multiplets according to quantum configurations of the two light quarks~\cite{Liu:2007fg,Yu:2021zvl,Wang:2021bmz,Matsuki:2017tqf,Wang:2017kfr,Mao:2017wbz,Cheng:2017ove,Yang:2020zrh,Chen:2018orb,Wang:2017goq,Wei:2016jyk,Lu:2016gev,He:2021xrh,Zhang:2008pm,Zhang:2008iz,Wei:2010zza,Guo:2008ns}.
The total wave function including flavour ($F$), spin ($s_{qq'}$) and orbital angular momentum ($l_{qq'}$) must be symmetric for the two light quarks $qq'$ to form an antisymmetric state
together with their antisymmetric colour configuration. Baryons with $l_{qq'}=s_{qq'}=0$ (antisymmetric in spin space) have a spin-parity of  $J^P=1/2^{+}$, and are grouped into a multiplet of three flavour-antisymmetric states for each heavy quark $Q$. While baryons with $s_{qq'}=1, l_{qq'}=0$
(symmetric in spin space)  have
$J^P=1/2^+$ or $J^P=3/2^+$,  and form a multiplet of six flavour-symmetric states for each $J^P$. These three different multiplets are shown in the bottom row of Fig.~\ref{sec:prod:fig:HbB}. Orbital and radial excitation can happen inside the two light quarks
($\rho$-mode) or between $Q$ and the $qq'$ system ($\lambda$-mode). The parity of a baryon is determined to be  $P=(-1)^{l_\rho+l_\lambda}$, where $l_\rho\equiv l_{qq'}$, and $l_\lambda$ is the orbital angular momentum between the $Q$ and $qq'$.
Beauty and charm baryons with $l_\lambda=l_{qq'}=s_{qq'}=0$ decay weakly and have been well established. However a chart of their excited states are far from being
complete. As a matter of fact only a few low lying states are observed, in particularly for beauty baryons as can be seen in Fig.~\ref{sec:prod:fig:HbB}.
Up to date no sign of $\rho$-mode states have been identified experimentally, probably because
they are too wide (hundreds of $\kern -0.125em\mev$) to be resolved from underlying background.

\begin{figure}[!tb]
\begin{center}
    \includegraphics[width=0.41\textwidth]{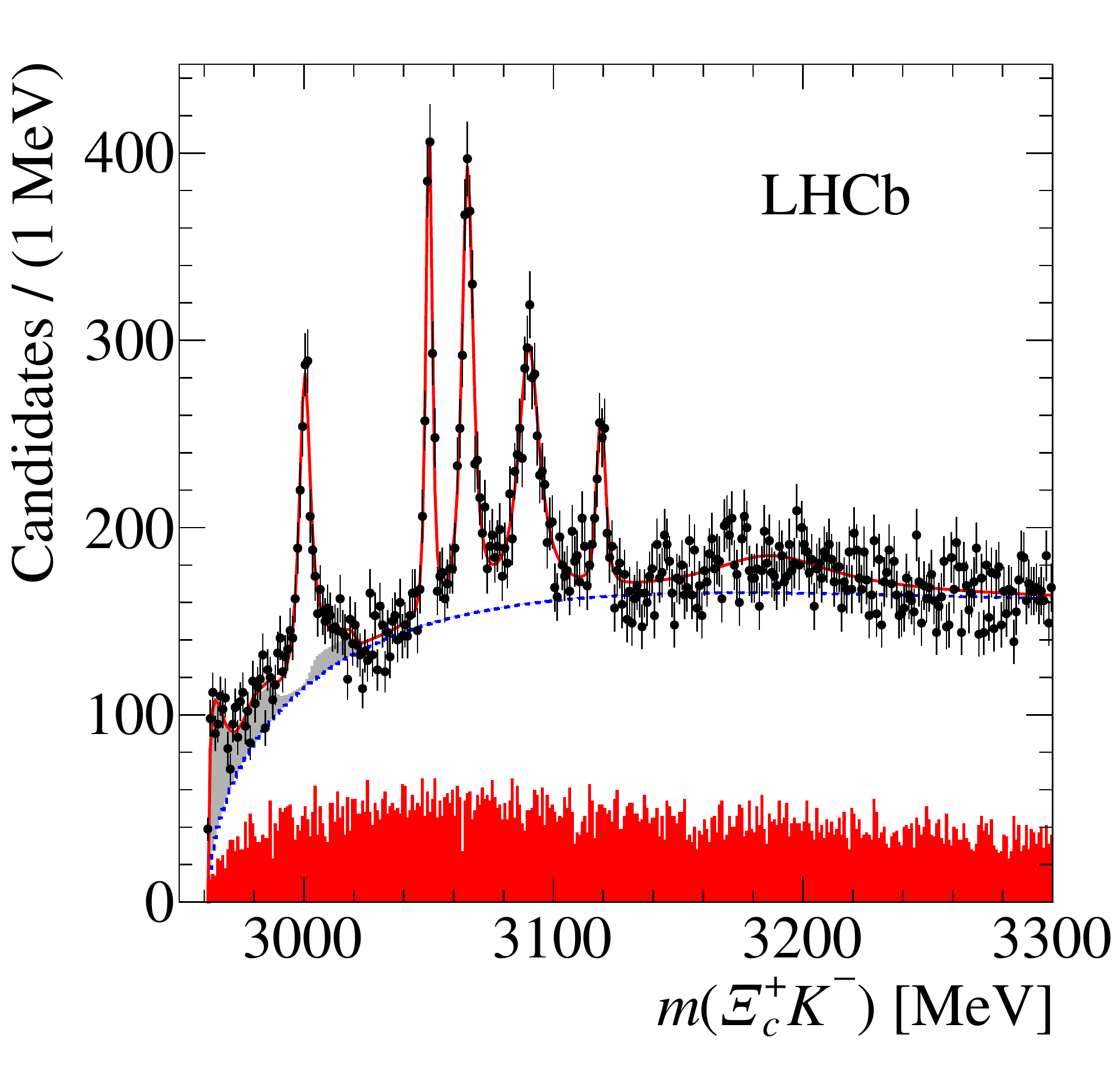}
    \includegraphics[width=0.52\textwidth]{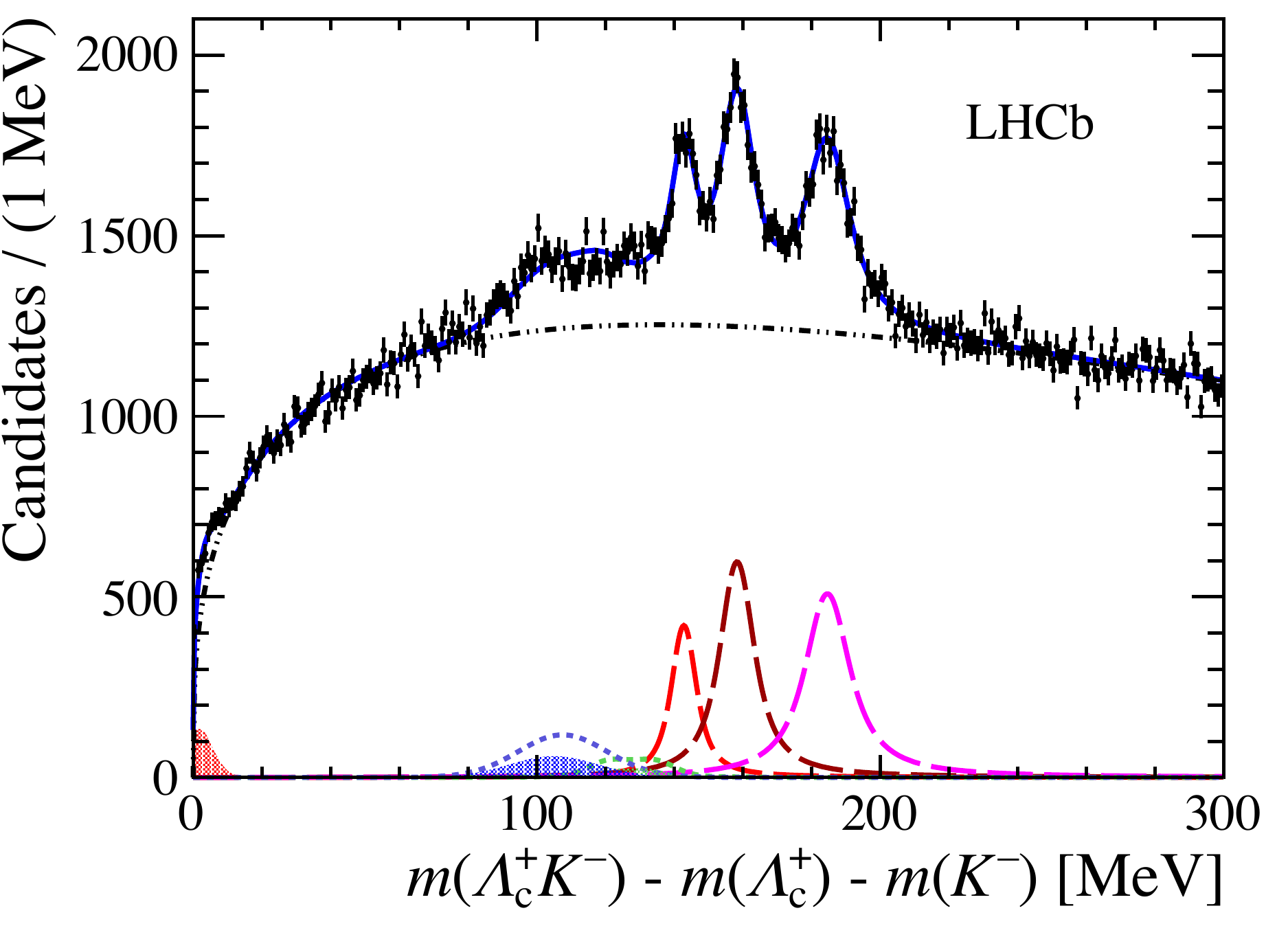}
    \caption{(Left) Invariant mass distribution of $\Xires^+_c\Km$, overlaid with the fit including five narrow $\Omegares_c$ states, taken from Ref.~\cite{LHCb-PAPER-2017-002}. (Right) Invariant mass distribution of $\Lc\Km$, overlaid with the fit including three narrow $\Xic$ states (red, brown and purple dashed lines) and partially reconstructed components, taken from Ref.~\cite{LHCb-PAPER-2020-004}.}
\label{sec:prod:fig:charmBaryons}
\end{center}
\end{figure}

\paragraph{Charm baryons} In the invariant mass spectrum of $\Xires^+_c\Km$ hadrons shown on the left of Fig.~\ref{sec:prod:fig:charmBaryons}, LHCb observed five states whose quark contents are considered to be $css$:  $\Omegares_c(3000)^0$, $\Omegares_c(3050)^0$,
$\Omegares_c(3066)^0$, $\Omegares_c(3090)^0$ and $\Omegares_c(3119)^0$~\cite{LHCb-PAPER-2017-002}. 
All these states have narrow widths,  below $10\mev$, and their mass differences are only tens of $\kern -0.125em\mev$.
The first four states are
confirmed by Belle in $\epem$ collisions~\cite{Belle:2017ext} and by LHCb in the exclusive $\Omegares_b^{-}\to \Xires^+_c\Km\pim$ decay~\cite{LHCb-PAPER-2021-012}. The spin assignments of the first four states favour $1/2, 3/2, 3/2, 5/2$, consistent with the expectations for $P$-wave $\lambda$ excitation~\cite{LHCb-PAPER-2021-012}. A determination of their parities will help to make firm conclusions. According to phenomenological models~\cite{Karliner:2017kfm,Wang:2017hej,Xu:2019kkt,Chen:2017gnu,Wang:2017vnc,Chen:2017sci,Yang:2017rpg}, one of the five $1P$ states with $l_\rho=0,l_\lambda=1$ in the mass region of observed states is missing, and the $\Omegares_c(3119)^0$ state may be a $2S$ or $D$-wave baryon. In the high mass region of LHCb data, a hint of a wide state $\Omegares_c(3188)^0$ is present,  to be confirmed in future analysis with additional statistics. It is noted that some of these states are considered to be exotic states of quark constituents $cssu\uquarkbar/d\dquarkbar$ rather than conventional $css$ baryons~\cite{Wang:2021cku,Wang:2018alb,Chen:2017xat,Wang:2017smo,Wang:2017xam}.

Similarly, excited $\Xires^0_c$ states are searched for by LHCb~\cite{LHCb-PAPER-2020-004} in the $\Lc\Km$ invariant mass spectrum shown on the right of Fig.~\ref{sec:prod:fig:charmBaryons}. A
new state $\Xic(2965)^0$ is observed, and the state $\Xic(2930)^0$ claimed by the Belle experiment~\cite{Belle:2017jrt} now splits into two structures, $\Xic(2923)^0$ and
$\Xic(2939)^0$. Separation of the $\Xic(2923)^0$ and $\Xic(2939)^0$ is recently confirmed in $B$ decay by LHCb~\cite{LHCb-PAPER-2022-028}. The widths of these three states are determined to be around $10\mev$. These states and previously known $\Xic(2790)^0$ and
$\Xic(2815)^0$ lie in the mass region of $1P$  excitation~\cite{Ebert:2007nw,Roberts:2007ni,Migura:2006ep,Yang:2021lce,Chen:2021eyk}. There are in total seven $1P$ $\Xires^0_c$ states of the $\lambda$-excitation. At least two of these $1P$ states are still missing; 
Besides, a complete and solid matching of these observed states to predicted spectrum is not resolved yet~\cite{Nieves:2019jhp}.

\paragraph{Bottom baryons} 
A summary of all bottom baryons with clear experiment evidences are shown in Fig.~\ref{sec:prod:fig:HbB}, organised in multiplets of $qq'$ flavour symmetry and $\lambda$-mode excitation when possible. The spin-parity quantum numbers for most of these states are not measured, so we rely on theoretical calculations as a guidance to make the classification. Actually, there is not always a consensus on the $J^P$ of each state, in particularly for $\Omegares_b$ baryons~\cite{Xu:2020ofp,Karliner:2020fqe,Xiao:2020gjo,Yang:2020zjl,Xiao:2020oif,Wang:2020pri,Liang:2020dxr,Chen:2020mpy,Liang:2020hbo,Lu:2019rtg,Chen:2018vuc,Wang:2020vwl,Mao:2015gya,Lu:2014ina,Yang:2018lzg,Huang:2018bed}.

In total, five states have been reported in the $\Lb\pip\pim$ mass spectrum: $\Lbsimple(5912)^0$ and $
\Lbsimple(5920)^0$ with widths below $1 \mev$~\cite{LHCb-PAPER-2012-012}, $\Lbsimple(6146)^0$ and $\Lbsimple(6152)^0$ with
widths of about $2\mev$~\cite{LHCb-PAPER-2019-025}, and $\Lbsimple(6072)^0$ with a width around
$70\mev$~\cite{LHCb-PAPER-2019-045}. Their masses match two $1P$, two $1D$ and $2S$ $\Lb$ $\lambda$-mode excitation respectively, though other assignments are also discussed~\cite{Chen:2019ywy,Yang:2019cvw,Wang:2019uaj,Wang:2018fjm,Mao:2020jln}. It is useful to note that intermediate
$\Sigmares^{(*)\pm}_b(\to\Lb\pi^{\pm})$ states are found to be present in the $\Lbsimple^{*(*)0}\to\Lb\pip\pim$ decays.

The ground
$\Sigmares^{\pm}_b$ and $\Sigmares^{*\pm}_b$ states were first detected in the $\Lb\pi^{\pm}$ mass spectrum  by CDF~\cite{CDF:2007oeq}.
In the same final state, two new ones, $\Sigmab(6097)^+$ and $\Sigmab(6097)^-$, are observed by LHCb~\cite{LHCb-PAPER-2018-032}, whose widths are about $30\mev$. These two new states
belong to the $P$-wave family, and many more of them are still missing, like for charm baryons. 

In analogy, excited $\Xib$ states are searched for by the
LHC experiments in the $\Xires^{0}_\bquark\pi^{-}/\Xires^{-}_\bquark\pi^{+}$ spectra. New states close to $\Xib\pi$ mass thresholds are observed, which include the low lying $\Xires^{*0}_\bquark$ baryon discovered
by CMS~\cite{CMS:2012frl} and $\Xires^{'-}_\bquark$, $\Xires^{*-}_\bquark$ states discovered by LHCb~\cite{LHCb-PAPER-2014-061}. 
The states $\Xires^{'-}_\bquark$ and $\Sigmares^{\pm}_b$ belong to the  flavour symmetric multiplet with $s_{qq'}=1, J^P=1/2^+$, while $\Xires^{*0}_\bquark$, $\Xires^{*-}_\bquark$ and $\Sigmares^{*\pm}_b$ belong to the flavour symmetric  multiplet with $s_{qq'}=1, J^P=3/2^+$. Going to the higher mass region, a state $\Xib(6227)^-$, with a width around $20 \mev$,  is found in both $\Xibz\pim$ and
$\Lb\Km$ final states~\cite{LHCb-PAPER-2018-013}, and its flavour partner $\Xib(6227)^0$  is found
in the $\Xibm\pip$ mass spectrum~\cite{LHCb-PAPER-2020-032}. They can be matched to $P$-wave
states or a mixture of several $P$-wave states with masses close to $6227\mevcc$. Very recently, two new states $\Xib(6327)^0$ and
$\Xib(6333)^0$, with widths below $2\mev$, are found in the $\Lb\Km\pip$ mass spectrum~\cite{LHCb-PAPER-2021-025}, consistent with the
$1D$ excitation of the $\Xibz$ baryon. These two states may also be present in the $\Xibz\pip\pim$ sample as well, demanding a future investigation of this decay mode by LHCb. In fact, in the $\Xibm\pip\pim$ spectrum, a $\Xib(6100)^-$ state is observed by CMS~\cite{CMS:2021rvl}, consistent with the $1P$ excitation of the flavour antisymmetric $\Xibm$ state with $J^P=3/2^-$. Apparently, the other $1P$ state with $J^P=1/2^-$ and a mass around $6100\mevcc$ is missing. No states with higher masses, for example flavour partners of $\Xib(6327)^0$ and $\Xib(6333)^0$,  are reported by CMS, which may be explained by lower production rate for these states.

Excited $\Omegab$ states are searched for in the $\Xibz\Km$ mass spectrum~\cite{LHCb-PAPER-2019-042}. Four narrow (width $<5\mev$) peaking
structures are identified with two of them having significance greater than five standard deviations ($5\sigma$), named as $\Omegares_b(6340)^-$ and
$\Omegares_b(6350)^-$ respectively. These states lie in the mass region of  $P$-wave excitation. More statistics in Run 3 will  allow for a further investigation of these states.

\begin{figure}[!tb]
\begin{center}
    \includegraphics[width=0.5\textwidth]{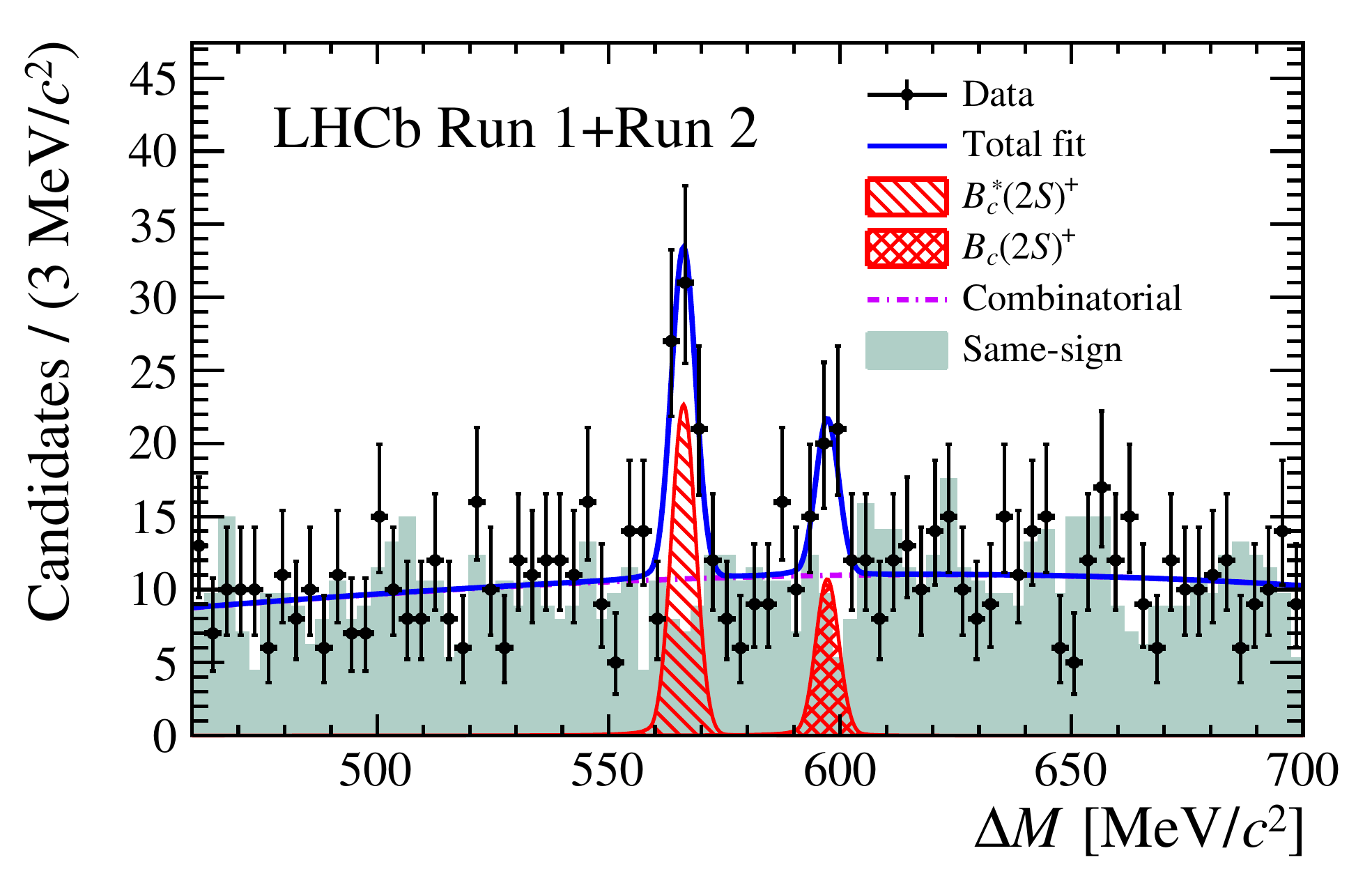}
    \includegraphics[width=0.43\textwidth]{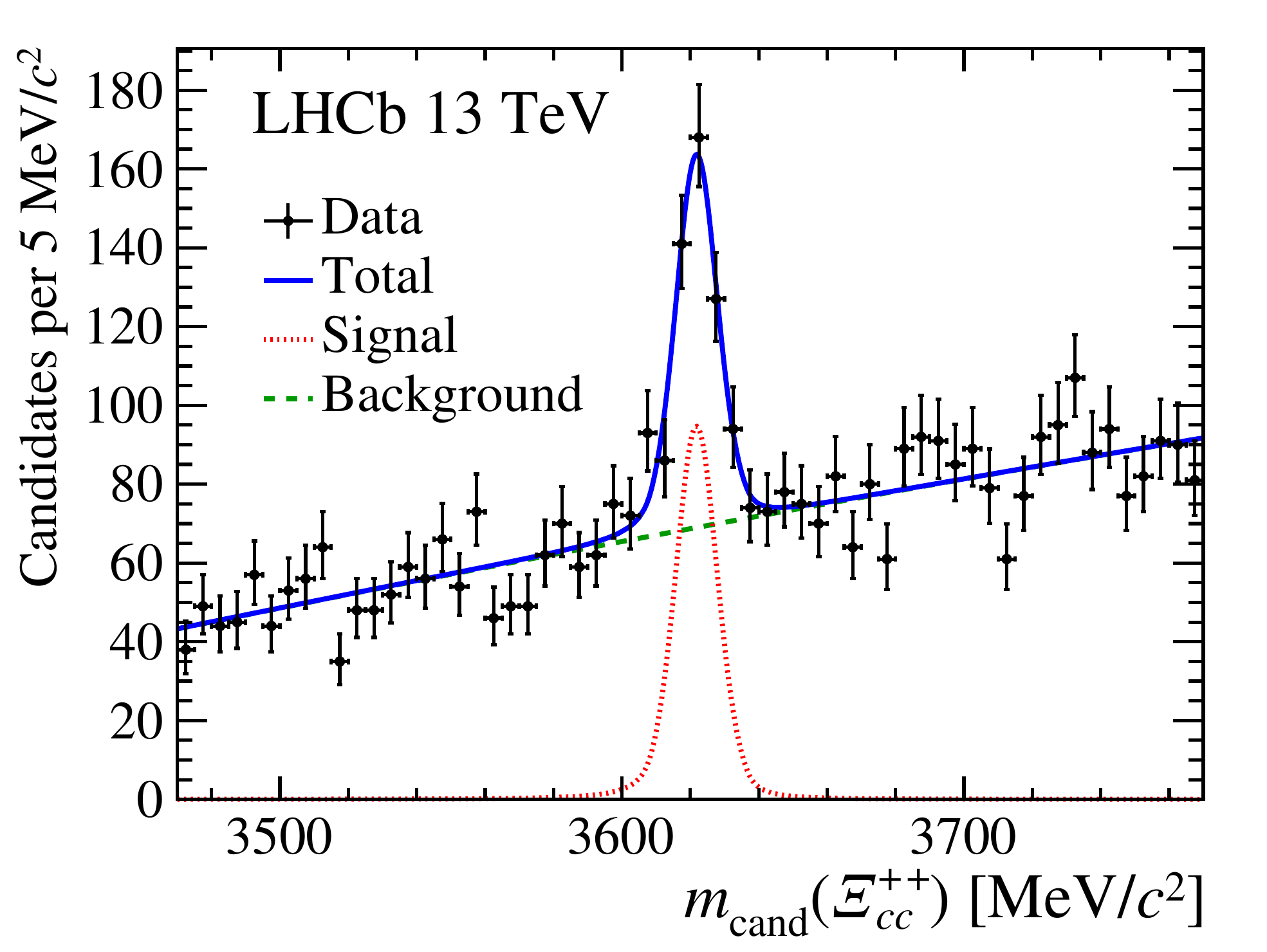}
    \caption{(Left) Invariant mass distribution of $\Delta M\equiv m_{\Bcp\pip\pim}-m_{\Bcp}$, overlaid with the fit including two narrow structures, corresponding to $\B_c^{*}(2S)^+$ and $\B_c(2S)^+$ respectively, taken from Ref.~\cite{LHCb-PAPER-2019-007}. (Right) Distribution of the reconstructed invariant mass $m_\mathrm{cand}(\Xiccpp)\equiv m(\Lc\Km\pip) -m_\mathrm{cand}(\Lc)+m_\mathrm{PDG}(\Lc)$, overlaid with the fit projections, taken from Ref.~\cite{LHCb-PAPER-2017-018}.}
\label{sec:prod:fig:BcXicc}
\end{center}
\end{figure}

\paragraph{Doubly heavy hadrons} LHCb has observed a few new hadrons with two heavy quarks. In the final states of $\Dz\Dzb$ and
$\Dp\Dm$ mesons produced promptly, a new particle $X(3842)$,  with a width of about $3\mev$, is discovered~\cite{LHCb-PAPER-2019-005}. It is consistent with the spin-3 conventional charmonium  $\psi_3({}^3D_3)$ with $J^{PC}=3^{--}$~\cite{Barnes:2005pb}.

As shown on the left of Fig.~\ref{sec:prod:fig:BcXicc}, two narrow structures are detected in the $\Bcp\pip\pim$ invariant mass spectrum of LHCb data~\cite{LHCb-PAPER-2019-007}. The left one corresponds to the $\B_c^{*}(2^3S_1)^+\to \B_c^{*}(1^3S_1)^+\pip\pim$ decay with the
photon in  decay of $\B_c^{*}(1^3S_1)^+\to\Bcp\gamma$ not detected. The peak on the right is consistent with the $\B_c(2^1S_0)^+\to
\Bcp\pip\pim$ decay. Almost in parallel, these two states are independently observed by CMS~\cite{CMS:2019uhm}. As the $\Bcp$ meson is  composed of two heavy quarks, its excitation mass spectroscopy can be calculated using models similar to those applied to heavy quarkonia~\cite{Gupta:1995ps,Chen:1992fq,Ding:2021dwh,Chen:2020ecu,Chang:2019wpt,Chang:2019eob} despite their hadroproduction mechanisms are very different~\cite{Chang:2003cq,Chang:2005hq}.

LHCb opens a new era in studies of doubly heavy baryons by observing
the $\Xiccpp$ baryon in the $\Xiccpp\to\Lc\Km\pip\pip$ mass spectrum~\cite{LHCb-PAPER-2017-018}, as shown on the right of Fig.~\ref{sec:prod:fig:BcXicc}. This discovery decay mode is predicted to have a relatively large branching fraction~\cite{Yu:2019lxw,Hu:2019bqj}. The $\Xiccpp$ state is later confirmed using the
$\Xiccpp\to\Xicp\pip$ decay~\cite{LHCb-PAPER-2018-026}. Its lifetime is measured to be about
$0.25\ps$~\cite{LHCb-PAPER-2019-037} and its mass is  precisely determined to be
$(3621.55\pm0.38)\mevcc$~\cite{LHCb-PAPER-2019-037}. Its SU(3) partners,  $\Xiccp$ and $\Omegacc$,  are  also searched for by LHCb, but with no
sign of observation yet~\cite{LHCb-PAPER-2019-029,LHCb-PAPER-2021-011,LHCb-PAPER-2021-019}. Theoretically, an important question is to understand the mass spectrum of the doubly charm  baryons and related systems, which depends on the binding energy between the two heavy quarks~\cite{Tong:2021raz,Li:2020uok,Wang:2021rjk,Liu:2019vtx,Dias:2019klk,Yu:2019yfr,Li:2019ekr,Chen:2017sbg,Wang:2017qvg,Guo:2021yws,Qin:2021dqo,Gao:2020ogo,Weng:2018mmf}. The successful discovery of $\Xiccpp$ has triggered wide theoretical work to understand the properties of baryons with more than a heavy quark~\cite{Han:2021gkl,Li:2021rfj,Han:2021azw,Wang:2020avt,Xiao:2017dly,Shi:2019hbf,Cheng:2019sxr,Zhang:2018llc,Jiang:2018oak,Zhao:2018zcb,Wang:2018utj,Cui:2017udv,Shi:2017dto,Meng:2017dni,Geng:2017mxn,Lu:2017meb,Xiao:2017udy,Li:2017pxa,Wang:2017azm,Wang:2017mqp,Yu:2017zst,Chen:2013upa,Wang:2017dcq}. In addtion, the $\Xiccpp$ baryon mass  is used to study the stability of tetraquark states with $QQ$ contents~\cite{Cheng:2020wxa,Lu:2020rog} with the assumption that $QQ$ form a heavy diquark~\cite{Shi:2020qde,An:2019idk,Meng:2018zbl,Zhou:2018bkn,Chen:2016ont}.


\subsubsection{Exotic hadrons} 

\def\zcOne{\ensuremath{T_{\psi 1}^{b}(4430)^+}\xspace}
\def\zcTwo{\ensuremath{T_{\psi 1}^{b}(3900)^+}\xspace}
\def\zcs{\ensuremath{T_{\psi s}^+}\xspace}
\def\zcsz{\ensuremath{T_{\psi s}^0}\xspace}
\def\XDKone{\ensuremath{T_{cs0}(2900)^0}\xspace}
\def\XDKtwo{\ensuremath{T_{cs1}(2900)^0}\xspace}

Many theoretical efforts have been placed to understand how quarks are combined to form a multi-body system~\cite{Dong:2021rpi,Ding:2021igr,Chen:2021crg,Yang:2020fou}. At the same time, more and more new states are observed experimentally which cannot fit into the conventional hadron spectra~\cite{Olsen:2017bmm}. As a result, the study on exotic hadrons has been a hot topic for the past decade. New results from LHCb are discussed below.

\paragraph{Tetraquark states}  LHCb provides essential information
for the understanding of previously known tetraquark states. For example LHCb determined the  quantum number of the $X(3872)$ state, first reported by Belle~\cite{Belle:2003nnu}, to
be $J^{PC}=1^{++}$ through a full amplitude analysis~\cite{LHCb-PAPER-2015-015}. LHCb also precisely measured the mass of the $X(3872)$ state  (referred to as $\chi_{c1}(3872)$ in Ref.~\cite{PDG2020}) to be  $m_{\chi_{c1}(3872)}-m_{\psitwos}=185.49\pm0.06\pm0.03\mevcc$~\cite{LHCb-PAPER-2020-008,LHCb-PAPER-2020-009}. Its Breit-Wigner (BW) width is determined by LHCb to be
$\Gamma_{\chi_{c1}(3872)}^\mathrm{BW}=0.96^{+0.19}_{-0.18}\pm0.21\mev$~\cite{LHCb-PAPER-2020-008,LHCb-PAPER-2020-009}. 
Evidence of its decay to $\psi(2S)\gamma$ is found and the branching fraction relative to $\jpsi\gamma$ is measured to $2.46 \pm 0.64 \pm 0.29$~\cite{LHCb-PAPER-2014-008}, disfavouring a pure $D\bar{D}^*$ molecule intepretation.
The $\rho^0$ and $\omega$ contributions are disentangled in its decay to the $\jpsi\pip\pim$ and a sizeable contribution from $\omega$ is confirmed~\cite{LHCb-PAPER-2021-045}.
The $\chi_{c1}(3872)$ state is the mostly studied exotic candidate, and its exotic behaviours include a extremely narrow width, isospin breaking decays and a mass close to the $D^*D$ threshold. Despite all the available information  we are still not sure whether it is a compact tetraquark state, a $\Dstarz\Dzb$ hadron molecule, a mixture of $\Dstarz\Dzb$ molecule with a $\chicone(2P)$ charmonium component or just caused by kinematic rescattering effect~\cite{Zhang:2020mpi,Meng:2021kmi,He:2021ihy,Maiani:2004vq,Wang:2013vex,Chen:2010ze,Li:2004sta,Guo:2014iya,Guo:2013zbw,Meng:2007cx,Guo:2016bjq,Luo:2017eub,Kang:2016jxw,Yuan:2005dr,Cui:2006mp,Meng:2005er,Zhang:2009vs,Zhang:2009bv,Meng:2013gga,Ding:2009vj,Chen:2016otp,Li:2012cs,Liu:2019stu,Chen:2013pya,Dong:2021juy,Guo:2014taa,Guo:2019qcn,Zhao:2014gqa,Liu:2008qb,Ke:2012gm,Geng:2017hxc,Meng:2014ota,Chen:2017vai,Shen:2010ky,Zhang:2007xa,Wang:2012cp,Wang:2015rcz,Yang:2017prf,Liang:2021fzr,Sun:2017wgf,Yu:2011wb,Cao:2020dlz,Qiao:2013xca}.

The exotic candidate \zcOne was first observed by Belle in the $\psitwos\pim$ mass spectrum in  $\Bz\to\psitwos\Kp\pim$ decays~\cite{Belle:2007hrb}.\footnote{The exotic hadron naming scheme is used~\cite{Gershon:2022xnn}.}
An amplitude analysis of the $\Bz\to\psitwos\Kp\pim$ decay is performed at LHCb, confirming the existence of the \zcOne state and determining it to be consistent
with a Breit-Wigner resonance with $J^P=1^+$~\cite{LHCb-PAPER-2014-014}. The quark contents of \zcOne, $c\cquarkbar\uquarkbar d$,  are the same as the \zcTwo state observed by the BESIII experiment~\cite{BESIII:2013ris}. Being charged, they are definitely not consistent with conventional charmonia and many phenomenological calculations are performed to explain their internal structure and properties~\cite{Wu:2021ezz,Wang:2021aql,Zhang:2020rqr,Xiao:2019spy,He:2017mbh,Chen:2015ata,Chen:2015fsa,Liu:2014eka,Ma:2014zva,Wang:2014vha,Ma:2014zua,Ke:2013gia,Zhao:2014qva,Gong:2016hlt,Gui:2018rvv,Wang:2019tlw,He:2014nxa,Liu:2009wb}.

\begin{figure}[!tb]
\begin{center}
    \includegraphics[width=0.8\textwidth]{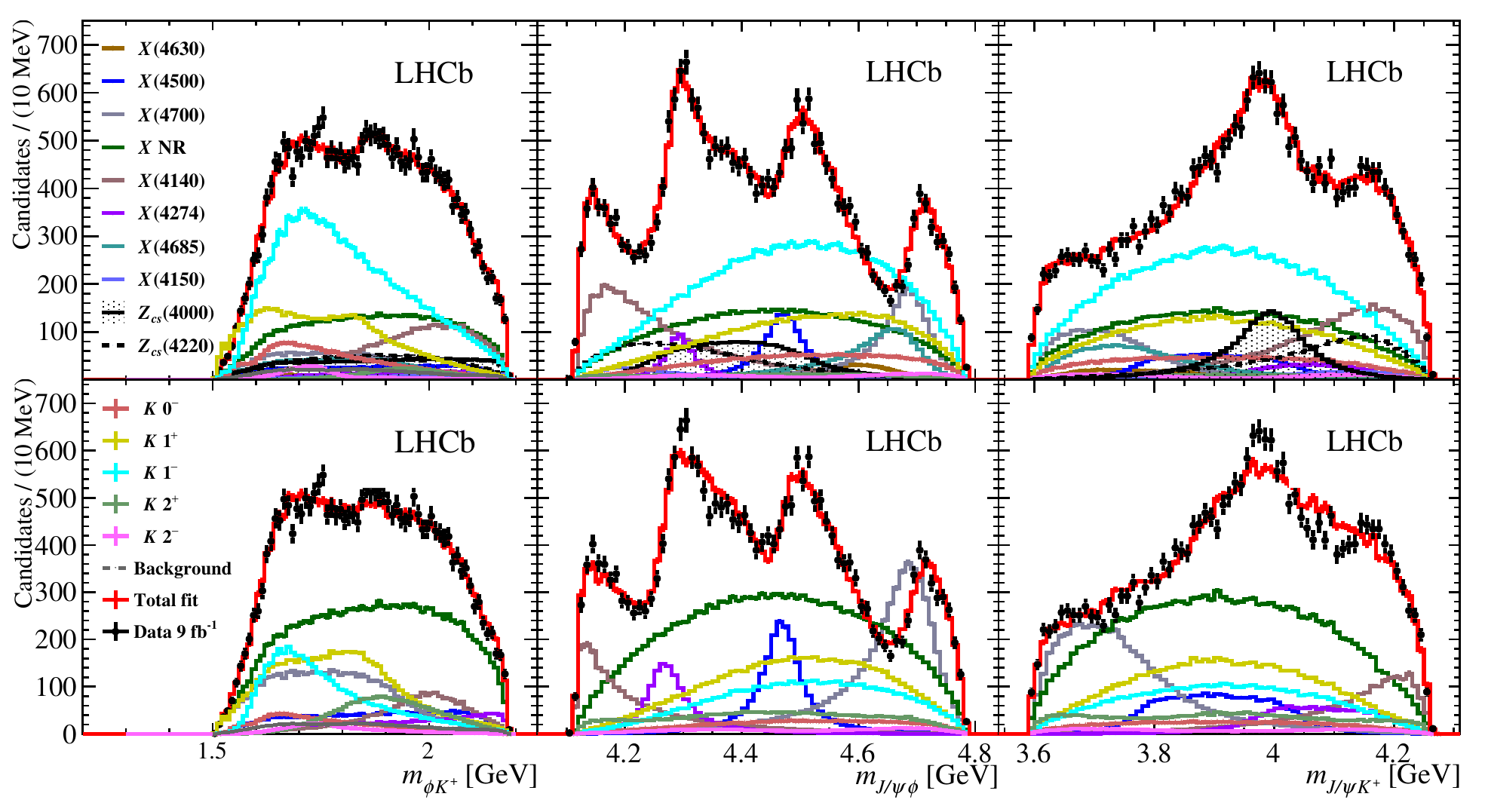}
    \caption{Distributions of (left) $\phi\Kp$, (middle)$\jpsi\phi$  and (right) $\jpsi\Kp$ invariant masses for (black data points) the $\Bp\to\jpsi\phi\Kp$ candidates  compared with  (red solid lines) the fit results using  (top row) the final model and (bottom row) the model of LHCb Run 1 analysis~\cite{LHCb-PAPER-2016-018}. Figure taken from Ref.~\cite{LHCb-PAPER-2020-044}.}
\label{sec:prod:fig:JpsiPhiK}
\end{center}
\end{figure}

The $\Bp\to\jpsi\phi\Kp$ decay is a zoo of exotic hadrons. In 2009, a narrow state $X(4140)$ was reported by CDF in the
$\jpsi\phi$ mass spectrum of the $\Bp\to\jpsi\phi\Kp$ decay~\cite{CDF:2009jgo,CDF:2011pep}, and is later confirmed by CMS~\cite{CMS:2013jru}. The quark
contents of the $X(4140)$ state is likely to be $\cquark\cquarkbar\squark\squarkbar$, consistent with an exotic hadron~\cite{Hu:2010fg,Meng:2019fan}, even though excited conventional
charmonia may have the chance to decay into $\jpsi\phi$ too. In the amplitude analysis by LHCb using Run 1 data, four
exotic candidates $X(4140)$, $X(4274)$, $X(4500)$ and $X(4700)$ are observed~\cite{LHCb-PAPER-2016-018}. Currently,
these $X$ states are considered to be either hadron molecules or compact tetraquark states or high-mass conventional
charmonia in various
calculations~\cite{Cao:2019wwt,Hao:2019fjg,Wang:2020dya,Chen:2016ncs,Lu:2016cwr,Wu:2016gas,Liu:2016onn,Wang:2016ujn,Wang:2014gwa,Zhang:2009st,Liu:2009iw,Chen:2016oma,Wang:2009wk,Wang:2017mrt,He:2013oma,Ferretti:2020civ,Deng:2019dbg,Wang:2021gml,Shi:2021jyr,Liu:2021xje,Ge:2021sdq,Yang:2021sue,Yang:2020nrt,Wang:2020rcx,Chen:2013wca,Jin:2020yjn,Liu:2021sxs}.
The LHCb analysis is updated recently with a sample that has six times more statistics, in which three more $X$ states are reported~\cite{LHCb-PAPER-2020-044}. In addition,  two \zcs structures are observed in the $\jpsi\Kp$
mass spectrum. The mass spectra and fit projections are shown in Fig.~\ref{sec:prod:fig:JpsiPhiK} and the properties of these exotic candidates are summarised in Table~\ref{sec:prod:tab:B2JpsiPhiK}.  The \zcs states mark the first observation of  exotic hadrons
with an $s$ quark through beauty decays. It is noted that another \zcs state is reported  by BESIII in the final state of
$D_s^-D^{*0}+D_s^{*-}\Dz$ pairs~\cite{BESIII:2020qkh}, with a mass and width different from those observed by LHCb. 
Very recently, evidence of \zcsz, isopin partner of \zcs, is found by LHCb in the $\Bp\to\jpsi\phi\KS$ decay through a
combined amplitude analysis of both $\Bp\to\jpsi\phi\KS$ and $\Bp\to\jpsi\phi\Kp$ decays~\cite{LHCb-PAPER-2022-040}. The
masses, widths of these two states and their contributions to the $B\to T_{\psi s} K$ decay are similar, confirming that
they are isospin partners.
There should be more states of $c\cquarkbar q\squarkbar$ quark contents, and their discovery will definitely help to understand the internal structure of strange tetraquark states~\cite{Cao:2021ton,Duan:2021pll}.

\begin{table}[!hbt]
\begin{center}
    \caption{Spin-parity, significance, masses, widths, and fit fractions of exotic candidates observed in the $\Bp\to\jpsi\phi\Kp$ decay
    by LHCb~\cite{LHCb-PAPER-2020-044}.}
    \label{sec:prod:tab:B2JpsiPhiK}
\vskip -0.2cm
\def\arraystretch{1.2}
\begin{tabular}{ccr@{\:$\pm$\:}lr@{\:$\pm$\:}lr@{\:$\pm$\:}l}
\hline
Contribution &Significance\,[$\times \sigma$]&            \multicolumn{2}{c}{$M_0$\,[$\kern -0.125em \mev$]}   & \multicolumn{2}{c}{$\Gamma_0$\,[$\kern -0.125em\mev$]}      & \multicolumn{2}{c}{FF\,[\%]}          \\
\hline \hline
$X(2^-)$ & & \multicolumn{2}{c}{} & \multicolumn{2}{c}{} &  \multicolumn{2}{c}{}  \\
$X(4150)$            & 4.8 (8.7) & $4146$&$18\pm33$ & $135$&$28\,_{-\,30}^{+\,59}$   & $2.0$&$0.5\,_{-\,1.0}^{+\,0.8}$ \\
\hline
$X(1^-)$ & & \multicolumn{2}{c}{} & \multicolumn{2}{c}{} &  \multicolumn{2}{c}{}  \\
$X(4630)$            & 5.5 (5.7) & $4626$&$ 16\,_{-\,110}^{+\,\phantom{0}18}$ & $174$&$27\,_{-\,\phantom{0}73}^{+\,134}$  & $2.6$&$0.5\,_{-\,1.5}^{+\,2.9}$ \\
\hline

All $X(0^+)$ & &\multicolumn{2}{c}{} &\multicolumn{2}{c}{} & $20$&$5\,_{-\,\phantom{0}7}^{+\,14}$  \\
$X(4500)$           & 20 (20) & $4474$&$3 \pm3 $ & $77$&$6\,_{-\,\phantom{0}8}^{+\,10}$ &  $5.6$&$0.7\,_{-\,0.6}^{+\,2.4}$ \\
$X(4700)$           & 17 (18) & $4694$&$ 4 \,_{-\,\phantom{0}3}^{+\,16}$ & $87$&$8\,_{-\,\phantom{0}6}^{+\,16}$ &  $8.9$&$1.2\,_{-\,1.4}^{+\,4.9}$ \\
\hline
All $X(1^+)$ & &\multicolumn{2}{c}{} &\multicolumn{2}{c}{} & $26$&$3\,_{-\,10}^{+\,\phantom{0}8}$ \\
$X(4140)$           & 13 (16) & $4118$&$ 11 \,_{-\,36}^{+\,19}$ & $162$&$21\,_{-\,49}^{+\,24}$  & $17$&$3\,_{-\,6}^{+\,19}$ \\
$X(4274)$           & 18 (18) & $4294$&$4 \,_{-\,6}^{+\,3}$     & $53$ &$5\pm5$                 & $2.8$&$0.5\,_{-\,0.4}^{+\,0.8}$ \\
$X(4685)$           & 15 (15) & $4684$&$ 7\,_{-\,16}^{+\,13}$   & $126$&$15 \,_{-\,41}^{+\,37}$ & $7.2$&$1.0 \,_{-\,2.0}^{+\,4.0}$ \\
\hline\hline
All $T_{\psi s}(1^+)$ & & \multicolumn{2}{c}{}& \multicolumn{2}{c}{}& $25$&$5\,_{-\,12}^{+\,11}$ \\
$T_{\psi s}(4000)$          & 15 (16) & $4003$&$ 6\,_{-\,14}^{+\,\phantom{0}4}$  & $131$&$15\pm26$  & $9.4$&$2.1\pm3.4$ \\
$T_{\psi s}(4220)$          & 5.9 (8.4) & $4216$&$24\,_{-\,30}^{+\,43}$ & $233$&$52\,_{-\,73}^{+\,97}$ & $10$&$4\,_{-\,\phantom{0}7}^{+\,10}$ \\
\hline\hline
\end{tabular}
\end{center}
\end{table}

\begin{figure}[!tb]
\begin{center}
\includegraphics[width=0.41\textwidth]{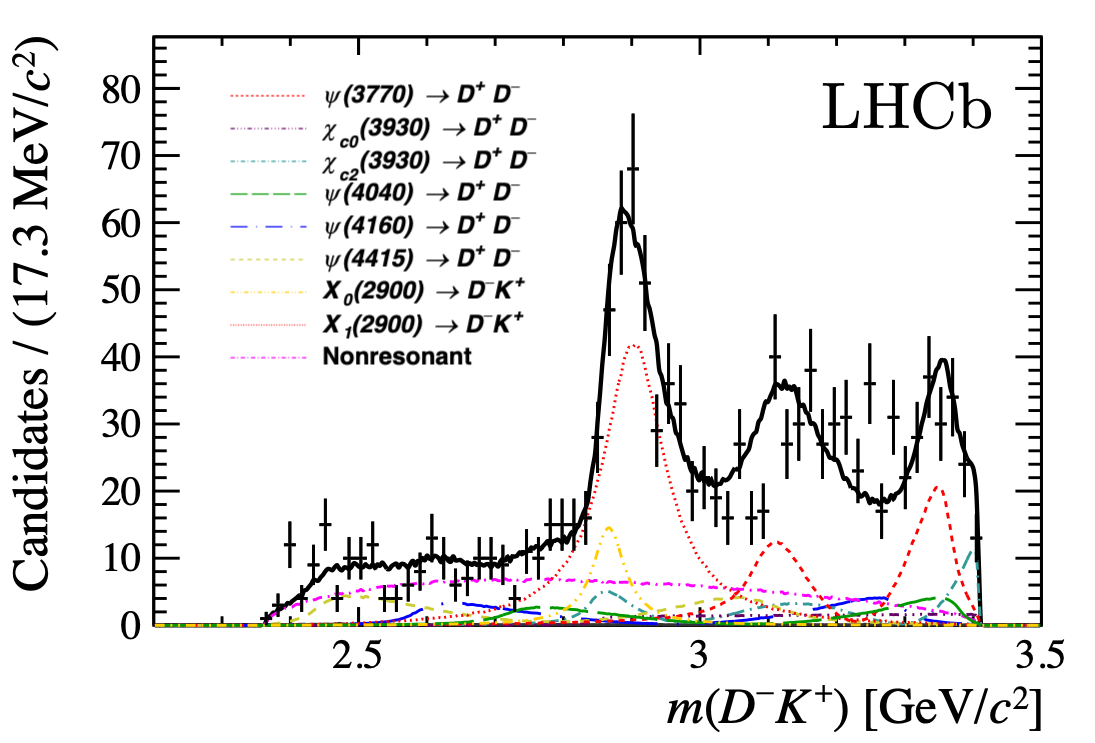}
\includegraphics[width=0.39\textwidth]{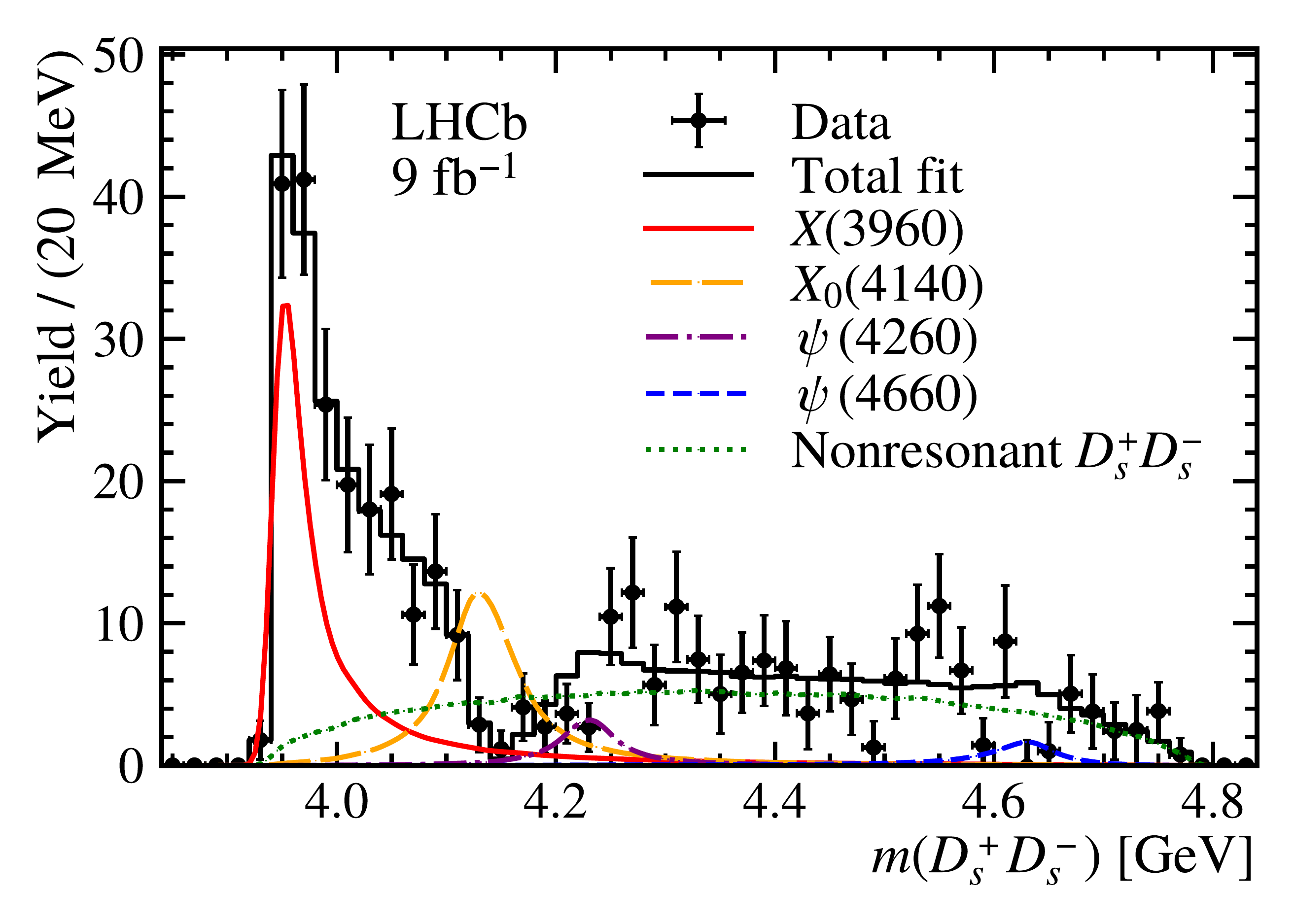}
\phantom{\includegraphics[width=0.15\textwidth]{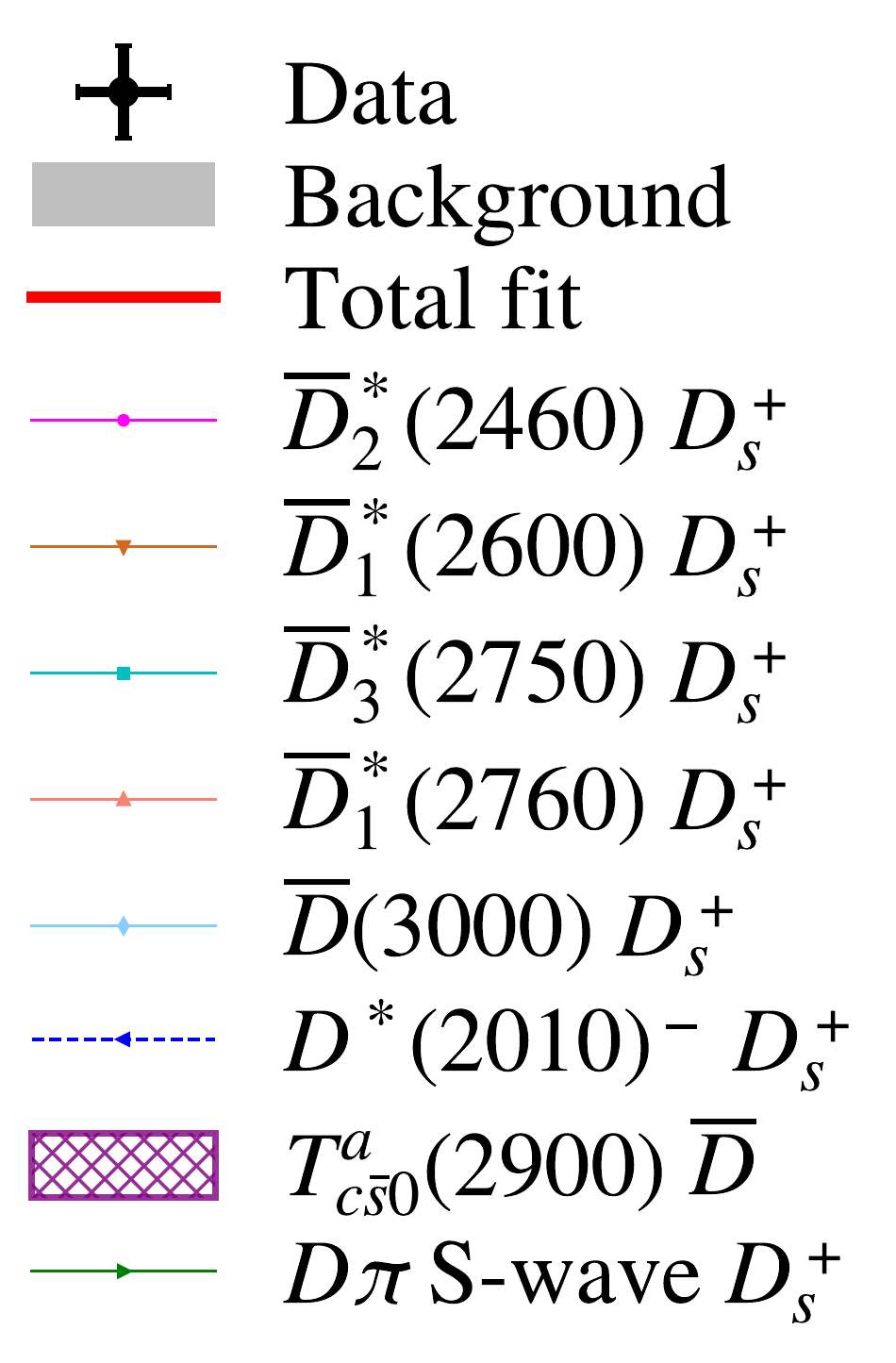}}
\includegraphics[width=0.4\textwidth]{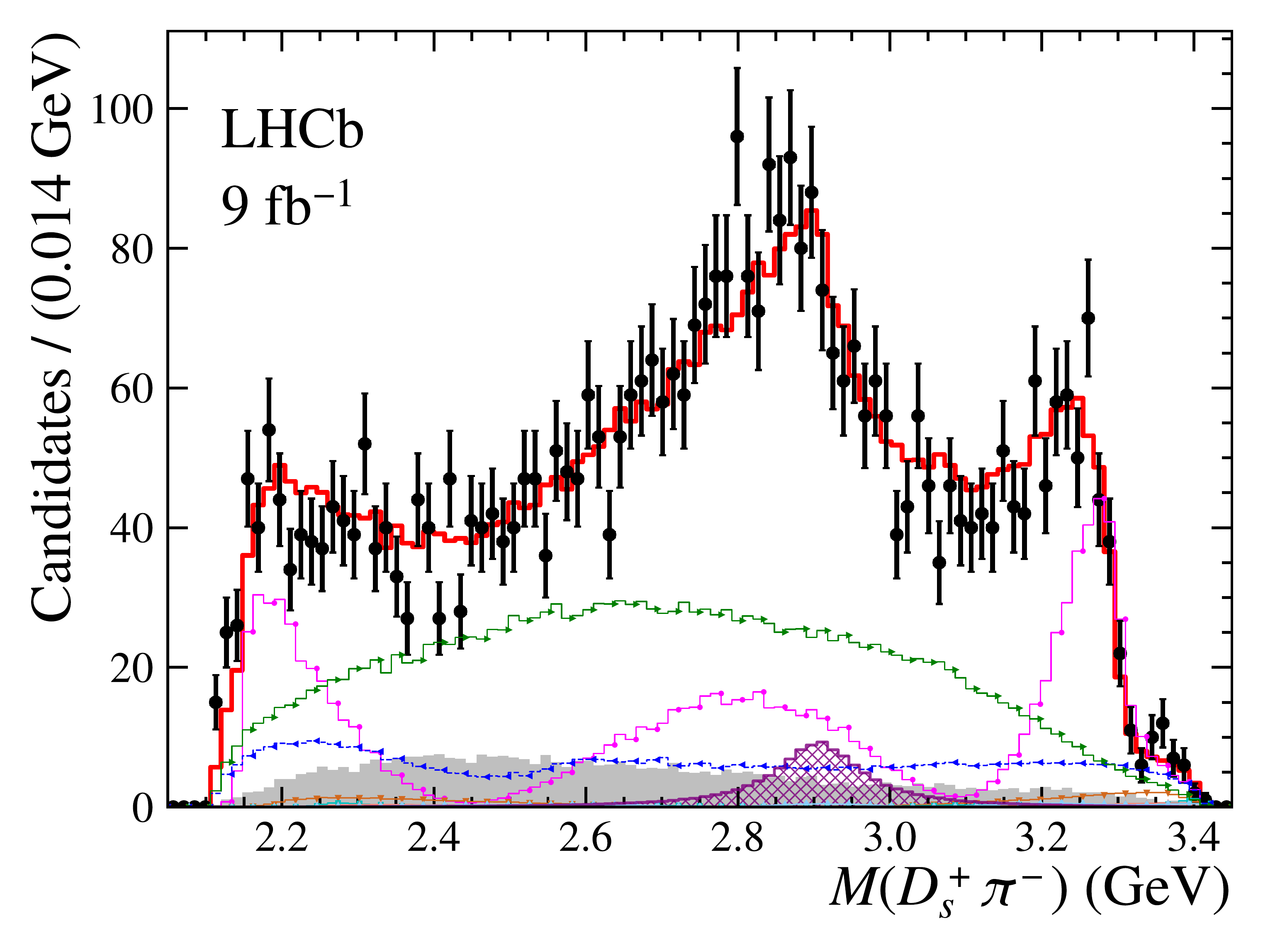}
\includegraphics[width=0.4\textwidth]{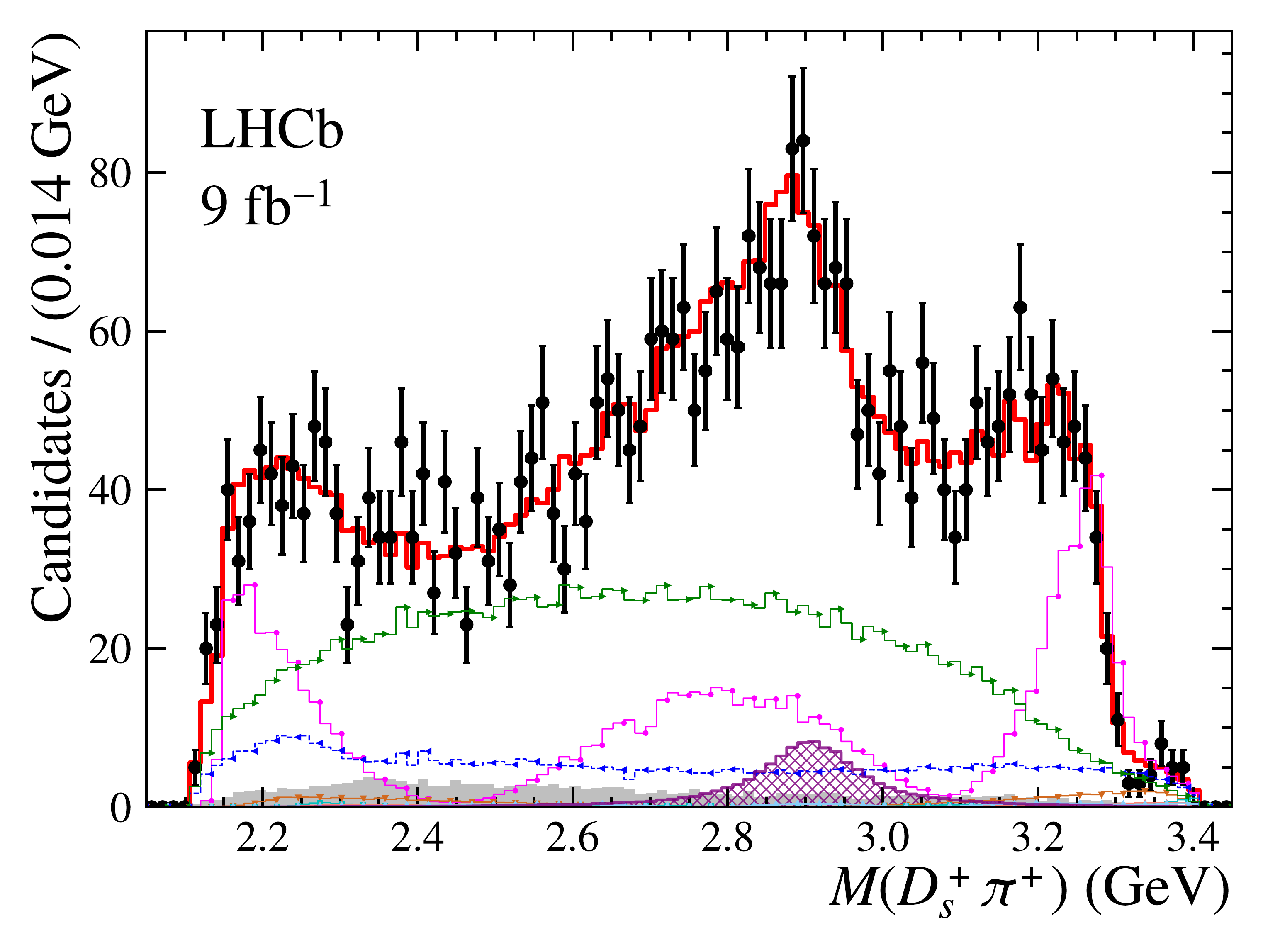} 
\includegraphics[width=0.15\textwidth]{figs/production/T2900leg.pdf}
\caption{
(Top left) Invariant-mass distribution of $\Dm\Kp$  in $\Bp\to\Dp\Dm\Kp$ decays in data and the fit projections, taken from Ref.~\cite{LHCb-PAPER-2020-025}. The yellow dot-dashed and red dotted line correspond to \XDKone and \XDKtwo, respectively, and more details can be found in Ref.~\cite{LHCb-PAPER-2020-025}.
(Top right) The distribution of the invariant mass of the $\Dsp\Dsm$ system in the $\Bp\to\Dsp\Dsm\Kp$ decay. A near-threshold structure, denoted as $X(3960)$, is marked in red~\cite{LHCb-PAPER-2022-018}. 
The distribution of the invariant mass of (bottom left) the $\Ds\pim$ pair in the $\Bz\to\Dzb\Ds\pim$ and (bottom right) $\Ds\pip$ system in the $\Bp\to\Dm\Ds\pip$ decay~\cite{LHCb-PAPER-2022-027}.
}
\label{sec:prod:fig:X3960}
\end{center}
\end{figure}

Exotic hadrons are also searched for in open charm final states using fully reconstructed beauty hadron decays. A Dalitz analysis of the $\Bp\to\Dp\Dm\Kp$ decay is performed by LHCb~\cite{LHCb-PAPER-2020-024,LHCb-PAPER-2020-025}, and two exotic states, \XDKone and \XDKtwo, are required to have a good
fit to the $\Dm\Kp$ invariant mass spectrum. The  $\Dm\Kp$  invariant-mass distribution and the fit projections are shown in Fig.~\ref{sec:prod:fig:X3960}. Their spin-parities are measured to be $J^P=0^+$ and $1^-$, and widths to be about $50\mev$ and $100\mev$ respectively. The quark contents of these $X$ states are
$\cquark\squark\uquarkbar\dquarkbar$.  These two $X$ states
contribute up to $35\%$ of the total $\Bp\to\Dp\Dm\Kp$ decay branching fraction, a magnitude similar to those of conventional charmonia in the decay. 
A final state rescattering effect is considered in order to explain  such a large branching fraction~\cite{Chen:2020eyu}. 
In a recent analysis of the $\Bz\to\Dzb\Ds\pim$ and $\Bp\to\Dm\Ds\pip$ decays by LHCb, two new states $T_{c\overline{s}0}^a(2900)^{++/0}$ are observed in
the $\Ds\pi^{+/-}$ systems~\cite{LHCb-PAPER-2022-026,LHCb-PAPER-2022-027}. The quark contents of these two
states are $\cquark\squarkbar\dquark\uquarkbar$ and $\cquark\squarkbar\uquark\dquarkbar$ respectively. The invariant mass
distributions of $\Ds\pim$ and $\Ds\pip$ are shown in Fig.~\ref{sec:prod:fig:X3960} superimposed with the
amplitude fit results. The masses and width of these two states are measured to be about $2.9\gevcc$ and   $0.15\gev$
respectively, and they contribute to about 3\% of the total $B\to\Dzb\Ds\pim$ decay.
The mass of the $T_{\cquark\squarkbar0}^a(2900)$ is consistent with the previously mentioned $\XDKone$ discovered in the $\Dm\Kp$ final state, but their widths and flavour contents are different.
The observation of $X\to D h$ ($D=\Ds, D$, $h=\pi, K$) states in the $B\to D \overline{D} h$ decay  opens a new avenue for studies of exotic hadrons composed of four different quark flavours~\cite{Lu:2020qmp,Liu:2020nil,Hu:2020mxp,Tan:2020cpu,Dong:2020rgs,Dai:2015bcc,Cheng:2020nho,Wang:2021lwy,Wang:2020xyc,Huang:2020ptc,He:2020jna,Chen:2020aos,Zhang:2020oze,Liu:2020orv}. 
Actually there are more than two dozens of $B\to D D K(\pi)$  decays, and also a few similar decays for $b$-baryons, and  it is promising that more $X$ states will be observed in these decays.

Recently, a Dalitz analysis for the $\Bp\to\Dsp\Dsm\Kp$ decay was performed by the \lhcb collaboration~\cite{LHCb-PAPER-2022-018,LHCb-PAPER-2022-019}. A near-threshold
structure in the $\Dsp\Dsm$ system is observed with a significance larger than $12\sigma$. The $\Dsp\Dsm$ invariant-mass
distribution is shown on the left of Fig.~\ref{sec:prod:fig:X3960}. The spin-parity of the $X(3960)$ state is determined to be $J^{PC} = 0^{++}$. The $X(3960)$ state is similar to the $\chi_{c0}(3930)$ state in Ref.~\cite{PDG2020}. If they are the same states, the partial width ratio is measured to be 
\begin{equation*}
    \frac{\Gamma\left(X \rightarrow D^{+} D^{-}\right)}{\Gamma\left(X \rightarrow D_{s}^{+} D_{s}^{-}\right)}=0.29 \pm 0.09 \pm 0.10 \pm 0.08,
\end{equation*}
where the first uncertainty is statistical, the second systematic, and the third external.
The ratio is smaller than unity. 
Since the creation of $\squark\squarkbar$ from vacuum is suppressed than $\uquark\uquarkbar$ or $\dquark\dquarkbar$ and the phase-space factor of $X \rightarrow D^{+} D^{-}$ is smaller than $X\to\Dsp\Dsm$, the $\Gamma\left(X \rightarrow D^{+} D^{-}\right)$ is expected to be larger than $\Gamma\left(X \rightarrow D_{s}^{+} D_{s}^{-}\right)$, which is inconsistent with the results from experiment. The inconsistency indicating the exotic nature of the $X(3960)$ states under the assumption of the $X(3960)$ and the $\chi_{c0}(3930)$ being the same states.

\begin{figure}[!tb]
\begin{center}
    \includegraphics[width=0.41\textwidth]{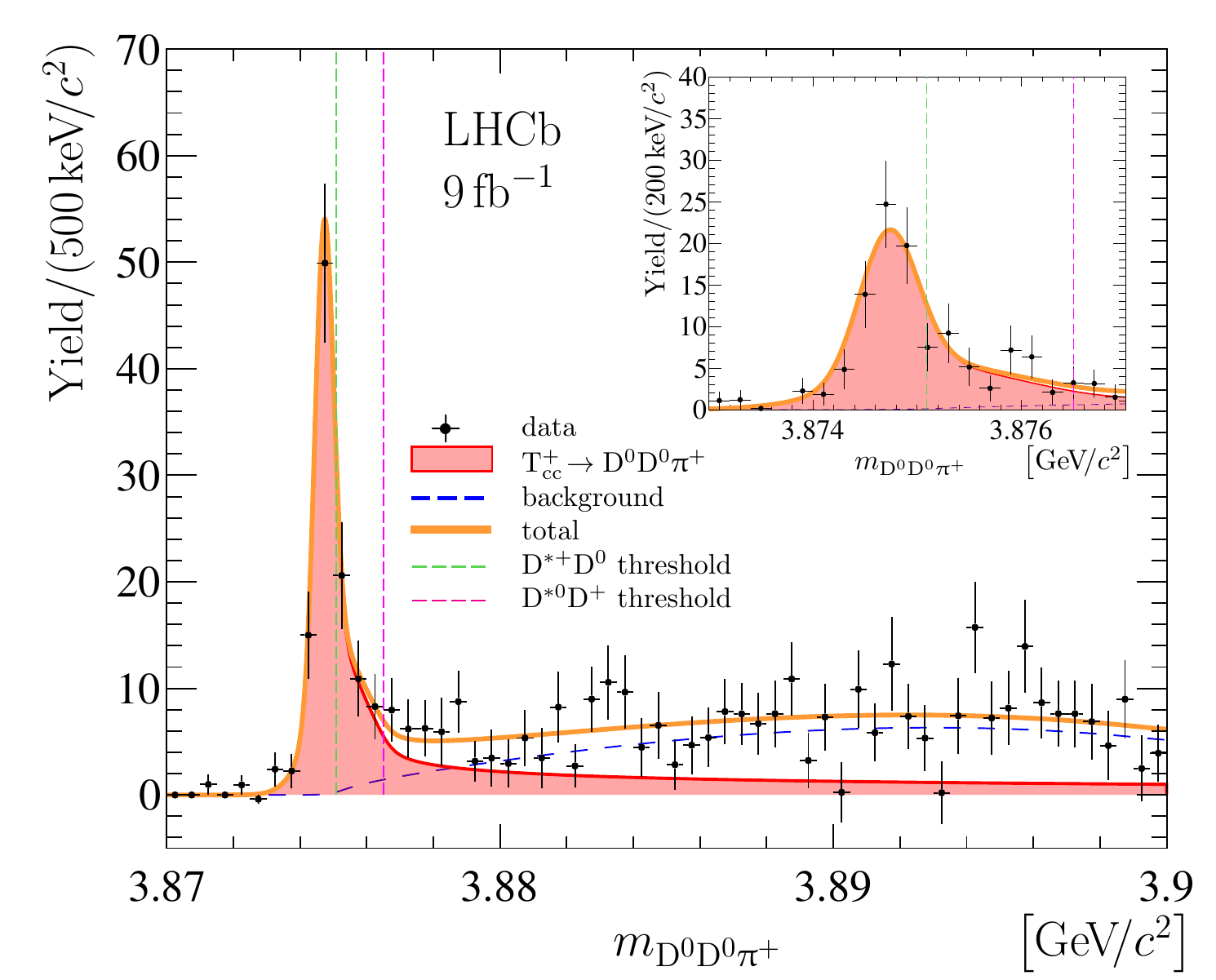}
    \includegraphics[width=0.5\textwidth]{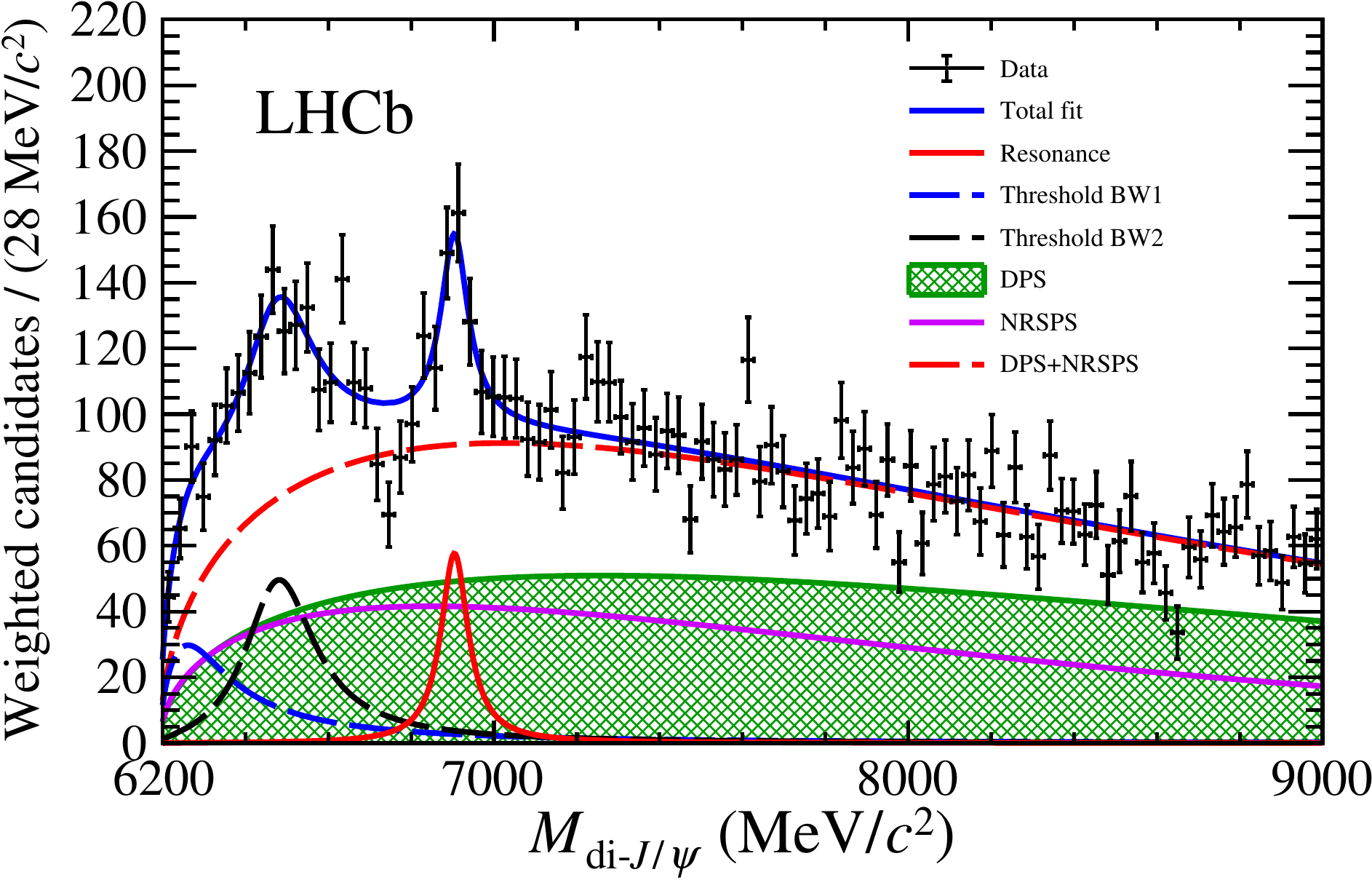}
    \caption{ 
    (Left) Distribution of $\Dz\Dz\pip$ invariant mass overlaid with fit projections.  The contribution of the non-$\Dz$ background has been statistically subtracted. Vertical dashed lines show the $D^{*+}\Dz$ and $\Dstarz\Dp$ mass thresholds respectively for a comparison. Figure taken from Ref.~\cite{LHCb-PAPER-2021-032}.
    (Right) Invariant mass distribution of the combination of two $\jpsi$ mesons overlaid with fit projections,  taken from Ref.~\cite{LHCb-PAPER-2020-011}. The fit model does not contain interference between any components.
    }
\label{sec:prod:figs1}
\end{center}
\end{figure}

Prompt production of di-charm hadrons has been suggested to search for exotic states containing multiple charm quarks~\cite{Yang:2009zzp}.
In the invariant mass spectrum of $\Dz\Dz\pip$, shown in Fig.~\ref{sec:prod:figs1}, a new structure is observed close to the $D^{*+}\Dz$ mass
threshold~\cite{LHCb-PAPER-2021-031, LHCb-PAPER-2021-032}. The structure is measured to be consistent with the ground state of a $T_{cc}^+$ isoscalar
tetraquark,  with $J^P=1^+$ and quark contents $cc\uquarkbar\dquarkbar$. Its Breit-Wigner mass is measured to be
$-273\pm61\pm5^{+11}_{-14}\kevcc$ below $m_{D^{*+}}+m_\Dz$, and  its Breit-Wigner width is $\Gamma_\mathrm{BW}=410\pm165\pm43^{+18}_{-38}\kev$.
The same state also appears in the $\Dz\Dz$ and $\Dz\Dp$ mass spectra, with a $\pip$, $\piz$ or
$\gamma$ in the $T_{cc}^+\to \Dz\Dz\pip$, $T_{cc}^+\to\Dz\Dp\piz$ or $T_{cc}^+\to\Dz\Dp\gamma$ decays undetected, respectively.
Dedicated studies of the $T_{cc}^+$ resonance lineshape are performed using a unitarised Breit-Wigner distribution, considering $T_{cc}^+$ decays in $\Dz\Dz\pip, \Dz\Dp\piz$ and $\Dz\Dz\gamma$ final states. The pole mass of the resonance in this advanced model is measured to be $-360\pm4^{+4}_{-0}\kevcc$
below $m_\Dstar+m_\Dz$, and the pole width is $\Gamma_\mathrm{pole}=-48\pm2^{+0}_{-14}\kev$. This extremely narrow width has attracted many theoretical interests~\cite{Ling:2021bir,Meng:2021jnw,Dai:2021wxi,Li:2021zbw,Guo:2021mja,Chen:2021kad}.  The $T_{cc}^+$ state is the first observed 
tetraquark candidates with two heavy quarks of the same flavour. Many theoretical models have been applied to explain the existence and structure of such a tetraquark state with a large fraction of them favouring a $D^{*+}\Dz$ hadron molecule interpretation~\cite{Xing:2021yid,An:2021vwi,Deng:2018kly,Wang:2017dtg,Karliner:2017qjm,Eichten:2017ffp,Zhu:2016arf,Yang:2019itm,Yan:2018gik,Tan:2020ldi,Kong:2021ohg,Hsiao:2021tyq,Chen:2021erj,Dong:2021lkh,He:2021oyy,Majarshin:2021hex,Wang:2021ajy,Du:2021zzh,Ling:2021asz,Baru:2021ldu,Ren:2021dsi,Wang:2017uld,Chen:2021htr}.


In addition to two open charm final states,  di-$\jpsi$ mass spectrum of prompt production is also studied using full LHCb data~\cite{LHCb-PAPER-2020-011}. Two peaking structures are observed in the mass range
$6.2<m_{\jpsi\jpsi}<7.4\gevcc$, where fully charmed tetraquark states are predicted~\cite{Iwasaki:1976cn,Chao:1980dv,PhysRevD.25.2370,PhysRevD.29.426,Badalian:1985es,
Berezhnoy:2011xn,Wu:2016vtq,Karliner:2016zzc,
Barnea:2006sd,Debastiani:2017msn,Liu:2019zuc,
Chen:2016jxd,
Wang:2019rdo,Bedolla:2019zwg,Lloyd:2003yc,
Chen:2020lgj,Wang:2018poa,
Anwar:2017toa,Esposito:2018cwh,
Becchi:2020mjz,Bai:2016int,Richard:2017vry,Vega-Morales:2017pmm,Chen:2019vrj,Berezhnoy:2012tu}. The first
structure (referred to as the threshold peak) covers the range between $6.2$ and $6.6\gevcc$ close to the di-$\jpsi$ mass threshold, and the
other one sits at $6.9\gevcc$, as shown on the left of Fig.~\ref{sec:prod:figs1}. The di-$\jpsi$ mass spectrum is modelled with a combination of 
BW functions for the two peaking structures and empirical smooth functions for SPS, DPS production of nonpeaking background. When no interference between BW and SPS is applied, the
threshold structure can be described by two BW functions and  the one at $6.9\gevcc$ is well described by a BW. 
The narrow structure, denoted as $T_{\psi\psi}(6900)$, is measured to have a mass of $m_{T_{\psi\psi}(6900)}=6905\pm11\pm7\mevcc$ and a width of
$\Gamma_{T_{\psi\psi}(6900)}=80\pm19\pm33\mev$.
Interpretations of the threshold peak using feeddown decays from excited quarkonium pairs are also possible. For various fit models without any interference, the dip around $6.8\gevcc$ can't be well described. Advance fit studies are performed introducing interference between SPS and resonant
structures. In one such fit, two BW functions are considered: a broad BW interfering with SPS used to describe the
threshold structure, and a stand-alone narrow one used to model the $6.9\gevcc$ peak. This new model could fit well the overall spectrum, and the
broad structure is now measured to have a mass around $6.7\gevcc$ and a width of about $0.3\gev$, while the $T_{\psi\psi}(6900)$ structure has
a mass consistent with the no-interference fit model, but its width becomes about twice larger. As the fit results for the broad
structure is not stable in different models, its nature is not fully resolved and more data are needed to provide
better information. 
The $T_{\psi\psi}(6900)$ and higher resonances have been recently confirmed by CMS~\cite{CMS:2022yhl} and ATLAS~\cite{ATLAS:2022hhx}
The $T_{\psi\psi}(6900)$ state is consistent with a genuine fully charmed tetraquark, however, in some models possible origins due to rescatterings of multiple charmonia or dibaryon molecules \etc are also discussed~\cite{Chao:2020dml,Weng:2020jao,An:2020vku,Lu:2020cns,Lee:2011rka,Jin:2020jfc,Lu:2017dvm,Meng:2017fwb,Wang:2020dlo,Tang:2019nwv,Wan:2020fsk,Yang:2020wkh,Li:2019uch,Liu:2020tqy,Zhao:2020zjh,Yan:2021glh,Liu:2021rtn,Wang:2021hdk,Li:2021ygk,Ke:2021iyh,Huang:2020dci,An:2020jix,Gong:2020bmg,Zhu:2020snb,Zhang:2020vpz,Cao:2020gul,Guo:2020pvt,Zhu:2020xni,Zhang:2020xtb,Faustov:2020qfm,Feng:2020riv,Ma:2020kwb,Dong:2020nwy,Karliner:2020dta,Wang:2020wrp,Yang:2020rih,Deng:2020iqw}. Other fully heavy tetraquarks, such as
$\bquark\bquarkbar\cquark\cquarkbar$ and $\bquark\bquarkbar\bquark\bquarkbar$,
can be searched for in $\PUpsilon\jpsi$ and $\PUpsilon\PUpsilon$ or similar final states, however current limited data are estimated to have small sensitivities for these states, demanding the increased luminosity in LHCb upgrades.

\begin{figure}[!tb]
\begin{center}
    \includegraphics[width=0.45\textwidth]{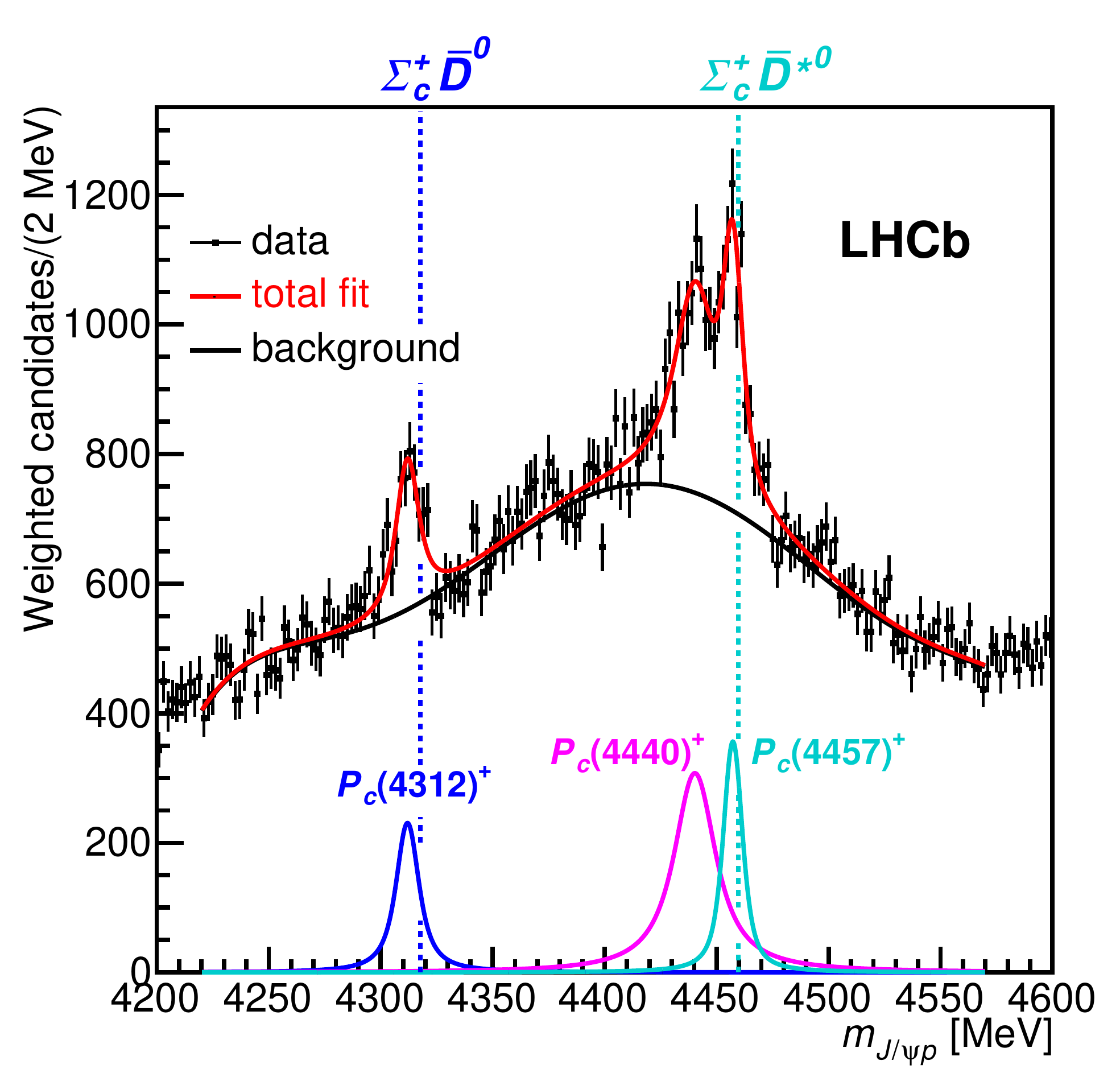}
    \includegraphics[width=0.42\textwidth]{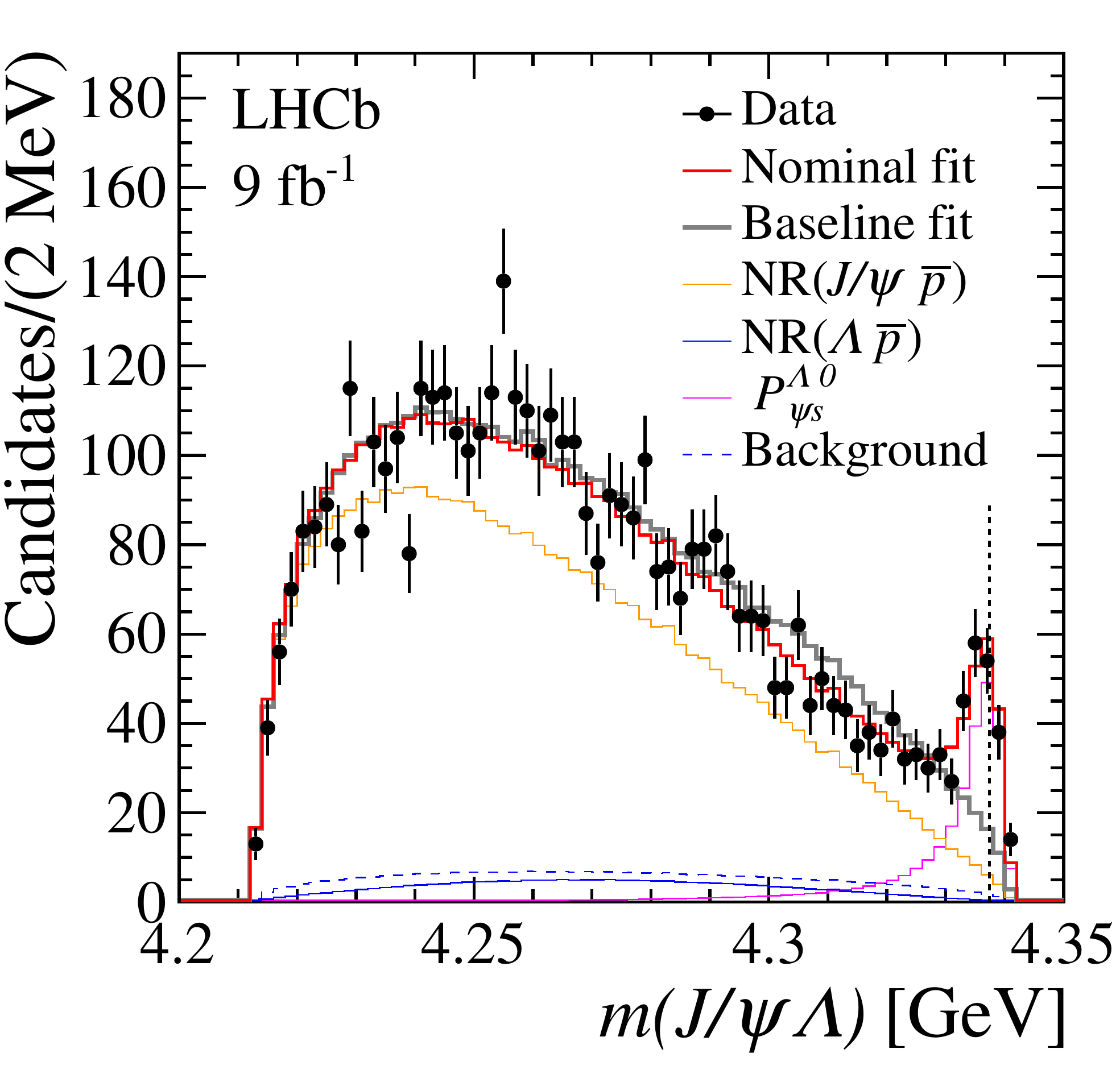}
    \caption{
    (Left) Invariant mass distribution of $\jpsi p$ in the $\Lb\to\jpsi\proton\Km$ decay fitted with three BW functions plus polynomial background, taken from Ref.~\cite{LHCb-PAPER-2020-011}. The vertical lines mark the $\Sigmares^+_c\Dzb$  and the $\Sigmares^+_c \Dstarzb$ mass thresholds respectively.
    (Right) Invariant mass distribution of the $\jpsi\Lz$ system in the $\Xib\to\jpsi\Lambdares\Km$ decay overlaid with fit projections, taken from Ref.~\cite{LHCB-PAPER-2022-031}. 
    }
\label{sec:prod:jpsiLambda}
\end{center}
\end{figure}

\paragraph{Pentaquark states} Following the successful discoveries of tetraquarks with $Q\overline{Q}$ contents, pentaquark states with $Q\overline{Q}$ were predicted, in the form of either meson-baryon hadron molecules or compact five-quark hadrons~\cite{Yang:2011wz,Zhu:2015bba,Li:2012bt,Liu:2018bkx}. Since 2015, several pentaquark candidates are found in LHCb, as summarised in Table~\ref{sec:prod:tab:Pc}.
The first observation of them is made in $\jpsi p$ final states in $\Lb\to\jpsi p\Km$ decays through
an amplitude analysis of Run 1 data~\cite{LHCb-PAPER-2015-029}. Two states were reported, $P_{\psi}^{N}(4380)^+$ and $P_{\psi}^{N}(4450)^+$, and their
evidence is also found in the $\Lb\to\jpsi p \pim$ decays with a similar amplitude study~\cite{LHCb-PAPER-2016-015}.  A model-independent moment analysis of the
$\Lb\to\jpsi p\Km$ decay concludes that
contributions of $\jpsi p$ exotics are essential, since only allowing $\Lambdares^*\to p\Km$ resonances in the decay are not sufficient to describe
data~\cite{LHCb-PAPER-2016-009}.  With full LHCb data, an amplitude analysis of $\Lb\to\jpsi p\Km$ decays becomes computationally very difficult. On the other hand, benefiting from the high statistics, one dimensional $\jpsi p$ mass spectrum is 
investigated to look for narrow pentaquark states~\cite{LHCb-PAPER-2019-014}. In this new analysis, the $P_{\psi}^{N}(4450)^+$ structure is found to consist of two narrow overlapping peaks
$P_{\psi}^{N}(4440)^+$ and $P_{\psi}^{N}(4457)^+$, and a new structure $P_{\psi}^{N}(4312)^+$ is observed. It is noted that the $P_{\psi}^{N}(4312)^+$ and $P_{\psi}^{N}(4457)^+$ 
states are close to the $\Sigmares^+_c\Dzb$ and $\Sigmares^+_c \Dstarzb$ mass
thresholds respectively, as shown in Fig.~\ref{sec:prod:jpsiLambda}
making them ideal candidates of meson-baryon molecules.
Recently, LHCb reported the evidence of a new pentaquark state, $P_{\psi}^{N}(4337)^+$, in the $\jpsi p(\antiproton)$ mass spectrum of
$\Bs\to\jpsi p\antiproton$ decays~\cite{LHCb-PAPER-2021-018}. This possible state is different from those observed in $\Lb$ decays, making beauty meson decays a new place to search for pentaquarks. 

Pentaquark candidates with strangeness are predicted in the $\Xibm\to\jpsi \Lz\Km$ decay~\cite{Wang:2019nvm,Chen:2015sxa}, which is an analogy of the $\Lb\to\jpsi p\Km$ channel by replacing the $d$ quark by the $s$ quark. An amplitude analysis of the $\Xibm\to\jpsi \Lz\Km$ decay is performed using full LHCb data, resulting in the evidence of a
new state, $P_{\psi s}^{\Lz}(4459)^0$, in the $\jpsi \Lz$ mass spectrum, with quark contents, $\cquark\cquarkbar\uquark\dquark\squark$. Its mass is measured to be
$4458.8\pm2.9^{+4.7}_{-1.1}\mevcc$  and width to be $17.3\pm6.5^{+8.0}_{-5.7}\mev$~\cite{LHCb-PAPER-2020-039}. According
to the prediction in Ref.~\cite{Wang:2019nvm}, there are two $\Xic\Dstar$ hadron molecules with masses within a few $\mevcc$ around the observed
structure. If the $P_{\psi s}^{\Lz}(4459)^0$ structure is found to be composed of two nearby states with LHCb upgrade data, it will be a strong proof of the molecular interpretation of such states.
Recently, an amplitude analysis of the $\Bm \to \jpsi \Lz \antiproton$ was performed~\cite{LHCB-PAPER-2022-031}. A narrow structure in the $\jpsi \Lz$ system, denoted as $P_{\psi s}^{\Lz}(4338)^0$, is observed with high significance, which is consistent with pentaquark state with strangeness. 
The invariant-mass distribution of the $\jpsi \Lz$ system is shown in Fig.~\ref{sec:prod:jpsiLambda}.
The mass and width of the state are measured to be $4338.2 \pm 0.7 \pm 0.4 \mev$ and $7.0 \pm 1.2 \pm 1.3 \mev$, respectively, where the first uncertainty is statistical and the second systematic. The spin-parity of the state is determined to be $J^P = \frac{1}{2}^-$.

Observations of pentaquark states with heavy-quark contents have triggered many studies from theorists, with the purpose of understanding their nature. In general, these states  are considered to either be compact tetraquarks, hadronic molecules or simple bumps due to kinematic rescatterings~\cite{Wang:2019nwt,Du:2019pij,Meng:2019ilv,Weng:2019ynv,Xu:2019zme,Ozdem:2017exj,Huang:2015uda,Du:2021bgb,Chen:2021cfl,Hu:2021nvs,Yalikun:2021bfm,Lu:2021irg,Du:2021fmf,Zhu:2021lhd,Nakamura:2021dix,Shi:2021wyt,Phumphan:2021tta,Nakamura:2021qvy,Xiao:2021rgp,Chen:2020kco,Wang:2020eep,Peng:2020gwk,Xu:2020gjl,Giachino:2020rkj,Chen:2019thk,Wang:2019ato,Pimikov:2019dyr,Wang:2019hyc,Yamaguchi:2019seo,Cheng:2019obk,Wang:2019got,Fernandez-Ramirez:2019koa,Xiao:2019mvs,Guo:2019kdc,He:2019ify,Guo:2019fdo,Chen:2019bip,Liu:2018nse,Ferretti:2018ojb,Hiyama:2018ukv,Qin:2018dqp,Richard:2017una,He:2016pfa,Eides:2017xnt,Yamaguchi:2017zmn,Dong:2017gaw,Lin:2017mtz,Chen:2016ryt,Guo:2016bkl,Santopinto:2016pkp,Shen:2016tzq,Shimizu:2016rrd,Lu:2016nnt,Oset:2016lyh,Chen:2016heh,Wang:2015ixb,Yang:2015bmv,Burns:2015dwa,Scoccola:2015nia,Wang:2015epa,Ghosh:2015xqp,Li:2015gta,Anisovich:2015cia,Mikhasenko:2015vca,Lebed:2015tna,Liu:2015fea,He:2015cea,Maiani:2015vwa,Guo:2015umn,Mironov:2015ica,Roca:2015dva,Chen:2015moa,Chen:2015loa,Karliner:2015ina}.
Measurements of their $J^P$ and production properties, and finding their flavour partners will shed light on the problem.
However, it is likely that debates on the nature of exotic hadrons will continue before we have a complete and coherent theory to explain all of them. In the past years, phenomenology models on the hadron spectroscopy evolved quickly and some patterns have been revealed, for example, dynamics close to two-hadron mass thresholds~\cite{Dong:2020hxe}. Besides, lattice QCD simulations have improved in computational performances and application scopes substantially over the years and will play more and more important roles in our understanding of low-energy QCD~\cite{Chiu:2006hd,Guo:2013nja,Bi:2015ifa,Liu:2016kbb,Leskovec:2014gxa,Gayer:2021xzv,Cheung:2020mql}. Hopefully, one day we can  predict low energy QCD phenomena as precise as other parts of the SM.

\begin{table}[!hbt]
\begin{center}
    \caption{Detection decay channels, experimental significance, masses and widths of pentaquark states reported by LHCb.}
    \label{sec:prod:tab:Pc}
    \begin{tabular}{c|cccc}
\hline
        State           & Decays     & Significance [$\sigma$] & Mass [$\mevcc$] & Width [$\mev$]\\
        \hline
        $P_{\psi}^{N}(4312)^+$   & $\jpsi p$  & $7.3\sigma$             & $4311.9\pm0.7^{+6.8}_{-0.6}$   & $\phantom{0}9.8\pm2.7^{+\phantom{0}3.7}_{-\phantom{0}4.5}$\\
        $P_{\psi}^{N}(4440)^+$   & $\jpsi p$  & $5.4\sigma$             & $4440.3\pm1.3^{+4.1}_{-4.7}$   & $20.6\pm4.9^{+\phantom{0}8.7}_{-10.1}$\\
        $P_{\psi}^{N}(4457)^+$   & $\jpsi p$  & $5.4\sigma$             & $4457.3\pm0.6^{+4.1}_{-1.7}$   & $0.53\pm2.0^{+\phantom{0}5.7}_{-\phantom{0}1.9}$\\
        
        $P_{\psi}^{N}(4337)^+$   & $\jpsi p$  & $3.1\sigma$             & $\phantom{0}4337\phantom{0}^{+\phantom{0}7}_{-\phantom{0}4}\phantom{0}\pm2$&  
        $\phantom{0}\phantom{0}29^{+\phantom{0}26}_{-\phantom{0}12}\pm14$\\
\hline
        $P_{\psi s}^{\Lz}(4459)^0$   & $\jpsi \Lz$  & $3.1\sigma$        & $4458.8\pm2.9^{+4.7}_{-1.1}$   & $17.3\pm6.5^{+\phantom{0}8.0}_{-\phantom{0}5.7}$\\
        $P_{\psi s}^{\Lz}(4338)^0$ & $\jpsi \Lz$ & $>15\sigma$& $4338.2 \pm 0.7 \pm 0.4$ & $7.0 \pm 1.2 \pm 1.3 $\\
        \hline
\hline
\end{tabular}
\end{center}
\end{table}

\clearpage

\section{Rare beauty hadron decays}
\label{sec:raredecay}

By rare decays we mainly refer to   flavour-changing-neutral-current (FCNC) processes, which are  highly suppressed in the SM by the Glashow-Iliopoulos-Maiani (GIM)  mechanism~\cite{Glashow:1970gm}. 
Rare decays of hadrons could receive significant contributions from new particles or new interactions beyond the SM. 
Precision measurements  of their properties play a special role in search of  physics beyond the SM. 

The LHCb collaboration has given priority to the study of FCNC $b\to s$ transitions, focusing on theoretically clean observables such as decay rates of purely leptonic $B$-meson decays, angular coefficients  in $b\to s \ellell$ decays, and ratio  of decay rates  between $b\to s \ellell$ processes with different lepton flavours. 
Analyses of $pp$ collision data  collected in the Run 1 and Run 2 periods have already led to some very important and interesting findings, including but not limited to the first observation of the purely leptonic decay $\Bs \to \mumu$, anomalous angular distribution in the decay  $\Bd \to \Kstarz \mumu$,  
and intriguing results from extensive tests of lepton flavour universality in  $B\to K^{(*)} \ellell$ decays.

\subsection{Effective field theory for \texorpdfstring{$b\to s$}{TEXT} transitions}

The effective Hamiltonian describing the quark-level $b\to s$ transitions is given by~\cite{Buras:1994dj, Buras:1998raa, Buchalla:1995vs, Altmannshofer:2008dz,Kruger:2005ep,Descotes-Genon:2011nqe,Lunghi:2006hc}  
\begin{equation}
   H_{\rm eff}(b\to s) =-{\frac{4G_F}{\sqrt{2}}} \Vtb\Vtss \sum_{i=1}^{10} C_i \Ope{i},  
   \label{eq:hamil}
\end{equation}
with $\mathcal{O}_{i}$ denoting the local operators in the SM and  $C_i$ indicating the corresponding Wilson coefficients.   
Of particular interest is the electromagnetic dipole operator corresponding to penguin diagrams mediated by photons,
\begin{equation}
   \Ope{7} = {\dfrac{e}{ 16 \pi^2}} m_b (\squarkbar \sigma_{\mu\nu} P_R b)  F^{\mu\nu}, 
   \label{eq:O7}
\end{equation}
and  semileptonic operators corresponding to loop diagrams mediated by $Z^0$ or $\Wpm$ bosons, 
\begin{equation}
   \Ope{9} ={\dfrac{e^2}{16 \pi^2}} (\squarkbar \gamma_{\mu} P_L b) (\overline{\ell} \gamma^{\mu} \ell), \,\,
   \Ope{10} ={\dfrac{e^2}{16 \pi^2}} (\squarkbar \gamma_{\mu} P_L b) (\overline{\ell} \gamma^{\mu} \gamma_5 \ell),
   \label{eq:O9O10}
\end{equation}
where $P_L=(1-\gamma_5)/2$ and $ P_R=(1+\gamma_5)/2$.
Contribution from  physics  beyond the SM can either alter the values of the Wilson coefficients and/or give rise to new operators that are absent or highly suppressed in the SM, such as the scalar and pseudo-scalar operators
\begin{equation}
   \Ope{S} ={\dfrac{e^2}{16 \pi^2}} (\squarkbar P_L b) (\overline{ell}  \ell), \,\,
   \Ope{P} ={\dfrac{e^2}{16 \pi^2}} (\squarkbar P_L b) (\overline{\ell} \gamma_5 \ell)\;,
   \label{eq:OsOp}
\end{equation}
and the  chirality-flipped operators $\mathcal{O}_{7,8,9,10,S,P}'$,  which are obtained by changing $P_{L(R)}$ to $P_{R(L)}$ in $\mathcal{O}_{7,8,9,10,S,P}$.

The Wilson coefficients $C^{(\prime)}_7$ can be probed in radiative $b$-hadron decays such as $\Bs \to \phi \gamma$ and $\Lb \to \Lz \gamma$, $C^{(\prime)}_{9,10}$ probed in semileptonic decays such as $B^0\to \Kp\ellell$ and $B^+\to \Kstarz  \ellell$, and $C^{(\prime)}_{S,P}$ probed in purely leptonic decays such as $B^0_s \to \ellell$.
Recent LHCb results on $b\to s$ transitions are summarised in the remainder of this section.


\subsection{Purely leptonic \texorpdfstring{$B$ meson decays}{TEXT}}
The decays $\Bds\to \ellell (\ell=\mu,\,e)$ are  among the most interesting probes of new physics. 
They are theoretically clean and are expected to be extremely rare  in the SM due to  helicity suppression in addition to the FCNC loop suppression. 
Their branching fractions in the SM are precisely predicted to be~\cite{Beneke:2019slt} 
\begin{equation}
\begin{aligned}
\BF(\Bs\to \mumu) &= (3.66\pm 0.14 )\times 10^{-9}, \\
\BF(\Bd\to \mumu) &= (1.03\pm 0.05 )\times 10^{-10}, \\
\BF(\Bs\to \epem) &= (8.60\pm 0.36 )\times 10^{-14},  \\
\BF(\Bd\to \epem) &= (2.41\pm 0.13 )\times 10^{-15}.
\end{aligned}
\label{eq:b2llSM}
\end{equation}
Note the  $\Bd \to \ellell$ decays proceed via $b\to d$ transitions, thus are further suppressed with respect to the $\Bs \to \ellell$ decays  
by a factor of $|\Vtd/\Vts|^2\sim \lambda^2$.   
The decay rates of $\Bds\to \ellell$ processes are highly sensitive to  (pseudo-)scalar interactions beyond the SM.

\begin{figure}[!tb]
\begin{center}
\includegraphics[width=0.45\textwidth]{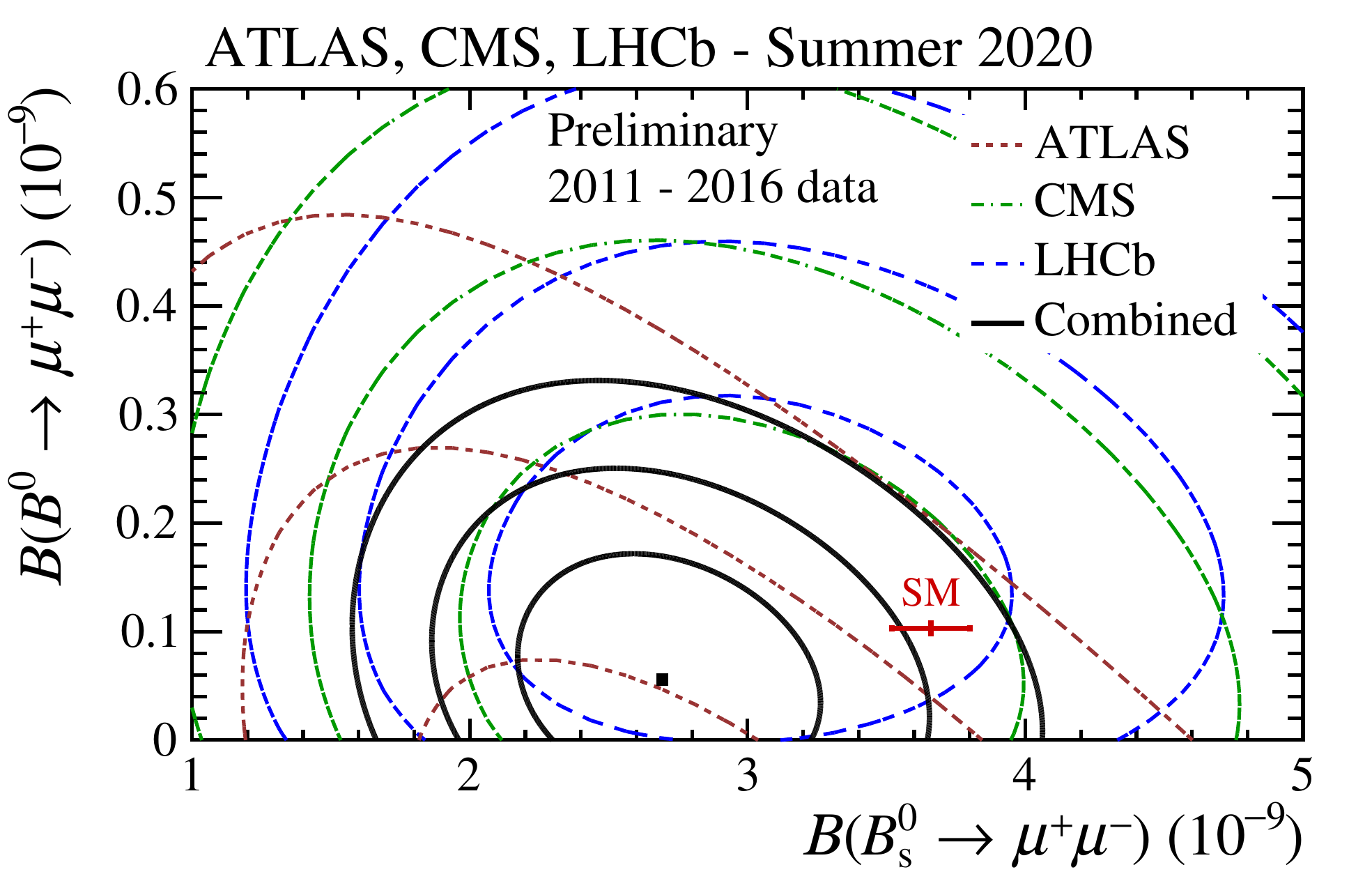}
\includegraphics[width=0.45\textwidth]{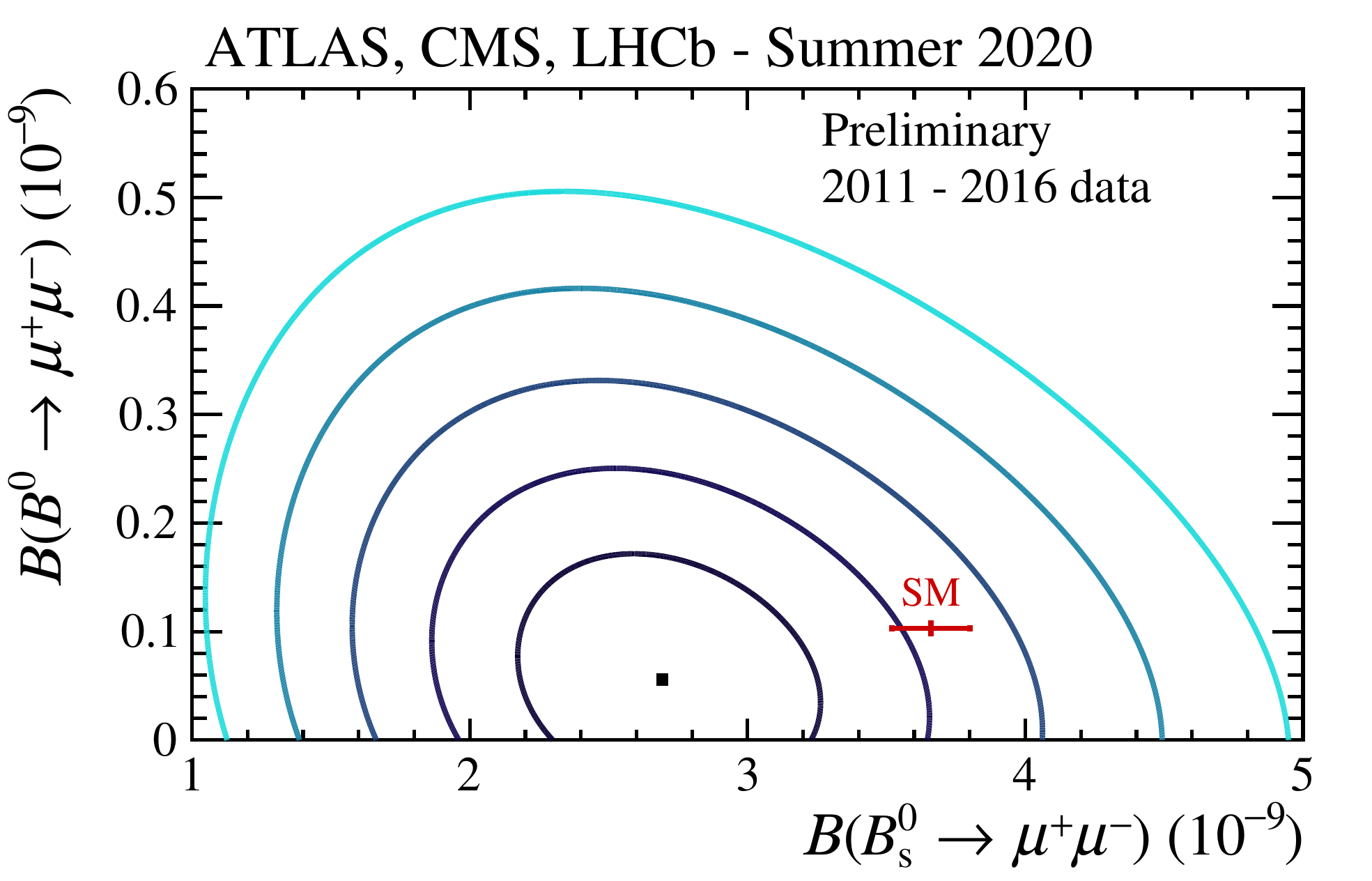}
\caption{
(Left) Likelihood contours in the $\BF(\Bs\to \mumu)$ - $\BF(\Bs\to \mumu)$ plane corresponding to $-2\Delta \ln L =2.3,\, 6.2$ and $11.8$, for ATLAS, CMS and LHCb experiments and  the combination. 
(Right) Likelihood contours for the combination
corresponding to $-2\Delta \ln L =2.3,\, 6.2,\, 11.8,\, 19.3$ and $30.2$. The data sets used were collected from 2011 to 2016.
Figures are extracted from Refs.~\cite{LHCb-CONF-2020-002}. 
}
\label{fig:b2mumu_comb}
\end{center}
\end{figure}

A joint analysis of data from the LHCb and CMS experiments  collected in Run~1 led to the observation of the $\Bs\to \mumu$ decay with a significance exceeding six standard deviations, and determined the branching fraction to be $\BF(\Bs\to \mumu)=(2.8^{+0.7}_{-0.6})\times 10^{-9}$~\cite{LHCb-PAPER-2014-049}.  
This result was later updated by LHCb~\cite{LHCB-PAPER-2017-001} and CMS~\cite{CMS:2019bbr}  by adding the 2016 data. 
The ATLAS collaboration  reported an evidence  
for the decay $\Bs\to \mumu$ with a significance of $4.6\sigma$  using the  data collected between 2011 and 2016~\cite{ATLAS:2018cur}. However, no significant signal for the decay $\Bd\to \mumu$ has been found by any experiment yet. 
A combination of the results from ATLAS, CMS and LHCb gives the branching fraction $\BF(\Bs\to \mumu)=(2.69^{+0.37}_{-0.35})\times 10^{-9}$, and sets an upper limit of $\BF(\Bd\to\mumu)<1.9\times 10^{-10}$ at 95\% confidence level~\cite{LHCb-CONF-2020-002}. 
Figure~\ref{fig:b2mumu_comb} shows the constraints on $\BF(\Bs\to \mumu)$ and $\BF(\Bs\to \mumu)$ from  the three experiments and the combined results, which  are compatible with the SM predictions~\cite{Beneke:2019slt}  within 2.1 $\sigma$. The difference is mainly driven by the ATLAS results,  which have an optimal solution  outside the physical region and are slightly in tension with the SM predictions. 

\begin{figure}
\begin{center}
\includegraphics[width=0.45\textwidth]{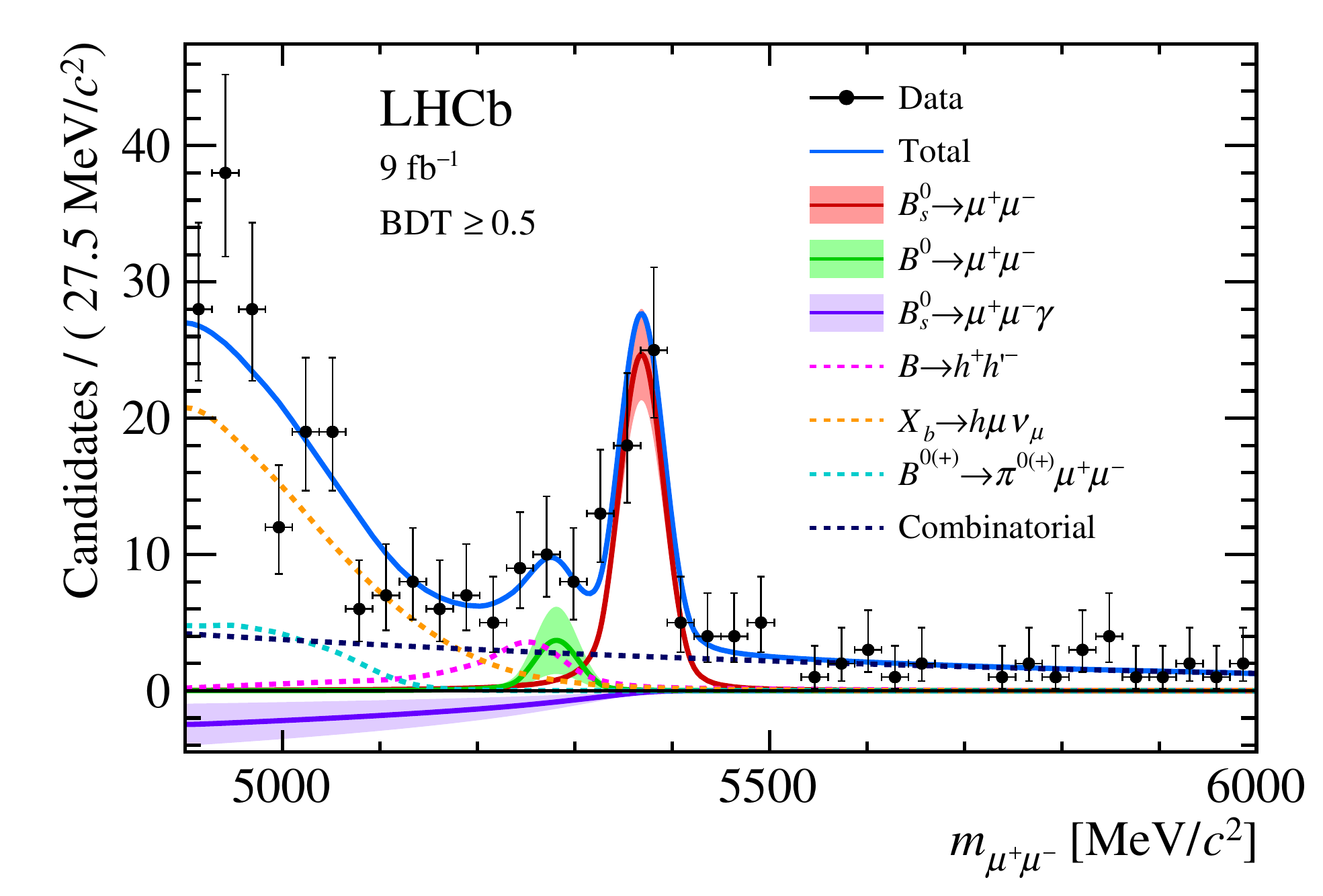}
\includegraphics[width=0.45\textwidth]{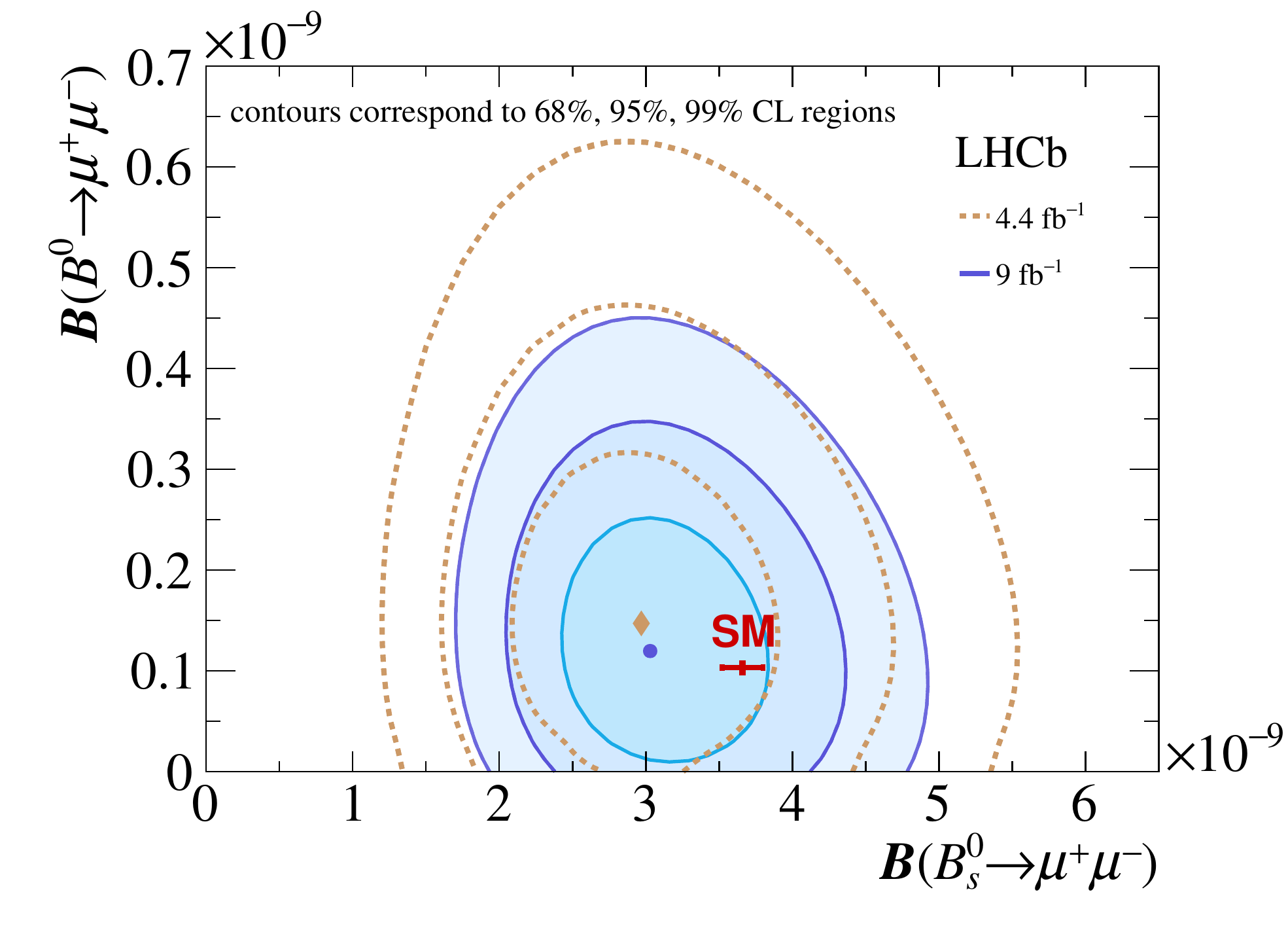}
\caption{ (Left) Invariant mass distribution of the selected $\Bds\to\mumu$ candidates with output of the used multivariate classifier above 0.5, superimposed with the fit result. (Right) Confidence intervals in the plane of the branching fractions of $\Bs\to \mumu$ and $\Bd\to\mumu$. 
The data sets used for the blue solid  (brown dotted)  contours were collected from 2011 to 2018 (2016). Figures are extracted from Refs.~\cite{LHCb-PAPER-2021-007,LHCb-PAPER-2021-008}. }
\label{fig:b2mumu2021}
\end{center}
\end{figure}

Very recently, LHCb reported updated results on the $\Bs\to \mumu$ and $\Bd\to \mumu$ decays  using all data collected in Run 1 and Run 2. 
The results are $\BF(\Bs\to \mumu)=(3.09^{+0.46+0.15}_{-0.43-0.11})\times 10^{-9}$, 
$\BF(\Bd\to\mumu)<2.6\times 10^{-10}$ at 95\% confidence level~\cite{LHCb-PAPER-2021-007,LHCb-PAPER-2021-008}, which are in good agreement with the SM expectations, as shown in Fig.~\ref{fig:b2mumu2021}. 

In addition to the decay rate, other interesting quantities, such as effective lifetime and CP asymmetry, can also be measured to search for possible non-SM contribution to the decay $\Bs\to \mumu$~\cite{DeBruyn:2012wk}.
Pioneering studies  of its  effective lifetime   using all   data collected in Run 1 and Run 2 have been performed at LHCb, leading to the result of  $\tau^{\rm eff}(\Bs\to\mumu) =2.07\pm 0.29 \pm 0.03$ \ps~\cite{LHCb-PAPER-2021-007,LHCb-PAPER-2021-008}. 
As a long-term goal, the effective lifetime and time-dependent CP violation of the decay $\Bs\to\mumu$ will be fully exploited in LHCb upgrade II, which aims to accumulate a $pp$ collision data sample of 300\invfb~\cite{LHCb-PII-Physics}. 

The decays  $\Bds\to\epem$ are even rarer than $\Bds\to \mumu$ and can provide powerful tests of lepton flavour universality. 
To date, the searches performed by the CDF and LHCb experiments have found no evidence for either $\Bs\to\epem$ or $\Bz\to\epem$. 
The most stringent upper limits  on their branching fractions are 
$\BF(\Bs\to\epem)<11.2\times 10^{-9}$ and 
$\BF(\Bd\to\epem)<3.0\times 10^{-9}$ at 95\% confidence level
set by LHCb~\cite{LHCb-PAPER-2020-001}.

In addition to decays to dileptons, the  LHCb experiment  has also searched for decays of neutral $B$ mesons to four leptons. 
A recent search using  the full Run~1 and Run~2 data sample found no hint of such decays, and  upper limits at 90\% confidence level for the nonresonant decays are determined to be $\BF(\Bs\to\mumu\mumu)<8.6\times10^{-10}$ and  $\BF(\Bd\to\mumu\mumu)<1.8\times10^{-10}$~\cite{LHCb-PAPER-2021-039}.
More stringent limits are set for decays involving the $\jpsi$ resonance or a promptly decaying intermediate scalar particle with a mass of $1\gevcc$.


\subsection{Semileptonic \texorpdfstring{$b\to s \ellell$}{TEXT} decays}
Semileptonic $\bquark\to\squark\ellell$ decays provide valuable insight into possible non-SM contributions that affect  the Wilson coefficients $C_9$ and $C_{10}$ of the electromagnetic operators. 
The presence of hadrons in the final state makes the search for new physics in semileptonic decays more complicated than that in purely leptonic decays. 
The challenges in hadronic form-factor calculations lead to significant uncertainties in the SM predictions of their decay rates.  
Fortunately, a number of relatively clean observables that are less affected by the form factors than the total decay rates have been identified, including  some special observables in angular distributions and observables for lepton universality test. A comprehensive study of these observables in a series of  $b\to s \ellell$ processes has been pursued by the LHCb collaboration and the results are summarized below.


\subsubsection{Differential decay rates with respect to $q^2$}
\label{dBdq2}

The differential branching fraction $d{\mathcal B}/dq^2$ can be measured in intervals of $q^2$, the invariant mass squared of the lepton pair, and compared with   SM predictions. 
Calculations of form factors are needed for making the SM predictions. 
Such calculations are challenging, and require different treatments depending on the $q^2$ regions. 
Light-cone sum rule calculations~\cite{Bharucha:2015bzk, Khodjamirian:2010vf,Gao:2019lta}  and lattice QCD calculations~\cite{Horgan:2013hoa, Horgan:2015vla} are often  used to determine the form factors in low- and high-$q^2$ regions, respectively. 
 
\begin{figure}[!tb]
\begin{center}
\includegraphics[width=0.45\textwidth]{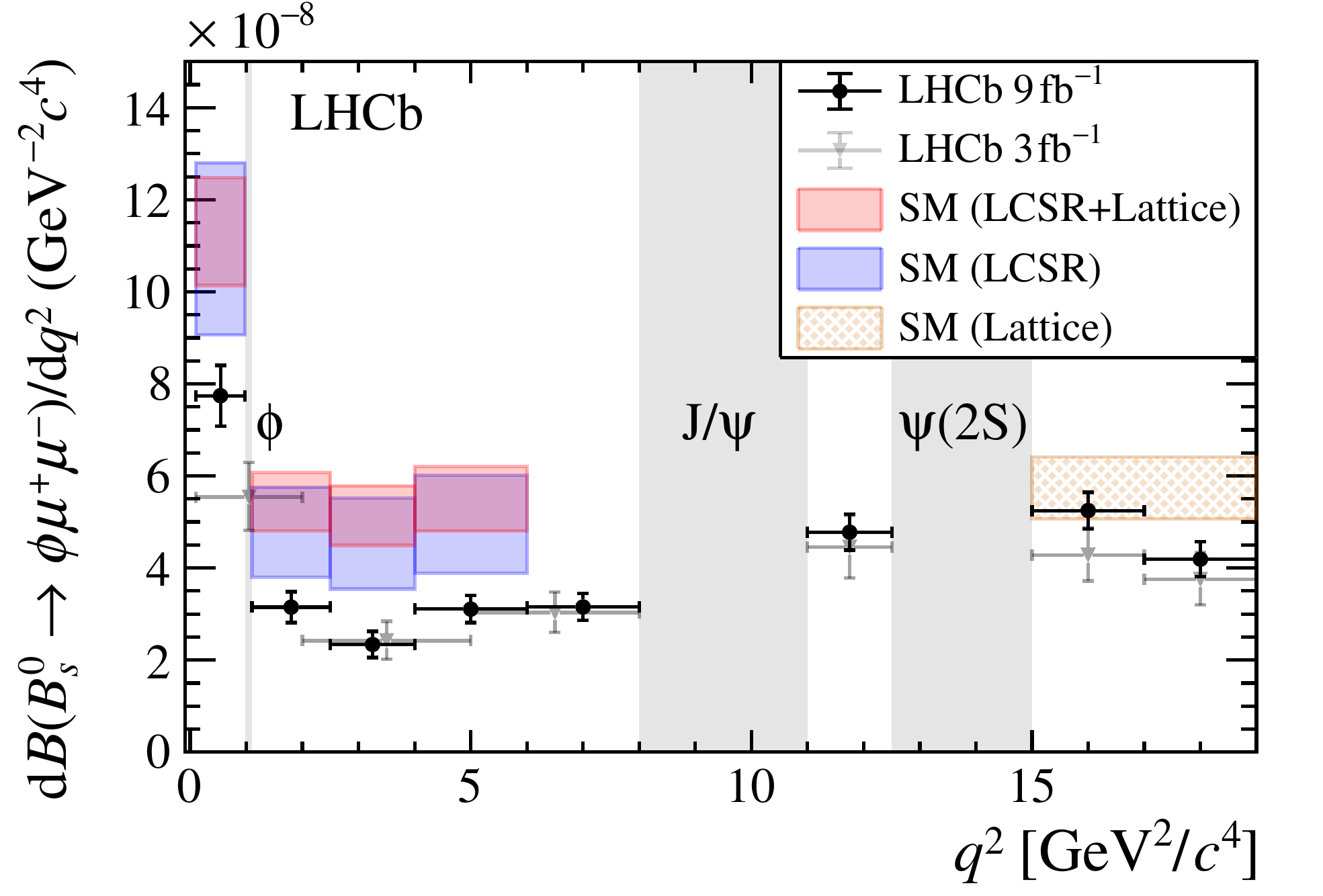}
\includegraphics[width=0.45\textwidth]{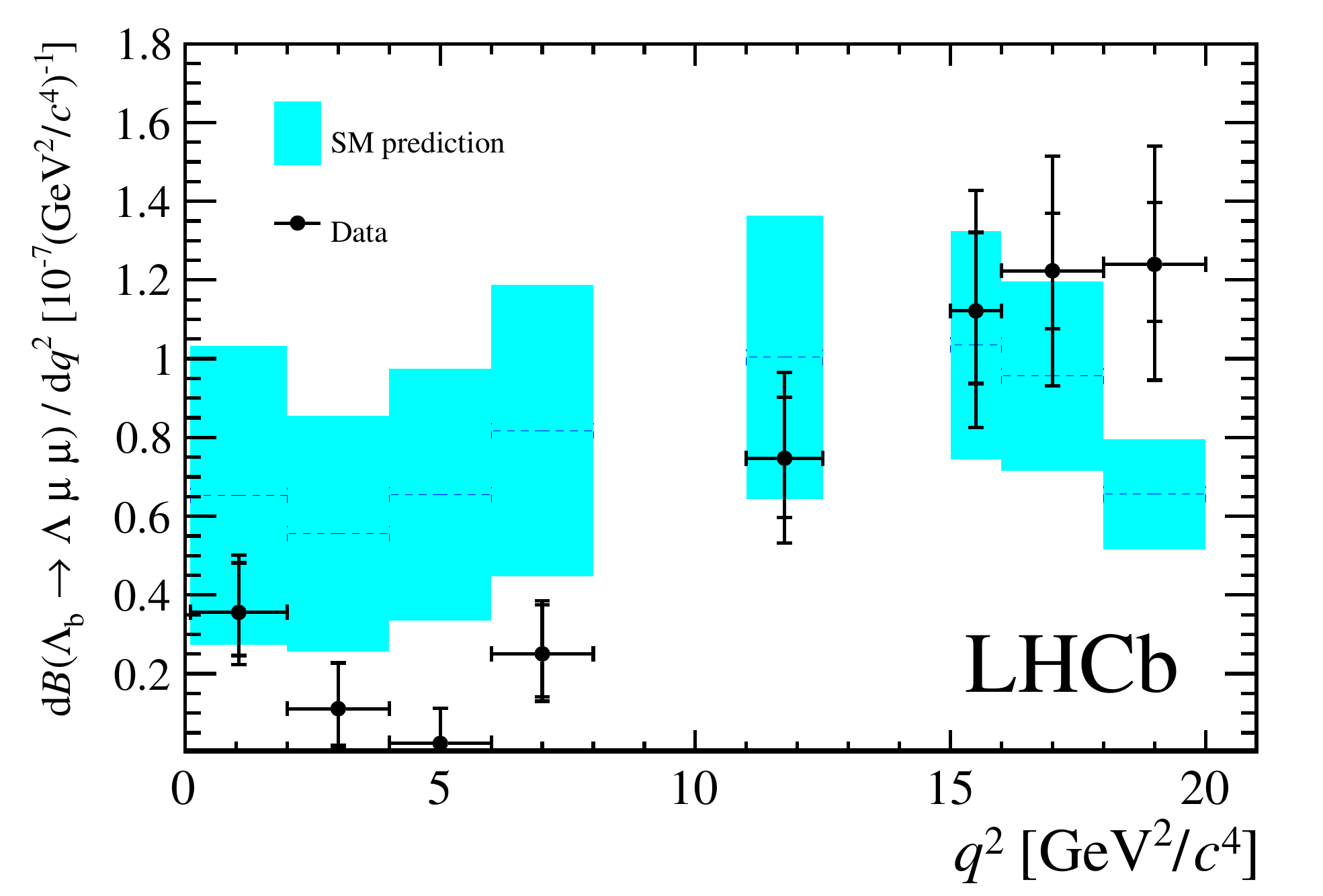}
\includegraphics[width=0.45\textwidth]{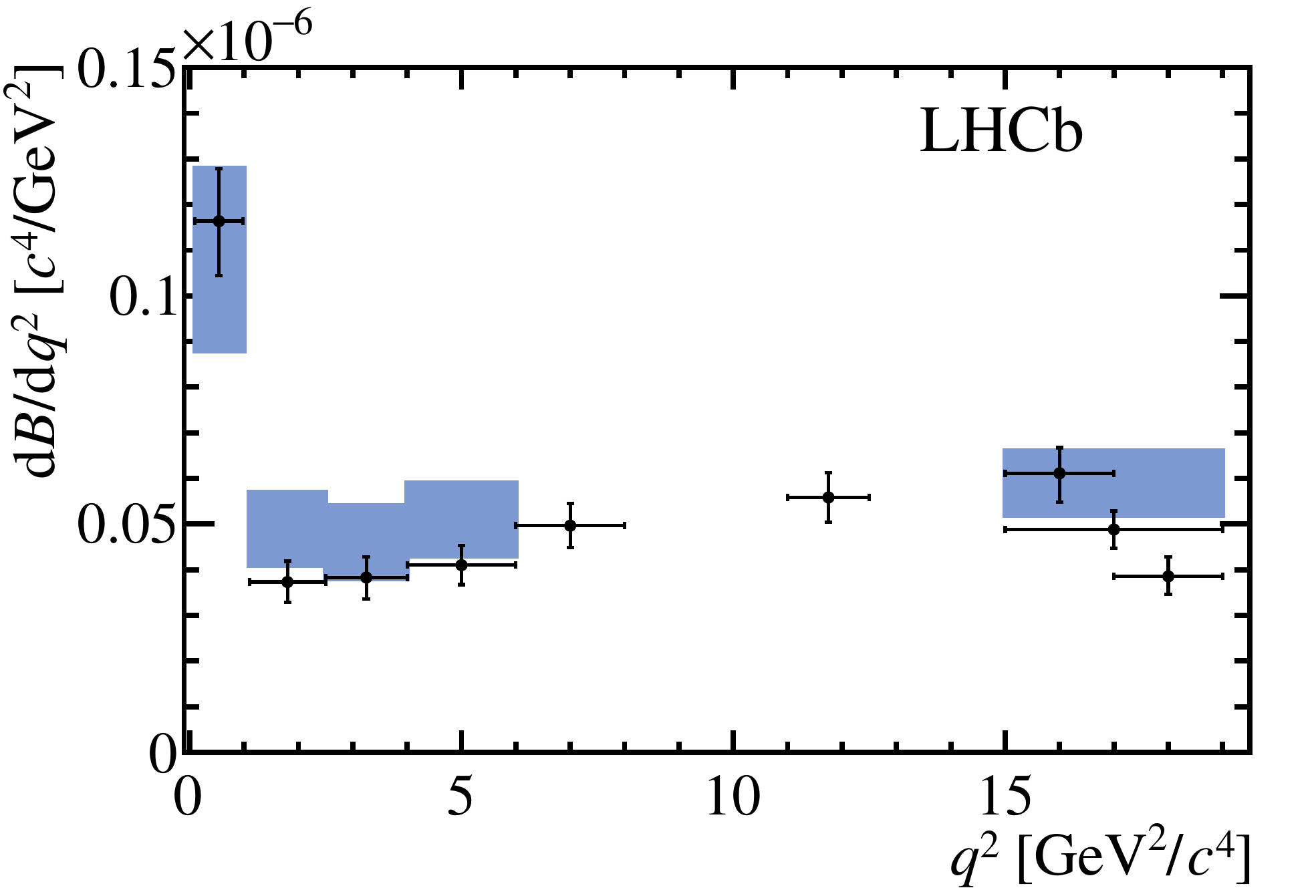}
\includegraphics[width=0.45\textwidth]{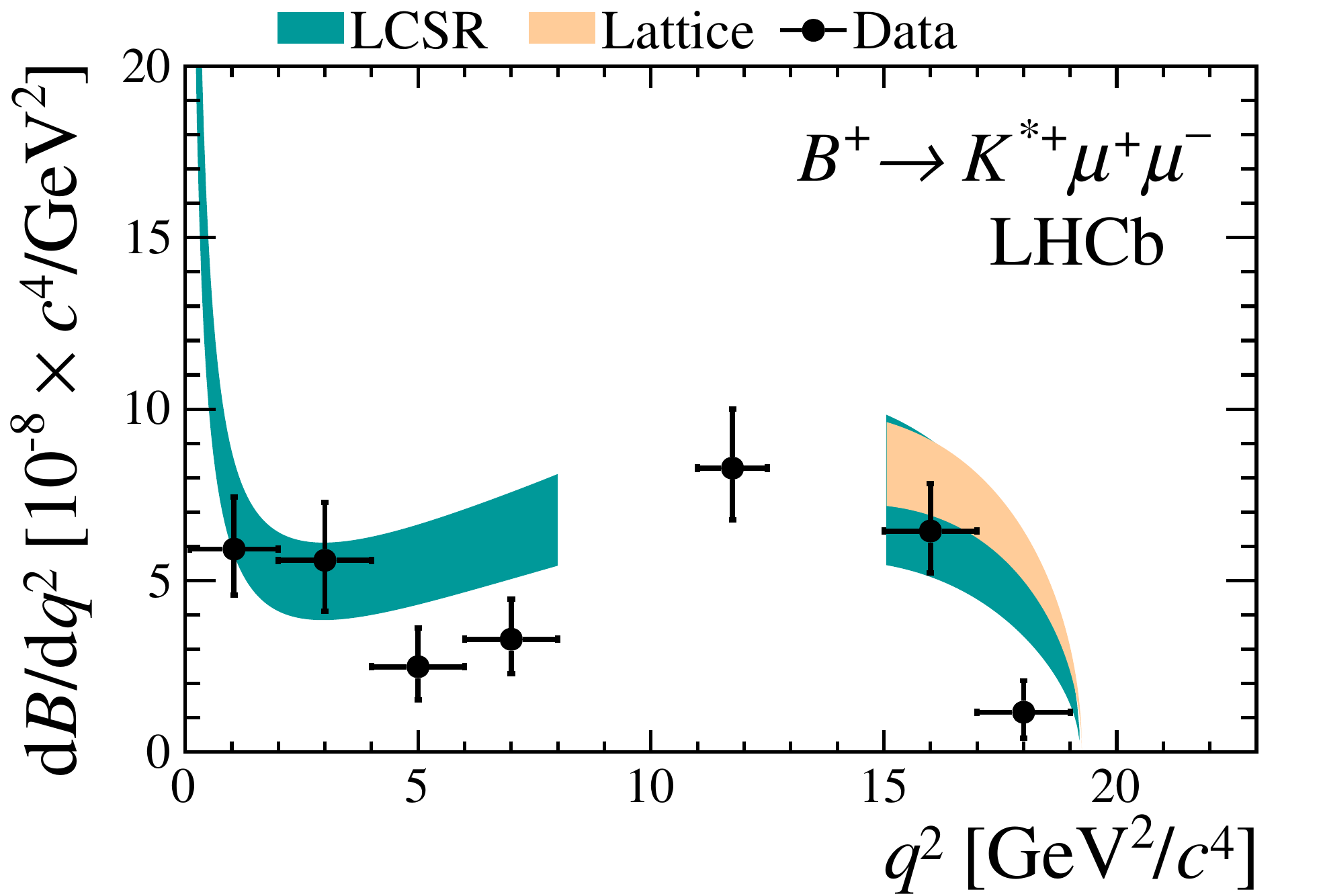}
\includegraphics[width=0.45\textwidth]{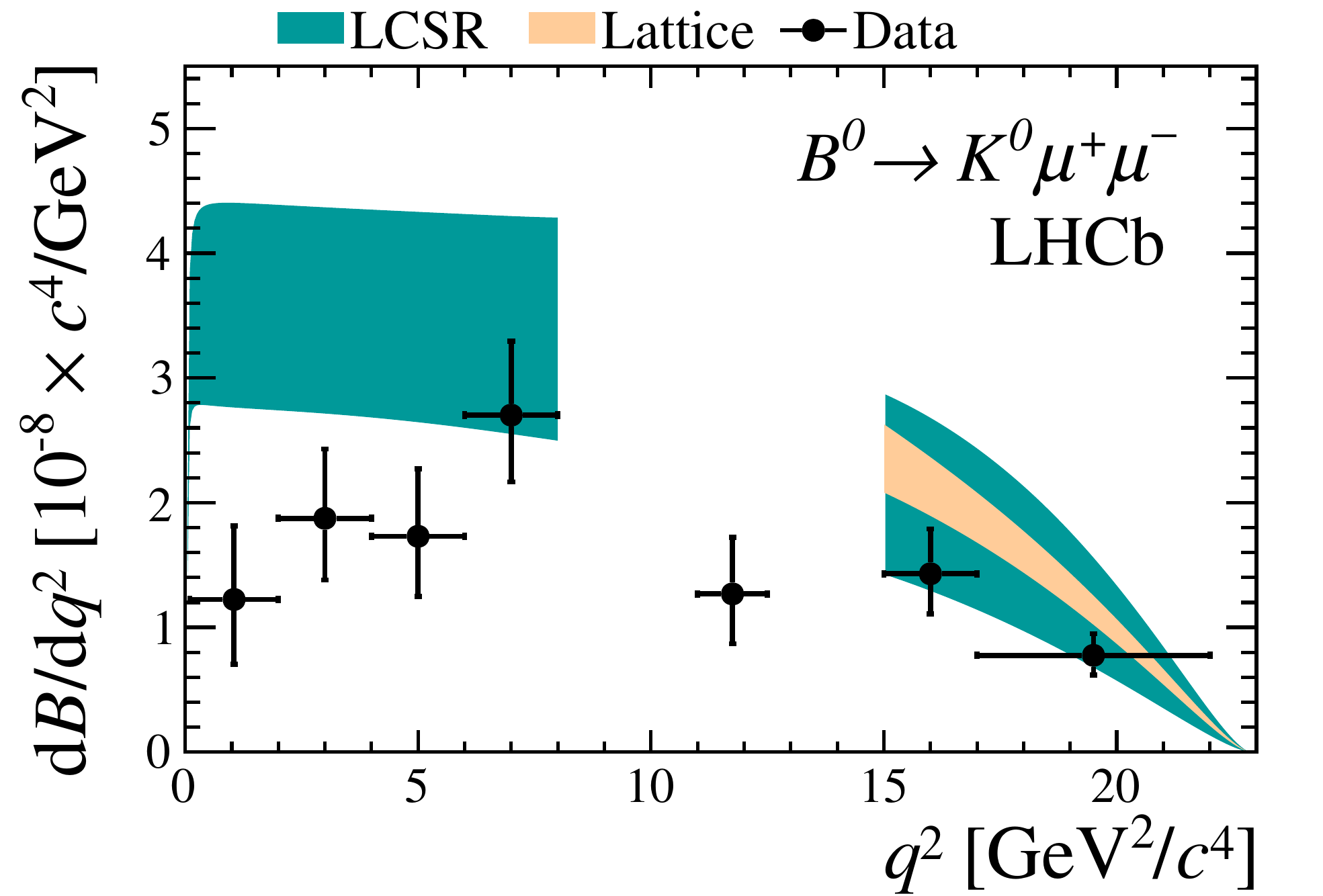}
\includegraphics[width=0.45\textwidth]{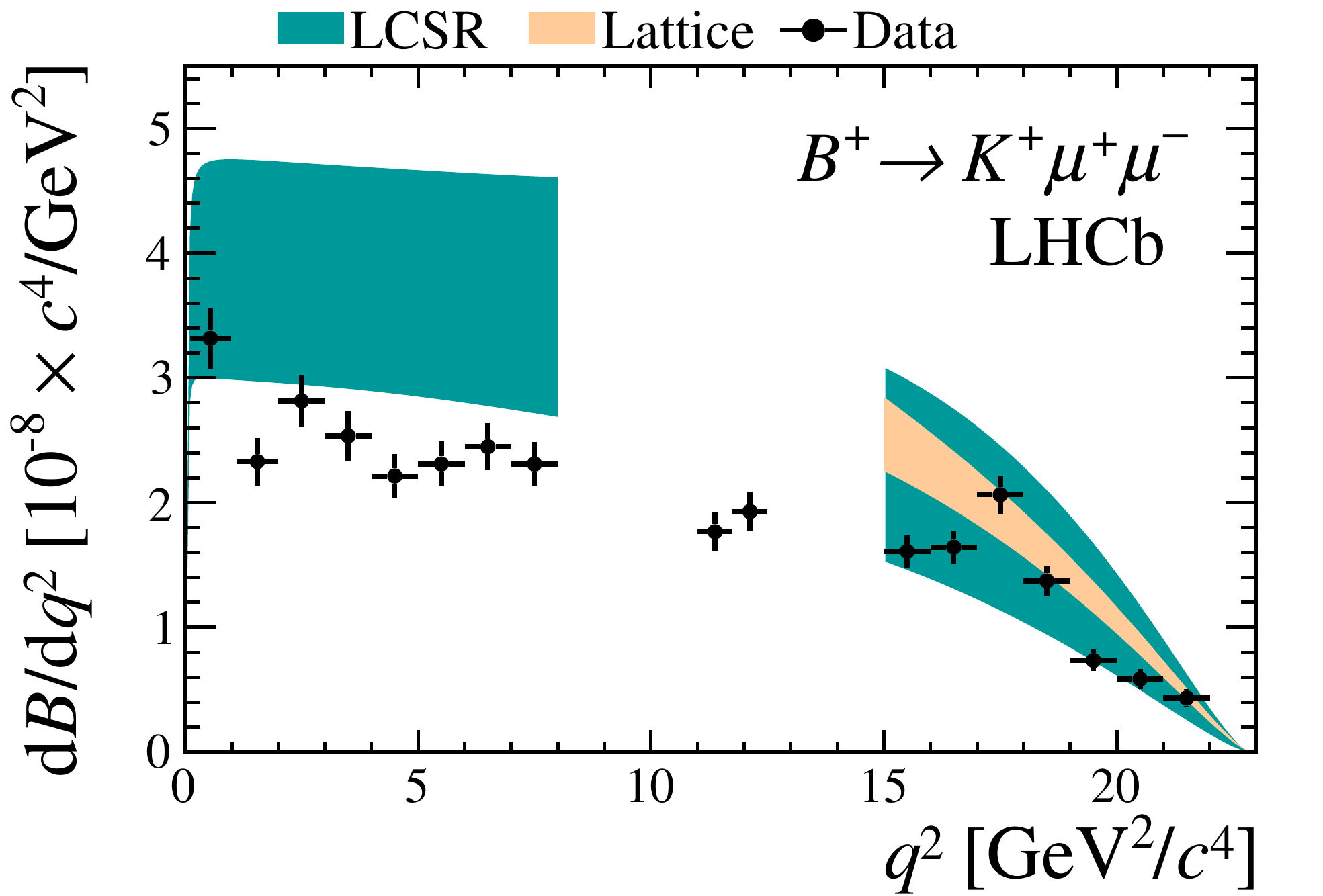}
\caption{ 
Top left: measured $d\BF(\Bs\to \phi \mumu)/dq^2$~\cite{LHCb-PAPER-2021-014} overlaid with the SM predictions based on light-cone sum rule calculations~\cite{Altmannshofer:2014rta,Altmannshofer:2015sma,Bharucha:2015bzk}  at low $q^2$ and lattice QCD calculations~\cite{Horgan:2013hoa, Horgan:2015vla} at high $q^2$;
top right: measured $d\BF(\Lb \to \Lz \mumu)/dq^2$~\cite{LHCb-Paper-2015-009} compared with the SM predictions using form factors from lattice QCD calculations~\cite{Detmold:2012vy};  
middle left: measured $d\BF(\Bd \to \Kstarz \mumu)/dq^2$~\cite{LHCb-Paper-2016-012} compared with the SM predictions using form factors from line-cone sum rule and lattice QCD calculations~\cite{Bharucha:2015bzk,Horgan:2013hoa};
middle right, bottom left and bottom right: 
measured $d\BF(\Bp \to \Kstarp \mumu)/dq^2$,
$d\BF(\Bd \to \KS \mumu)/dq^2$ and
$d\BF(\Bp \to \Kp \mumu)/dq^2$~\cite{LHCb-Paper-2014-006} 
compared with the SM predictions using form factors from light-cone sum rule and lattice QCD calculations~\cite{Bobeth:2011gi, Bobeth:2011nj}.
}
\label{fig:b2smumu_br}
\end{center}
\end{figure}

After measuring the branching fraction of the decay $\Bs \to \phi \mumu$ to be about $3\sigma$ below the SM expectation value~\cite{LHCb-PAPER-2015-023}, the LHCb experiment further studied its differential branching fraction  as a function of $q^2$. The top left plot in Fig.~\ref{fig:b2smumu_br} shows the latest results of  $d\BF(\Bs\to \phi \mumu)/dq^2$ obtained using Run 1 data~\cite{LHCb-PAPER-2021-014}, where the $\jpsi$ and $\psitwos$ regions are excluded. 
A puzzle appears in the  range $1<q^2<6\gevgevcccc$.  
 The  branching fraction integrated over this range is measured to be $(2.88\pm 0.22 ) \times 10^{-8}$.  Currently, the most precise SM prediction for this range is   $(5.37\pm 0.66 ) \times 10^{-8}$, obtained from a combination of light-cone sum rule and lattice QCD calculations. A discrepancy of  $3.6\,\sigma$ is observed. Similar patterns are also seen in the LHCb measurements of differential branching fractions in the decays $\Lb\to \Lz \mumu$~\cite{LHCb-Paper-2015-009},  
  $\Bd \to \Kstarz \mumu$~\cite{LHCb-Paper-2016-012},
$\Bp\to\Kp\mumu$,
$\Bz \to \KS \mumu$ and
$\Bp\to\Kstarp \mumu$~\cite{LHCb-Paper-2014-006}, as shown in Fig.~\ref{fig:b2smumu_br}.

\subsubsection{Angular distributions}

Angular distributions in $b\to s \ellell$ decays contain rich information about  interference  between the SM and non-SM contributions that may not  be accessible via  decay rates integrated over angular variables. A set of  $q^2$-dependent angular coefficients   can be extracted from the angular distributions and used as probes for new physics, which are complementary to  branching fractions and $d\BF/dq^2$. 
Based on these  coefficients,  we can define some theoretically clean observables with reduced dependency on the form factors.

Of particular interest is the angular distribution of the $\mbox{\Bd\to \Kstarz(\to \Kp\pim) \mumu}$ decay, which  has been extensively studied by \babar~\cite{BaBar:2006tnv}, \belle~\cite{Belle:2016fev}, ATLAS~\cite{ATLAS:2018gqc}, CMS~\cite{CMS:2017rzx} and LHCb~\cite{LHCb-PAPER-2011-020,LHCb-PAPER-2013-019, LHCb-PAPER-2015-051, LHCb-PAPER-2020-002}. 
Following the definitions in Ref.~\cite{Altmannshofer:2008dz},
the CP-averaged angular distribution of the decay $\Bd\to \Kstarz \mumu$ with $\Kstarz\to \Kp\pim$ is given by
\begin{equation}
\begin{aligned}
    \frac{1}{d(\Gamma + \bar{\Gamma})/dq^2}
    \frac{d^4(\Gamma +\bar{\Gamma})}{dq^2 d\vec{\Omega}} 
    =  \frac{9}{32\pi}&  \left[ \tfrac{3}{4}(1-F_{\rm L})\sin^2\theta_K 
    +F_{\rm L}\cos^2\theta_K\right. \\
    & +  \tfrac{1}{4} (1-F_{\rm L})\sin^2\theta_K \cos2\theta_\ell 
    -F_{\rm L}\cos^2\theta_K \cos 2\theta_\ell \\
    &+ S_4 \sin 2\theta_K \sin 2\theta_\ell \cos\phi + S_5 \sin 2\theta_K \sin \theta_\ell \cos \phi \\
    &+ \tfrac{4}{3} A_{\rm FB} \sin^2\theta_K \cos \theta_\ell + S_7 \sin 2\theta_K \sin\theta_\ell \sin \phi \\
    &+ \left. S_8 \sin 2\theta_K \sin 2 \theta_\ell \sin \phi + S_9 \sin^2\theta_K \sin^2\theta_\ell \sin 2\phi \right],
\end{aligned}
\label{eq:kstarmumu}
\end{equation}
where $\vec{\Omega}=(\cos\theta_K, \cos \theta_\ell, \phi)$, $\theta_K$ is the angle between the directions of the $\Kp$ ($\Km$) and  $\Bz$ ($\Bzb$) in the rest frame of the $\Kstarz$ ($\Kstarzb$) system, $\theta_\ell$ is the angle between the direction of the  $\mup$ ($\mun$) and the opposite direction of the $\Bz$ ($\Bzb$) in the rest frame of the $\mumu$ system, $\phi$ is the angle between the plane defined by the muon pair and the plane defined by the kaon and pion in the $\Bz$ ($\Bzb$) rest frame.
Eight observables  can be extracted, including  the fraction of the longitudinal polarisation of the $\Kstarz$ meson ($F_{\rm L}$),  
the forward-backward asymmetry of the $\mumu$ system ($A_{\rm FB}$),  and six other angular coefficients ($S_i$, $i=3,4,5,7,8,9$).  
Using the $S_i$ coefficients,
new  observables less sensitive to form factor uncertainties are defined, such as  $P^{\prime}_i =S_i/\sqrt{F_{\rm L}(1-F_{\rm L})}$ $(i=3,4,5)$~\cite{Descotes-Genon:2014uoa}. 

\begin{figure}[!tb]
\begin{center}
\includegraphics[width=0.45\textwidth]{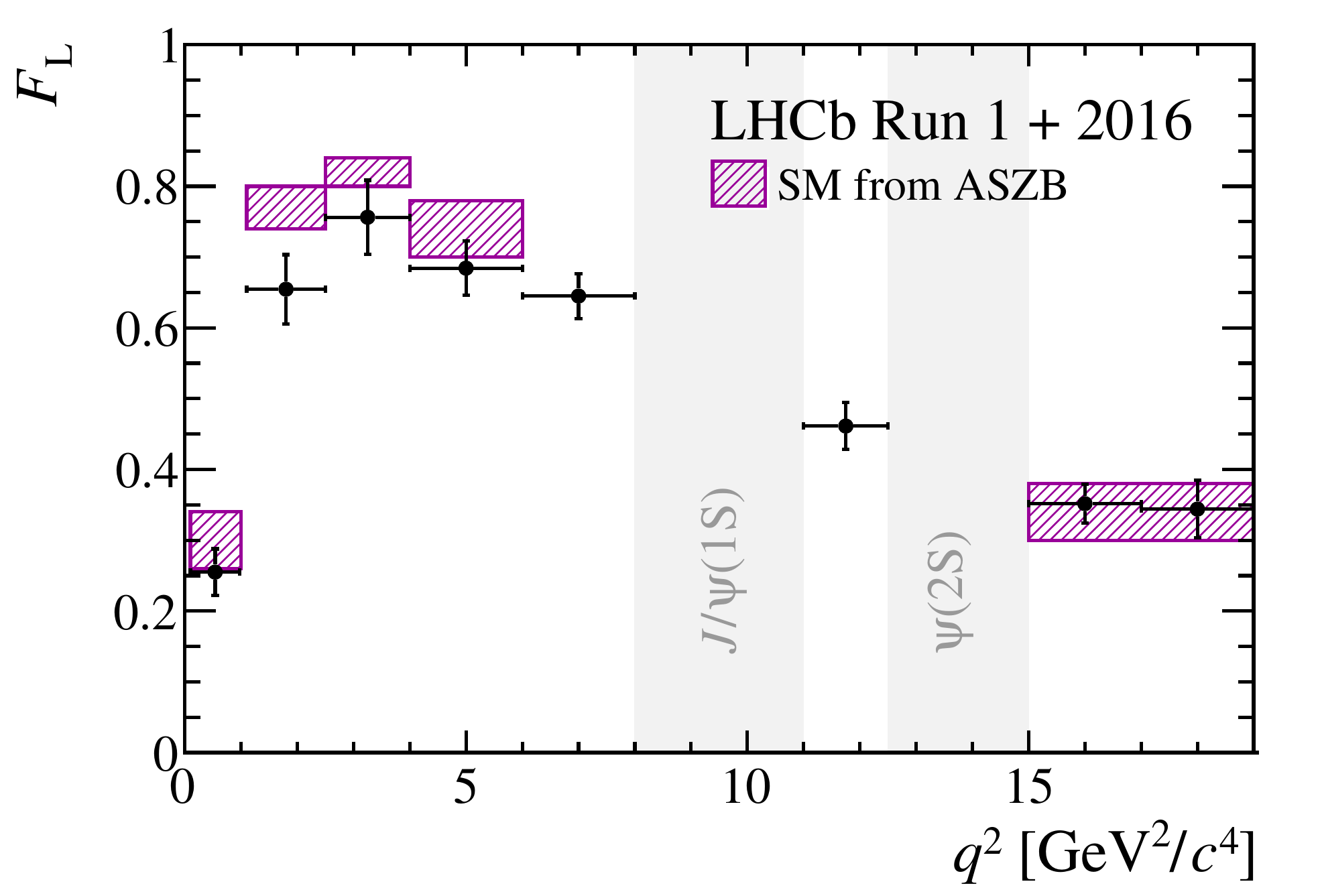}
\includegraphics[width=0.45\textwidth]{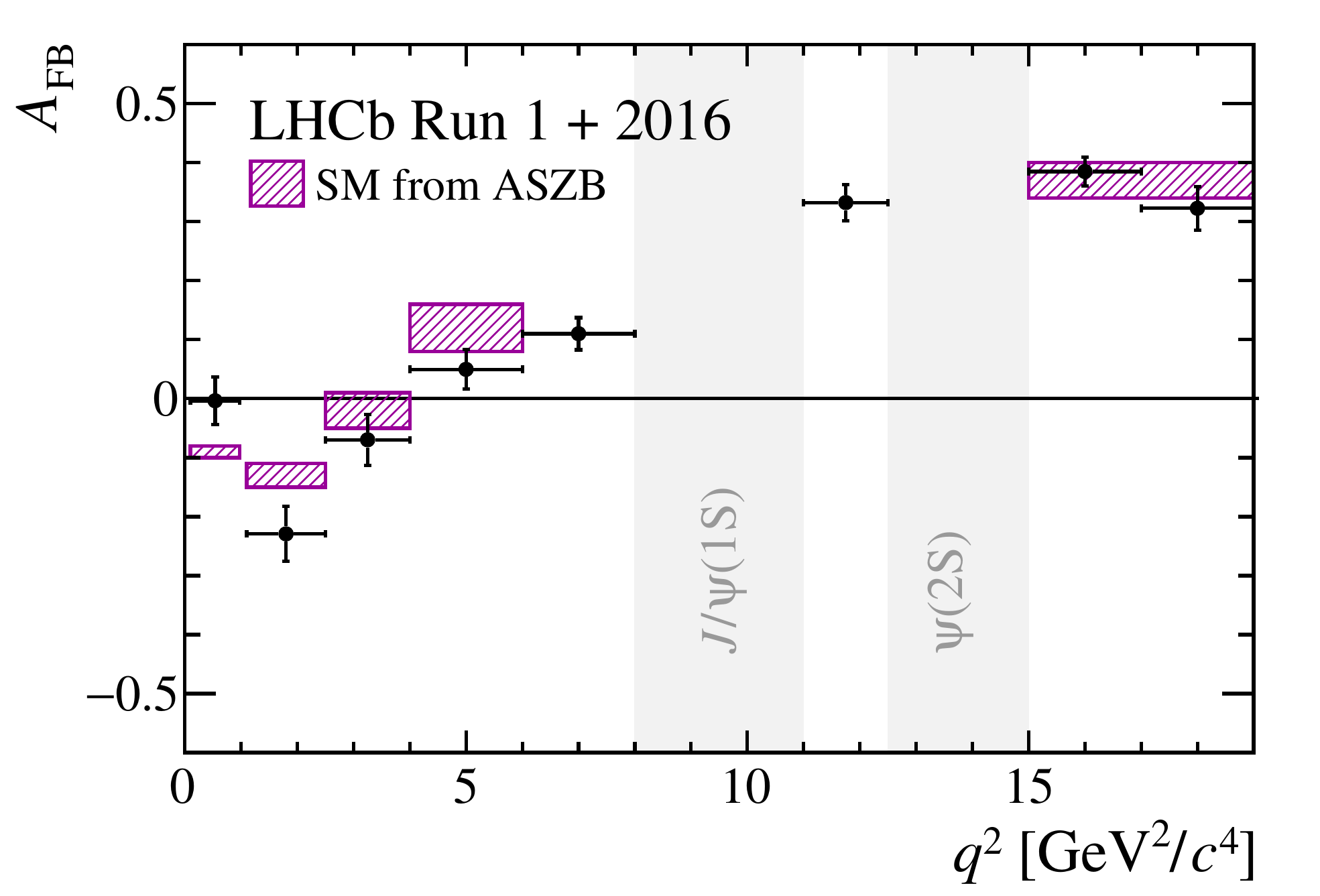}
\includegraphics[width=0.45\textwidth]{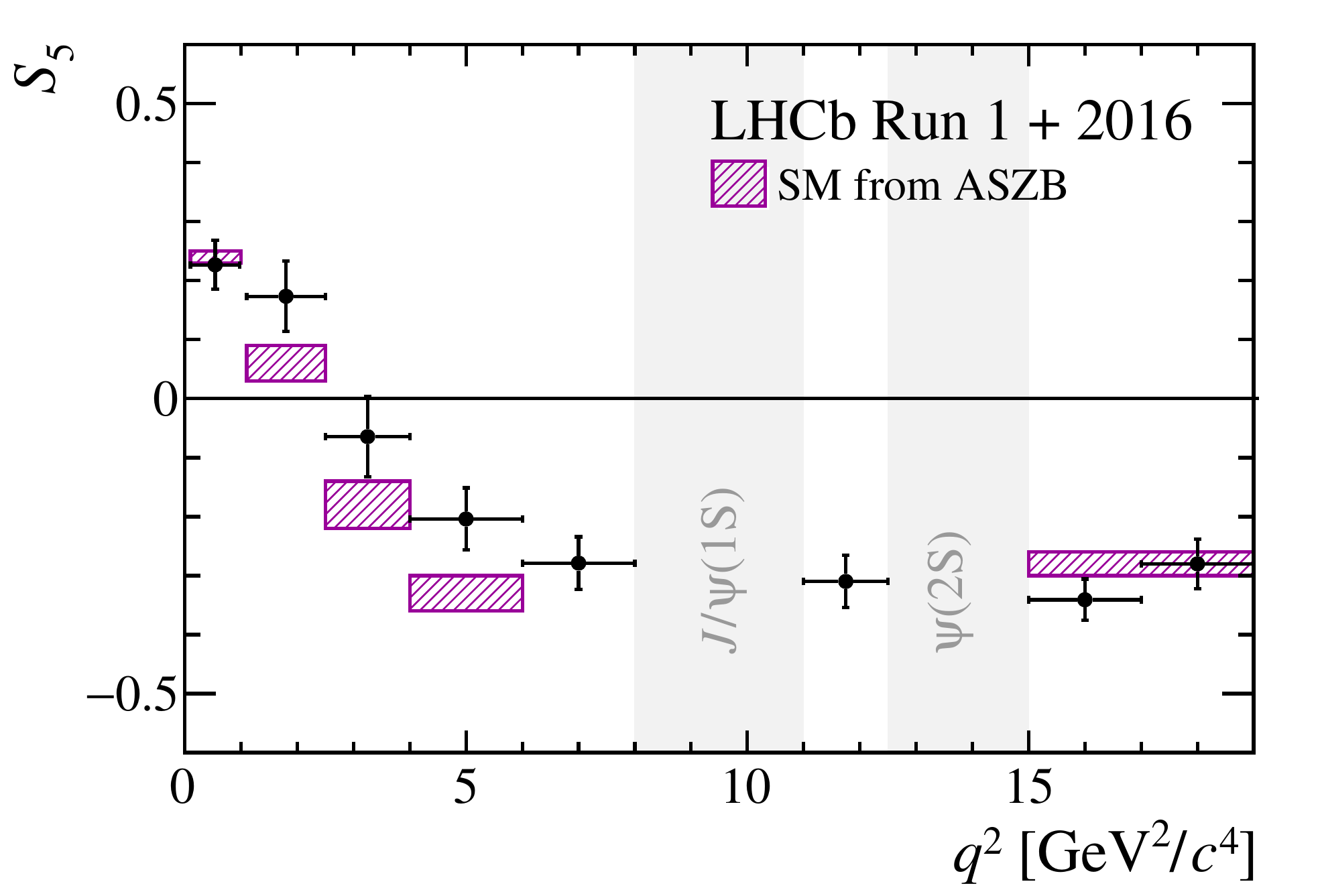}
\includegraphics[width=0.45\textwidth]{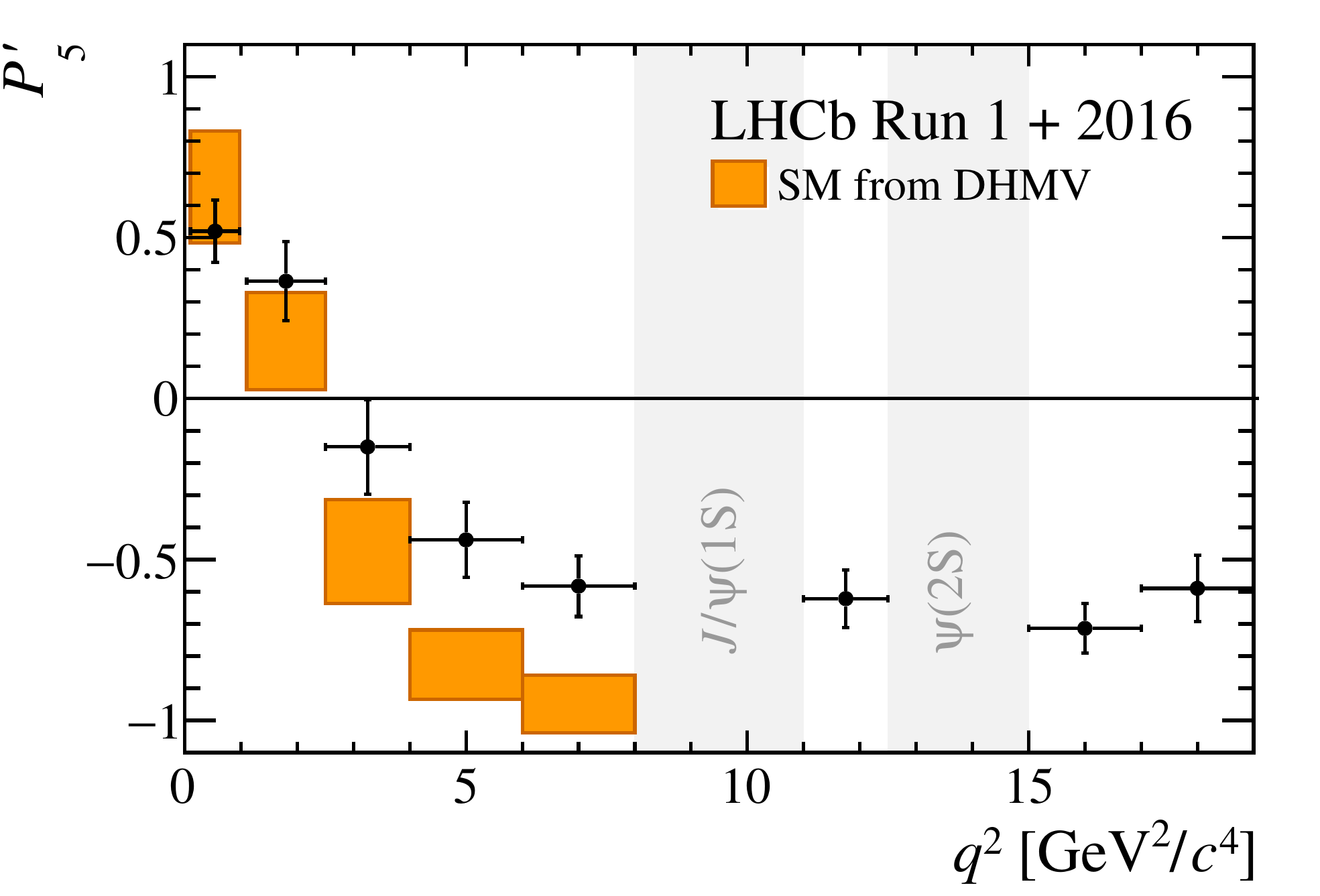}
\caption{ Results of the observables $F_{\rm L}$, $A_{\rm FB}$, $S_5$ and $P^{\prime}_5$ in bins of $q^2$, compared with the SM predictions. 
Figures are extracted from Ref.~\cite{LHCb-PAPER-2020-002}. }
\label{fig:p5pr_puzzle}
\end{center}
\end{figure}

The latest LHCb results on  angular analysis of the $\Bd\to \Kstarz \mumu$ decay are obtained  using data collected in 2011, 2012 and 2016~\cite{ LHCb-PAPER-2020-002}. The majority of the angular  observables
are  consistent with the SM predictions~\cite{
Altmannshofer:2014rta} based on form factors obtained from a combination of  light-cone sum rule calculations~\cite{Bharucha:2015bzk} for low-$q^2$ regions and lattice QCD calculations~\cite{Horgan:2013hoa, Horgan:2015vla}  for high-$q^2$ regions. 
A clear exception is seen with the robust observable  $P^{\prime}_5$ defined using $S_5$, as shown in  Fig.~\ref{fig:p5pr_puzzle}. 
The measured values of $P^{\prime}_5$ in the intervals $4.0<q^2<6.0\gevgevcccc$ and $6.0<q^2<8.0\gevgevcccc$ are found to be higher than the SM predictions~\cite{Descotes-Genon:2014uoa,Khodjamirian:2010vf} by $2.5\,\sigma$ and $2.9\,\sigma$, respectively.  
These results confirm the  discrepancy in $P^{\prime}_5$ observed in an earlier LHCb analysis with Run 1 data~\cite{LHCb-PAPER-2015-051}.  According to model-independent fits using the FLAVIO software package~\cite{Straub:2018kue},  the overall tension with the SM  is increased  from $3.0\,\sigma$ to $3.3\,\sigma$.
The fits reveal that the current measurements of the angular observables in $\Bd\to \Kstarz \mumu$  can be accommodated by shifting the real part of the Wilson coefficient $C_9$ from its SM value by $0.99^{+0.25}_{-0.21}$~\cite{ LHCb-PAPER-2020-002}.

\begin{figure}
\begin{center}
\includegraphics[width=0.45\textwidth]{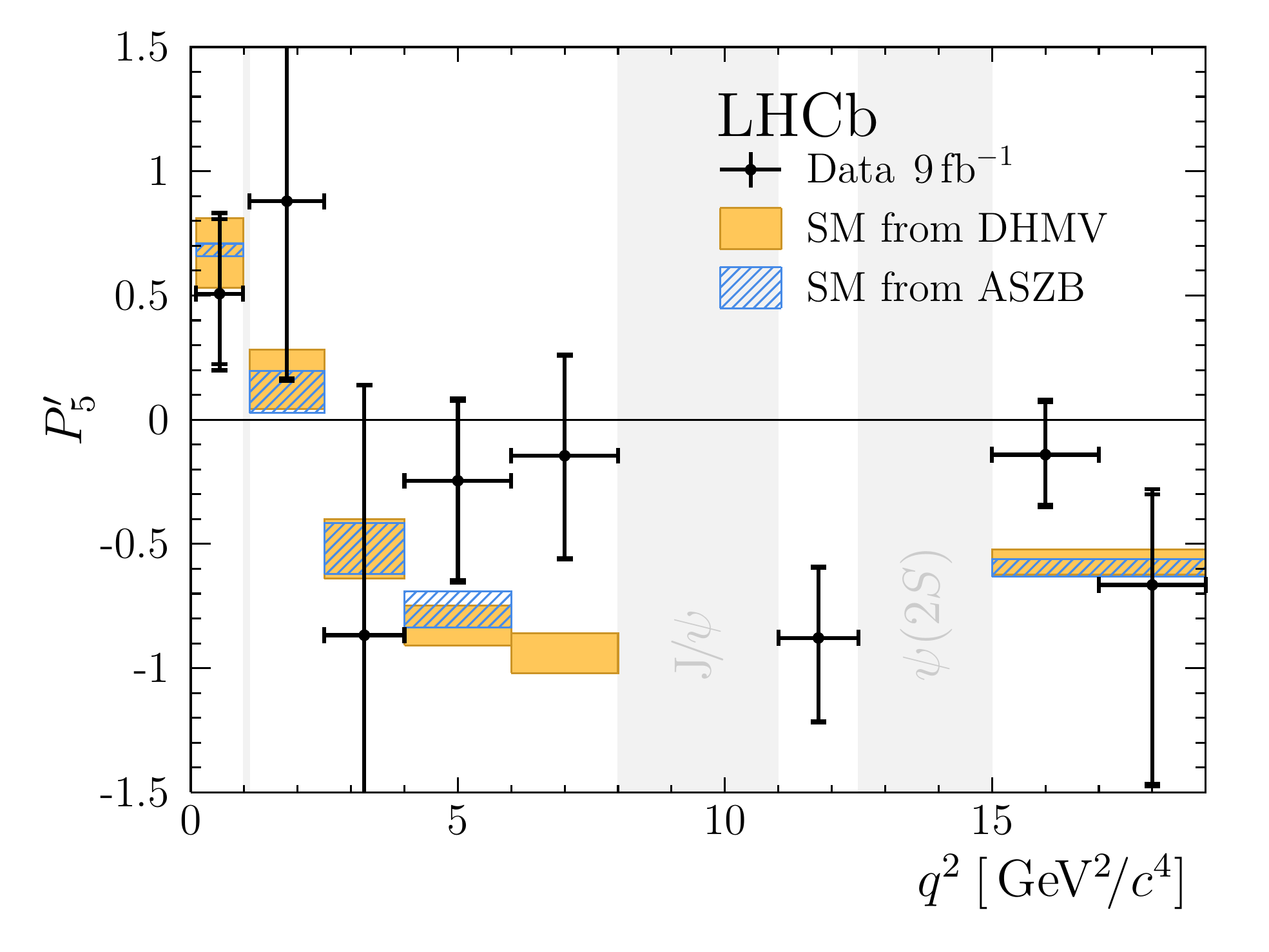}
\includegraphics[width=0.45\textwidth]{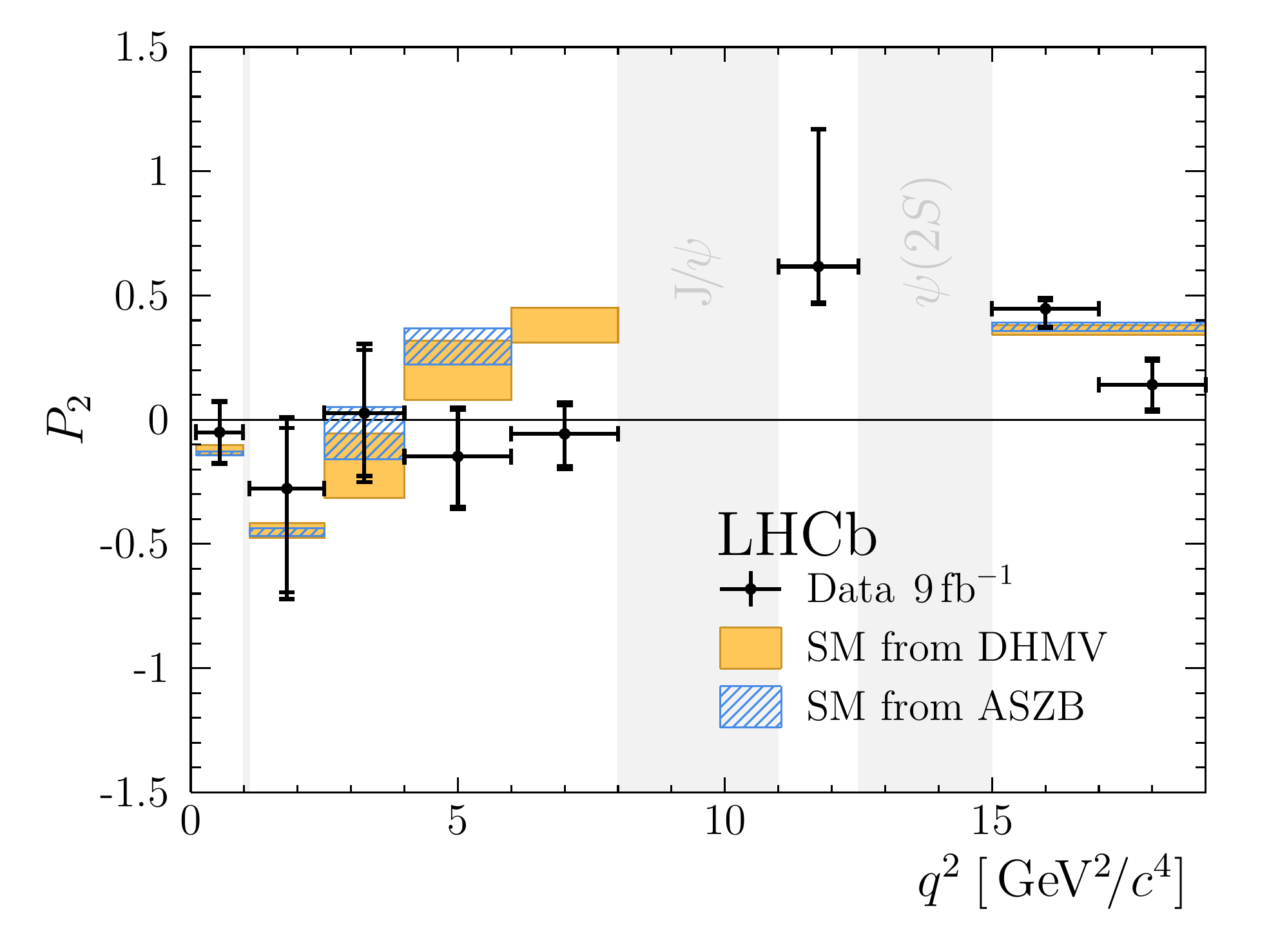}
\caption{ Results for the observables  $P^{\prime}_5$ and $P_2$ in $B^+\to K^{*+} \mumu$,
 compared with SM predictions. Figures are extracted from Ref.~\cite{LHCb-PAPER-2020-041}. }
\label{fig:Bp2Kstmumu}
\end{center}
\end{figure}

Recently, LHCb reported  results of  angular analysis of the $B^+\to K^{*+}(\to \KS\pip) \mumu$ decay using Run~1 and Run~2 data\cite{LHCb-PAPER-2020-041}. 
A trend of deviations from the SM predictions in~$P^{\prime}_5$, similar to that in the isospin partner decay $B^0\to \Kstarz \mumu$, is shown in the left of Fig.~\ref{fig:Bp2Kstmumu}. 
Meanwhile,  a large  discrepancy in the measurement of  $P_2=\frac{2}{3}A_{\rm FB}/(1-F_{\rm L})$  has also been observed in the  $6.0<q^2<8.0\gevgevcccc$ region, where the measurement deviates from its SM prediction~\cite{Altmannshofer:2014rta, Descotes-Genon:2014uoa} by $3.0\,\sigma$ (Fig.~\ref{fig:Bp2Kstmumu} right ).

\begin{figure}[!tb]
\begin{center}
\includegraphics[width=0.45\textwidth]{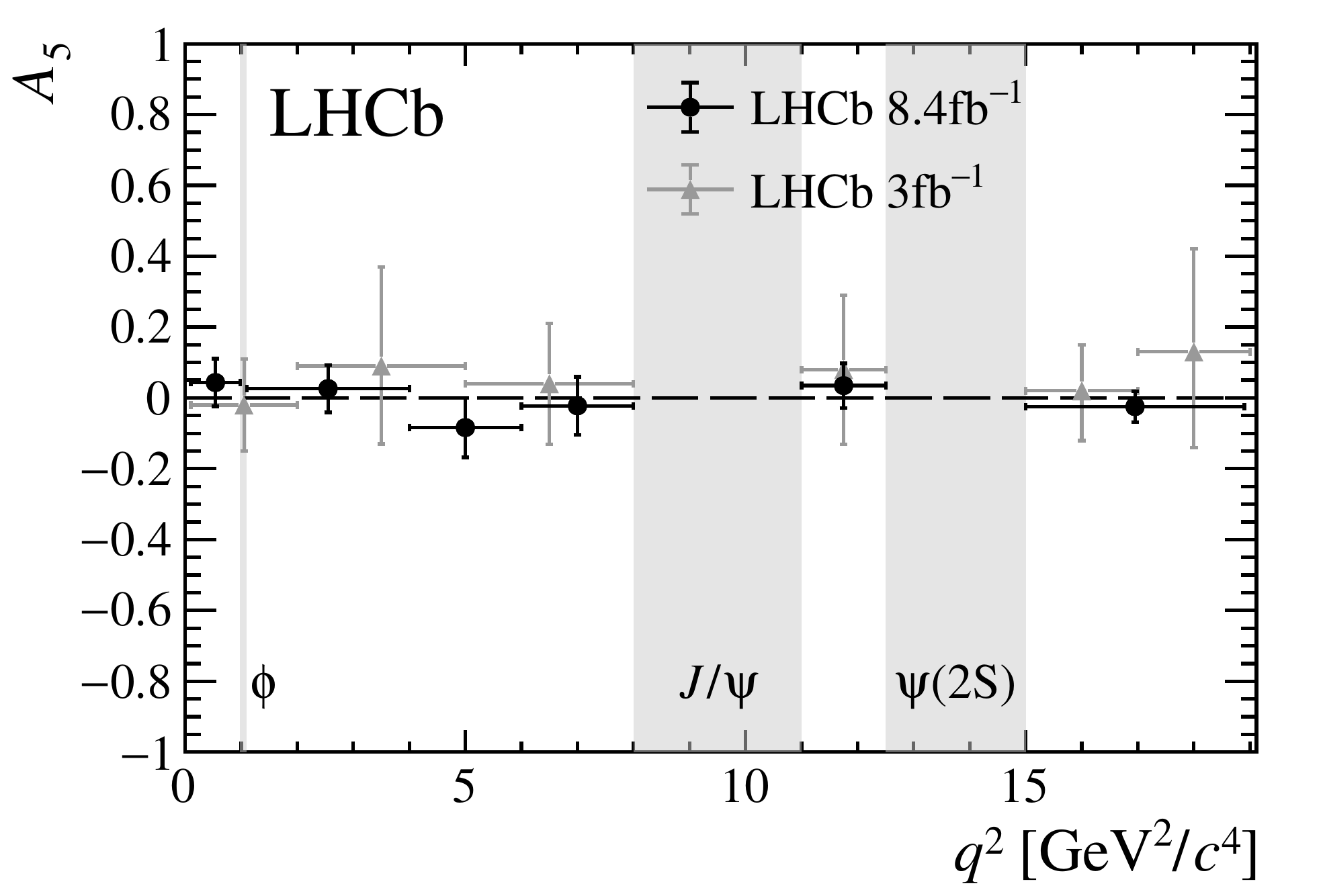}
\includegraphics[width=0.45\textwidth]{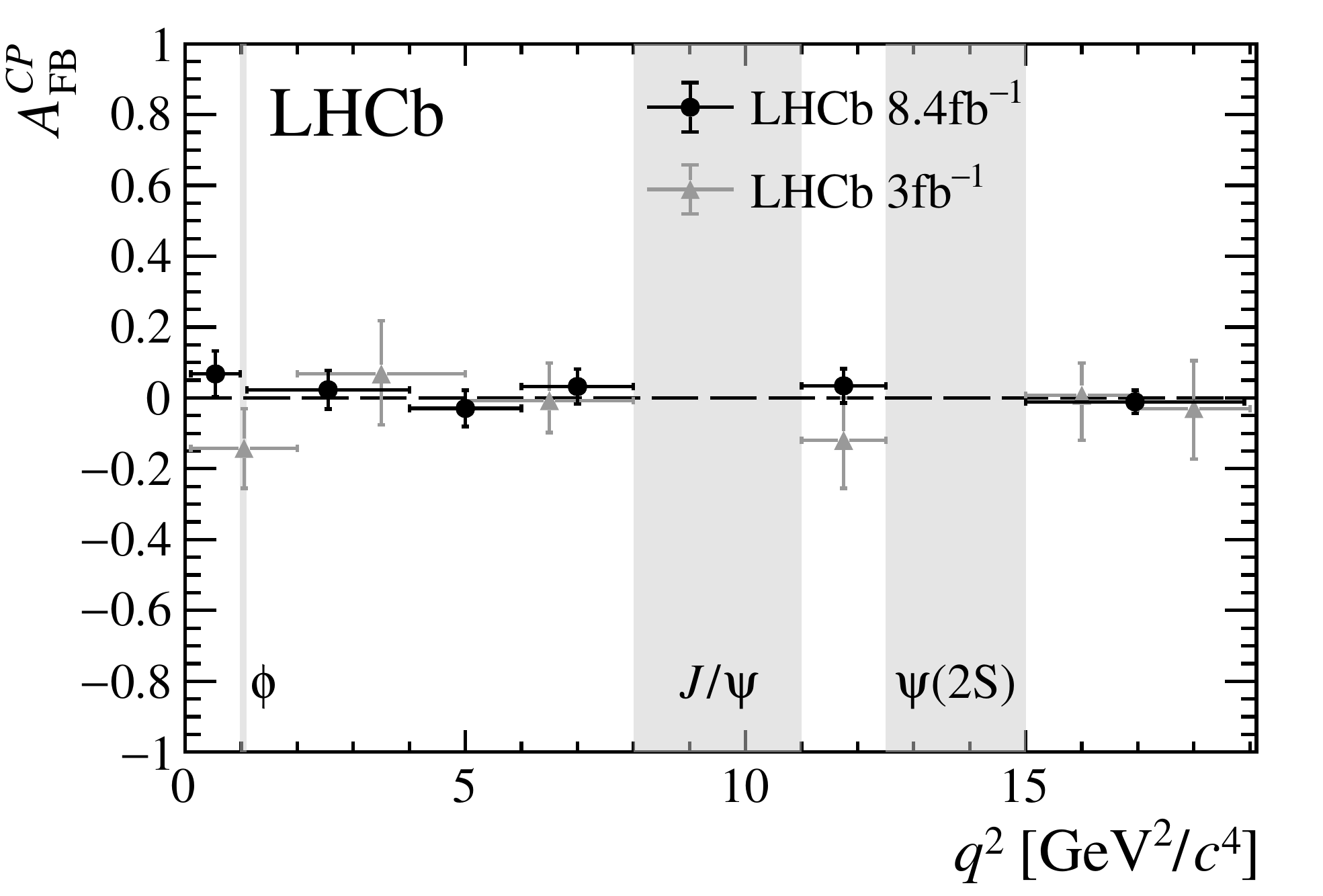}
\includegraphics[width=0.45\textwidth]{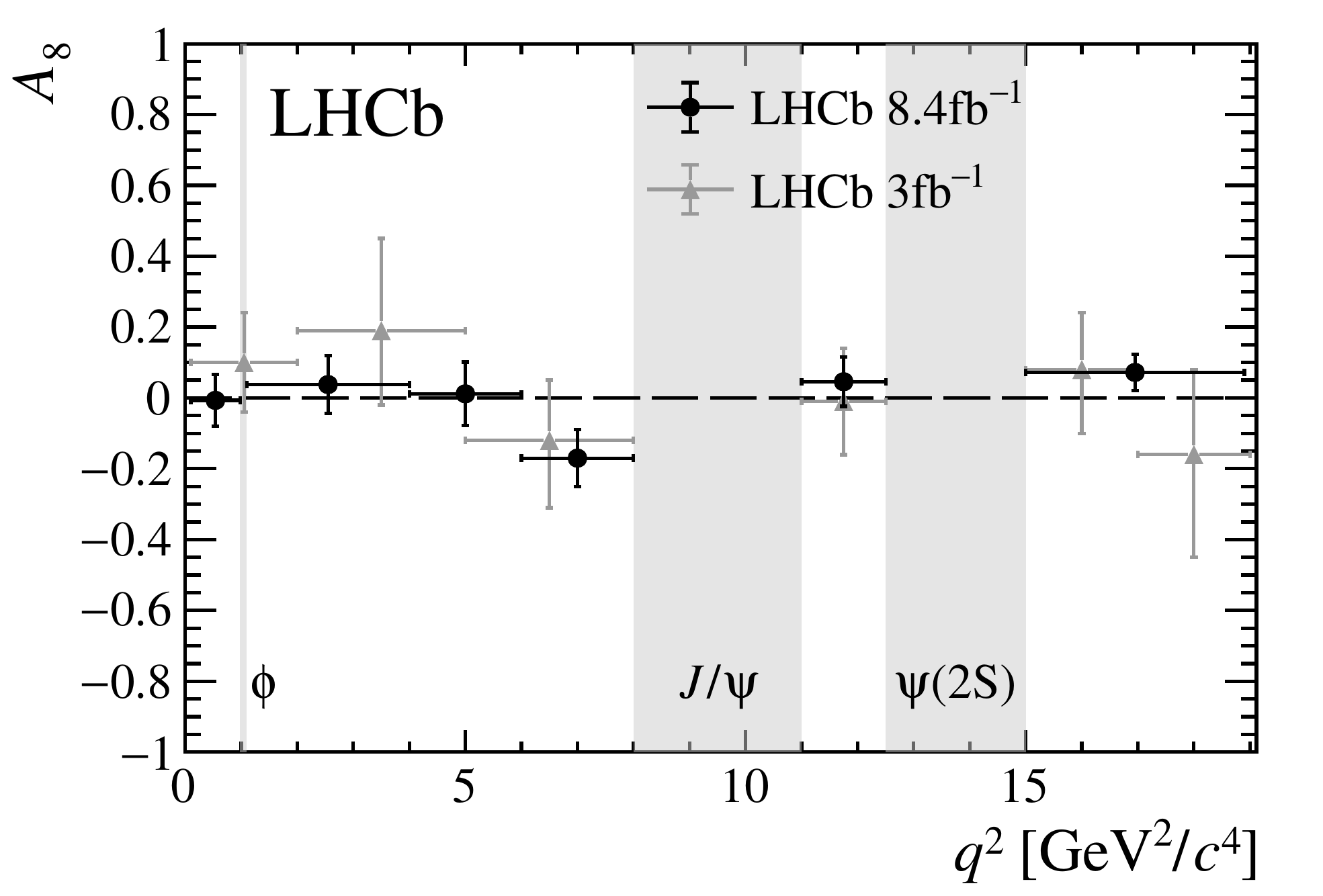}
\includegraphics[width=0.45\textwidth]{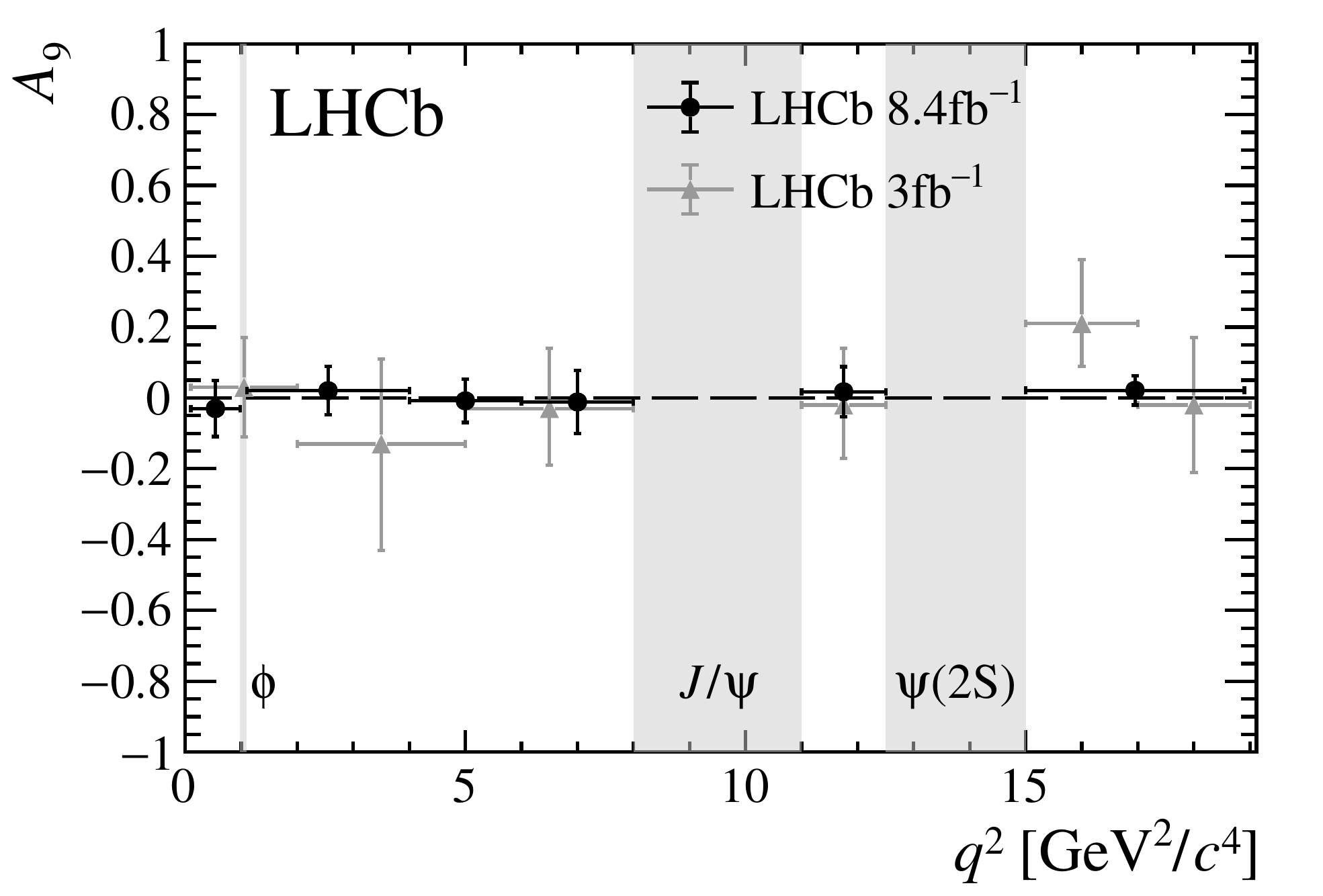}
\caption{ CP asymmetries $A^{CP}_{\rm FB}$ and $A_{5,8,9}$ in intervals of $q^2$ in the decay $\Bs \to \phi \mumu$ measured using Run 1 and Run 2 data (black) in Run 1 data only (grey). Figures are extracted from Ref.~\cite{LHCb-PAPER-2021-022}. }
\label{fig:Bs2phimumu}
\end{center}
\end{figure}

An untagged time-integrated angular analysis of the decay $\Bs \to \phi \mumu$ has  been performed by the LHCb collaboration using data collected in 2011 to 2012 and 2016 to 2018~\cite{LHCb-PAPER-2021-022}.  
In this channel, the same particles ($\mu^+$ and $\Kp$) are used to define the angular variables for both $\Bs$ and $\Bsb$ decays, since the final state is not self-tagging as in the $\Bd\to \Kstarz\mumu$ case.
With this convention, the coefficients of the terms in the CP-averaged  time-integrated angular distributions corresponding to the interference between CP-even ($0$ or $||$) and CP-odd ($\perp$ or $S$) amplitudes  are CP asymmetries, $A^{\CP}_{\rm FB}$ and $A_{5,8,9}$, rather than the CP-average observables $A_{\rm FB}$ and $S_{5,8,9}$ in Eq.\ref{eq:kstarmumu}.  
These asymmetries can arise from either direct CP violation or  nonzero effective mixing phase,  with the latter contribution suppressed by the small value of $\Delta\Gamma_s/\Gamma_s$. 
They are  predicted to be close to  zero in the SM but has some sensitivity to new physics contributions~\cite{Bobeth:2008ij}. 
The measurements of $A^{CP}_{\rm FB}$ and $A_{5,8,9}$ in intervals of $q^2$ are shown in Fig.~\ref{fig:Bs2phimumu}, which are consistent with CP invariance.
Much more information on  CP violation can be obtained from time-dependent angular analysis of tagged $\Bs\to \phi \mumu$ decays~\cite{Descotes-Genon:2015hea}, which may become feasible with the huge amount of  data  that will be collected with the upgraded LHCb detector in the coming data-taking periods. 

Angular coefficients in  the $B^+ \to K^+ \mumu$ decay 
have  been measured by LHCb~\cite{LHCb-PAPER-2014-007} and CMS~\cite{CMS:2018qih} using Run 1 data. 
Angular observables in $\Lb \to p\Km \mumu$ decays have been determined by LHCb~\cite{LHCb-PAPER-2018-029} from a moment analysis using data collected between 2011 and 2016.  
These results are  consistent with the SM predictions but limited by statistical uncertainties.




\subsubsection{Lepton flavour  universality tests}
In the SM, the couplings of  the three generations of leptons to the electroweak gauge bosons $Z^0$ and $\Wpm$
are assumed to be identical. This is known as lepton flavour  universality (LFU). Under this assumption,  processes involving the three flavours of charged leptons, $e$, $\mu$ and $\tau$, have equal  rates up to corrections caused by different lepton masses, which can be trivially taken into account. Contributions of new particles or new interactions may lead to violation of LFU, particularly in FCNC processes such as $b\to s \ellell$ decays.
Stringent tests of LFU can be performed by measuring the ratio of the branching fractions between $B\to X \mumu$ and $B\to X \epem$ decays~\cite{Hiller:2003js,Bordone:2016gaq,Isidori:2020acz} outside the charmonium regions in the dilepton mass spectrum, with $X$ indicating the hadron(s) in the decays. 
The ratio is denoted by
\begin{equation}
    R_{X} \equiv\frac{\BF(B\to X \mumu)}
                {\BF(B \to X \epem)}\;.
\end{equation}
In the SM, $R_X$ is expected to be very close to unity with negligible theoretical uncertainty,
due to the small and precisely known difference between the  muon and electron masses. 
On the experimental side, reconstruction of electrons is challenging due to the Bremsstrahlung radiation. The decays $B\to X \jpsi(\to \mumu)$ and $B\to X \jpsi(\to \epem)$ are used as control channels for cancellation of systematic uncertainties associated with electron reconstruction. 
Practically, the ratio $R_X$ is measured using a double-ratio technique following the equation
\begin{equation}
    R_X =\frac{\BF(B\to X \mumu)/\BF(B\to X \jpsi(\to \mumu))} {\BF(B \to X \epem)/\BF(B\to X \jpsi (\to e^+e^-))}\;,
\end{equation} 
where $\BF(B\to X \jpsi(\to \mumu))/\BF(B\to X \jpsi(\to \epem))$ is known to be very close to unity~\cite{PDG2020}.

\begin{figure}[!tb]
\begin{center}
\includegraphics[width=0.45\textwidth]{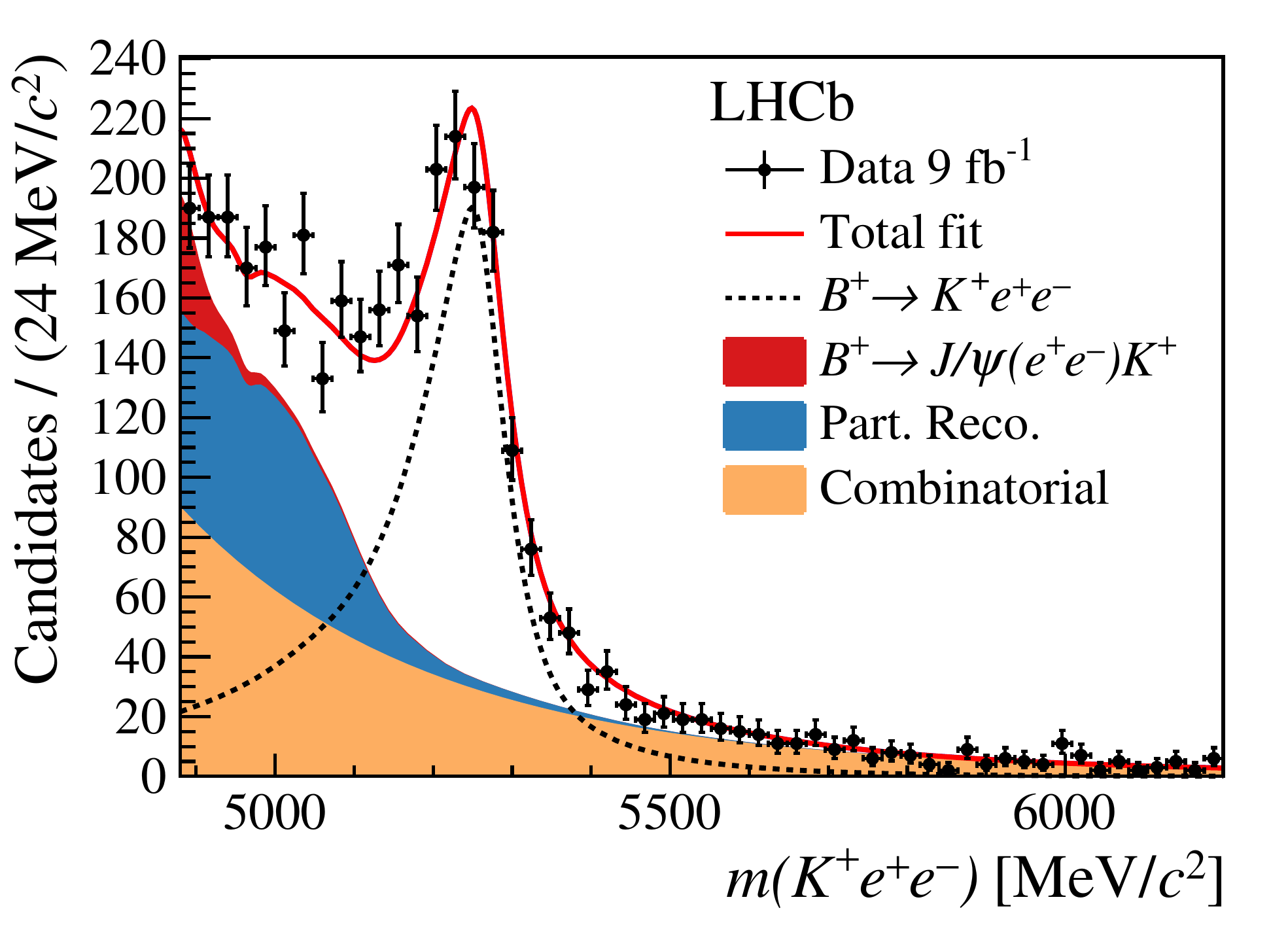}
\includegraphics[width=0.45\textwidth]{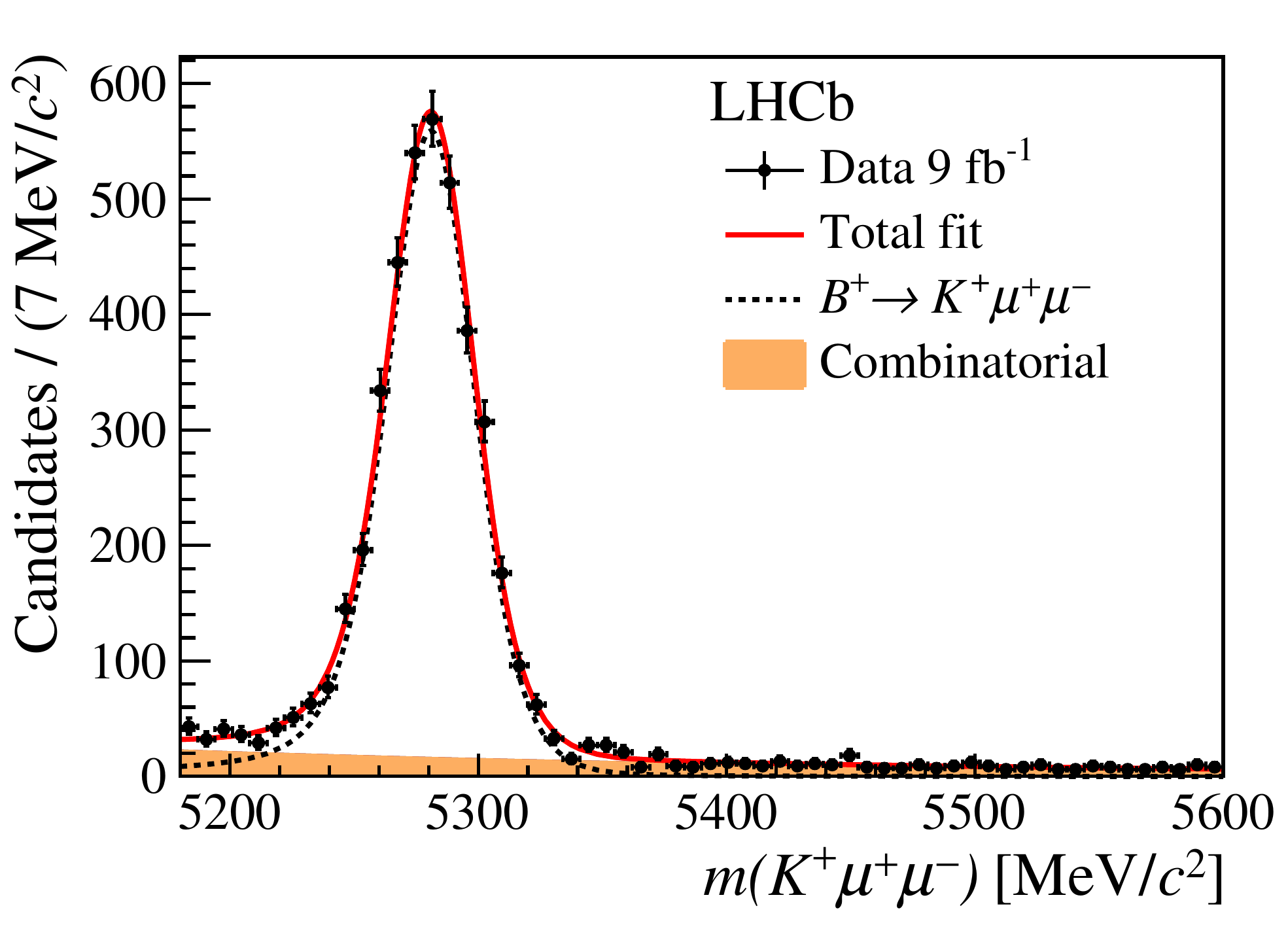}
\caption{ Invariant mass distributions of the selected (left) $\Bp\to\Kp \epem$  and (right) $\Bp\to \Kp \mumu$ candidates, superimposed by the fit results. Figures are extracted from Ref.~\cite{LHCb-PAPER-2021-004}.
}
\label{fig:RK1}
\end{center}
\end{figure}

\begin{figure}[!tb]
\begin{center}
\includegraphics[width=0.45\textwidth]{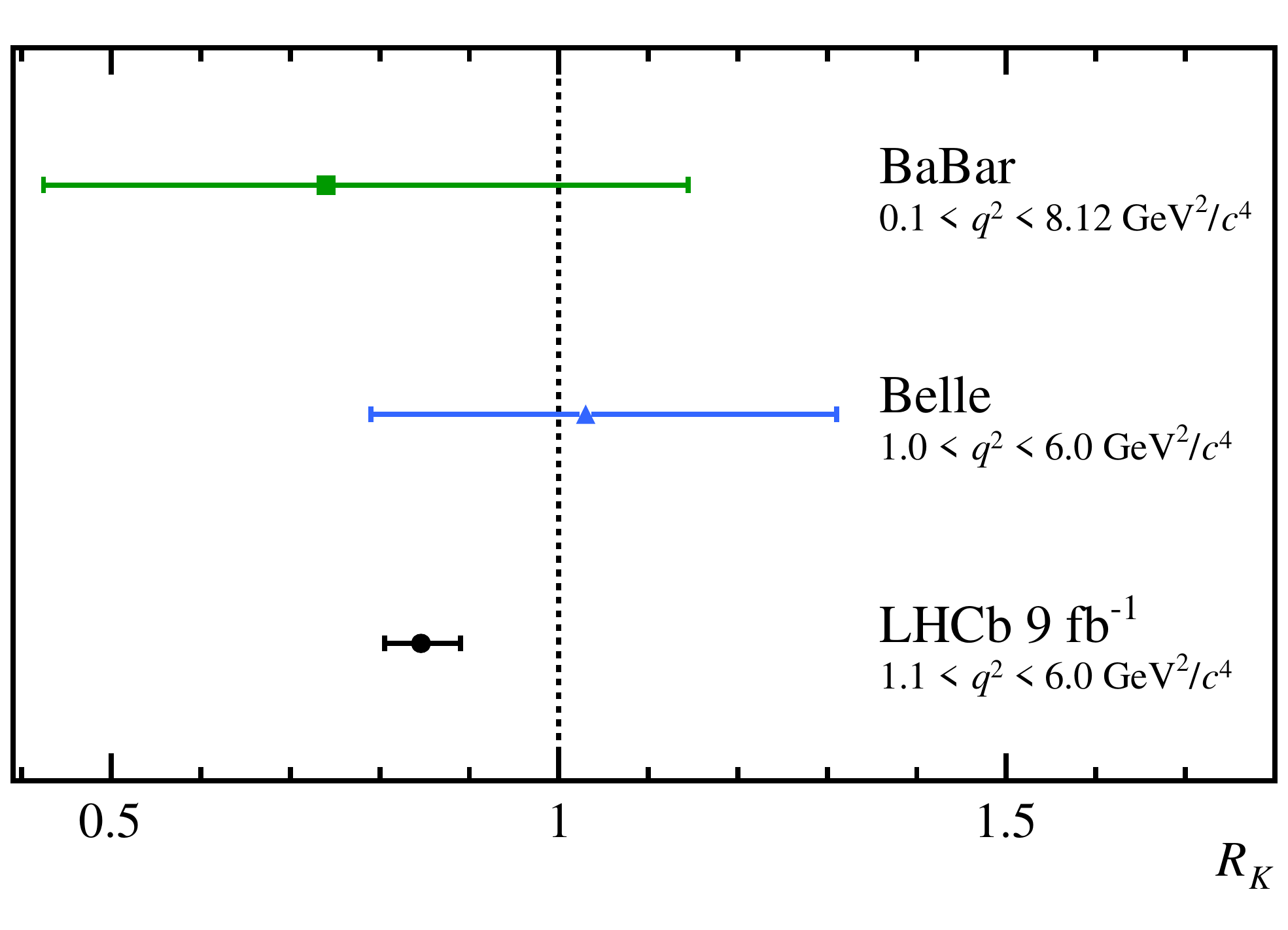}
\caption{ Comparison of the LHCb $R_K$ result with the measurements by B-factories~\cite{BaBar:2012mrf,Belle:2009zue}. 
The vertical line indicates the SM expected value. Figures are extracted from Ref.~\cite{LHCb-PAPER-2021-004}.
}
\label{fig:RK2}
\end{center}
\end{figure}

The LHCb collaboration  previously  measured  $R_K=0.846^{+0.060}_{-0.054}\stat{}^{+0.016}_{-0.014}\syst$ 
in $\Bp\to\Kp\ellell\; (\ell=e,\,\mu)$ decays in the dilepton  
mass-squared range $1.1<q^2<6.0\gevgevcccc$
using Run 1 and part of Run 2 data~\cite{LHCb-PAPER-2019-009}.
The result was below the SM expectation~\cite{Straub:2018kue} by $2.5\,\sigma$.
Recently, LHCb  updated the $R_K$ measurement 
using the full Run 1 and Run 2 sample~\cite{LHCb-PAPER-2021-004}.
The mass distributions of the $\Bp\to\Kp\ellell$ candidates are shown in Fig.~\ref{fig:RK1}.  
The $R_K$ value is measured to be
\begin{equation*}
R_K(1.1<q^2<6.0\gevgevcccc)=
0.846^{+0.042}_{-0.039}\stat{}^{+0.013}_{-0.012}\syst\;,
\end{equation*}
 which are lower than the SM prediction,  $1.00\pm 0.01$~\cite{Descotes-Genon:2015uva, Bobeth:2007dw, Bordone:2016gaq,van_dyk_danny_2016_159680, Straub:2018kue}, 
 by $3.1\,\sigma$.
A comparison of the LHCb $R_K$ result with the values measured by  \babar~\cite{BaBar:2012mrf} and \belle~\cite{Belle:2009zue} is shown in Fig.~\ref{fig:RK2}.

Tests of LFU have also been performed 
in other $b\to s \ellell\;(\ell=e,\mu)$ decays. 
Based on the Run 1 data sample, the LHCb collaboration has determined 
the ratios of branching fractions of 
$B^0\to \Kstarz \ellell\; (\ell=e,\mu)$  decays
in two regions of dilepton mass-squared below the $\jpsi$ resonance to be~\cite{LHCb-PAPER-2017-013} 
\begin{equation*}
\begin{aligned}
R_{\Kstarz}(0.045<q^2<1.1\gevgevcccc) & = 0.66^{+0.11}_{-0.07} \stat \pm 0.03 \syst\;, \\
R_{\Kstarz}(1.1<q^2<6.0\gevgevcccc) & =  0.69^{+0.11}_{-0.07} \stat\pm 0.05 \syst\;.
\end{aligned}
\end{equation*}
These results are in tension with the SM predictions~\cite{Bordone:2016gaq,Descotes-Genon:2015uva,Capdevila:2016ivx,Capdevila:2017ert,Serra:2016ivr,van_dyk_danny_2016_159680, Straub:2018kue,Bharucha:2015bzk,Jager:2014rwa} at the level 
of $2.1-2.3\,\sigma$ and $2.4-2.5\,\sigma$, respectively. 
LHCb has also measured 
\begin{equation*}
R_{pK}(0.1<q^2<6.0\gevgevcccc) =0.86^{+0.14}_{-0.11} \stat\pm 0.05 \syst
\end{equation*}
in $\Lb\to p\Km\ellell\;(\ell=e,\mu)$ decays
using Run 1 and part of Run 2 data~\cite{LHCb-PAPER-2019-040}. 

Very recently, LHCb reported the observation of 
the decays $\Bz\to \KS \epem$ and  $\Bp\to \Kstarp \epem$ and 
the measurements of  LFU observables $R_{\KS}$ and $R_{\Kstarp}$
 using the full Run 1 and Run 2 data samples~\cite{LHCb-PAPER-2021-038}. 
 The obtained results of $R_{\KS}$ and $R_{\Kstarp}$  are
\begin{equation*}
    \begin{aligned}
    R_{\KS}(1.1<q^2<6.0\gevgevcccc) & = 0.66^{+0.20}_{-0.14} \stat{}^{+0.02}_{-0.04} \syst\;, \\
R_{\Kstarp}(0.045<q^2<6.0\gevgevcccc) & = 0.70^{+0.11}_{-0.07} \stat{}^{+0.03}_{-0.04} \syst\;,
    \end{aligned}
\end{equation*}
which are lower than but consistent with the SM predictions at $1.5\,\sigma$ and $1.4\,\sigma$, respectively. 

The anomalous results that the LHCb collaboration has obtained in the study of  LFU  and  angular distributions in  $b \to s \ellell$ decays are highly interesting but not well understood yet. 
These results have prompted extensive theoretical studies of   potential new physics effects in  $b \to s \ellell$ transitions~\cite{Alguero:2019ptt,Ciuchini:2020gvn, Aebischer:2019mlg,Alguero:2021anc,Altmannshofer:2021qrr,Geng:2021nhg,Cornella:2021sby,Isidori:2021vtc,Isidori:2021tzd,Hurth:2021nsi}. Particularly, new physics scenarios that mainly affect  $b \to s \mumu$ transitions are preferred, according to  
  global analysis in the framework of SM effective field theory~\cite{Aebischer:2019mlg,Alguero:2021anc,Altmannshofer:2021qrr,Geng:2021nhg,Isidori:2021tzd,Hurth:2021nsi}.  
  
  It should be noted with caution  that
  the current measurements of these rare decays   are largely limited by the statistical precision, and the  non-trivial background contamination and reconstruction efficiency in   $b \to s \epem$ decays may also need to be further scrutinized. 
  An improved understanding of  FCNC $b$-hadron decays  can  be achieved using the huge amount of data that will be recorded following Upgrade I and Upgrade II. 
  This will not only significantly increase the accuracy of the benchmark measurements in $b \to s$ transitions, but also provide great opportunities to explore new observables and new rare decay modes that are currently inaccessible, such as time-dependent observables in $\Bz \to \KS \mumu$~\cite{Descotes-Genon:2020tnz,Kosnik:2021owl} and $\Bs \to \phi \mumu$~\cite{Descotes-Genon:2015hea}   decays,
 and   lepton universality ratios and angular observables in heavily suppressed  $b \to d \ellell$ transitions~\cite{Bordone:2021olx,Rusov:2019ixr,Soni:2020bvu,Kindra:2018ayz}.

\subsection{Radiative \texorpdfstring{$b\to s \gamma$}{TEXT} decays}

The effective Hamiltonian for  $b\to s \gamma$ transitions can be approximately written as
\begin{equation}
   H_{\rm eff}(b\to s \gamma) =-{\dfrac{4G_F}{\sqrt{2}}} V_{tb}V^*_{ts} ( C_7 \Ope{7} + C^{\prime}_7 \mathcal{O}^{\prime}_7),  
   \label{eq:hamil2}
\end{equation}
where only  the leading operator $\mathcal{O}_{7}$ and its chirality-flipped counterpart  $\mathcal{O}^{\prime}_7$ are included. In the SM, the   coefficient $C^{\prime}_7$ is given by $C^{\prime}_7=\frac{m_s}{m_b} C_7$ due to the chiral $V-A$ structure of the weak interaction, where $m_s$ ($m_b$) indicates the mass of the $s$ ($b$) quark. 
Consequently, the photons emitted in radiative $b$-hadron ($\bquarkbar$-hadron) decays
 are predominantly left-handed (right-handed).  
 Amplitudes with  right-handed photons, $A_R$, are suppressed by the ratio $m_s/m_b$  compared with those with left-handed photons, $A_L$, but could be enhanced in new physics scenarios with right-handed charged current, such as supersymmetric grand unified theories and  left-right symmetric models~\cite{Atwood:1997zr, Everett:2001yy,Grinstein:2004uu, Becirevic:2012dx,Kou:2013gna,Haba:2015gwa,Paul:2016urs}.

Rich information on photon polarisation can be obtained from time-dependent analysis of  $B_q^0 (\bar{B}^0_q) \to f_{\rm CP} \gamma$ decays~\cite{Atwood:1997zr, Atwood:2004jj, Muheim:2008vu}, where $f_{\rm CP}$ is a CP eigenstate with eigenvalue $\eta$. The time-dependent decay rates summing over  left-handed and right-handed photons are expressed as 
\begin{equation}
  P(t) =P_0 e^{{-\Gamma_q t}} 
  \left[\cosh(\Delta\Gamma_q t/2)-A^{\Delta} \sinh(\Delta\Gamma_q t/2) +\xi  C\cos(\Delta m_qt)-\xi S \sin(\Delta m_q t) \right]\;,
   \label{eq:b2sg_rate}
\end{equation}
where $\xi$ takes the value of $+1$ ($-1$) for an initial $B^0_q$ ($\bar{B}^0_q$) meson. 
The coefficient $C$ quantifies CP violation in the decay. This type of CP violation  has been constrained to be small in radiative $B$ meson decays by \babar, \belle and \lhcb~\cite{BaBar:2012fqh,Belle:2017hum,LHCb-PAPER-2012-019}. 
Assume no CP violation in the decay for simplicity, it is convenient to write 
\begin{equation}
\begin{aligned}
&\mathrm{Favoured:}\,\,\,\phz    A(\bar{B}^0_q \to f_{\rm CP} \gamma_{L}) = a_{L} e^{i \delta_{L}} e^{i \phi_{L}} \,\;\Rightarrow\;
  A(B^0_q \to f_{\rm CP} \gamma_{R}) =\eta \, a_{L} e^{i \delta_{L}} e^{-i\phi_{L}}\,,\\
&\mathrm{Suppressed:\,}   A(\bar{B}^0_q \to f_{\rm CP} \gamma_{R}) = a_{R} e^{i \delta_{R}} e^{i \phi_{R}}\; \Rightarrow\;A(B^0_q \to f_{\rm CP} \gamma_{L}) =\eta \,  a_{R} e^{i \delta_{R}} e^{-i\phi_{R}}\,,
\end{aligned}
\label{eq:}
\end{equation}
where $a_{L(R)}$, $\delta_{L(R)}$ and $\phi_{L(R)}$ are the size, strong phase and weak phase of \mbox{$A(\bar{B}^0_q \to f_{\rm CP} \gamma_{L(R)})$}, respectively.
The terms $\sin(\Delta m_q)$ and $\cosh(\Delta \Gamma_q t/2)$ in Eq.~\ref{eq:b2sg_rate}
arise from  interference of the amplitudes of direct decay, $A(\bar{B^0_q} \to f_{\rm CP} \gamma_L)$ or $A(B^0_q \to f_{\rm CP} \gamma_R)$, and the decay via $B^0_q$-$\bar{B}^0_q$ mixing, $p/q A(B^0_q \to f_{\rm CP} \gamma_L)$ or $q/p A(\bar{B}^0_q \to f_{\rm CP} \gamma_R)$. 
The mixing-induced observables $S$ and $A^{\Delta}$ are  given by~\cite{ Muheim:2008vu}
\begin{equation}
\begin{aligned}
 S &\approx  \frac{2\eta\, r}{1+r^2} \cos (\delta_L-\delta_R)  \sin (\phi_q-\phi_L-\phi_R) \,,\\
 A^{\Delta } &\approx  \frac{2\eta\, r}{1+r^2} \cos (\delta_L-\delta_R)  \cos (\phi_q-\phi_L-\phi_R) \;,
\end{aligned}
\end{equation}
where  $r\equiv|a_R/a_L|\approx|C^{\prime}_7/C_7|$, $\phi_q$ is the $B^0_q$-$\bar{B}^0_q$ mixing phase. 
 The values of $S$ and $A^{\Delta}$ are expected to be small  in the SM due to the suppression by the ratio $r\approx m_s/m_b$. 
 Since   $S$ and $A^{\Delta}$ are approximately linearly dependent on  $r$, they are sensitive to even a small increase of right-handed photons. 
 
 Currently, the observables $S$ and $A^{\Delta}$ are only weakly constrained. 
 The $\Bd \to \KS \piz \gamma$ decay is a golden channel to study photon polarisation at $B$ factories.
The mixing-induced CP asymmetry in this channel has been measured to be
 $S_{\KS \piz \gamma} = -0.10\pm 0.31 \stat \pm 0.07 \syst$~\cite{Belle:2006pxp} and 
 $S_{\KS \piz \gamma} = -0.78\pm 0.59 \stat \pm 0.09 \syst$~\cite{BaBar:2008okc} by the \belle and \babar collaborations, respectively, both consistent with the SM expectation value of roughly $\frac{m_s}{m_b} \sin 2\beta  $.  
 The coefficient of the $\sinh{\Delta\Gamma_d t/2}$ term,
 $A^{\Delta}$, 
 is inaccessible  in  $\Bd$ decays  due to the tiny value of the $\Bd$ width difference, $\Delta\Gamma_d$. 
 
\begin{figure}[!tb]
\begin{center}
\includegraphics[width=0.45\textwidth]{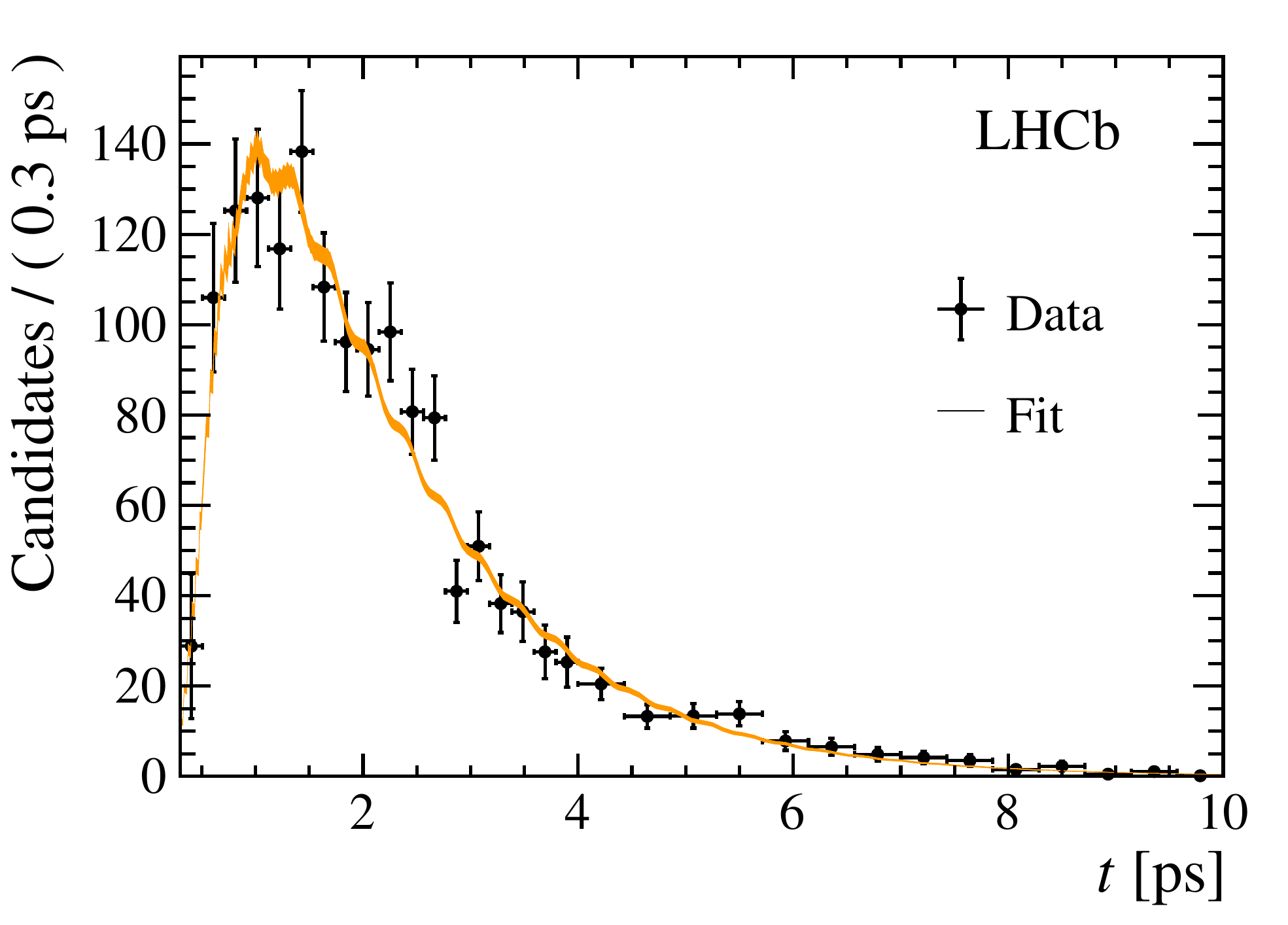}
\includegraphics[width=0.45\textwidth]{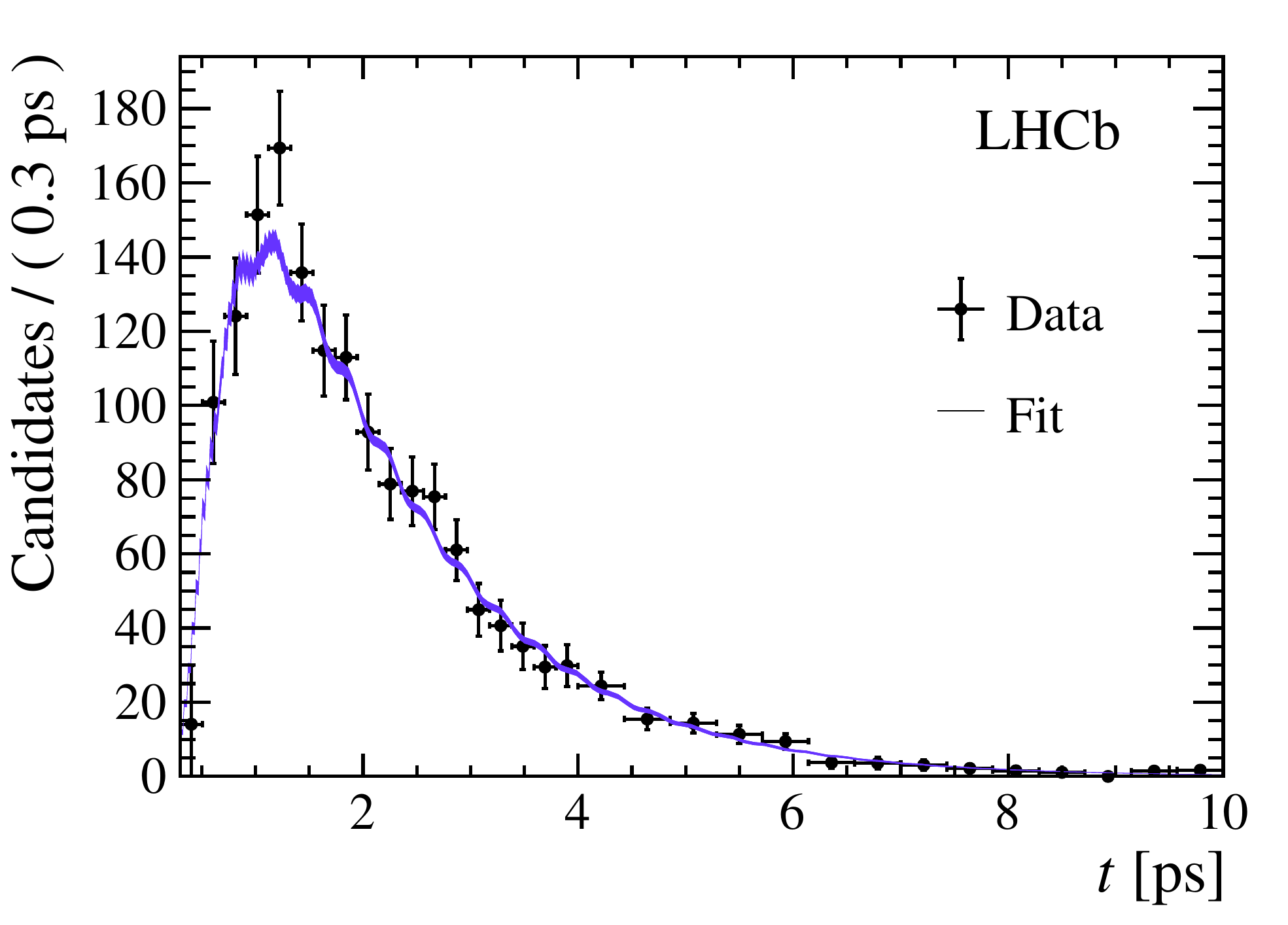}
\caption{ Decay-time distributions of the tagged (left) $\Bs\to \phi \gamma$ and (right) $\Bsb \to \phi \gamma$ candidates, superimposed 
by the fit projections. Plots are taken from Ref.~\cite{LHCb-PAPER-2019-015}.
}
\label{fig:phig}
\end{center}
\end{figure}

Reconstruction of  $\Bd \to \KS \piz \gamma$ decays is challenging at the LHCb  experiment. Alternatively, LHCb can measure mixing-induced CP violation in $B^0\to \KS \pip \pim \gamma$ decays through a time-dependent amplitude analysis~\cite{LHCb-PII-Physics}. A more promising channel to probe right-handed NP
is the  decay $\Bs \to \phi \gamma$. Both  $S$ and   $A^{\Delta}$ can be measured in this channel and they are predicted to be close to zero in the SM~\cite{Muheim:2008vu}:
\begin{equation*}
S_{\phi \gamma} ({\rm SM}) = 0.000 \pm 0.002\;,\; 
A^{\Delta}_{\phi \gamma } ({\rm SM}) =0.047\pm 0.039\;.
\end{equation*}
Using data collected in Run 1, the LHCb collaboration studied the tagged time-dependent decay rates of $\Bs \to \phi \gamma$, which are shown in Fig.~\ref{fig:phig}. The mixing-induced observables are measured to be~\cite{LHCb-PAPER-2019-015}
\begin{equation*}
 S_{\phi \gamma } =
0.43 \pm 0.30 \stat \pm 0.11 \syst\;,\;
A^{\Delta}_{\phi \gamma}=-0.67 ^{+0.37}_{-0.41} \stat \pm 0.17 \syst\;,
\end{equation*}
 which are  in agreement with the SM expectations.

The  $\Bz\to \Kstarz \epem$ decay in the low-$q^2$ region  offers a powerful probe of right-handed new physics. 
In the vicinity of the  photon pole, the decay amplitudes are dominated by contributions from the electromagnetic Wilson coefficients $C^{(\prime)}_7$. 
An  angular analysis  can be performed in 
a similar way as that in the $\Bz \to \Kstarz \mumu$ case. 
In the angular distribution, there are two terms arising from interference of the left-handed and right-handed decay amplitudes that  are proportional to $C_7$ and $C^{\prime}_7$, respectively. 
For small values of $r$ ($\equiv|C_7^\prime/C_7)|$), 
the coefficients of these two terms, denoted 
 $A^{(2)}_{\rm T}$ and $A^{\rm Im}_{\rm T}$,
 are approximately expressed as~\cite{Becirevic:2011bp,Paul:2016urs}
\begin{equation}
   A^{(2)}_{\rm T}\approx r\,\cos (\phi_L-\phi_R)\;,\;
   A^{\rm Im}_{\rm T}\approx r\,\sin (\phi_L-\phi_R)\;,
\end{equation}
where $\phi_{L}$ and $\phi_R$ represent the phases of $C_7$ and $C^{\prime}_7$, respectively. 
Like the  mixing-induced observables in $\Bs \to \phi \gamma$, $A^{(2)}_{\rm T}$ and $A^{\rm Im}_{\rm T}$ depend approximately linearly on $r$, 
thus can provide high sensitivity  to right-handed currents in the small $r$ region.
Using data collected in Run 1 and Run 2, the LHCb collaboration has measured    $A^{(2)}_{\rm T}$ and $A^{\rm Im}_{\rm T}$  to be~\cite{LHCb-PAPER-2020-020}
\begin{equation*}
 A^{(2)}_{\rm T} =
0.11 \pm 0.10 \stat \pm 0.02 \syst\;,\;
A^{\rm Im}_{\rm T}=0.02 \pm 0.10 \stat \pm 0.01 \syst\;.
\end{equation*}
These results are compatible with the following SM predictions calculated using the FLAVIO software package~\cite{Straub:2018kue}:
\begin{equation*}
 A^{(2)}_{\rm T} ({\rm SM}) =
0.033 \pm 0.020 \;,\;
A^{\rm Im}_{\rm T} ({\rm SM})=-0.00012\pm0.00034\;,
\end{equation*}
and provide the most stringent constraint on the $b\to s \gamma$ photon polarisation. 


Photon polarisation in $b \to s \gamma$ transitions
can also be probed by exploiting the angular correlations 
in radiative decays of $b$ baryons or charged $b$ mesons.
 Since current detection technology cannot distinguish left-handed and right-handed photons, the final states with both left-handed and right-handed photons  are summed together. The left-handed  amplitude $A_L$ and right-handed amplitude $A_R$ add incoherently in the form of  $|A_L|^2+|A_R|^2$, without any interference.
 In certain  cases, the angular distributions allow for determining a parity violation parameter, $A_{\rm parity}$, which is proportional to the photon polarisation~\cite{Gronau:2002rz, Kou:2010kn},
 \begin{equation}
A_{\rm parity} \propto  \lambda_{\gamma}\equiv \frac{|A_L|^2-|A_R|^2}{|A_L|^2+|A_R|^2}\approx \frac{1-r^2}{1+r^2}\;.
\end{equation}
 This approach is powerful  in probing  large right-handed currents  but has limited sensitivity to any small  right-handed component. 
 
 The LHCb collaboration observed a significantly non-zero up-down asymmetry of  the photons in $B^-\to K^-\pip \pim \gamma$ decays in the range $m_{K\pi\pi} =[1.1,\,1.3]\gevcc$~\cite{LHCb-PAPER-2014-001} with respect to the plane defined by the three final-state hadrons in their rest frame. 
 This observation demonstrates that the photons are indeed polarised.
 However, it is nontrivial to translate the measured asymmetry into a constraint on the polarisation parameter $\lambda_{\gamma}$, due to currently limited knowledge of the
 structure and decay dynamics of the  intermediate  resonances involved in this process.
 A recent theoretical study pointed out that this $m_{K\pi\pi}$ range is dominated by the $K_1(1270)$  resonance and proposed to exploit  the charm  decay $D\to K_1 e\nu$ to quantify the hadronic effects in  $K_1 \to K^-\pip \pim$~\cite{Wang:2019wee}, which can be studied at a future Super $\tau$-charm factory~\cite{Cheng:2022tog}.

 The baryonic decay  $\Lb\to \Lz \gamma$, observed by  the LHCb experiment using data collected in 2016~\cite{LHCb-PAPER-2019-010},
 provides a more convenient way  to measure the photon polarisation  in $b \to s \gamma$ transitions~\cite{Gremm:1995nx, Mannel:1997xy, Hiller:2001zj}. 
 The angular distribution of this process is given by the differential rate
\begin{equation}
    \frac{d\Gamma}{d\cos\theta_p} \propto 1-\alpha_{\Lz} \lambda_{\gamma} \cos \theta_p \;,
\end{equation}
where $\theta_p$  is the helicity angle of the proton in the $\Lz$ rest frame with respect to the opposite direction of the photon, $\alpha_{\Lz}$ is the   decay parameter of the weak process $\Lz \to p \pim$. The photon polarisation parameter has recently been measured to be  
$\lambda_{\gamma}=0.82^{+0.17}_{-0.26}(\rm stat){}^{+0.04}_{-0.13}(\rm syst)$~\cite{LHCb-PAPER-2021-030} by the LHCb experiment using all data from Run~2 and the average of the decay parameter values of $\Lz$  and 
$\Lbar$ measured by \mbox{BESIII}, $\alpha_{\Lz}=0.754\pm 0.004$~\cite{BESIII:2018cnd}. This result is in agreement with the SM predictions from Refs.~\cite{Wang:2008sm,Mannel:2011xg,Gutsche:2013pp}.
The LHCb experiment also searched for the decay $\Xi_b^-\to \Xi^- \gamma$ using Run 2 data and found no signal~\cite{LHCb-PAPER-2021-017}.

\subsection{Other rare decays of beauty hadrons}

Besides the  $b\to s \ellell$ decays discussed above, the LHCb experiment has also performed studies of other  rare decay processes of beauty hadrons. These include: lepton-flavour violating decays $\Bd\to\Kstarz\tau^{\pm}\mu^{\mp}$~\cite{LHCb-paper-2022-021}, $\Bd\to \Kstarz \mu^{\pm}e^{\mp}$ and $\Bs\to\phi\mu^{\pm}e^{\mp}$ ~\cite{LHCb-paper-2022-008}, $\Bp\to K^+\mun\tau^{+}$~\cite{LHCb-paper-2019-043}, $\Bp\to K^+\mu^{\pm}e^{\mp}$~\cite{LHCb-paper-2019-022},
$B^0_{(s)}\to\tau^{\pm}\mu^{\mp}$~\cite{LHCb-PAPER-2019-016}, 
$B^0_{(a)}\to e^{\pm}\mu^{\mp}$~\cite{LHCb-PAPER-2017-031};
lepton- and baryon-number violating  decays $B^0_{(s)}\to p\mun$ ~\cite{LHCb-paper-2022-022};
$b\to d \ellell$ decays $\Bs\to \Kstarz \mumu$~\cite{LHCb-paper-2018-004}, $\Lb\to p\pim\mumu$~\cite{LHCb-paper-2016-049}, $\Bp\to\pip\mumu$~\cite{LHCb-paper-2012-020};
annihilation-type decays $\Bd\to\phi\mumu$~\cite{LHCb-paper-2021-042}, $\Bd\to\jpsi \phi$~\cite{LHCb-paper-2020-033}.
Due to the limited space, the results of these studies are not included in this review.







\clearpage

\section{CP violation in beauty and CKM parameters}
\label{sec:beautyCPV}

CP violation is a necessary condition to explain the matter-dominated universe. 
 While the SM with the CKM mechanism can account for the current experimental results on CP violation, it fails to explain  the cosmological matter–antimatter imbalance. Searching for new sources of CP violation  is one of the primary goals of flavour physics. This can be done by overconstraining the CKM matrix using measurements of the matrix  elements in many different processes. 

The decays of $b$-hadrons provide a number of key measurements to access the five  CKM matrix elements  related to the $b$ or $t$ quark. 
Taking advantages of the intense source of b-hadrons at the LHC and a detector designed to probe CP violation in heavy-flavour decays,  the LHCb experiment  has been the leading experiment in the field of $B$ physics  in the past ten years, and achieved  some of the most precise measurements of CP violation and mixing of $B$ mesons. Particularly, the  precision of the CKM angle $\gamma$ is now approaching that of the indirect determination;
the CP violation parameter $\phi_s$ and mixing parameter  $\dms$ of the \Bs system, which are key observables for NP searches, have been pinned down with unprecedented precision. 


This section describes the  key measurements of the CKM elements in the beauty sector by the LHCb experiment using the data taken in Run 1 and Run 2. The parameters $\gamma$, $\phi_s$ and $\dms$, which have received the most significant improvements, are discussed in detail.
Other observables related to CKM global fit will also be discussed briefly while many other interesting topics, such as  CP violation in $b$-baryon decays, are not mentioned.

\subsection{CKM angle \texorpdfstring{$\gamma$}{TEXT}}
The angle $\gamma$, defined as $\arg[-(V_{ud}V_{ub}^*)/(V_{cd}V_{cb}^*)]$, is one of the key observables related to the CKM matrix. As can be seen from Fig.~\ref{fig:rhoeta_large}, one of the main limitations of global constraints comes from the angle $\gamma$. To be noted, the fits have already included recent $\gamma$ measurements from LHCb which improves the sensitivity on $\gamma$ from $14^{\circ}$, established at the era of $B$-factories, to around $5^{\circ}$. Future improvements on the sensitivities on $\gamma$ can be foreseen with the upgrade of LHCb and running of the Belle II experiment. In the following sections, we  briefly overview the main developments on $\gamma$ measurements in the past several years from the LHCb experiment.
 
The direct determination of the angle $\gamma$ is obtained through interference between $b \to c$ and $b \to u$ tree-level processes, where new physics hardly enters~\cite{Brod:2014bfa}. 
The hadronic parameters of the system are all determined from experimental data and related theoretical uncertainty is negligible~\cite{Brod:2013sga}. The direct $\gamma$ measurements thus serve as key inputs for SM predictions, which can be compared with other NP sensitive measurements to search for physics beyond the SM. 

Several methods have been proposed to measure the angle $\gamma$, based on the types of $D$ decays. In this paper, when not specified, $D$ means a mixture state of $D^0$ and $\overline{D}^0$. The GLW method~\cite{Gronau:1991dp,Gronau:1990ra} refers to those decays with $D$ into a CP eigenstate or multi-body $D$ decays which can be effectively considered as a CP eigenstate using a CP-even fraction $F_+$. 
The ADS method~\cite{Atwood:1996ci} refers to two-body $D$ decays or multi-body $D$ decays where the detailed structures over phase space are considered by introducing a global coherent factor $R_D$ and an effective strong phase $\delta_D$. In this case, the interference happens between $b\to c$ transition, with $D$ decaying into doubly-Cabibbo-suppressed final states, and $b\to u$ transition, with $D$ decaying into Cabibbo favoured final state.  The BPGGSZ method~\cite{Giri:2003ty,Bondar:2005ki,Bondar:2008hh} refers to $D$ decaying into multi-body final state where the phase space is binned to make full use of the statistic power of the decay. In addition, the angle $\gamma$ can also be extracted from time-dependent CP violation measurements of $\Bs$ decays. The measurements from the LHCb experiments with these methods are discussed in detail in the following section. 

\subsubsection{GLW and ADS measurements}\label{sec:GLWADS}
The GLW and ADS channels are usually considered together for their similarities in their final states and experimental treatments. The GLW and ADS measurements have been performed in many decay channels to obtain the best sensitivity on $\gamma$. The full list of the measured channels can be found in Ref.~\cite{LHCb-PAPER-2021-033}. 

The decay rate of the GLW method is
\begin{equation}
\label{EQ_GLW}
    \Gamma(B^- \to D[\to f_{\textrm{GLW}}]h^{-}) \propto 1+r_B^2  + 2\kappa_B r_B(2F_+ -1) \cos(\delta_B - \gamma),
\end{equation}
where $F_+ = 1 (0)$ means a pure CP-even (CP-odd) state while decays with $F_+ =0.5$ does not have any sensitivity on $\gamma$. The value of $F_+$ can be determined from quantum-coherent data from BESIII and CLEO-c experiments in a model-independent way. For example, the decays $B^- \to DK^-$, $D \to \pi^+ \pi^- \pi^0$ and $D \to K^+ K^- \pi^0$ have been used to measure the angle $\gamma$ in LHCb using the measured $F_+^{\pi^+\pi^-\pi^0}$ and $F_+^{K^+K^-\pi^0}$ of $0.973 \pm 0.017$ and $0.73 \pm 0.06$, respectively~\cite{Malde:2015mha}. One can see that almost full sensitivity can be achieved in the $\pi^+\pi^-\pi^0$ channel without considering structures over the phase space while further binning the phase space of $K^+K^-\pi^0$ will help get more sensitivity due to small $F_+^{K^+K^-\pi^0}$. In fact, the amplitude analyses of $D^0 \to \pi^+ \pi^- \pi^0$ and $D^0 \to K^+ K^- \pi^0$  channels have both been performed~\cite{BaBar:2007dro,CLEO:2005uoz,BaBar:2007soq,CLEO:2006tjv}. The relative strong phase between $D^0 \to \rho^+ \pi^-$ and $D^0 \to \rho^- \pi^+$ is $(-2.0 \pm 0.8)^{\circ}$ while that between $D^0 \to K^{*}(892)^+K^-$ and $D^0 \to K^{*}(892)^-K^+$ is $(-37.0 \pm 2.9)^{\circ}$, this is consistent with the fact that $F_+^{\pi^+\pi^-\pi^0}$ is close to 1 and is much larger than $F_+^{K^+K^-\pi^0}$.
The parameters $r_B$ and $\delta_B$ are the amplitude ratio and phase difference between $b \to u$ and $b \to c$ processes, $\kappa_B$ is the coherent factor to take into account the sensitivity lost due to contamination of other contributions when the bachelor $h$ is a broad resonant structure, e.g. $K^*(892)$. The coherent factor $\kappa_B$ is obtained based on amplitude models used to describe multi-body $B$ decays. In Eq.~\ref{EQ_GLW} only $B^-$ is written; however, it also applies for $B^0$ decays. The sensitivity of $\gamma$ is directly linked to  the size of $r_B$ and $\kappa_B$, larger $r_B \times \kappa_B$ gives better sensitivity on $\gamma$. The CP measurements from the GLW channels are only sensitive to $\cos(\delta_B - \gamma)$ which has four-fold ambiguity on determination of angle $\gamma$.

In the GLW modes, as $D$ decays into CP eigenstates, in case of no CP violation in $D$ decays, the decay amplitudes and their phases are the same for $D^0$ and $\overline{D}^0$. However, LHCb has discovered direct CP violation in $D^0 \to K^+ K^-$ and $D^0 \to \pi^+ \pi^-$ by looking at the difference of CP violation between the two decay channels~\cite{LHCb-PAPER-2019-006}. This indicates small difference of amplitudes and phases between the two channels, which affects $\gamma$ determination. Studies~\cite{Wang:2012ie} show that the effect on $\gamma$ determination is smaller than 0.5$^{\circ}$. The CP violation measured between $D^0 \to K^+ K^-$ and $D^0 \to \pi^+ \pi^-$ has been considered in the LHCb $\gamma$ combinations shown later. 

\begin{figure}
\begin{center}
\includegraphics[width=0.45\textwidth]{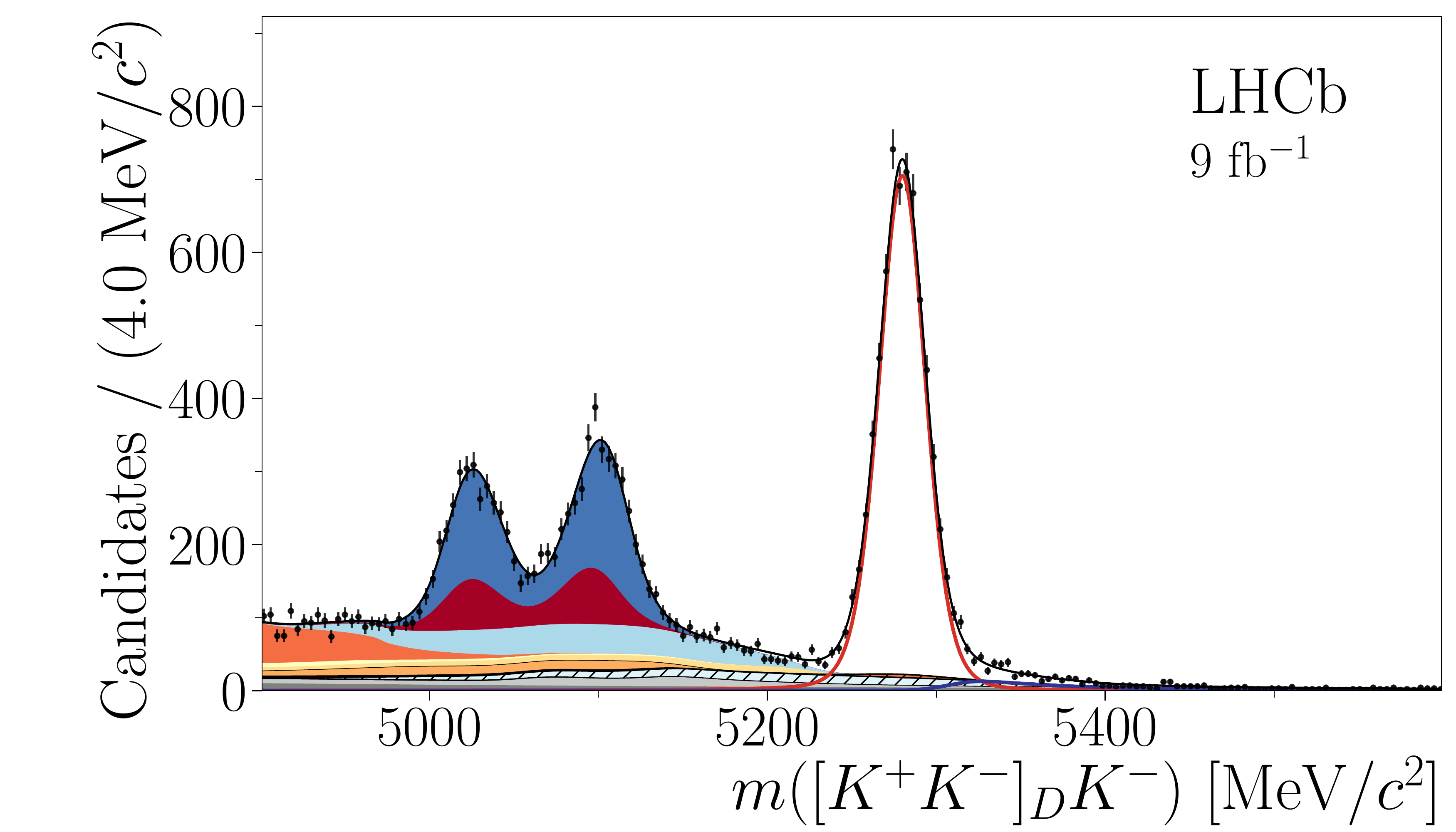}
\includegraphics[width=0.45\textwidth]{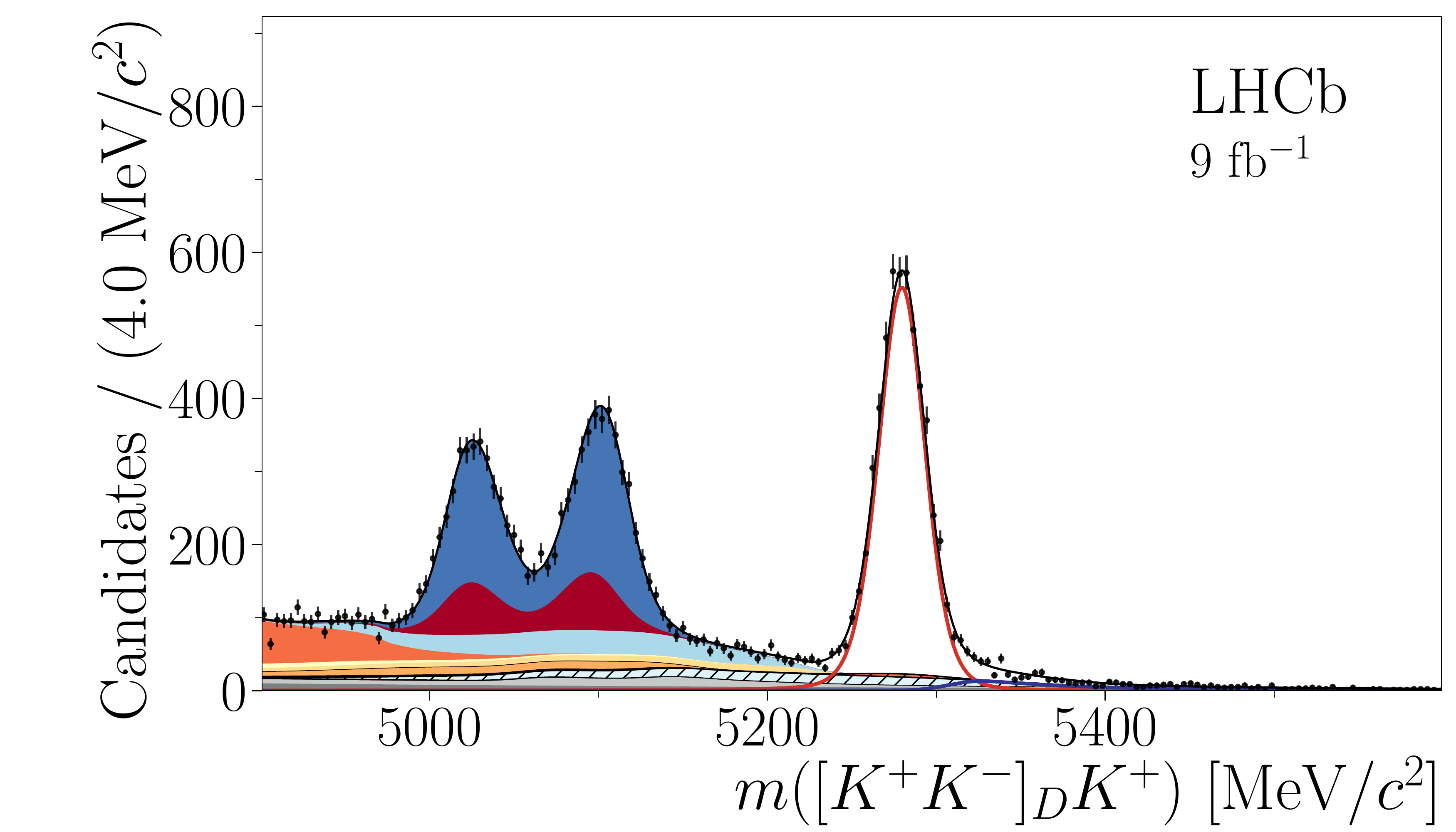}
\caption{ The invariant mass distributions of $(K^+K^-)_DK^+$ (left) and $(K^+K^-)_DK^-$ (right) from \cite{LHCb-PAPER-2020-036}.
}
\label{fig:GLW}
\end{center}
\end{figure}

Based on Eq.~\ref{EQ_GLW}, the angle $\gamma$ can be accessed by measuring the relative decay rate difference for $B^+$ and $B^-$ mesons. An example of the decay rate difference can be seen directly from the raw yields of the invariant mass distributions of the decays $B^{\pm} \to D(\to K^+K^-)K^{\pm}$ as shown in Fig.~\ref{fig:GLW}. The figures are from the latest LHCb measurements using $9\invfb$ data~\cite{LHCb-PAPER-2020-036}. By further considering the production asymmetry of $B^+$ and $B^-$ and detection efficiency difference between $K^+$ and $K^-$, the size of the CP violation of the decay can be determined. 

\begin{figure}
\begin{center}
\includegraphics[width=0.45\textwidth]{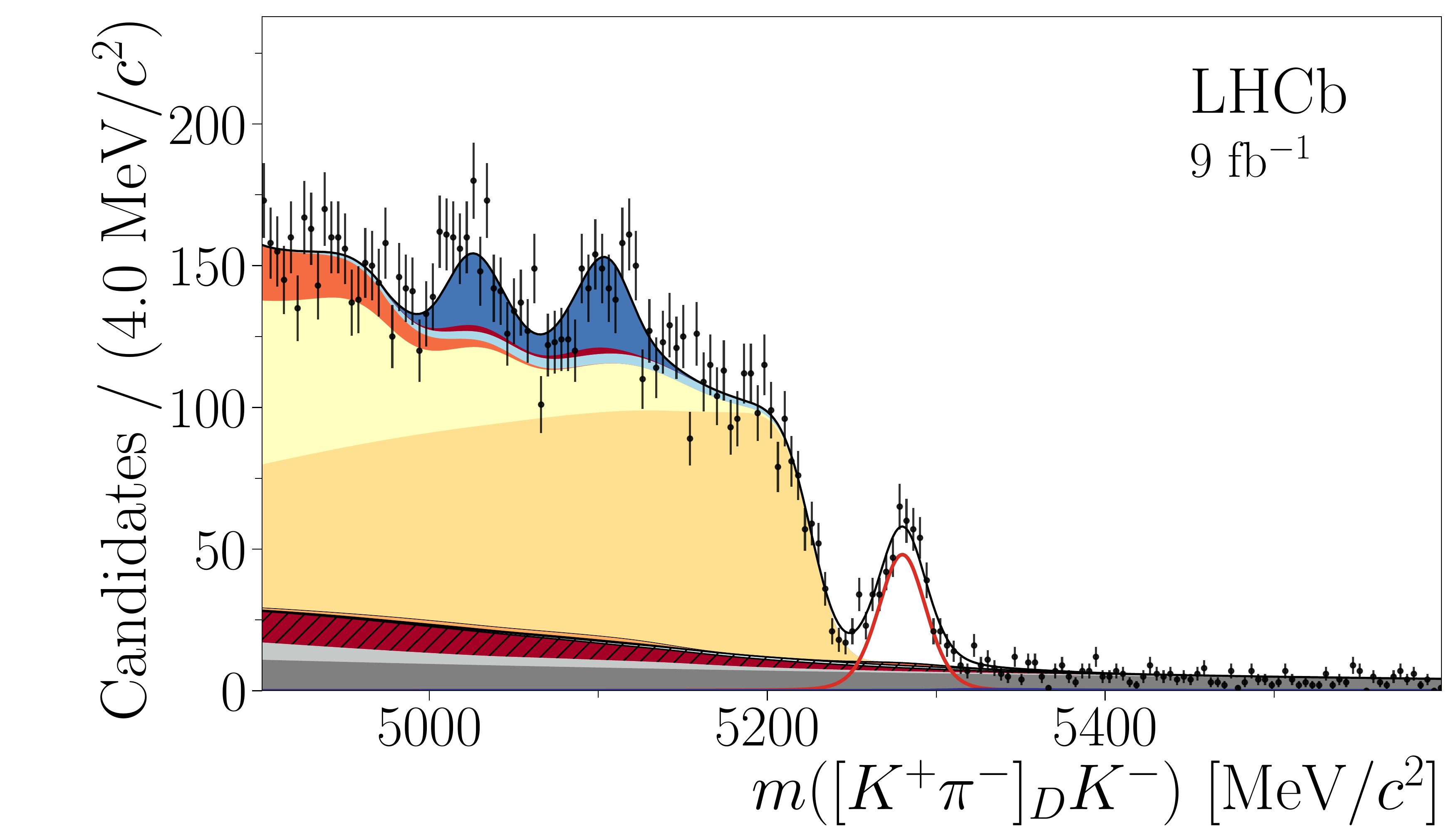}
\includegraphics[width=0.45\textwidth]{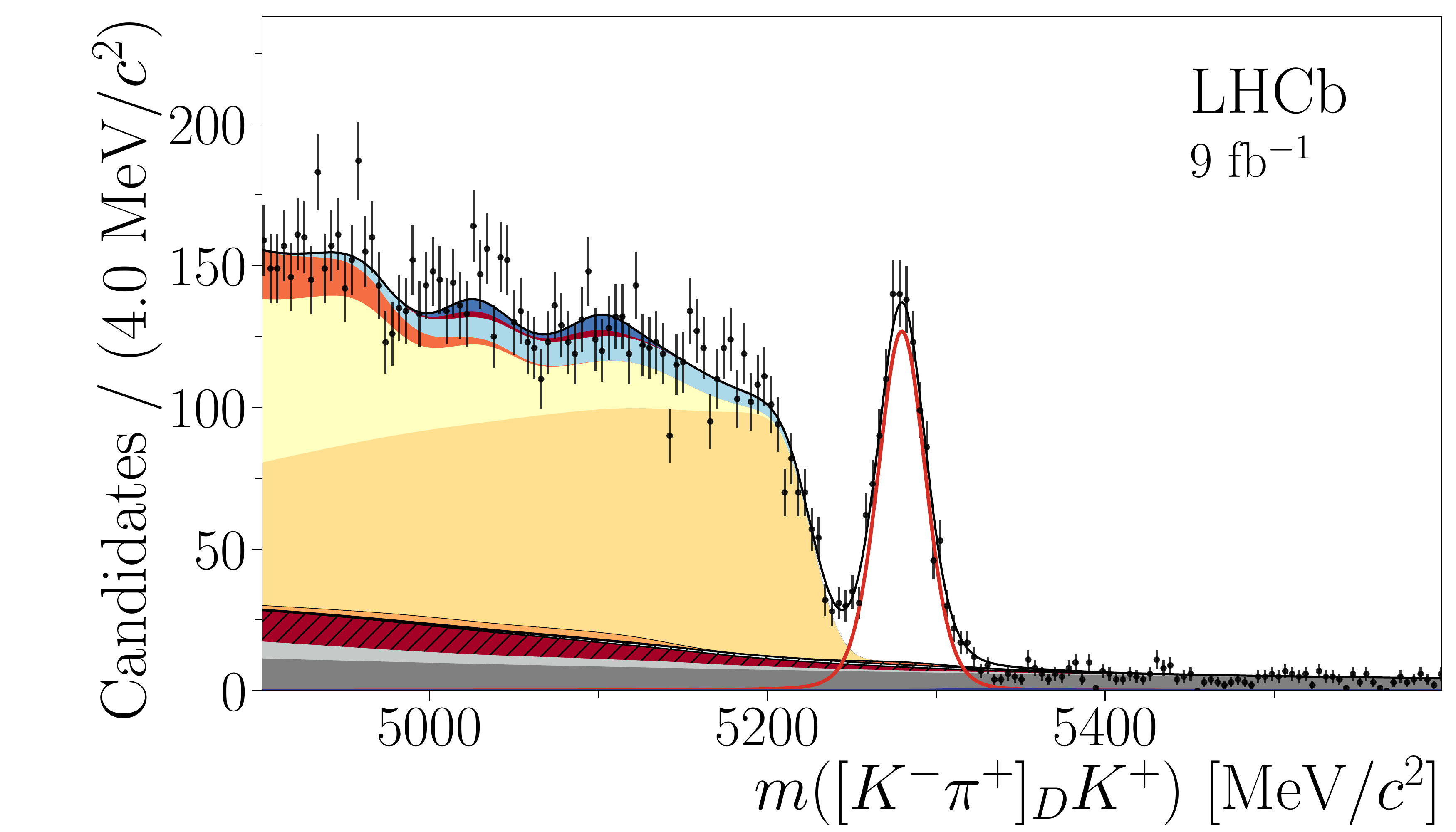}
\caption{ The invariant mass distributions of $(K^+\pi^-)_D K^+$ (left) and $(\pi^+K^-)_D K^-$ (right) from \cite{LHCb-PAPER-2020-036}.
}
\label{fig:ADS}
\end{center}
\end{figure}

The decay rate of the ADS channel has a similar form of 
\begin{equation}
 \Gamma(B^- \to D[\to f_{\textrm{ADS}}]h^-) \propto r_D^2+r_B^2  + 2\kappa_B r_B r_D R_D \cos(\delta_D + \delta_B - \gamma),
\end{equation}
where $r_D$ and $\delta_D$ are the average amplitude ratio and phase difference between doubly Cabibbo-suppressed and Cabibbo-favoured $D$ decays. The value of $r_D$ is at similar magnitude as $r_B$ and thus leads to larger CP violation. However, the statistics of the ADS channel is also suppressed. This can be seen from the measurements done by the LHCb experiment using $B^{\pm} \to (K^{\mp}\pi^{\pm})K^{\pm}$ decays as show in Fig.~\ref{fig:ADS}~\cite{LHCb-PAPER-2020-036}.
The measurements also have four-fold ambiguity, however, as $\delta_D$ is not zero, combining with GLW mode gives two-fold ambiguity on determination of angle $\gamma$. 

The coherent factor $R_D$ equals to one for two-body final states and is less than one for multi-body final states to take into account dilutions due to different resonant contributions from Cabibbo-favoured and doubly-Cabibbo-suppressed $D$ decays. 
The values of $r_D$, $R_D$ and $\delta_D$ can also be determined using the quantum-coherent data from BESIII and CLEO-c experiments. For example, the coherent factor of $\Dz(\Dzb) \to K^+\pi^+ \pi^- \pi^-$, $R_{K3\pi}$ is measured by the two experiments to be $0.43^{+0.17}_{-0.13}$~\cite{Evans:2016tlp}, which is significantly lower than one. Further binning the phase space according to the variation of strong phase determined from amplitude analysis can improve the sensitivity on $\gamma$~\cite{Evans:2019wza}, and a first measurement using this approach is recently published~\cite{LHCb-PAPER-2022-017}.

The GLW and ADS measurements have been performed by the LHCb experiment with $D$ decaying mainly into charged final states. However, final states with neutral particles are also studied, either by reconstructing the neutral particles $\pi^0$ or $\gamma$ as in recent study of $B^{\pm}\to D h^{\pm}, D \to h^{\pm}h^{'\mp}\piz$~\cite{LHCb-PAPER-2021-036}, or in a partially reconstructed method, where $\pi^0$ and $\gamma$ from $D^{*0}$ are not reconstructed~\cite{LHCb-PAPER-2020-036}. In the second case, the reconstruction efficiency is much higher, however, the sensitivity of $\gamma$ is limited by the background contributions, especially those from $B^0 \to D^{*-}K^+$ decays which has similar line-shapes as signal.

\subsubsection{Measurements using the GGSZ method}
As has discussed in the above section, for multi-body $D$ decays, if CP even fraction $F_+$ for self-conjugated decay or coherent factor $R$ for semi-flavour tagged decay is significantly smaller than one, further sensitivity can be achieved by considering the variation of $r_D$ and $\delta_D$ over the phase space. One can model the $D$ decays with an amplitude model which provides the information of strong phase $\delta_D$ over the phase space. However, this may suffer from large systematic uncertainties due to modelling of amplitude distributions. 

\begin{figure}
\begin{center}
\includegraphics[width=0.45\textwidth]{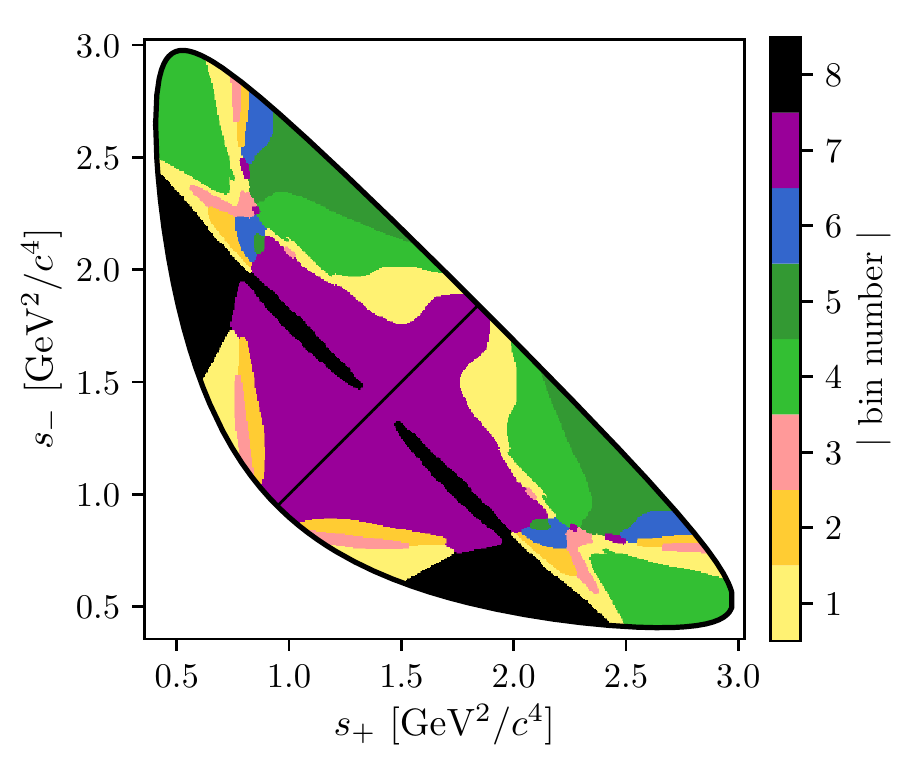}
\includegraphics[width=0.45\textwidth]{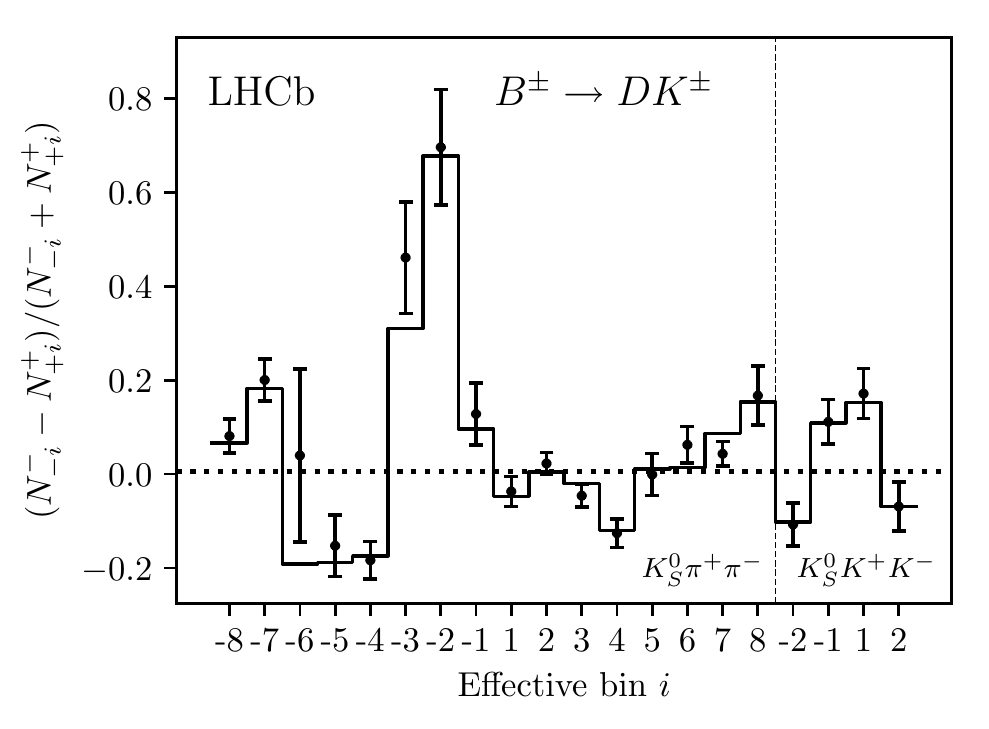}
\caption{ 
(Left) Binning scheme of $D \to \KS \pi^+ \pi^-$ decays used for $\gamma$ measurements; (right) Observed raw CP asymmetries and the predicted values using obtained CP parameters in different bins~\cite{LHCb-PAPER-2020-019}. 
}
\label{fig:GGSZ}
\end{center}
\end{figure}

An alternative method~\cite{Bondar:2005ki, Bondar:2008hh} is to bin the phase space and the effective $r_D^{i}$ and $\delta_D^{i}$ (or $c_i$ and $s_i$) in bin $i$ are defined as \begin{equation}
    c_i + i s_i \equiv R_D^{i} e^{i\delta_D^{i}}  \equiv \frac{\int_{\textrm{bin i}} A_f(p) \overline{A}_f(p)^* dp}{\sqrt{\int_{\textrm{bin i}} |A_f(p)|^2 dp} \sqrt{\int_{\textrm{bin i}} |\overline{A}_f(p)|^2dp}}.
\end{equation}
The GGSZ mode refers to decays into self-conjugated final states. BPGGSZ is used when referring to the analysis method using a binned strategy. In the BPGGSZ analysis, $c_i$ and $s_i$ are used instead of effective $R_D$ and $\delta_D$ for the benefit of better statistical performance. 
The values of $c_i$ and $s_i$ can be determined using quantum coherent data collected by BESIII and CLEO-c experiments, where a mixture of $D^0$ and $\overline{D}^0$ can be achieved from $\psi(3770) \to D^0 \overline{D}^0$. As $\psi(3770)$ is a parity-odd state, the $D^0$ and $\overline{D}^0$ are in quantum-correlated state of $(|D^0 \overline{D}^0> - |\overline{D}^0 D^0>)/\sqrt{2}$~\cite{CLEO:2010iul,BESIII:2020hlg, BESIII:2020khq, BESIII:2020hpo}.
Using the measured $c_i$ and $s_i$ values, the angle $\gamma$ and strong parameters $r_B$ and $\delta_B$ can be extracted. The method has been applied to $B^- \to DK^-$, $D \to \KS \pi^+ \pi^- $ and $D \to \KS K^+ K^-$ decays by LHCb~\cite{LHCb-PAPER-2018-017, LHCb-PAPER-2020-019}, where 16 bins and 4 bins are used according to the statistics of the decays, respectively. In order to optimise the sensitivity, binning schemes are chosen according to the strong phase variation over the Dalitz plot considering possible background contamination and efficiency effects. Symmetry between $\KS \pi^+ (\KS K^+)$ and  $\KS \pi^- (\KS K^-)$ is also used to increase the sensitivity on $c_i$ and $s_i$. The binning scheme optimised for the $\KS \pi^+ \pi^-$ decays is shown in Fig.~\ref{fig:GGSZ}~\cite{BESIII:2020hlg, BESIII:2020khq}.

Each bin of the BPGGSZ method can offer constraints to the angle $\gamma$ and to the strong parameters $r_B$ and $\delta_B$. The number of measurements is much more than the number of unknown parameters, thus global production asymmetry and detection asymmetry can be treated as fit variables and related systematic uncertainties  are reduced. While for the ADS and GLW methods, production asymmetry and detection asymmetry have to be considered using control channels.   
The measured yield difference in each bin is shown in Fig.~\ref{fig:GGSZ} for the $\KS\pi^+\pi^-$ and $\KS K^+K^-$ decays. Clear CP violation can be found. By combining the statistical power of these bins, the angle $\gamma$ is measured to be $(68.7^{+5.2}_{-5.1})^{\circ}$~\cite{LHCb-PAPER-2020-019}. 

\subsubsection{Multi-body $B$ decays}
Similar to multi-body $D$ decays, multi-body $B$ decays can also be used where $r_B$ and $\delta_B$ now is a function of $B$ decay Dalitz plot. However, in this case, one can only use a model to describe different resonant contributions, where some can only be obtained through $b \to c$ process and some can be obtained through both $b \to c$ and $b \to u$ processes and the interference between them gives sensitivity to the angle $\gamma$. The measurements have been performed by the LHCb collaboration in $B^0 \to D K^+ \pi^-$ decays with $D \to K^+K^- (\pi^+\pi^-)$~\cite{LHCb-PAPER-2015-059} using 3 fb$^{-1}$ Run 1 data. However, due to limited statistics, the sensitivity on $\gamma$ is still low. Further measurements with all the data collected by the LHCb experiment will be very interesting. 

\subsubsection{Time-dependent $\Bs$ decays}
The angle $\gamma$ can be measured through time-dependent $\Bs$ and $B^0$ decays where the weak phases extracted are $(\gamma - 2\beta_s)$ and $(\gamma + 2\beta)$, respectively. In hadron collider experiments like LHCb, the golden channels are $\Bs \to D_s^{\pm} K^{\mp}$, $\Bs \to D_s^{\pm} K^{\mp} \pi^+\pi^-$ and $\Bs \to D \phi$ decays. The time-dependent analyses have been performed for the first two channels~\cite{LHCb-PAPER-2020-030,LHCb-PAPER-2017-047}, while only branching fraction has been measured for the $\Bs \to \overline{D}^0 \phi$ decay~\cite{LHCb-PAPER-2013-035,LHCb-PAPER-2018-015}. 

As $\Bs$ mixing is involved, a time-dependent analysis is needed to extract CP parameters. The time-dependent decay rate of the $\Bs$ decay into a final state $f$ is given by
\begin{equation}
    \frac{d\Gamma_{\Bs\to f}(t)}{d t} \propto e^{-\Gamma_s t} [\cosh(\frac{\Delta \Gamma_s t}{2}) + A_f^{\Delta \Gamma} \sinh(\frac{\Delta \Gamma_s t}{2}) + C_f \cos(\Delta m_s t) - S_f\sin(\Delta m_s t)],
\end{equation}
where $\Delta \Gamma_s = \Gamma_{B_L} - \Gamma_{B_H}$ and $\Delta m_s = m_{B_H} - m_{B_L}$ are the decay-width and mass differences between the light ($B_L$) and heavy ($B_H$) $\Bs$ mass eigenstates and $\Gamma_s$ is the average $\Bs$ decay width. For the decays to the CP-conjugated final states, the CP violation parameters $C_f$, $S_f$ and $A_f^{\Delta \Gamma}$ are replaced with $C_{\overline{f}}$, $S_{\overline{f}}$ and 
$A_{\overline{f}}^{\Delta \Gamma}$. These CP violation parameters are related with $\gamma$ through
\begin{eqnarray}\label{Eq:TD_Gamma}
    C_f &=& -C_{\overline{f}} =  \frac{1-r_B^2}{1+r_B^2}, \\\nonumber
    S_f &=& \frac{2r_B \sin(\delta_B-(\gamma-2\beta_s))}{1+ r_B^2}, S_{\overline{f}} = \frac{-2r_B \sin(\delta_B+(\gamma-2\beta_s))}{1+ r_B^2}, \\\nonumber
    A_f^{\Delta \Gamma} &=& \frac{-2r_B\cos(\delta_B - (\gamma-2\beta_s))}{1+ r_B^2}, A_{\overline{f}}^{\Delta \Gamma} = \frac{-2r_B\cos(\delta_B + (\gamma-2\beta_s))}{1+ r_B^2}
\end{eqnarray}
following the definitions in Ref.~\cite{LHCb-PAPER-2014-038}.
In the formulae above, we have assumed no CP violation in either the mixing and in the decay amplitude. 

The time-dependent CP violation has been measured using 3.0 fb$^{-1}$ and 9.0 fb$^{-1}$ data for $\Bs \to D_s^{\pm} K^{\mp}$~\cite{LHCb-PAPER-2017-047} and $\Bs \to D_s^{\pm} K^{\mp} \pi^+ \pi^-$~\cite{LHCb-PAPER-2020-030}, respectively. The measured values of the angle $\gamma$, using the world-average value of $-2\beta_s$, are $\gamma = (128^{+17}_{-22})^{\circ}$ and $\gamma = (44 \pm 12)^{\circ}$, respectively (modulo 180$^{\circ}$). The $\Bs \to D_s^{\pm} K^{\mp} \pi^+ \pi^-$ is complicated due to multiple bachelor particles, and an amplitude analysis is needed. However, as discussed before, multi-body $B$ decays effectively introduce a dilution factor $\kappa$, as Eq.~\ref{Eq:TD_Gamma} have five constraints while together with $\kappa$, there are four unknown variables. The measurement is performed in the same paper in a model-independent way, which leads to $\gamma = (44^{+20}_{-13})^{\circ}$. The sensitivity is worse than model-dependent results as expected.

Sensitivity studies with $\Bs \to D \phi$ using a time-integrated method has also been performed~\cite{Ao:2020cwh}, the expected statistical sensitivity of $\gamma$ is about $(8-19)^{\circ}$ using the \mbox{9\invfb} $pp$ collision data collected by the LHCb experiment. It is pointed out that additional sensitivity on $\gamma$ can be achieved using the longitudinal polarised part of $\Bs \to D^* \phi$ decays with a partially reconstructed technique~\cite{LHCb-PAPER-2018-015}. Besides, further sensitivities on $\gamma$ can also be obtained by using other quasi-two-body decays in $\Bs \to D K^+K^-$ decays~\cite{Wang:2011zw}.

\subsubsection{Combination on $\gamma$}
\label{sec:04:01:05gammaCombination}
A combination of the parameter $\gamma$ was recently performed using all available results of $\gamma$ measurements from the LHCb experiment~\cite{LHCb-PAPER-2021-033}.
The determination of $\gamma$ relies on the inputs from charm decays; on the other hand, precise $\gamma$ measurements and strong parameters of $B$ decays can offer valuable constraints on the charm parameters, which in turn helps constraining the mixing parameters of $\Dz$ mesons. 
Therefore, the measurements that are sensitive to the charm mixing parameters are also used in the combination. The CKM angle $\gamma$ and charm mixing parameters are simultaneously determined with significant improvements.
Here we focus on the $\gamma$ combination results, leaving those on the charm mixing parameters to be discussed later in Section~\ref{sec:05:01:CharmMixing}.

\begin{figure}
\begin{center}
\includegraphics[width=0.6\textwidth]{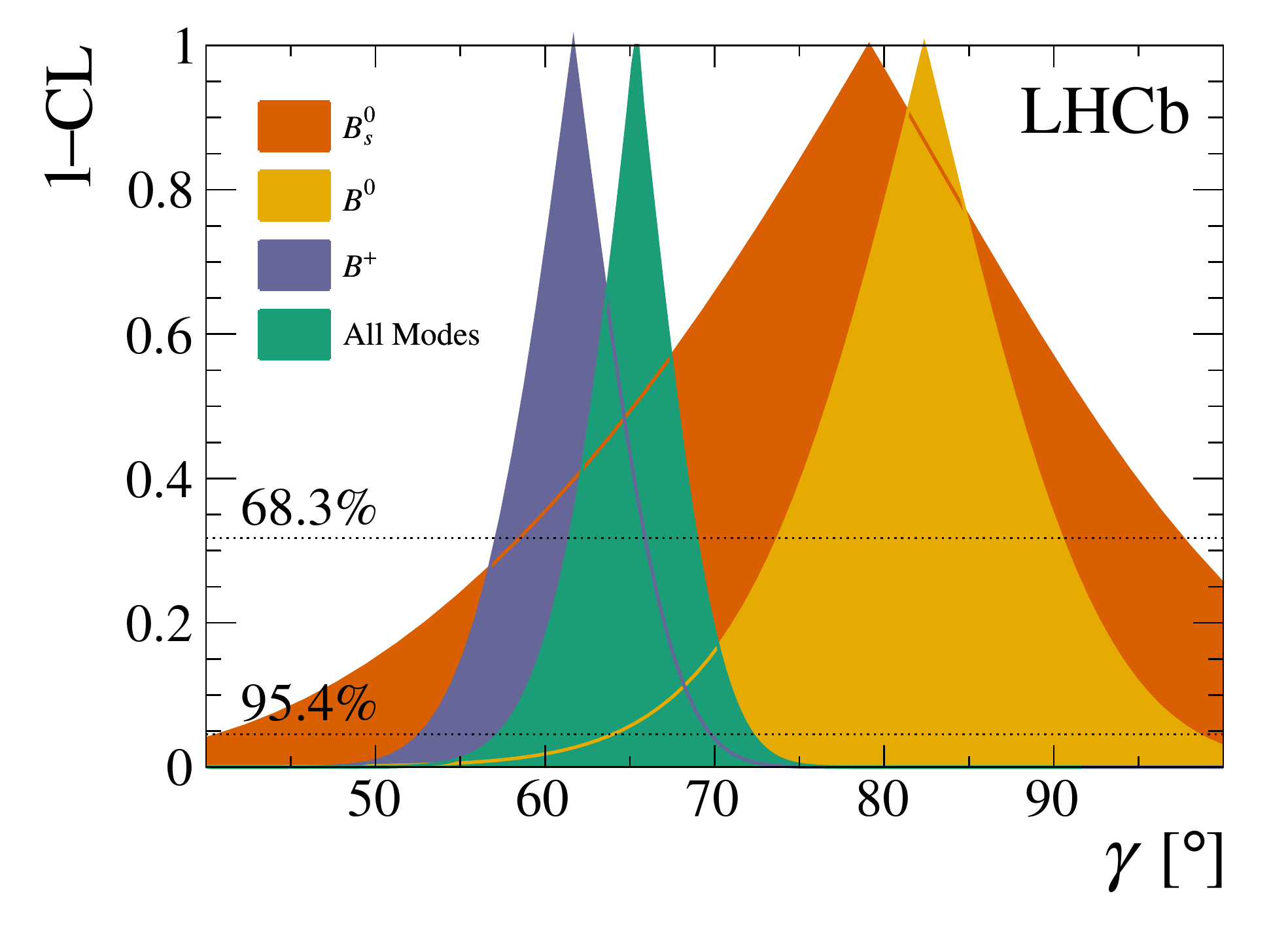}
\caption{$\gamma$ combinations from different $B$ hadrons~\cite{LHCb-PAPER-2021-033}. 
}
\label{fig:Comb_B}
\end{center}
\end{figure}
Figure~\ref{fig:Comb_B} shows the $\gamma$ contributions from different $B$ decays.
The sensitivity on $\gamma$ mainly comes from $B^+$ decays. 
For $B^0$ and $\Bs$ decays, where multi-body decays are usually involved, only a few channels have been studied due to the complications in analysis procedures.
The small sample sizes and small number of available measurements limit their sensitivity on $\gamma$.
The central values of $\gamma$ determination from $B^0$ and $\Bs$ are around $20\degrees$ higher, which motivates further measurements with the $B^0$ and $\Bs$ mesons to check overall consistency between different $B$ mesons.
Special efforts to the $\Bs$ meson are well worth, since the $\Bs$-$\Bsb$ mixing is involved and new physics contributions can easily enter inside. 

\begin{figure}
\begin{center}
\includegraphics[width=0.45\textwidth]{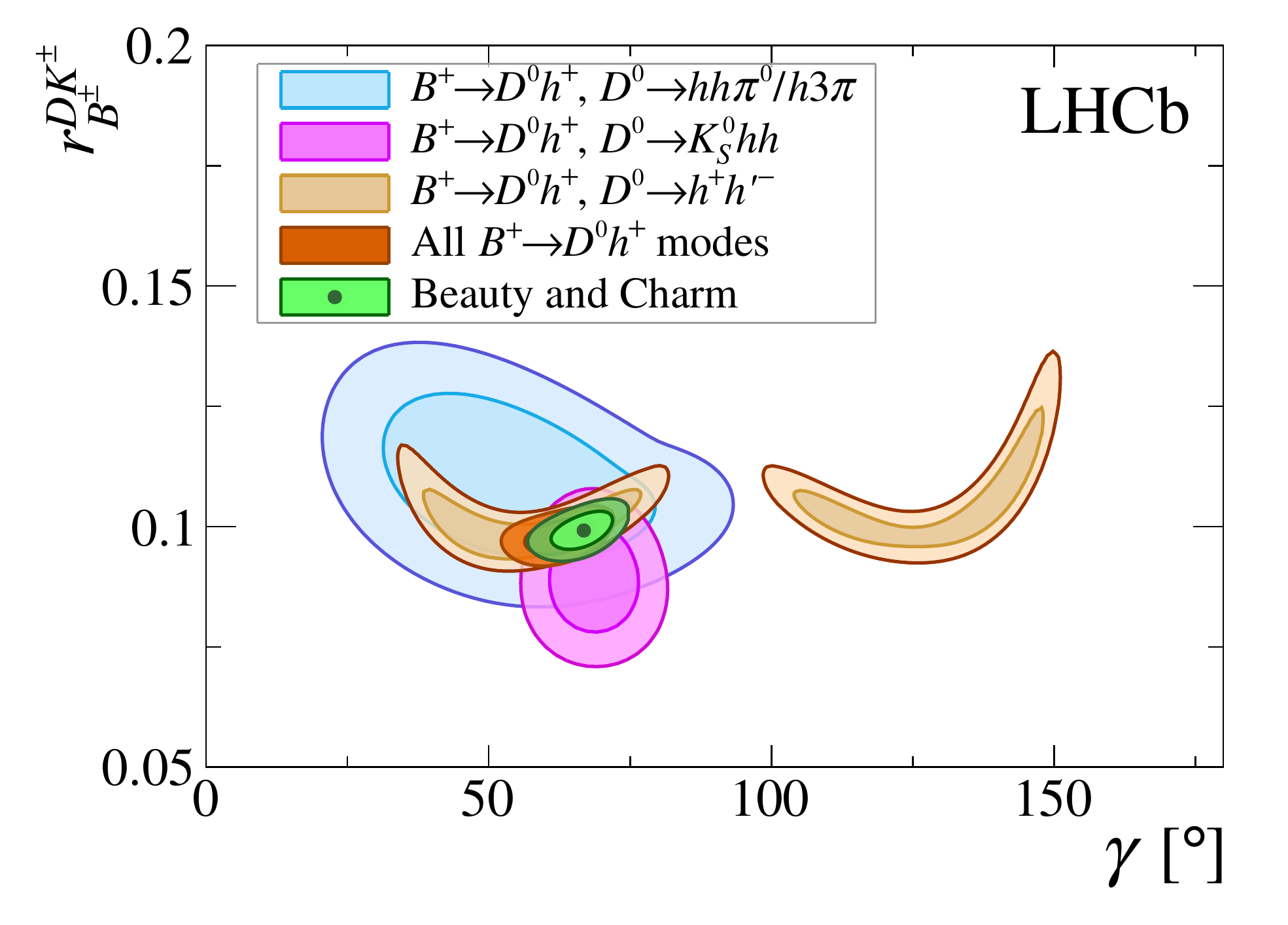}
\includegraphics[width=0.45\textwidth]{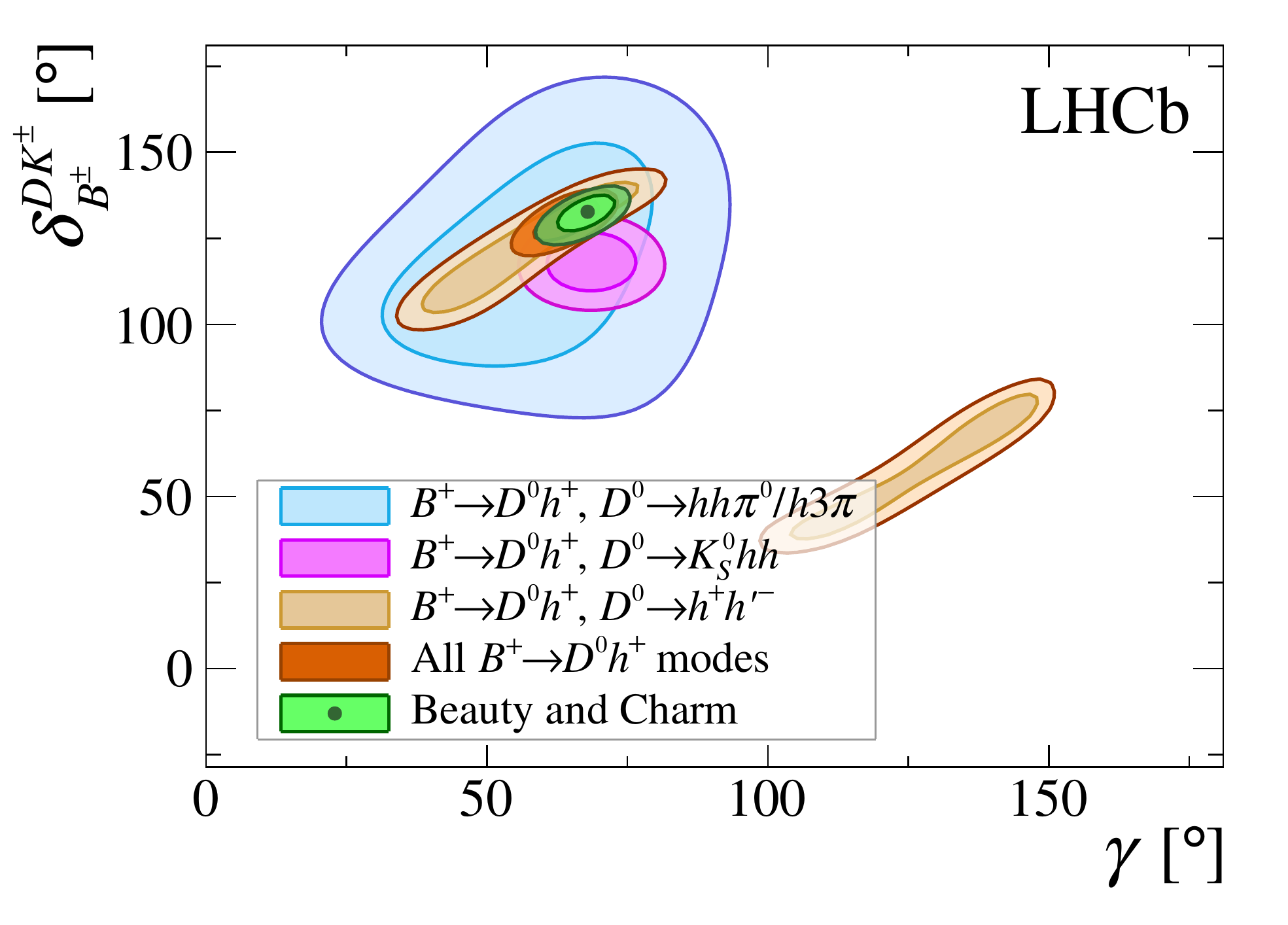}
\caption{Combinations of $\gamma$ from all LHCb measurements, 2D contours of $r_B$ vs $\gamma$ (left) and $\delta_B$ vs $\gamma$ (right) have been shown at 68.3\% confidence level~\cite{LHCb-PAPER-2021-033}.
}
\label{fig:Comb_rB_deltaB}
\end{center}
\end{figure}

The 1$\sigma$ contour of the constraint on $\gamma$ and strong parameters of $B^+$ decays are shown in Fig.~\ref{fig:Comb_rB_deltaB}. Ambiguities on determination of $\gamma$ from two-body $D$ decays can be seen as has been discussed in sec.~\ref{sec:GLWADS}, however, they constrain the parameter space into a very narrow region and together with the unique determination of $\gamma$ from the BPGGSZ channel, $\gamma$ can be precisely obtained.
The combined $\gamma$ is found to be $\gamma = (65.4^{+3.8}_{-4.2})^{\circ}$ and is the most precise determination from a single experiment. 

\subsection{CKM angle \texorpdfstring{$\beta$}{TEXT}}
The angle $\beta$, defined as arg$(-V_{cd}V_{cb}^*/V_{td}V_{tb}^*)$, is approximately the phase of $V_{td}^*$ in the Wolfenstein parameterisation~\cite{Wolfenstein:1983yz}.
 It enters the  decay time distributions  of \Bd  and $\Bzb$ meson decays   due to $B^0$-$\Bzb$ oscillation. 
 The effective value of
 $\sin(2\beta)$, which could have been altered by NP contributions in $B^0$-$\Bzb$  mixing , can be extracted from the time-dependent CP asymmetries of  \Bd  decays  via $b \to \ccbar s$ transitions  following the relation 
\begin{equation}
    A_{\textrm{CP}}(t) = \frac{\Gamma(\Bzb(t)\to f_{\rm CP}) - \Gamma(B^0(t)\to f_{\rm CP})}
    {\Gamma(\Bzb(t)\to f_{\rm CP}) + \Gamma(B^0(t)\to f_{\rm CP})} \approx \eta_{f} \sin(2\beta)\sin(\Delta mt)\;,
\end{equation}
where $f_{\rm CP}$ is a CP eigenstate with eigenvalue $\eta_f$, 
and the approximation assumes no  CP violation in the mixing or decay. Any significant deviation of the measured $\sin(2\beta)$ value from the indirect determination of $\sin(2\beta)$ through a global CKM fit excluding  $\sin(2\beta)$  measurements is a clear sign of NP.

The $\ccbar$ pair could appear either 
in a charmonium meson or in two charmed mesons
in the final state. 
The precision of the $\beta$  is mainly driven by \Bd decays to charmonium final-states due to their large decay rates and distinct characteristics for identification. 
The world-average of $\sin(2\beta)$ of all the charmonium measurements is  sin$(2\beta)=0.699 \pm 0.017$~\cite{HFLAV18}. The LHCb experiment has performed measurements of $\sin(2\beta)$
in the decays $\Bd \to \jpsi(\to \mumu) K^0_S$~\cite{LHCb:2015ups}, $\Bd \to \jpsi(\to e^+e^-) K^0_S$  and  $\Bd \to \psi(2S)(\to \mumu) K^0_S$~\cite{LHCb:2017mpa}. The combined  value is $\sin(2\beta)=0.760\pm 0.034$, 
the precision of which is already comparable to that of the  \babar result  $\sin(2\beta)=0.09\pm 0.03 \pm 0.01$~\cite{BaBar:2009byl} and the Belle result $\sin(2\beta)=0.67\pm 0.02 \pm 0.01$~\cite{Belle:2012paq}.
An improvement  by a factor of two is expected from measurements including  LHCb Run 2 data. 

The presence of small penguin contributions in $b \to \ccbar s$ processes  may shift the measured values of $\sin(2\beta)$ by up to few percent~\cite{Ciuchini:2005mg}. The decays via tree-level $b \to \cquark\uquarkbar\dquark$ transitions, though having smaller signal yields due to the small branching fractions of $D$ decays, are free of the  penguin effects and thus theoretically clean. 
A recent joint analysis  of the decay
$\overline{B}^0 \to Dh^0$ with  $D\to \KS h^+ h^-$ by  the \babar and Belle  experiments
measured $\sin(2\beta)=0.80 \pm 0.14\pm 0.04\pm 0.03$ and $\cos(2\beta)=0.91 \pm 0.22 \pm 0.09 \pm 0.07$, which ruled out the other solution of $\beta$ at 7.3$\sigma$~\cite{BaBar:2018agf,BaBar:2018cka}.
Analysis of this decay is challenging at LHCb due to the presence of  $\KS$ and $\pi^0$ mesons in the final state. On the other hand, 
the decay $\overline{B}^0 \to D\pi^+\pi^-$ followed by $D\to K^+K^- (\pi^+\pi^-)$ only involves charged particle  and thus is ideal for LHCb to pursue.

\subsection{CKM angle \texorpdfstring{$\beta_s$}{TEXT}}
 The angle $\beta_s$, defined as $-$arg$(-V_{cb}V_{cs}^*/V_{tb}V_{ts}^*)$, is approximately the phase of $V_{ts}$ in the Wolfenstein parameterisation.
 The effective value of $-2\beta_s$  can be measured in the time-dependent CP asymmetries of $\Bs$ decays to  CP eigenstates via $b \to \ccbar s$ transitions, and is  denoted $\phi_s^{\ccbar s}$.
In contract to the angle $\beta$, $\beta_s$ is very small. The SM prediction for $\phi_s $ is $\phi_s^{\rm SM} = -2\beta_s=-0.03696 \pm 0.0004$~\cite{CKMfitter2015}, which is 
subject to small corrections due to the neglected penguin contributions in  $b \to \ccbar s$ decays. 
Presence of new particles  in $\Bs$-$\Bsb$ mixing diagrams may have a sizeable effect on  $\phi_s$, making it an sensitive probe of physics beyond the SM.

The  LHCb experiment has performed measurements of $\phi_s^{\ccbar s}$ in the decays $\Bs \to \jpsi(\to \mumu) \phi$~\cite{LHCb-PAPER-2019-013}, $\Bs \to \jpsi(\to \mumu) \pi^+ \pi^-$~\cite{LHCb-PAPER-2019-003}, $\Bs \to \jpsi(\to \mumu) K^+ K^-$ with $m(K^+K^-)>1.05\gevcc$~\cite{LHCb-PAPER-2017-008}, $\Bs \to \psitwos(\to \mumu)\phi$~\cite{LHCb-PAPER-2016-027}, $\Bs \to D_s^+ D_s^-$~\cite{LHCb-PAPER-2014-051} and more recently in $\Bs \to \jpsi(\to \epem) \phi$~\cite{LHCb-PAPER-2020-042}. A combination of the measurements  in $\Bs \to \jpsi(\to \mu^+\mu^-) \phi$ and  $\Bs \to \jpsi(\to \mu^+\mu^-) \pi^+ \pi^-$ obtained using data taken during 2011-2016 gives $\phi_s^{\ccbar s}= -0.042 \pm 0.025$ rad~\cite{LHCb-PAPER-2019-013}. 

\begin{figure}
\centering
\includegraphics[width=0.45\textwidth]{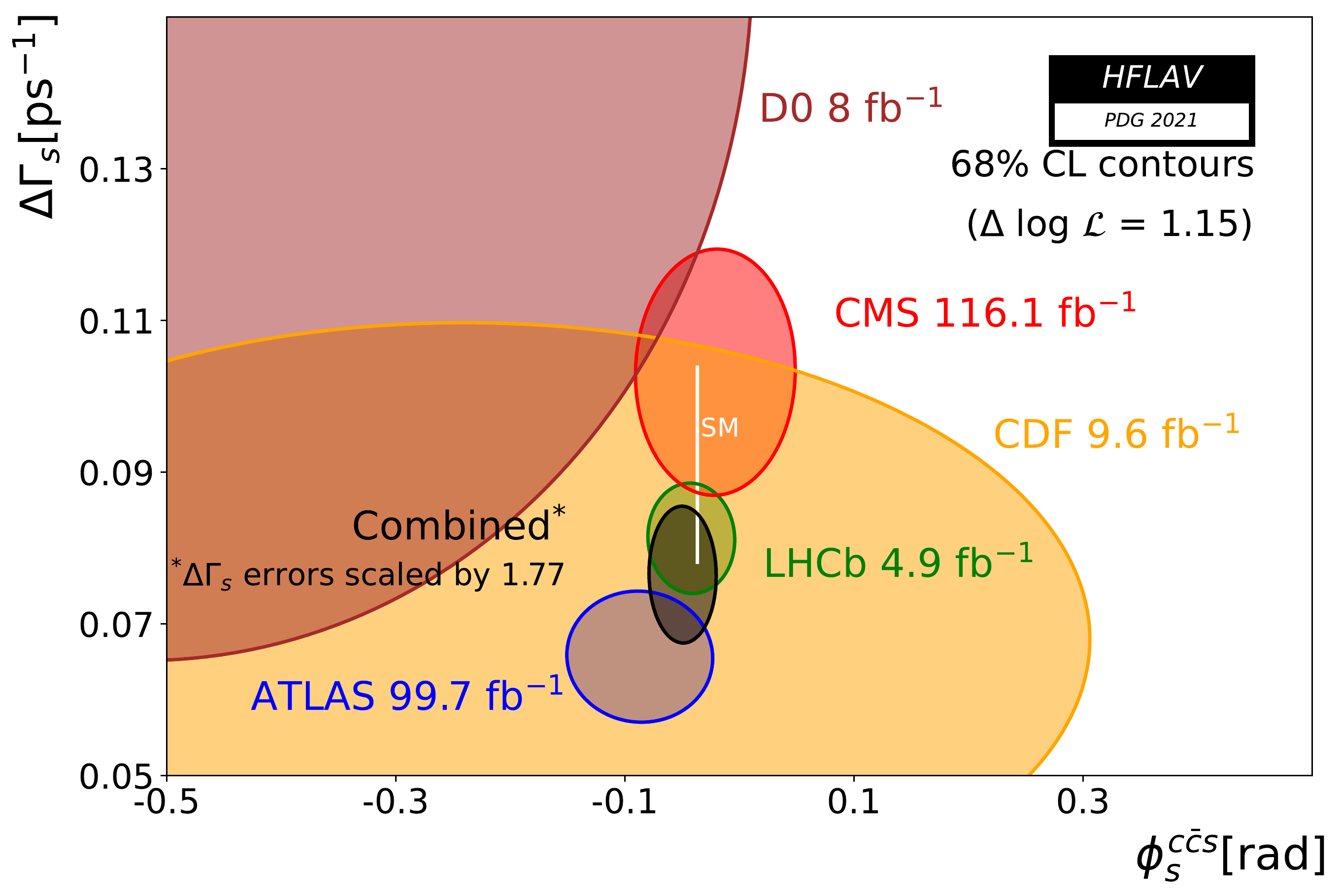}
\includegraphics[width=0.46\textwidth]{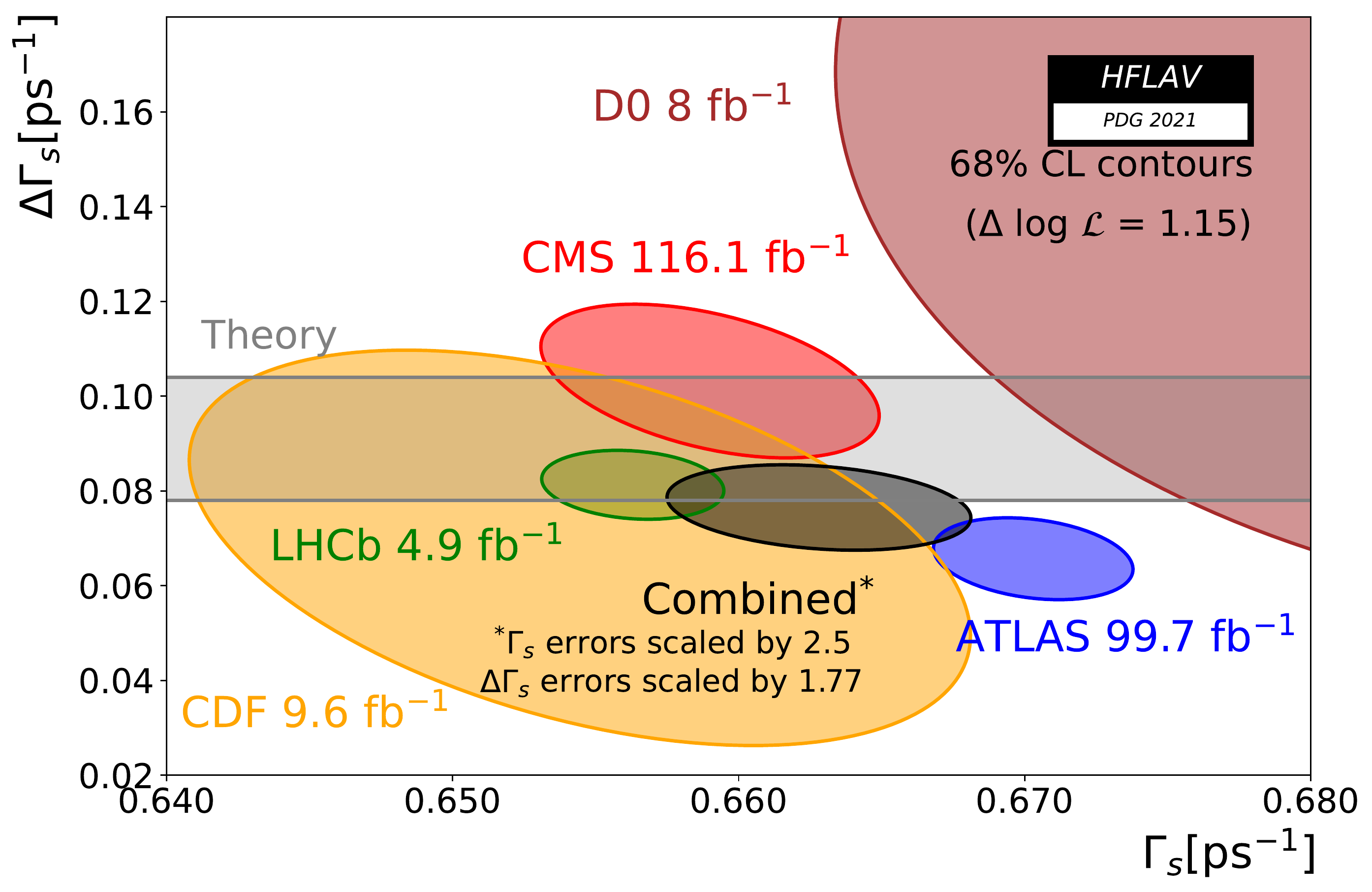}
\caption{ Combination  of $\phi_s^{\ccbar s}$ measurements by HFLAV~\cite{HFLAV18}, where 2D contours of (left) $\phi_s^{\ccbar s}$ vs $\Delta\Gamma_s$ and (right) $\Gamma_s$ vs $\Delta \Gamma_s$ are displayed at 68\% confidence level.
}\label{fig:PhisDGs}
\end{figure}

The average of all LHCb measurements of  $\phi_s^{\ccbar s}$ 
is compared with the ATLAS~\cite{ATLAS:2020lbz} and  CMS~\cite{CMS:2020efq} results in Fig.~\ref{fig:PhisDGs}, where 2-dimensional contours in the plane of $\phi_s^{\ccbar s}$ versus the \Bs decay width difference ($\Delta \Gamma_s$) are displayed at 68\% confidence level~\cite{HFLAV18}. 
Note ATLAS and CMS have only performed measurements in $\Bs \to \jpsi(\to \mumu) \phi$ decays, due to constraints from their trigger systems,. 
The current world-average value is $\phi_s^{\ccbar s}= -0.050 \pm 0.019$~rad~\cite{HFLAV18}. 
In Fig~\ref{fig:PhisDGs}, one can see a good agreement in the $\phi_s^{\ccbar s}$ measurements
from different measurements. However, some tension is observed for  $\Delta \Gamma_s$ and $\Gamma_s$. Factors of 2.5 and 1.77 have been applied to scale up the uncertainties of $\Gamma_s$ and $\Delta \Gamma_s$ in the combination. Further investigations by the relevant experiments are needed to solve this problem.
With the uncertainty of $\phi_s^{\ccbar s}$
well below its SM value, 
the study of CP violation in \Bs decays enter an era of precision test, and control of the penguin pollution in $\phi_s^{\ccbar s}$ using data-driven methods is essential for identification of NP signals in  $\Bs$-$\Bsb$ mixing~\cite{Liu:2013nea, Faller:2008zc, Nagahiro:2008cv, DeBruyn:2010hh, Jung:2012mp, DeBruyn:2014oga, Frings:2015eva, Barel:2020jvf}. 
In this regard, the LHCb experiment has measured CP violation in the penguin-enhanced $b\to \ccbar d$ decays $\Bd \to \jpsi \rho^0$~\cite{LHCb-PAPER-2014-058} and $\Bs \to \jpsi \Kstarzb$~\cite{LHCb-PAPER-2015-034} decays. The measurements  are used to estimate the penguin shift of $\phi_s^{\ccbar s}$ measured in $\Bs \to \jpsi \phi$,  assuming SU(3) flavour symmetry. The shift is found to be compatible  with zero~\cite{LHCb-PAPER-2015-034}, with an uncertain well below the statistical uncertainties  of the current $\phi_s^{\ccbar s}$ measurements.

Similar CP-violating phases can also be measured in  decays of \Bs mesons via
$b \to \ssbar s$ and $b \to \ddbar s$ transitions,  denoted $\phi_s^{\ssbar s}$ and $\phi_s^{\ddbar s}$, respectively.
Since  these decays are dominated by penguin diagrams with internal top quarks, 
the phases $\phi_s^{\ssbar s}$ and $\phi_s^{\ddbar s}$ receive  contributions from the decay amplitudes that cancel out the contribution of $-2\beta$  from the $\Bs$ - $\Bsb$ mixing, resulting vanishing values in the SM predictions. Measurements of these quantities can  probe NP in these FCNC decays. 
 LHCb has measured  $\phi_s^{\ssbar s}$ and $\phi_s^{\ddbar s}$ in the $\Bs \to \phi \phi$ and $\Bs \to \Kstar(892)\Kstar(892)$ decays using data taken during 2011-2016, and the results are
  $\phi_s^{\ssbar s} = -0.073 \pm 0.115 \pm 0.027$~rad~\cite{LHCb-PAPER-2019-019} and $\phi_s^{\ddbar s} = -0.10 \pm 0.13 \pm 0.14$~rad~\cite{LHCb-PAPER-2017-048}, respectively, where the first uncertainties are statistical and the second systematic.

\subsection{CKM elements \texorpdfstring{$V_{ub}$}{TEXT} and \texorpdfstring{$V_{cb}$}{TEXT}}
\begin{figure}
\centering
\includegraphics[width=0.6\textwidth]{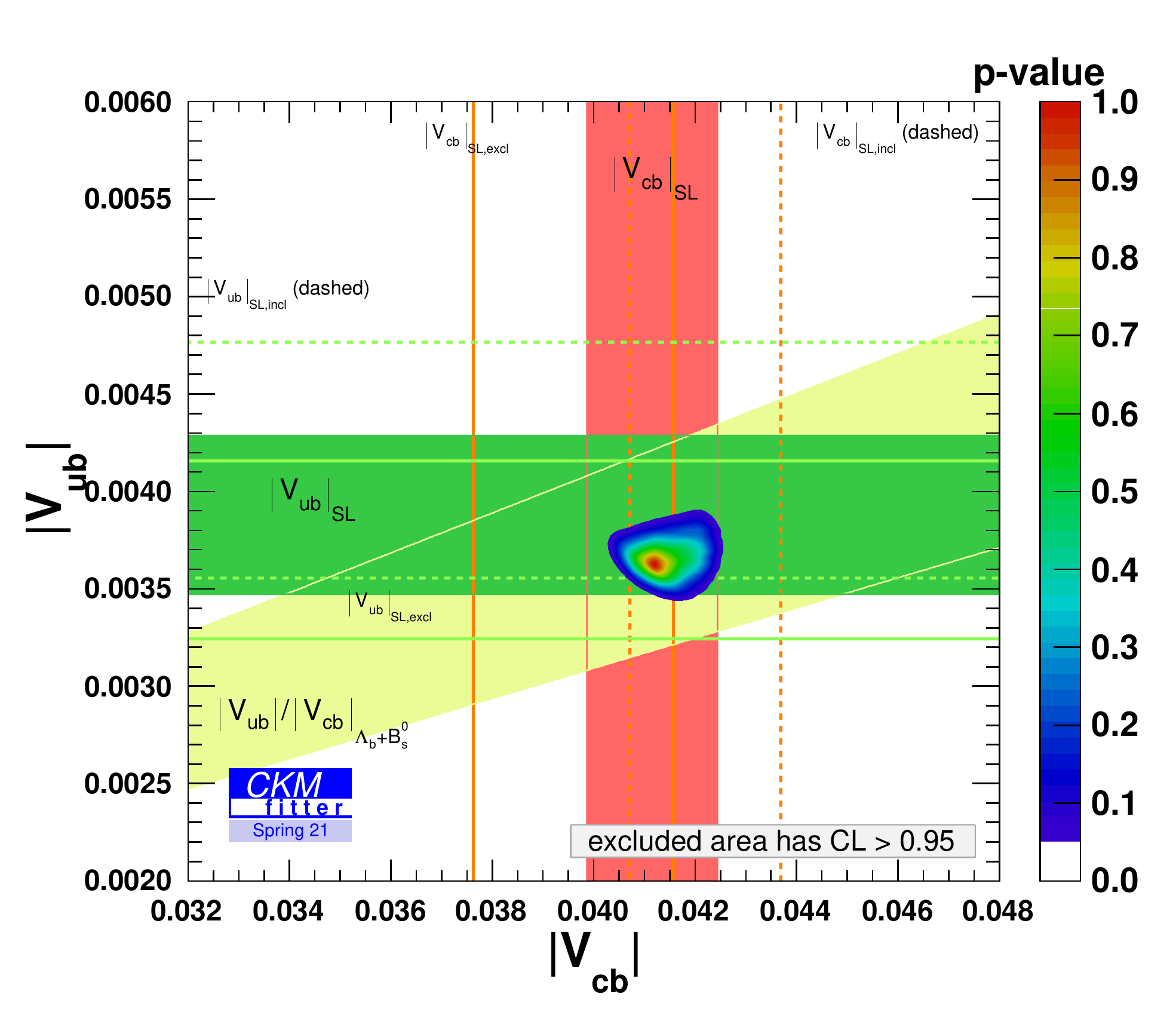}
\caption{ Global fits of $|V_{ub}|$ and $|V_{cb}|$ from the CKMfitter group~\cite{CKMfitter2005} where measurements using different methods have been displayed.
}\label{fig:VubVcb}
\end{figure}

The amplitudes of the CKM matrix elements $V_{ub}$ and $V_{cb}$ are measured through semi-leptonic transitions of $b \to u \ell\nu$ and $b \to c \ell \nu$. They have been extensively studied previously in $B$-factories using the so-called exclusive and inclusive methods, which infer to whether a specific decay channel is used or not. The two approaches suffer from different theoretical and experimental uncertainties and offer important cross-checks between each other. In the inclusive measurements, the Heavy Quark Expansion (HQE) is used as $\Lambda_{\textrm{QCD}}/m_b$ is small and the Operator Product Expansion (OPE) calculates non-perturbative contributions involved.
For the exclusive measurements, parameterisation of the form factor of the corresponding decay is needed, where inputs are obtained from light-cone sum rules (LCSR)~\cite{Li:2012nk, Wang:2015vgv, Wang:2017jow, Lu:2018cfc, Gao:2019lta, Khodjamirian:2011jp} or from lattice QCD (LQCD). However, tensions have been found between the inclusive and exclusive results as can be seen in Fig.~\ref{fig:VubVcb}~\cite{CKMfitter2005}. 
Efforts from both experimental and theoretical sides are needed to understand the discrepancy. 

Unlike $B$-factories, where full kinematics can be obtained, LHCb can not obtain kinematic information of neutrinos from energy-momentum conservation as it covers only forward region. In addition, background contributions from other $b$ and $c$ hadrons, and also huge combinatorial backgrounds randomly combined from tracks other than signal makes the analyses of semi-leptonic decays very complicated. Despite of these difficulties, the LHCb experiment has successfully measured the ratio of $|V_{ub}|$ and $|V_{cb}|$ using $\Lb \to p \mun \neum$ ($q^2 > 15\gevgevcccc$) and $\Lb \to \Lc \mu^- \neum$ ($q^2 > 7\gevgevcccc$)~\cite{LHCb-PAPER-2015-013}. The usage of the control channel $\Lb \to \Lc \mu^- \neum$ not only cancels out common systematic uncertainties between the two channels, but also offers a global scale needed to determine $|V_{ub}|$ from branching fractions. Using the updated branching fraction measurement of $\Lc \to pK^-\pi^+$, benefiting from the $\Lc\Lcbar$ data collected by the BESIII experiment~\cite{BESIII:2015bjk}, the ratio is determined to be $|V_{ub}|/|V_{cb}| = 0.079 \pm 0.009$. 

Using the same approach, the ratio of $|V_{ub}|$ and $|V_{cb}|$ is also determined using $\Bs \to K^- \mu^+ \nu_{\mu}$ and $\Bs \to D_s^- \mu^+ \neum$~\cite{LHCb-PAPER-2020-038}. In the measurement, two $q^2$ regions are used, \mbox{$q^2 > 7\gevgevcccc$} and $q^2 < 7\gevgevcccc$ where the form factors are obtained from LQCD and LCSR, respectively. However, the ratios of $|V_{ub}|$ and $|V_{cb}|$ obtained from the two methods differ significantly, $|V_{ub}|/|V_{cb}|_{q^2 < 7\gevgevcccc} = 0.0607 \pm 0.0015 \pm 0.0013 \pm 0.0008 \pm 0.0030$ and $|V_{ub}|/|V_{cb}|_{q^2 > 7\gevgevcccc} = 0.0946 \pm 0.0030^{+0.0024}_{-0.0025} \pm 0.00013 \pm 0.0068$, where the first uncertainty of each result is statistical, the second systematic, the third due to $D_s^-$ branching fraction, and the last one from the form factor.  The discrepancy between the two 
results clearly indicates that more efforts from theoretical side are needed to resolve the tension on $|V_{ub}|$ and $|V_{cb}|$ measurements. 

In addition to the ratio between the two CKM matrix elements, the LHCb experiment is also exploring its potential in determine the $|V_{cb}|$ alone using $\Bs \to D_s^{(*)-}\mu^+ \nu_{\mu}$~\cite{LHCb-PAPER-2019-041}. The branching fraction of $\Bs \to D_s^{(*)-}\mu^+ \nu_{\mu}$ is needed to set the global scale for the determination of $|V_{cb}|$. This is obtained using the control channel $B^0 \to D^{(*)-}\mu \nu_{\mu}$, where the ratios of 
\begin{eqnarray}
    R &=& \frac{\Bs \to D_s^{-}\mu^+ \nu_{\mu}}{B^0 \to D^{-}\mu \nu_{\mu}}, \\
    R^* &=& \frac{\Bs \to D_s^{*-}\mu^+ \nu_{\mu}}{B^0 \to D^{*-}\mu \nu_{\mu}},
\end{eqnarray}
are determined. 
The data from LHCb offer $q^2$ dependence needed to extract $|V_{cb}|$ together with non-perturbative inputs.
Using form factor parameterisations from Caprini, Lellouch and Neubert~\cite{Caprini:1997mu} or from Boyd, Grinstein and Lebed~\cite{Boyd:1994tt,Boyd:1997kz}, the measured values of $|V_{cb}|$ are $(41.4 \pm 0.6 \pm 0.9 \pm 1.2) \times 10^{-3}$ and $(42.3 \pm 0.8 \pm 0.9 \pm 1.2) \times 10^{-3}$, respectively, where the first uncertainty is statistical, the second systematic, and the last one due to external inputs such as theoretical inputs on form factors, branching fractions of $D_s^-$ or $D^-$ decays, $\Bs$ lifetime etc. 


\subsection{\texorpdfstring{$\Delta m_d$}{TEXT} and \texorpdfstring{$\Delta m_s$}{TEXT}}
The parameters $\Delta m_d$ and $\Delta m_s$ 
denote the mass differences between the heavy and light mass eigenstate of the $\Bd$ and \Bs systems, and define   
 the oscillation frequencies of $B^0$ mixing and $\Bs$ mixing, respectively.
Currently, The  most precise determination of $\Delta m_d$ comes from the LHCb measurements in semileptonic decays with a $D^-$ or $D^{*-}$ meson using 3 fb$^{-1}$ of data. Combining the results obtained in the two decay modes yields $\Delta m_d = (0.5050 \pm 0.0021 \pm 0.0010)$ ps$^{-1}$~\cite{LHCb-PAPER-2015-031}, where the first uncertainty is statistical and the second systematic.
The world-average is
$\Delta m_d = 0.5065 \pm 0.0019$ ps$^{-1}$~\cite{HFLAV18}. 
The determination of  $\Delta m_s$ is also led by the LHCb experiment. Combining the recent 
 measurements in the decays  $\Bs \to D_s^- \pi^+$~\cite{LHCb-PAPER-2021-005} and $\Bs \to D_s^- \pi^+ \pi^- \pi^+$~\cite{LHCb-PAPER-2020-030} and earlier measurements  yields   $\Delta m_s =17.7656 \pm 0.0057$ ps$^{-1}$.

However, the constraints on the CKM matrix elements provided by the $\Delta m_d$ and $\Delta m_s$ measurements rely on  the decay constants and  Bag parameters of the the $B^0$ and $\Bs$ mesons, which are obtained from Lattice QCD calculations~\cite{FlavourLatticeAveragingGroup:2019iem}. The precision of these hadronic parameters
is much worse than the experimental precision, thus limits the constraining  power of $\Delta m_d$ and $\Delta m_s$ on the CKM  global fit. Further improvements in Lattice QCD calculations
are eagerly awaited.

\subsection{Global fit}

 The four parameters of the
 CKM  matrix, namely $A$, $\lambda$, $\rho$ and $\eta$ in the Wolfenstein parameterization~\cite{Wolfenstein:1983yz}, are measured  in different processes and some of the key observables have been discussed in the above sections.  A global fit is needed to get the best sensitivity and to probe NP effects. 
The $\chi^2$ value of the fit provides a measure of 
the overall consistency between the different measurements,
while  the pull value for each measurement quantifies the difference between the measured value and  value predicted by the fit results.
Clues for new physics can be identified from large $\chi^2$ or pull values. 

\begin{figure}
\begin{center}
\includegraphics[width=0.6\textwidth]{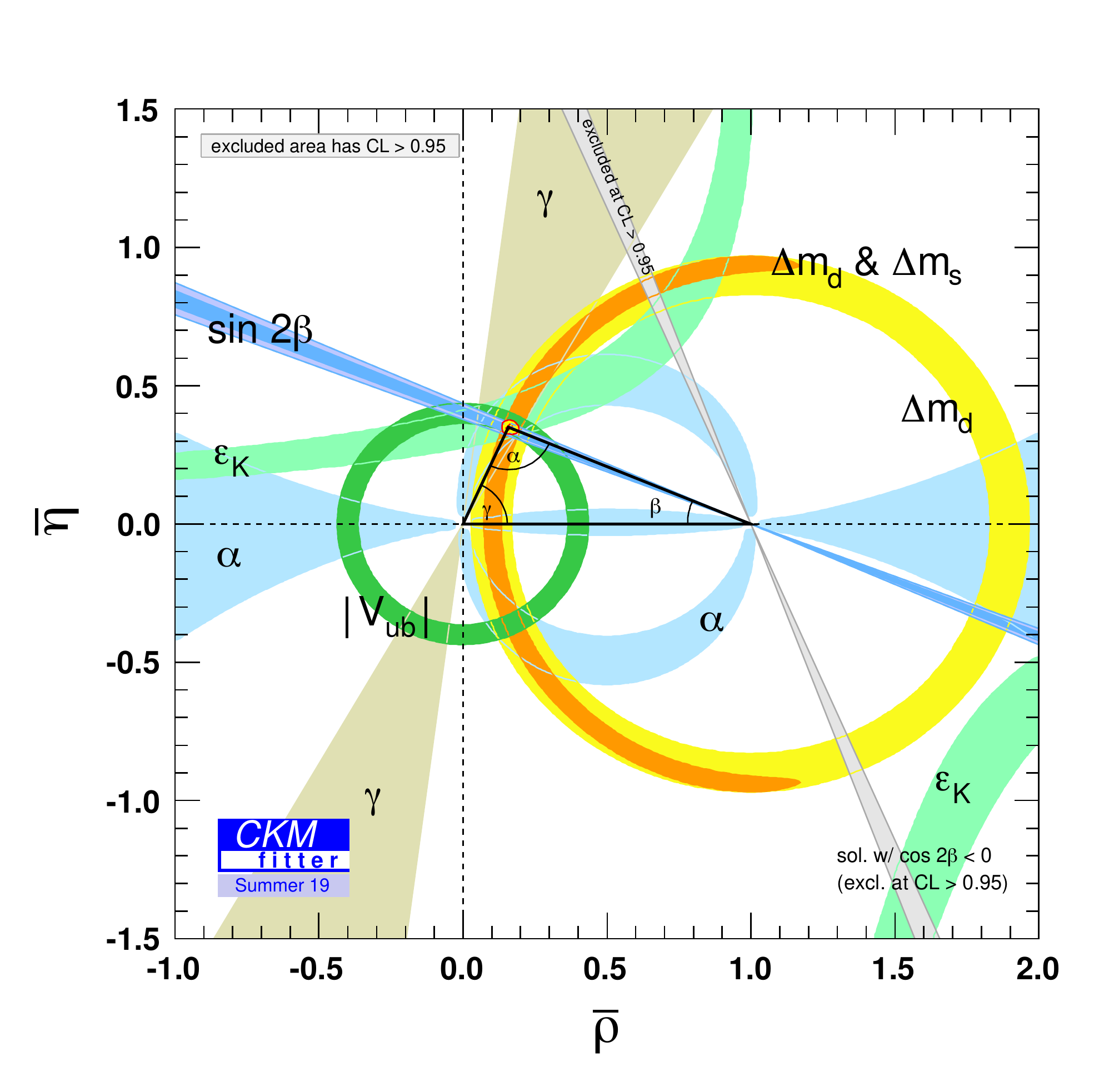}
\caption{Global fit results of the CKM matrix from different measurements, provided by the CKMfitter group~\cite{CKMfitter2005}. Different colours and labels on the plot indicate constraints from measurements of different observables.
}
\label{fig:rhoeta_large}
\end{center}
\end{figure}

The constraining of the CKM matrix is usually illustrated in  complex planes using  triangles defined using unitarity relations of the CKM matrix. The most commonly quoted CKM triangle corresponds to the relation $\Vud\Vubs + \Vcd\Vcbs + \Vtd\Vtbs =0$, where $V_{ij}$ is the CKM matrix element 
between the quarks of the flavours $i$ and $j$. 
The results of the state-of-the-art global fit  are shown in Fig.~\ref{fig:rhoeta_large} provided by the CKMfitter group~\cite{CKMfitter2015}. Inside, $\bar{\rho}$ and $\bar{\eta}$~\cite{Buras:1994ec} are used to ensure the relationship $\rhobar + i \etabar = - (V_{ud}V_{ub}^*)/(V_{cd}V_{cb}^*)$. 

Within current precision, different measurements cross on a single point and give an overall consistent picture. However,  the argument of matter-antimatter asymmetry suggests  that CP violation  from sources beyond the CKM matrix may break the consistency. One of the main efforts of the LHCb experiment is to search for such a discrepancy by further improving measurement precision in the beauty and charm sectors.

In the following decade, both the LHCb  and Belle II experiments will accumulate much more data to further constrain the CKM matrix. With data collected till 2025, either experiment will be able to reduce the uncertainty of $\gamma$ to around $1.5\degrees$, and  further improve it to $0.3\degrees$ after 2030s. Other CKM angles and matrix elements will also be significantly improved. 
Details on the future outlook
can be found in section~\ref{sec:upgrade-prospect}. Together with improvements of other measurements and lattice calculations, NP may be observed from inconsistency between different measurements. 

\clearpage

\section{Charm mixing and CP violation}
\label{sec:charmCPV}

Charm physics covers the studies of hadrons containing charm quarks.
CP violation in the charm sector is expected to be incredibly small in the SM, of the order $\mathcal{O}(10^{-3})$ or less~\cite{PDG2020}. 
However, the presence of new physics may enhance the amount of CP violation,
which can be probed using  the enormously large sample of charmed hadrons at LHCb. Particularly, the study of mixing and CP violation of neutral $D$ mesons can provide unique probes of  NP in FCNC transitions in the up-type quark sector, complementary to the study of  mixing and CP violation in neutral $B$ and $K$ mesons, which are sensitive to NP in FCNC transitions of down-type quarks. For this reason, 
this section mainly focuses on  results in  mixing and CP violation of \Dz mesons from the LHCb experiment.

\subsection{Neutral \D meson mixing}
\label{sec:05:01:CharmMixing}
\begin{figure}[!b]
\begin{center}
\includegraphics[width=0.49\textwidth]{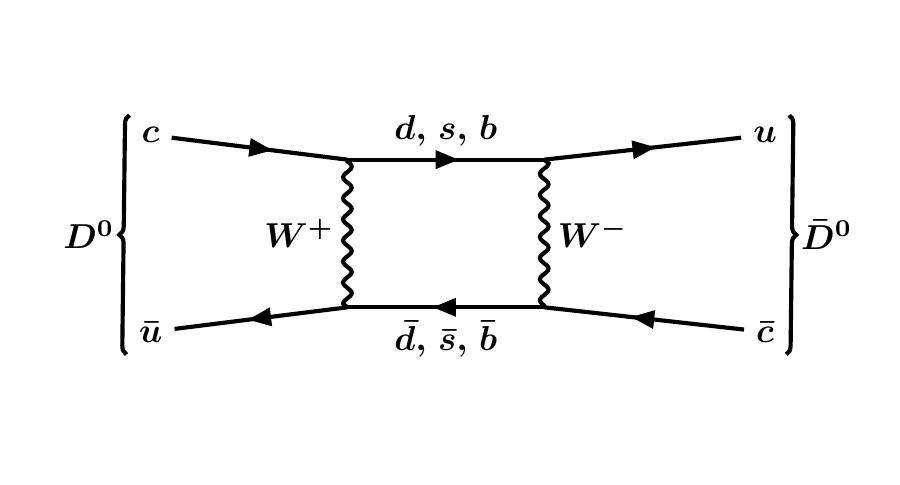}
\includegraphics[width=0.49\textwidth]{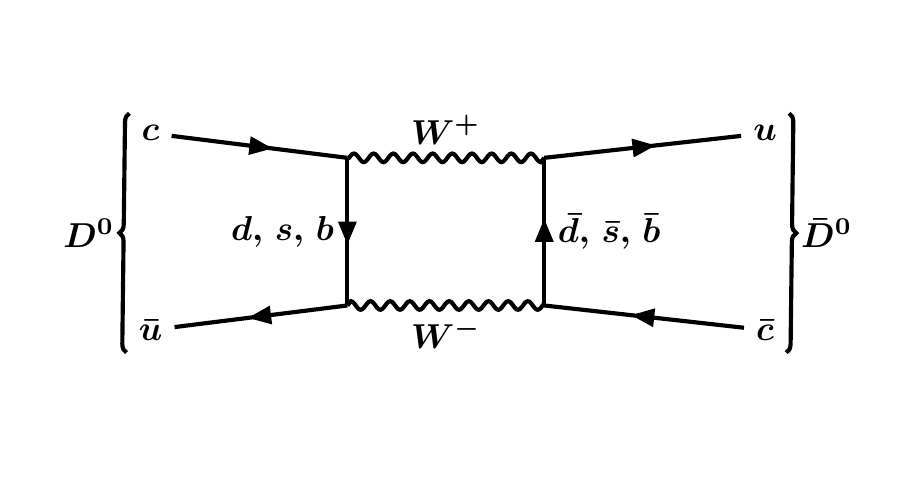}\\
\includegraphics[width=0.49\textwidth]{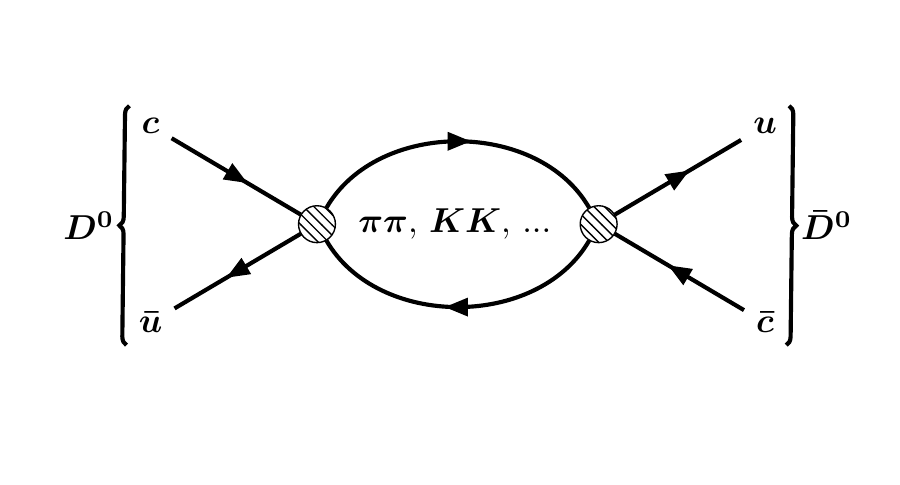}
\end{center}
\caption{Oscillation of $\Dz$ mesons via (top) $\Wpm$ exchanges or (bottom) rescattering effect.
}
\label{fig:charm:00}
\end{figure}

Similar to  neutral $\Kz$ and $\Bds$ mesons, the neutral charmed meson, $\Dz$,
can oscillate to its antiparticle partner, $\Dzb$, via the short-distance $\Wpm$ exchange or long-distance rescattering diagrams, as shown in Fig.~\ref{fig:charm:00}.
This phenomenon of oscillation or mixing can be characterised by 
the normalised (dimensionless) mixing parameters $x$ and $y$, defined as
\begin{equation}
x\equiv \frac{\Delta M}{\Gamma},\quad y\equiv \frac{\Delta \Gamma}{2\Gamma},
\end{equation}
where $\Delta M$ ($\Delta\Gamma$) is the mass (decay width) difference of the heavy and light mass eigenstates, and $\Gamma$ is the average decay width.
Unlike in the case of the $\Kz$ or $\Bds$ system, both  $x$ and $y$ in the \Dz system are significantly smaller than unity, 
thus  very large data samples are required to observe \Dz mixing and determine the  tiny values of $x$ and $y$.  

Evidence of $\Dz$-$\Dzb$ mixing was first reported by \babar~\cite{BaBar:2007kib} and \belle~\cite{BELLE:2007dgh} in 2007, and later also seen by \cdf~\cite{CDF:2007bdz} in 2008.
Subsequent measurements by \babar~\cite{BaBar:2008xkf,BaBar:2009kmh} with different \Dz decay channels provided more evidences of the mixing.  
The combination of these measurements confirmed the existence of charm mixing with a significance more than $5\sigma$.
The first observation of $\Dz$-$\Dzb$ mixing in a single measurement was achieved by LHCb~\cite{LHCb-PAPER-2012-038} in 2012 by using the data taken in 2011 to study the time-dependent ratio of $\Dz\to\Kp\pim$ (doubly Cabibbo-suppressed, DCS) to $\Dz\to\Km\pip$ (Cabibbo favoured, CF) decay rates.
The \Dz candidates are selected from the  $\Dstarp\to\Dz\pip$ decays, where the charge of the pion directly from each \Dstarp decay is used to determine the \Dz flavour at its production time.

\begin{figure}[!tb]
\begin{center}
\includegraphics[width=0.5\textwidth]{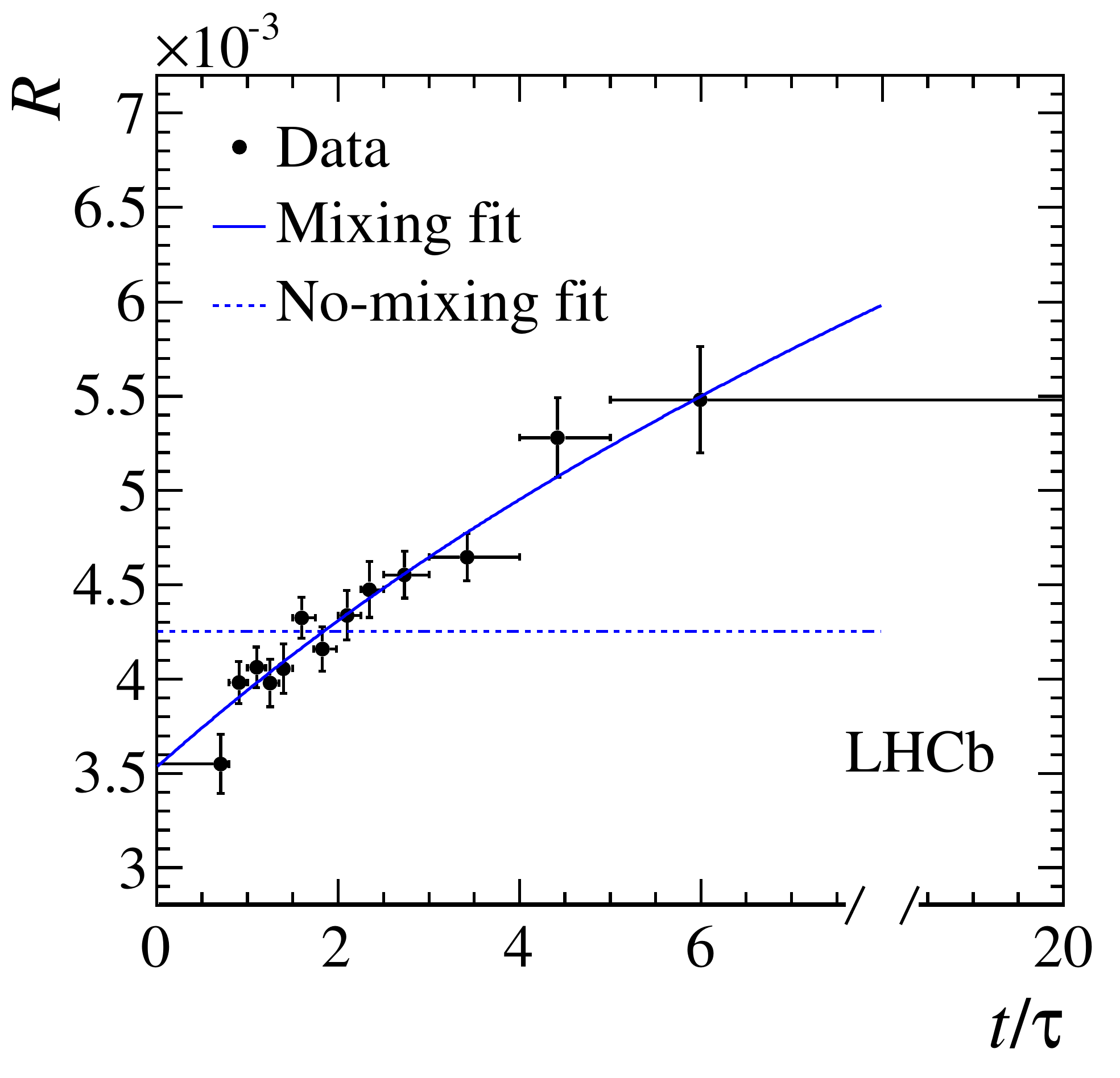}
\caption{ 
Ratio of the $\Dz\to\Kp\pi$ to $\Dz\to\Km\pip$ decay rates, $R$, as a function of the \Dz decay time normalised by the average $\Dz$ lifetime~\cite{LHCb-PAPER-2012-038}. 
The solid (dashed) line indicates the projection of the mixing  (no-mixing) fits. 
}
\label{fig:charm:03}
\end{center}
\end{figure}

The $\Dstarp\to\Dz(\to\Km\pip)\pip$ process, referred to as the right-sign (RS) process, 
is dominated by a CF decay, contaminated with a small contribution from the \Dz-\Dzb mixing followed by the DCS decay; 
the $\Dstarp\to\Dz(\to\Kp\pim)\pip$ process, referred to as wrong-sign (WS) process, 
includes contributions from both the DCS decay and the \Dz-\Dzb mixing followed by the CF decay.
Under the assumption of small mixing and negligible CP violation, the time-dependent ratio of the WS to the RS decay rates, $R$, is given by~\cite{Bianco:2003vb}
\begin{equation*}
R(t)
\approx R_D + \sqrt{R_D}\,\yprime\,\dfrac{t}{\tau}+\dfrac{{\xprime}^2+{\yprime}^2}{4}\left(\dfrac{t}{\tau}\right)^2,
\end{equation*}
where $t/\tau$ is the decay time normalised to the average $\Dz$ lifetime, $R_D$ is the ratio between the DCS and CF decay rates, and $\xprime$ and $\yprime$ are the mixing parameters `rotated' by the strong phase difference $\delta$ between the DCS and CF amplitudes: 
$x'=x\cos \delta+y\sin\delta$ and $y'=y\cos \delta-x\sin\delta$.
The time evolution of the ratio $R$ is shown in Fig.~\ref{fig:charm:03}.
Further studies with larger data samples have  also  been performed by LHCb~\cite{LHCb-PAPER-2012-038,LHCb-PAPER-2013-053,LHCb-PAPER-2016-033,LHCb-PAPER-2017-046} and the results are summarized in Table~\ref{sec:charm:tab:1}. 

\begin{table}[!t]
\begin{center}
    \caption{ Summary of the charm mixing parameters measured by LHCb  using  $\Dz\to\Kp\pim$ decays.
    Results under the assumptions of CP invariance and CP violation are given.
    }
    \label{sec:charm:tab:1}
\vskip -0.2cm
{\small
\begin{tabular}{lcccc}
\hline
 & \multicolumn{2}{c}{no CPV} &  \multicolumn{2}{c}{CPV allowed}\\
Data sample & $x'^2$ ($\times 10^{-3}$)  & $y'$ ($\times 10^{-3}$) &   $x'^2$ ($\times 10^{-3}$) & $y'$ ($\times 10^{-3}$) \\ 
\hline
$1.0\invfb$, $D^*$\ tag~\cite{LHCb-PAPER-2012-038} & $-0.09\pm 0.13$ & $7.2\pm 2.4$  & -  &  - \\
 \multirow{2}*{$3.0\invfb$, $D^*$\ tag ~\cite{LHCb-PAPER-2013-053}}&   \multirow{2}*{$0.055\pm 0.049$} &  \multirow{2}*{ $2.8\pm 1.0$}  & $\Dz$: $\phz 0.049\pm 0.070$  & $5.1\pm 1.4$   \\
 & & & $\Dzb$: $\phz 0.060\pm 0.068$ & $4.5\pm 1.4$  \\
 \multirow{2}*{$3.0\invfb$, $B$\ tag ~\cite{LHCb-PAPER-2016-033}}&   \multirow{2}*{$0.028\pm 0.310$} &   \multirow{2}*{$4.6\pm 3.7$} & $\Dz$: $-0.019\pm 0.447$ & $5.81\pm 5.26$  \\
 & & &$\Dzb$: $\phz 0.079\pm 0.433$ & $3.32\pm 5.23$  \\
 \multirow{2}*{$5.0\invfb$, $D^*$\ tag ~\cite{LHCb-PAPER-2017-046}}&   \multirow{2}*{$0.039\pm 0.027$} &  \multirow{2}*{$5.28\pm 0.52$}  & $\Dz$: $\phz 0.061\pm 0.037$  & $5.01\pm 0.74$  \\
 & & &$\Dzb$: $\phz 0.016\pm 0.039$ & $5.54\pm 0.74$ \\
\hline
\end{tabular}
}
\end{center}
\end{table}

The measurements in the $\Dz\to\Km\pip$ decay are sensitive to the normalised decay-width difference $y$ and the sum $x^2+y^2$ (under the assumption of negligible CP violation), but not to the sign of the normalised mass difference $x$. One approach to solve this problem is to study the Dalitz distributions of  three-body decays. 
The `golden channel' at LHCb for such studies is the decay $\Dz\to\KS\pip\pim$, where the decay to the  $\KS\pip\pim$  final state  proceed mainly via the following three  processes with different intermediate  resonances:
1) the $\KS\rhoz$ process with $\rhoz\to\pip\pim$, which is common for both $\Dz$ and $\Dzb$ mesons; 
2) the $\Kstarm\pip$ process with $\Kstarm\to\KS\pim$, which is a CF decay;
and 3) the $\Kstarp\pim$ process with $\Kstarp\to\KS\pip$, which is either a DCS decay or  \Dz-\Dzb oscillation followed by a CF decay. 

In the Dalitz phase space, the DCS and CF decay amplitudes of the $\Dz\to\KS\pip\pim$ decay populate the same space and interfere. Therefore, the parameters $x$ and $y$ can be determined by measuring the strong phase difference between the contributing amplitudes in an amplitude analysis, or by importing the average strong-phase difference in  regions of phase space obtained by  $\epem$ experiments operating at the energy of the $\psi(3770)$ resonance. 
The latter approach is employed in several LHCb measurements~\cite{LHCb-PAPER-2015-042,LHCb-PAPER-2019-001,LHCb-PAPER-2021-009}, and the most recent measurement~\cite{LHCb-PAPER-2021-009} led to the first observation of a nonzero mass difference between the two mass eigenstates in the $\Dz$-\Dzb system.
The results of these measurements using  $\Dz\to\KS\pip\pim$ decays are summarised in Table~\ref{sec:charm:tab:2}.
\begin{table}[!tb]
\begin{center}
    \caption{ 
    Summary of the charm mixing parameters measured by LHCb using  $\Dz\to\KS\pip\pim$ decays.
    }
    \label{sec:charm:tab:2}
\vskip -0.2cm
\begin{tabular}{ccc}
\hline
 & \multicolumn{2}{c}{ CP-averaged parameters} \\
Data sample & $x$ ($\times 10^{-3}$)  & $y$ ($\times 10^{-3}$) \\
\hline
$1.0\invfb$, $D^*$\ tag~\cite{LHCb-PAPER-2015-042} & $-8.6\pm5.3\pm1.7$  & $0.3\pm4.6\pm1.3$  \\
$3.0\invfb$, $B$\ tag~\cite{LHCb-PAPER-2019-001} & $2.7\pm1.6\pm0.4$ & $7.4\pm3.6\pm1.1$ \\
$5.4\invfb$, $D^*$\ tag~\cite{LHCb-PAPER-2021-009} & $3.97 \pm 0.46 \pm 0.29$ & $4.59 \pm 1.20 \pm 0.85$ \\
\hline
 &  \multicolumn{2}{c}{ CP-violating parameters}\\
Data sample &  $\Delta x$ ($\times 10^{-3}$) & $\Delta y$ ($\times 10^{-3}$)\\
\hline
$3.0\invfb$, $B$\ tag~\cite{LHCb-PAPER-2019-001} & $-0.53\pm0.70\pm0.22$ & $0.6\pm1.6\pm0.3$ \\
$5.4\invfb$, $D^*$\ tag~\cite{LHCb-PAPER-2021-009} & $-0.27 \pm 0.18 \pm 0.01$ & $0.20 \pm 0.36 \pm 0.13$ \\
\hline
\end{tabular}
\end{center}
\end{table}

As mentioned in Section~\ref{sec:04:01:05gammaCombination}, different from the past LHCb $\gamma$ combinations, the recent combination exploited the LHCb measurements that are sensitive to the CKM angle $\gamma$ and to the charm mixing parameters, and the $\gamma$ angle and charm mixing parameters are simultaneously determined~\cite{LHCb-PAPER-2021-033}.
The motivation for the simultaneous combination is as follows:
\begin{itemize}
\item The $\gamma$ angle and the strong phase difference between the interfering $B$ decays are now so precisely constrained by the large $B$-meson samples that the strong phase difference, $\delta_{D}^{K\pi}$, between the decays $\Dz\to\Km\pip$ and $\Dzb\to\Km\pip$ can achieve a precision of about a factor of  two better than the previous world average~\cite{HFLAV18}.
This improvement can then be used to improve the precision of the charm mixing parameters $x$ and $y$.
\item Due to non-negligible effects originating from charm-meson mixing, a simultaneous combination is needed to obtain an unbiased determination of the $\gamma$ angle and the charm mixing parameters $x$ and $y$.
\end{itemize}

\begin{figure}[!tb]
\begin{center}
\includegraphics[width=0.5\textwidth]{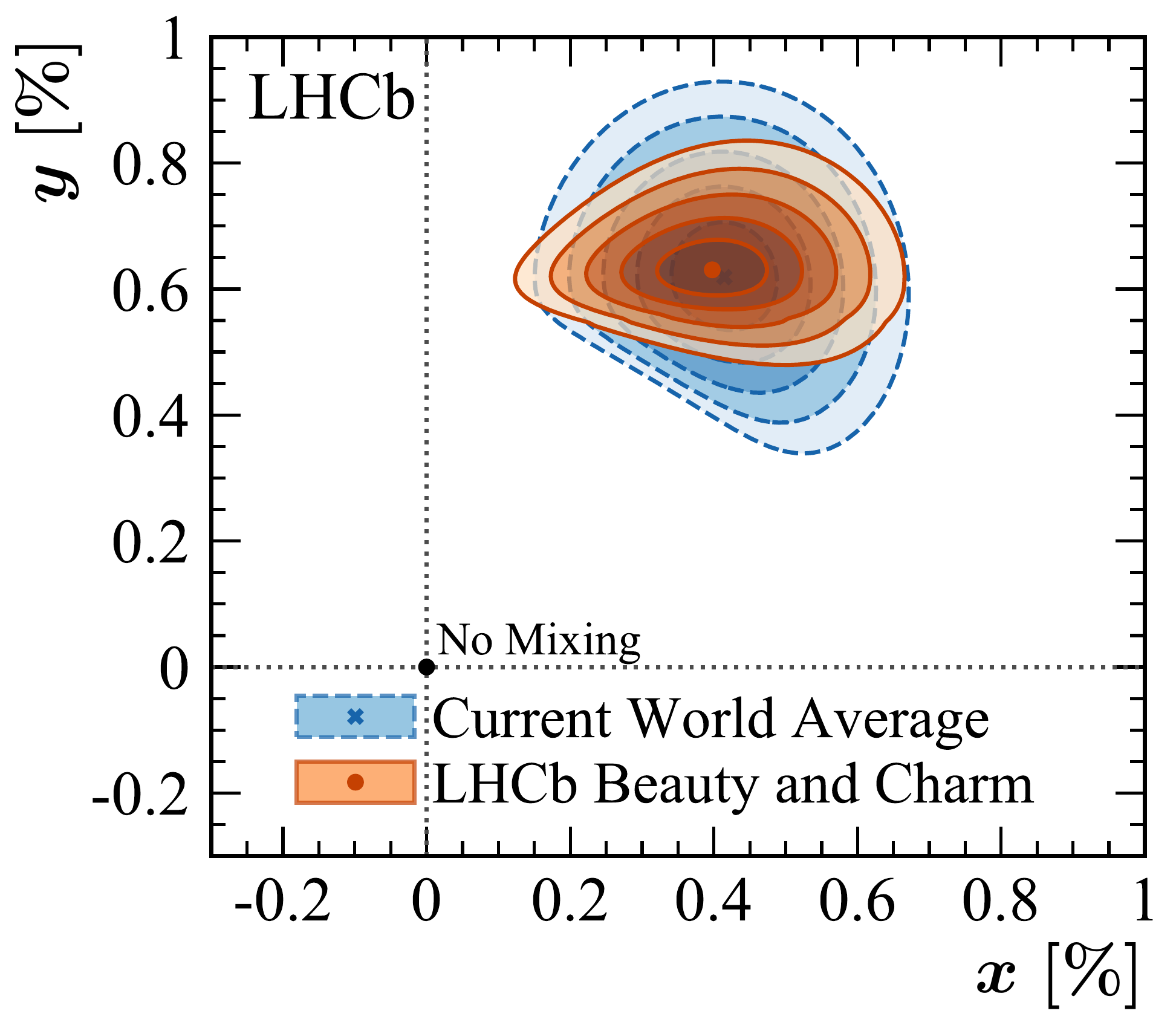}
\caption{ 
Profile likelihood contours (from 1 to 5 $\sigma$) of the charm mixing parameters $x$ and $y$.
The solid (brown) contours are determined from the simultaneous combination~\cite{LHCb-PAPER-2021-033}, while the dashed (blue) indicate the current world average from Ref.~\cite{HFLAV18}. 
}
\label{fig:charm:05:01:combinedXY}
\end{center}
\end{figure}

In the charm sector, the inputs used in the combination are obtained from the time-dependent measurements of $\Dz\to h^+ h^-$, $\Dz\to\Kp\pim$, $\Dz\to\Kpm\pimp\pip\pim$, and $\Dz\to\KS\pip\pim$  decays performed by LHCb~\cite{LHCb-PAPER-2019-006,LHCb-PAPER-2014-013,LHCb-PAPER-2015-055,LHCb-PAPER-2018-038,LHCb-PAPER-2014-069,LHCb-PAPER-2016-063,LHCb-PAPER-2019-032,LHCb-PAPER-2020-045,LHCb-PAPER-2016-033,LHCb-PAPER-2017-046,LHCb-PAPER-2015-057,LHCb-PAPER-2019-001,LHCb-PAPER-2021-009,LHCb-PAPER-2015-042}.
Figure~\ref{fig:charm:05:01:combinedXY} shows the two-dimensional profile likelihood contours in the $x$-$y$ plane.
The values of $x$ and $y$, determined in the simultaneous combination, are found to be
\begin{equation*}
x=(0.400{}_{-0.053}^{+0.052})\%,\quad
y=(0.630{}_{-0.030}^{+0.033})\%.
\end{equation*}
These results provide the most precise determinations of the parameters $x$ and $y$. 
Particularly, the precision of $y$ is improved by a factor of two with respect to the current world average~\cite{HFLAV18}.

\subsection{CP violation}


\subsubsection{Time-integrated CP violation}

The time-integrated CP asymmetry, $\mathcal{A}_{CP}$,  in the  decay $D\to f$ is dominated by the direct CP asymmetry. 
Its measurement follows the formula
\begin{equation}
    \mathcal{A}_{CP}=\mathcal{A}_{raw}-\mathcal{A}_{prod}-\mathcal{A}_{det},
\end{equation}
where $\mathcal{A}_{prod}$ denotes the 
 meson production asymmetry between the $c$-hadron and its antiparticle, 
$\mathcal{A}_{det}$ represents the detection asymmetry, and $\mathcal{A}_{raw}$ is the raw asymmetry between the yields of $D\to f$ 
and  $\overline{D}\to\overline{f}$ decays.
Often the difference of  CP asymmetries between two different decay processes  are measured, which is defined as 
\begin{equation}
 \Delta \mathcal{A}_{CP}\equiv \mathcal{A}_{CP}(D\to f_1) -\mathcal{A}_{CP}(D\to f_2),    
\end{equation}
 where $f_1$ and $f_2$ are two different final states with similar topologies. The effects of production asymmetry and CP asymmetries in mixing as well as part of the detection asymmetries  on $\Delta \mathcal{A}_{CP}$
are largely cancelled.






Two-body decays of $D$ mesons are particularly interesting due to their super large sample sizes,  
 which are crucial for probing the tiny CP violating effects.
The first observation of CP violation in the charm sector was reported by LHCb in 2019 using the Run 2 data~\cite{LHCb-PAPER-2019-006}.
The difference of $\mathcal{A}_{CP}$ between the $\Dz\to\Kp\Km$ and $\Dz\to\pip\pim$ decays, $\Delta \mathcal{A}_{CP}\equiv  \mathcal{A}_{CP}(\Kp\Km)-\mathcal{A}_{CP}(\pip\pim)$, was measured 
with a deviation from zero corresponding to a significance of $5.3\,\sigma$.
Table~\ref{sec:charm:tab:3} summarises the $\Delta \mathcal{A}_{CP}$ results of  a series of LHCb measurements~\cite{ LHCb-PAPER-2011-023,LHCb-PAPER-2013-003,LHCb-PAPER-2014-013,LHCb-PAPER-2015-055,LHCb-PAPER-2019-006}. 

Several $\mathcal{A}_{CP}$ measurements using two-body decays have been performed by LHCb as well~\cite{LHCb-PAPER-2012-052,LHCb-PAPER-2014-018,LHCb-PAPER-2015-030,LHCb-PAPER-2016-041,LHCb-PAPER-2018-012,LHCb-PAPER-2019-002,LHCb-PAPER-2020-047,LHCb-PAPER-2021-001,LHCb-PAPER-2021-051,LHCb-PAPER-2022-024}, and the first evidence of direct CP violation in a specific charm hadron decay was reported in Ref.~\cite{LHCb-PAPER-2022-024}.

\begin{table}[!tb]
\begin{center}
    \caption{ 
    Summary of the $\Delta \mathcal{A}_{CP}$ results from the LHCb measurements in the charm sector.
    }
    \label{sec:charm:tab:3}
\vskip -0.2cm
\begin{tabular}{lc}
\hline
Data sample &  $\Delta \mathcal{A}_{CP} $ ($\times 10^{-3}$) \\
\hline
$0.62\invfb$, $D^*$\ tag~\cite{LHCb-PAPER-2011-023} &  $-8.2\pm 4.1\pm0.6$\\
$1.0\invfb$, $B$\ tag~\cite{LHCb-PAPER-2013-003} &  $4.9\pm3.0\pm1.4$\\
$3.0\invfb$, $B$\ tag~\cite{LHCb-PAPER-2014-013} &  $1.4\pm1.6\pm0.8$\\
$3.0\invfb$, $D^*$\ tag~\cite{LHCb-PAPER-2015-055} & $-1.0\pm0.8\pm0.3$ \\
$5.9\invfb$, $B$ or $D^*$\ tag~\cite{LHCb-PAPER-2019-006} & $-1.54\pm0.29$ \\
\hline
\end{tabular}
\end{center}
\end{table}

While  multi-body charm decays often have much smaller sample sizes compared to two-body charm decays, they can  provide excellent opportunities for CP violation measurements. The presence of intermediate resonances can lead to large variation of the strong phase difference between the interfering amplitudes, which can lead to sizeable local CP asymmetries.
Besides the $\Delta \mathcal{A}_{CP}$ method, several techniques to search for CP violation in multi-body charm decays are exploited by LHCb, including amplitude analysis~\cite{LHCb-PAPER-2018-041}, the binned $\chisq$ technique~\cite{LHCb-PAPER-2011-017, LHCb-PAPER-2013-041, LHCb-PAPER-2013-057, LHCb-PAPER-2014-046, LHCb-PAPER-2019-026}, and an unbinned technique called the energy test~\cite{LHCb-PAPER-2014-054, LHCb-PAPER-2016-044}. 
For the latter two methods, model-dependent analyses are eventually required to pin down the source in case significant CP violation were observed.

\begin{figure}
\begin{center}
\includegraphics[width=\textwidth]{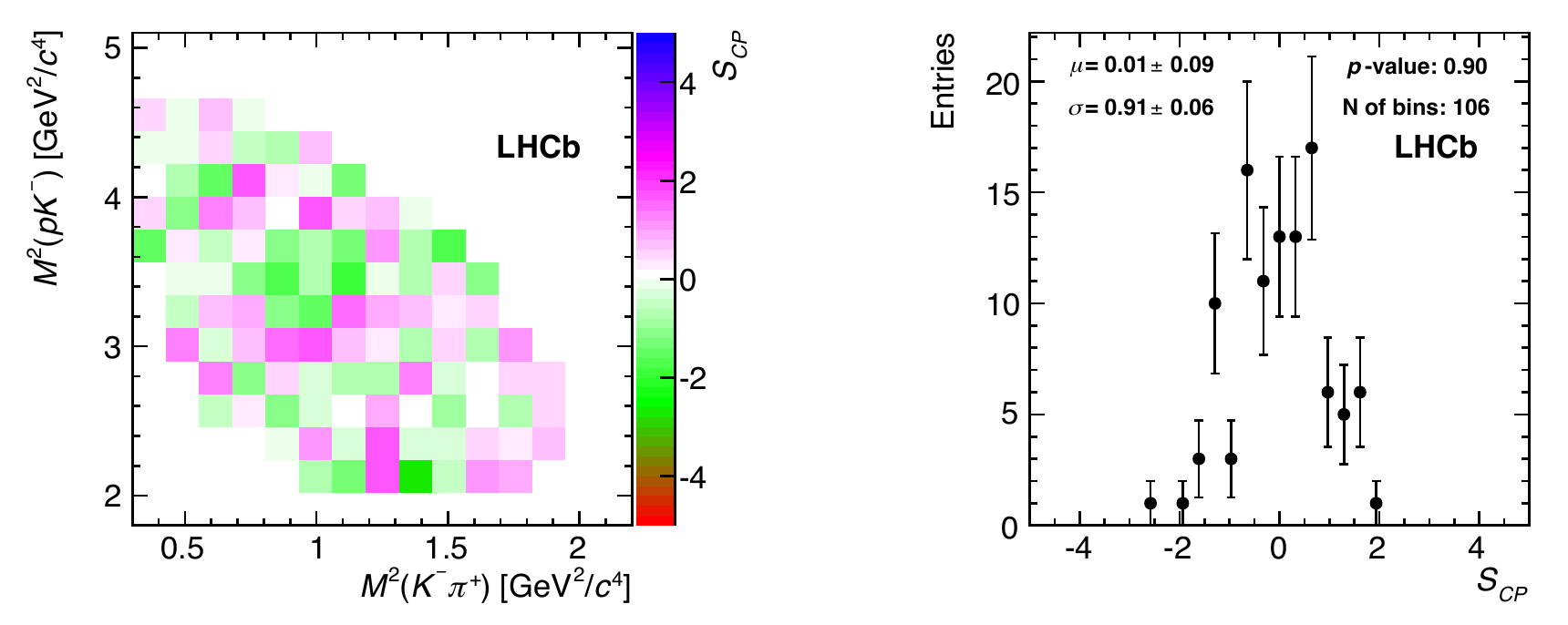}
\caption{(left) Distributions of local per-bin asymmetry significance $S_{CP}$ and (right) corresponding one-dimensional distributions obtained by the binned $\chisq$ method for $\Xires_c^+\to p \Km \pip$ decays. Figures taken from Ref.~\cite{LHCb-PAPER-2019-026}.
}
\label{fig:charm:Scp}
\end{center}
\end{figure}

The binned $\chisq$ technique computes  the distribution of local asymmetries and compare it with a normal distribution to judge if CP violation were observed. 
An example of binned $\chisq$ distribution in a Dalitz plot is shown in Fig.~\ref{fig:charm:Scp}. 
This method relies on the optimal choice of the binning scheme. 
Wide bins across resonances can lead to the cancellation of real CP asymmetries within a bin.

\begin{figure}
\begin{center}
\includegraphics[width=0.49\textwidth]{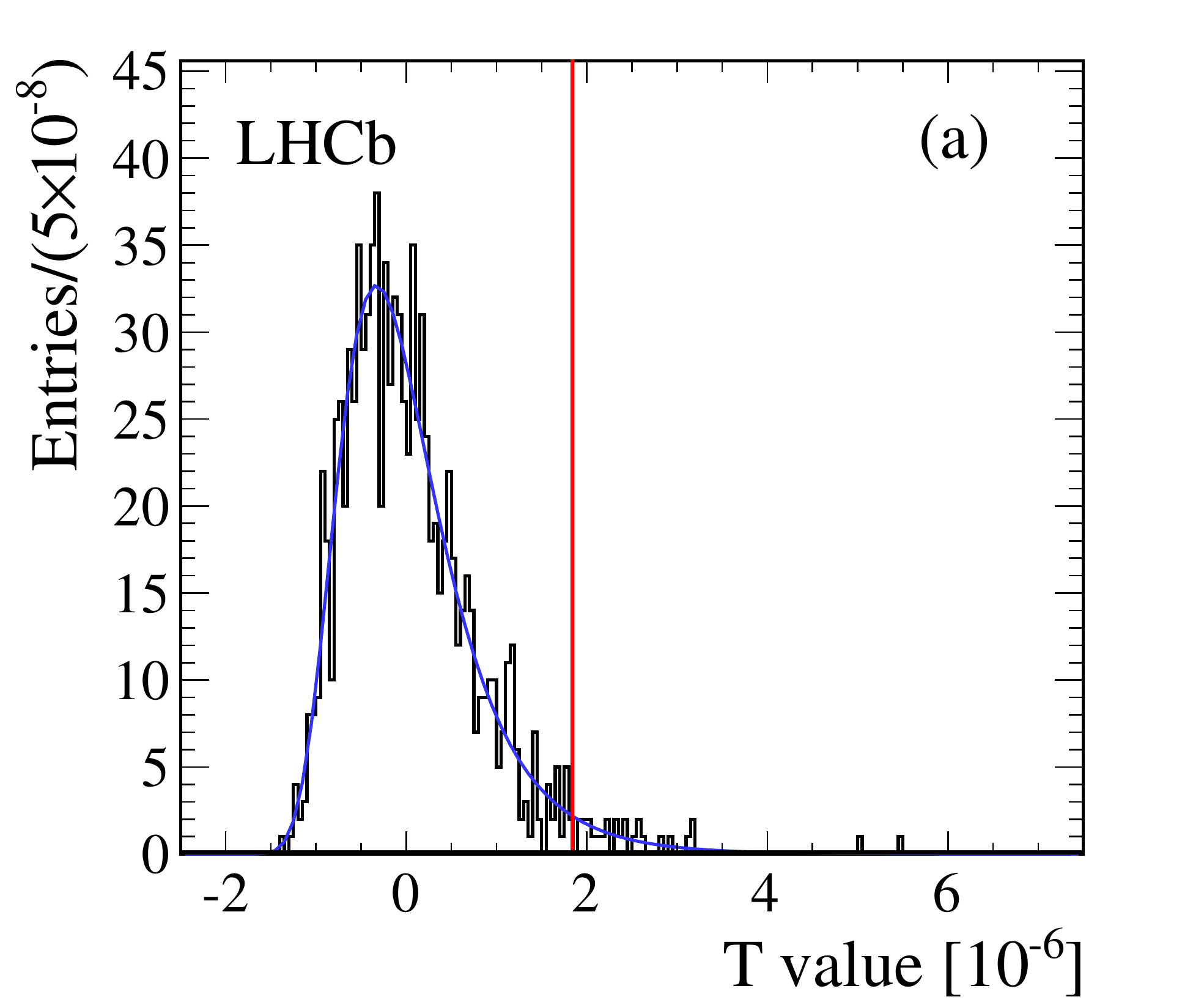}
\includegraphics[width=0.49\textwidth]{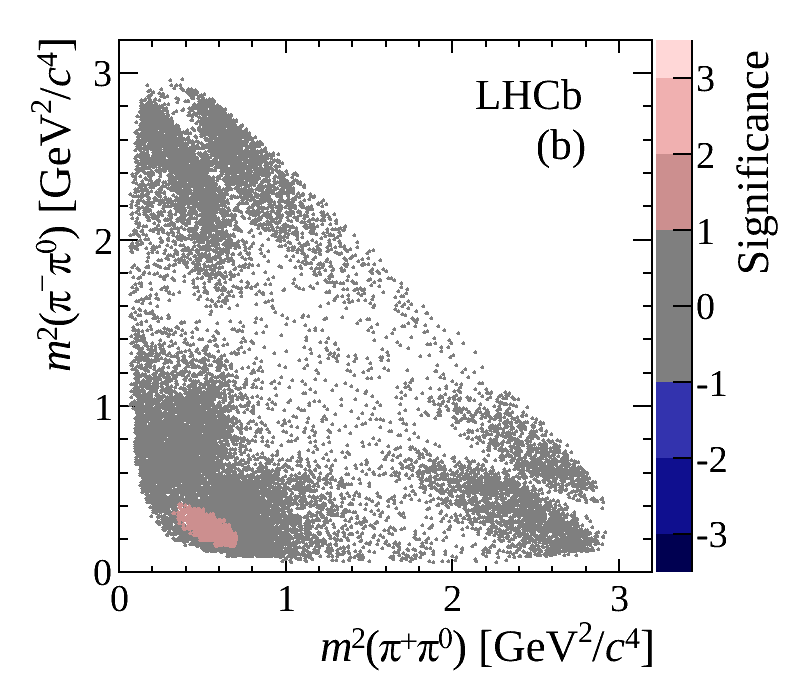}
\caption{ (a) The global  test statistic compared with the distribution of the statistic calculated from a large number of random permutations obtained from the energy test applied in the CP violation searches in $\Dz\rightarrow \pim\pip\pi^0$ decays. (b) Dalitz plot distribution of significance of local test statistics.  Figures are taken from Ref.~\cite{LHCb-PAPER-2014-054}.
}
\label{fig:charm:ET}
\end{center}
\end{figure}

The LHCb collaboration has developed a novel unbinned method, energy test~\cite{Williams:2011cd, Parkes:2016yie},  to perform model-independent 
search for CP violation in many-body decays. With this method,  a test statistic, $T$,  is defined.  For a given data sample, 
   a $p$-value for the hypothesis of CP invariance is assigned by   comparing the observed value of $T$
to the distribution of $T$ obtained from many random permutations of the data.  
This method  has been applied to search for  CP violation in decays of charm mesons and beauty baryons. As an example, Fig.~\ref{fig:charm:ET} shows the global  test statistic compared with the distribution of the statistic  from many random permutations, and  
the Dalitz plot distribution of significance of local test statistics
in $\Dz \to \pim\pip\piz$ decays.
Despite the many efforts made by LHCb and the significant improvements in the measurement precision, no  evidence of CP violation in multi-body charm decays  has ever been found  to date.  
Table~\ref{sec:charm:tab:4} summarises the searches for direct CP violation in phase space of charm decays by LHCb.

\begin{table}[!hbt]
\begin{center}
    \caption{ Summary of LHCb direct CP violation searches in phase space of charm decays.}
    \label{sec:charm:tab:4}
\vskip -0.2cm
\begin{tabular}{lcc}
\hline

Decay channel & Data sample & Method  \\
\hline
$D^+\rightarrow \Km\Kp\pip$~\cite{LHCb-PAPER-2011-017} & $35\invpb$ & binned $\chi^2$  \\ 
$\Dz\rightarrow \Km\Kp\pim\pip$~\cite{LHCb-PAPER-2013-041} & $1.0\invfb$, $D^*$\ tag & binned $\chi^2$ \\
$\Dz\rightarrow \pim\pip\pim\pip$~\cite{LHCb-PAPER-2013-041} & $1.0\invfb$, $D^*$\ tag & binned $\chi^2$ \\
$D^+\rightarrow \pim\pip\pip$~\cite{LHCb-PAPER-2013-057} &  $1.0\invfb$ & binned $\chi^2$ \\ 
$\Dz\rightarrow \Km\Kp\pim\pip$~\cite{LHCb-PAPER-2014-046} & $3.0\invfb$, $B$ tag & binned $\chi^2$ \\
$\Dz\rightarrow \pim\pip\pi^0$~\cite{LHCb-PAPER-2014-054} & $2.0\invfb$, $D^*$\ tag& energy test \\
$\Dz\rightarrow \pim\pip\pim\pip$~\cite{LHCb-PAPER-2016-044} & $3.0\invfb$, $D^*$\ tag& energy test \\
$\Lambda_c^+\rightarrow p h^- h^+$~\cite{LHCb-PAPER-2017-044}  & $3.0\invfb$ &  $\Delta \mathcal{A}_{CP}$  \\
$\Dz\rightarrow \Km\Kp\pim\pip$~\cite{LHCb-PAPER-2018-041} & $3.0\invfb$, $B$ tag& amplitude analysis \\
$\Xi_c^+\rightarrow p \Km \pip$~\cite{LHCb-PAPER-2019-026} & $3.0\invfb$&  binned $\chi^2$ \\

\hline
\end{tabular}
\end{center}
\end{table}


 \subsubsection{ $\ycp$ and $\agamma$ measurements}
 
The amplitudes of
the direct decay  of \Dz to a CP eigenstate and the decay after mixing can interfere and lead to indirect CP asymmetry. Its contribution to the time-integrated CP asymmetry is denoted $\mathcal{A}_{CP}^{ind}$.

Due to the  $\Dz$-$\Dzb$ mixing, the effective decay width of $\Dz$ decays to a CP-even final state (e.g. $f$=$\Kp\Km$ or $\pip\pim$),  $\Gamma_{CP+}$, differs from the average decay width $\Gamma$. 
We can define  the  parameter $\ycp\equiv \Gamma_{CP+}/ \Gamma-1$ to represent the amount of CP-violation in mixing. The quantity $\ycp$ is related to $x$ and $y$, $|q/p|$, and $\phi\equiv arg(q\overline{A}/pA)$,
$$
\ycp\approx\dfrac{1}{2}\left(\left|\dfrac{q}{p}\right|+\left|\dfrac{p}{q}\right|\right)y\cos\phi-\dfrac{1}{2}\left(\left|\dfrac{q}{p}\right|-\left|\dfrac{p}{q}\right|\right)x\sin\phi\;.
$$
Only if CP is conserved, $\ycp$ is equal to $y$. 
The decay rate asymmetry is defined as
\begin{align}
    \agamma\approx\dfrac{1}{2}\left(\left|\dfrac{q}{p}\right|-\left|\dfrac{p}{q}\right|\right)y\cos\phi-\dfrac{1}{2}\left(\left|\dfrac{q}{p}\right|+\left|\dfrac{p}{q}\right|\right)x\sin\phi \nonumber\;.
\end{align}
The asymmetry \agamma is related to the 
indirect CP asymmetry $\mathcal{A}_{CP}^{ind}$  through
$\mathcal{A}_{CP}^{ind}=-\agamma$.

The quantities $\ycp$ and $\agamma$ can be determined by measuring the ratio of the effective lifetimes of $\Dz$ and $\bar{\Dz}$  decays to the same CP eigenstate:
\begin{align}
  \ycp&=\dfrac{2\tau(\Dz\rightarrow f_{\not{C}\not{P}})}{\tau(\Dzb\rightarrow f_{CP})+\tau(\Dz\rightarrow f_{CP})}-1, \\
\agamma&=\dfrac{\tau(\Dzb\rightarrow f_{CP})-\tau(\Dz\rightarrow f_{CP})}{\tau(\Dzb\rightarrow f_{CP})+\tau(\Dz\rightarrow f_{CP})},  
\end{align}
where   $f_{\not{C}\not{P}}$ denotes a non-CP-eigenstate, such as $\Km\pip$.

\begin{figure}
\begin{center}
\includegraphics[width=0.49\textwidth]{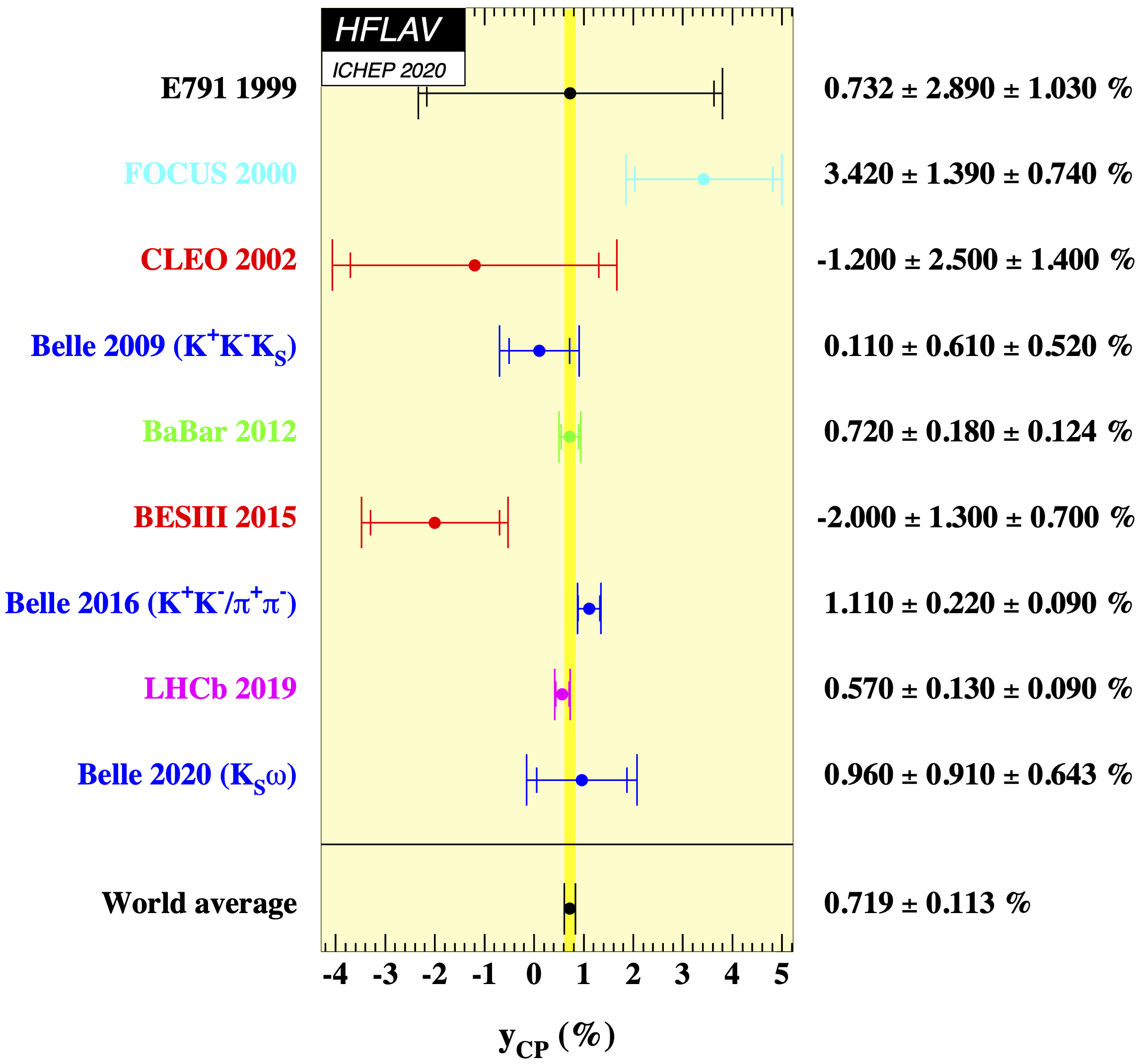}\includegraphics[width=0.49\textwidth]{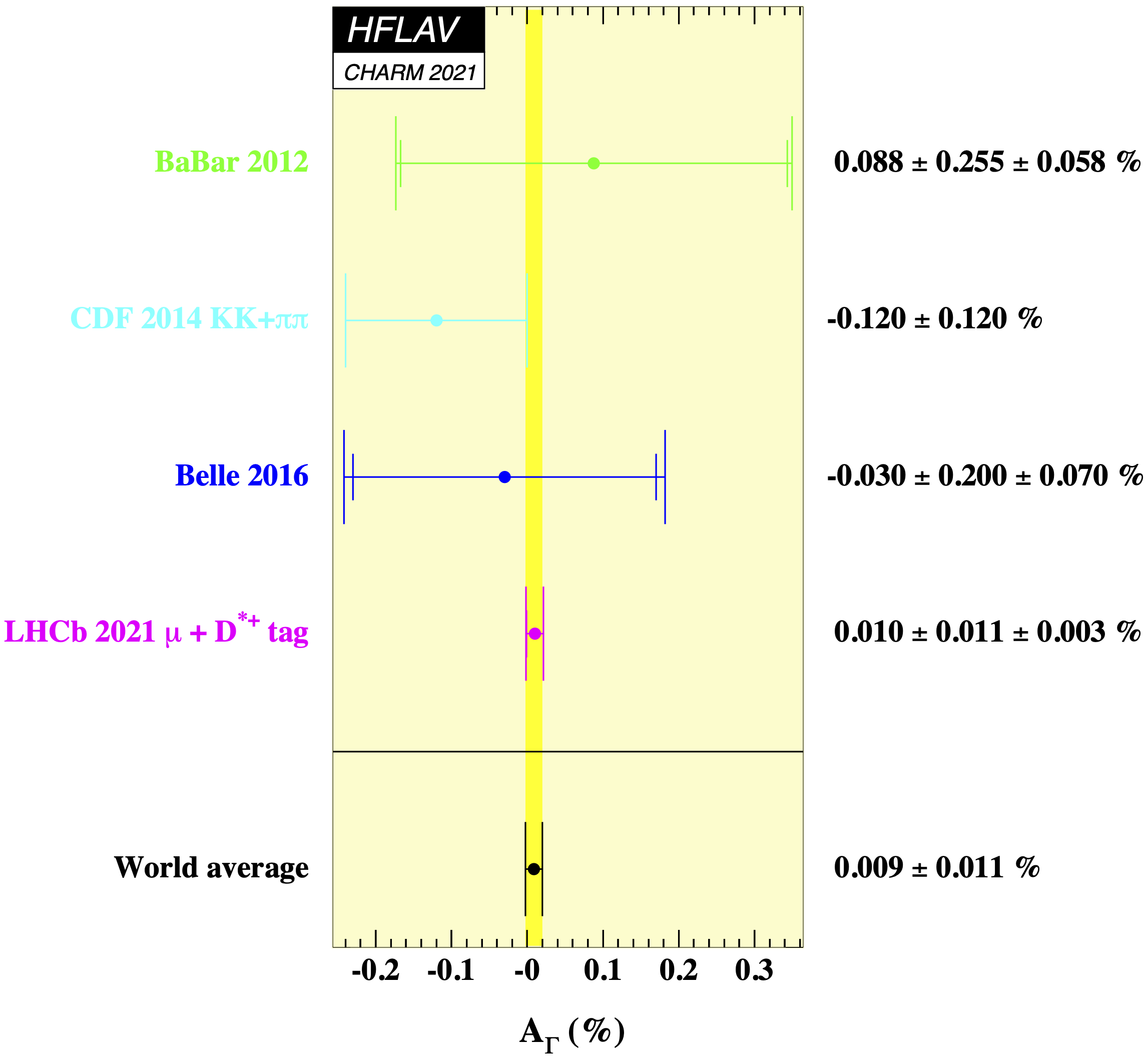}
\caption{The world averages of  $\ycp$ and $\agamma$. Figures taken from Ref.~\cite{HFLAV18}
}
\label{fig:charm:average}
\end{center}
\end{figure}

In recent years, LHCb has preformed several measurements of $\ycp$ and $\agamma$, which are summarized in
Table~\ref{sec:charm:tab:ycp}~\cite{LHCb-PAPER-2011-032,LHCb-PAPER-2013-054,LHCb-PAPER-2014-069,LHCb-PAPER-2018-038,LHCb-PAPER-2016-063,LHCb-PAPER-2019-032,LHCb-PAPER-2021-041}. 
A recent study shows that using the average decay width of $\Dz\to\Km\pip$ and $\Dzb\to\Kp\pim$ decays as a proxy to the   average decay width of the neutral charm meson mass eigenstates $D_1$ and $D_2$ does not give direct access to $\ycp$
but rather corresponds to $\ycp-\ycp^{K\pi}$~\cite{Pajero:2021jev}, where $\ycp^{K\pi}$ is approximately equal to $-0.4\times 10^{-3}$\cite{LHCb-PAPER-2021-041}. 
In Ref~\cite{LHCb-PAPER-2020-045}, an LHCb legacy result of $\agamma$ combined with both $D^*$ and $B$ flavour tag using 2011-2012 and 2015-2018 data sample is obtained.
None of these measurements  shows any indication of CP violation in $\Dz$-$\Dzb$ mixing or in the interference between mixing and decay. Figure~\ref{fig:charm:average} compares the $\ycp$ and $\agamma$ measurements performed by different experiments, and the  averages provided by the Heavy Flavour Averaging Group~\cite{HFLAV18} . The world averages are dominated by the measurements  by the LHCb experiment.

\begin{table}[!t]
\begin{center}
    \caption{ Summary of LHCb $\ycp$ and $\agamma$ measurements.}
    \label{sec:charm:tab:ycp}
\vskip -0.2cm
    \resizebox{\textwidth}{!}{%
\begin{tabular}{lccc}
\hline
Data sample & Final state(s) & $\ycp$ (\%)  & $\agamma$ ($\times 10^{-3}$) \\
\hline
$29\invpb$, $D^*$\ tag~\cite{LHCb-PAPER-2011-032} & $\Kp \Km$  & $0.55\pm0.63\pm0.41$ & $-5.9\pm 5.9\pm 2.1$ \\
$1.0\invfb$, $D^*$\ tag~\cite{LHCb-PAPER-2013-054} &  $\pip \pim$  & - & $0.33\pm1.06\pm 0.14$ \\
$1.0\invfb$, $D^*$\ tag~\cite{LHCb-PAPER-2013-054} &$\Kp \Km$  &  - & $-0.35\pm0.62\pm0.12$ \\
$3.0\invfb$, $B$\ tag~\cite{LHCb-PAPER-2014-069} &  $\pip \pim$ & - & $-0.92\pm2.6^{+0.25}_{-0.33}$\\
$3.0\invfb$, $B$\ tag~\cite{LHCb-PAPER-2014-069} & $\Kp \Km$  & - & $-1.34\pm 0.77^{+0.26}_{-0.34}$\\
$3.0\invfb$, $B$\ tag~\cite{LHCb-PAPER-2014-069} & $\pip \pim$ \& $\Kp \Km$  & - & $-1.25\pm0.73$  \\
$3.0\invfb$, $B$\ tag~\cite{LHCb-PAPER-2018-038} & $\pip \pim$ \& $\Kp \Km$  & $0.57\pm 0.13\pm 0.09$ & -  \\
$3.0\invfb$, $D^*$\ tag~\cite{LHCb-PAPER-2016-063} &  $\pip \pim$ & - & $0.46\pm 0,58\pm 0.12$\\
$3.0\invfb$, $D^*$\ tag~\cite{LHCb-PAPER-2016-063} & $\Kp \Km$  & - & $-0.30\pm 0.32\pm 0.10$\\
$3.0\invfb$, $D^*$\ tag~\cite{LHCb-PAPER-2016-063} & $\pip \pim$ \& $\Kp \Km$  & - & $-0.13\pm 2.0\pm 0.7$  \\
$5.4\invfb$, $B$\ tag~\cite{LHCb-PAPER-2019-032} &   $\pip \pim$ & - & $0.22\pm 0.70\pm 0.08$   \\
$5.4\invfb$, $B$\ tag~\cite{LHCb-PAPER-2019-032} &  $\Kp \Km$ & - & $-0.43\pm 0.36\pm 0.05$ \\
$6\invfb$, $D^*$\ tag~\cite{LHCb-PAPER-2020-045} &  $\pip \pim$ & - & $0.4\pm 0.28\pm 0.04$ \\
$6\invfb$, $D^*$\ tag~\cite{LHCb-PAPER-2020-045} &  $\Kp \Km$ & - & $0.23\pm 0.15\pm 0.03$ \\
$8.4\invfb$, $D^*$\ or\ $B$\ tag~\cite{LHCb-PAPER-2020-045} &  $\pip \pim$ & - & $0.36\pm 0.24\pm 0.04$ \\
$8.4\invfb$, $D^*$\ or\ $B$\ tag~\cite{LHCb-PAPER-2020-045} &  $\Kp \Km$ & - & $0.03\pm 0.13\pm 0.03$ \\
$8.4\invfb$, $D^*$\ or\ $B$\ tag~\cite{LHCb-PAPER-2020-045} &  $\pip \pim$ \& $\Kp \Km$ & - & $0.10\pm0.11\pm0.03$ \\
$6\invfb$, $D^*$\ tag~\cite{LHCb-PAPER-2021-041} &  $\pip \pim$ & $0.657\pm 0.053\pm 0.016$~\footnotemark[1]  & - \\
$6\invfb$, $D^*$\ tag~\cite{LHCb-PAPER-2021-041} &  $\Kp \Km$ &  $0.708\pm 0.030\pm 0.014$~\footnotemark[2] & - \\
$6\invfb$, $D^*$\ tag~\cite{LHCb-PAPER-2021-041} &  $\pip \pim$ \& $\Kp \Km$ &  $0.696\pm 0.026\pm 0.013$~\footnotemark[3] & - \\
\hline
\end{tabular}
}
\end{center}
{\footnotesize
\footnotetext[1] 01.  $\ycp^{\pi\pi}-\ycp^{K\pi}$ is measured in this analysis. \\
\footnotetext[2] 02. $\ycp^{KK}-\ycp^{K\pi}$ is measured in this analysis. \\
\footnotetext[3] 03. $\ycp-\ycp^{K\pi}$ is measured in this analysis. \\
}
\end{table}

\clearpage

\section{Prospects and summary}
\label{sec:upgrade}
\newcommand{\invcmsqs}{\,\mathrm{cm}^{-2}\mathrm{s}^{-1}}

\subsection{Upgrade plan of LHCb}
\label{sec:upgrade-plan}
\begin{figure}[!b]
\begin{center}
\includegraphics[width=0.8\textwidth]{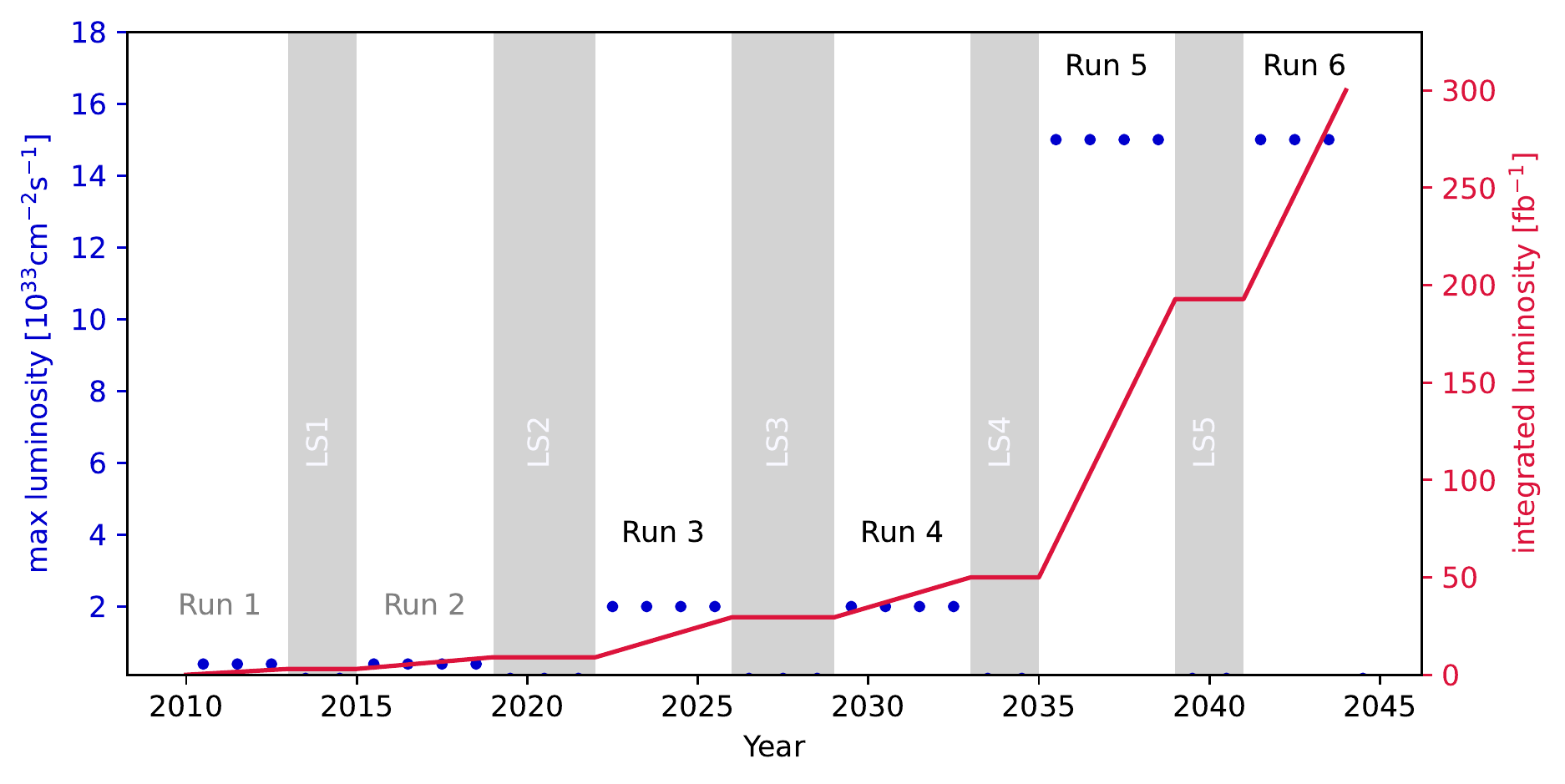}
\caption{Instantaneous and integrated luminosities of LHCb as functions of year.}
\label{fig:upgrade-timeline}
\end{center}
\end{figure}

The physics output discussed in this review shows that LHCb has successfully
deepened our understanding of flavour physics with experimental data taken
up to the year of 2018. Most of the the key flavour observables are measured to
an unprecedented precision, yet it is generally true that the uncertainties
are still dominated by statistical fluctuation. To further increase the availability
of high-quality collision data, the LHCb detector is currently under a major
upgrade~\cite{LHCb-TDR-012}, known as Upgrade Ia or simply Upgrade~I.
The installation has almost completed by the end of the second Long Shutdown (LS2)
of the LHC, and the upgraded detector is starting to take data in  2022
with an instantaneous luminosity of $2 \times 10^{33} \invcmsqs$,
five times the value achieved so far.
Fig.~\ref{fig:upgrade-timeline} shows the plan for LHCb operation after upgrade.
LHCb aims to accumulate an integrated luminosity of approximately $23 \invfb$
by the end of Run 3 around 2025, and a total of $50 \invfb$ by the end of Run 4.
Note that during the Long Shutdown 3 (LS3) between Run 3 and 4, intensive work
will be done on the machine configuration to prepare the High Luminosity Large Hadron Collider (HL-LHC)~\cite{Aberle:2749422}.
During this period consolidation work (Upgrade~Ib) will be carried out at LHCb
with only minor change on the detector configuration or performance.
In order to fully exploit the HL-LHC potential in flavour physics, the collaboration
plans another major upgrade, Upgrade II~\cite{LHCb-PII-EoI}, to enable
the detector to operate at luminosity as high as $1.5 \times 10^{34} \invcmsqs$.
This will allow for an integrated luminosity of $\sim 300\invfb$ to be achieved
in the lifetime of the (HL-)LHC.

To put the LHCb prospects in context, other players in the field of
flavour physics study have to be mentioned. The Belle II
experiment~\cite{Hashimoto:2004sm}, the $B$-factory at the superKEKB~\cite{Ohnishi:2013fma},
has started to take collision data of $\epem\to\FourS\to \B\Bbar$ since 2018, 
aiming to collect an integrated luminosity of $50 \invab$ by 2025.
Belle II and LHCb are expected to be competitive in pushing the
measurement precision, though with very different systematic uncertainties,
of a number of key flavour observables such as the CKM unitarity angles,
the Wilson coefficients and the tests on lepton flavour universality.
Given the difference in production mechanism and corresponding detector features,
their complementarity should be more appreciated.
With the beam energy constraints Belle II will be generally good at treating
final states with neutral particles ($\gamma$, $\piz$) or missing energy (neutrinos);
It has unique advantage in studying the $\tau$ leptons
through $\epem \to \tautau$ process.
With the huge cross section of heavy hadrons, LHCb will have larger yield in
most of the final states especially those with charged tracks.
The beauty hadrons produced at high-energy $pp$ collisions are highly boosted,
hence their decay vertices could be well separated from the primary vertices.
With an excellent vertex detector LHCb fully exploits this phenomenon
to suppress background for most $B$ signatures to an  extremely low level.
In addition to $B$ mesons, the studies at LHCb extend to all type of heavy hadrons
like $\Bs$, $\Bc$ and beauty baryons. 
Future electron positron colliders proposed primarily for Higgs study, such as
CEPC~\cite{CEPCStudyGroup:2018ghi} and FCC-ee~\cite{FCC:2018evy}, 
also plan to operate at $Z$ pole. With yields of $B$-mesons comparable or higher
than Belle II depending on the machine design, future $Z$ factories produce
all spectrum of beauty hadrons with large boost and efficient trigger, 
hence will also contribute to heavey flavour physics with unique advantages.
At the HL-LHC era ATLAS and CMS will keep pushing the high-energy frontier
by studying the Higgs and searching for signature of new physics beyond
the Standard Model, meanwhile the yield of beauty hadrons will be so considerable
that measurement of a few key flavour observables can be made precisely,
especially those with a pair of muons in their final state.
They are designed to perform well at high $\pt$ and central rapidity,
perfectly complementing the kinematic range of LHCb.
The BESIII experiment~\cite{BESIII:2009fln} will continue to 
operate for another 10 years~\cite{BESIII:2020nme} and accumulate a sample of 
charm mesons corresponding to $\sim 20\invfb$, which will not only allow 
more mixing and CP violation study in the charm sector but also help to 
reduce the uncertainties related to the charm strong phase in measurement of the CKM $\gamma$ angle~\cite{Wilkinson:2021tby}.
Even larger $D\bar{D}$ samples at the order of $\sim \mathrm{ab}^{-1}$ are 
expected at proposed future tau-charm factories SCTF~\cite{Epifanov:2020elk}
and STCF~\cite{Peng:2020orp}.

\subsubsection{Upgrade I}
\label{sec:upgrade-UI}
The ongoing Upgrade I aims to increase the instantaneous luminosity of LHCb
from the current value of $4 \times 10^{32}\invcmsqs$ to $2 \times 10^{33}\invcmsqs$.
Na\"{i}vely it would mean a five-fold increase in all signal yield, which will be
roughly true for final states with muons. Actually the gain for all-hadronic
final state will be more than that due to a major change in the trigger
system~\cite{LHCb-TDR-016}.
The hardware trigger L0, which reduces the data rate from \mbox{40 MHz} to
\mbox{1 MHz}, will be completely removed after Upgrade I,
allowing a more flexible full software trigger.
Generally the trigger efficiencies for all-hadronic final states
are expected to be doubled, as taken in simulation study of upgrade
performances, however this number could vary depending on the individual channel.

The increased pile-up causes much higher combinatorial background and
more challenging track reconstruction, therefore the tracking
systems~\cite{LHCb-TDR-013,LHCb-TDR-015} have been completely redesigned
with higher granularity and better radiation tolerance so as to provide
uncompromised tracking performance at higher pile-up.
Components of the particle identification systems~\cite{LHCb-TDR-014}
will be reused as much as possible, yet the readout electronics
will be replaced in accordance with the 40 MHz readout rate. 
As a result of higher luminosity, improved trigger rate and larger number of
output channels, the data volume to be treated either in real time or offline
will be substantially higher, hence new software infrastructure
and computing models have been developed
correspondingly~\cite{LHCb-TDR-017,LHCb-TDR-018,LHCb-TDR-021} to ensure
physics data to be processed and stored in a timely manner.
A new subsystem has been installed to enhance the detector's 
capability in fixed target and heavy-ion studies without
disturbing the main physics program~\cite{LHCb-TDR-020}.
The simulation study shows that detector performance after upgrade
will be at least as good as before, with improvement at some areas.

\subsubsection{Upgrade II}
\label{sec:upgrade-UII}
By the end of Run 4 LHCb will have accumulated 50 $\invfb$ $pp$ collision data,
with many subdetectors reaching end of lifetime.
Operation at the same condition beyond that point would be less attractive.
To fully exploit the HL-LHC potential in flavour physics study, 
the collaboration proposed Upgrade II towards an integrated luminosity of
$300 \invfb$~\cite{LHCb-PII-EoI}.
The physics cases with the luminosity an order of magnitude higher than
before HL-LHC time have been studied extensively by the collaboration
and summarised in a document in 2018~\cite{LHCb-PII-Physics}.
A few benchmarks will be discussed below.
Note that the HL-LHC baseline design assumes LHCb running condition
to be the same as in Run 3, the HL-LHC experts recently released a report on
the upgrade feasibility from the machine side~\cite{Efthymiopoulos:2319258}
showing that possible solution of operating at a luminosity of
$1.5 \times 10^{34}\invcmsqs$ will allow the target of $300 \invfb$ to be met.

The seven-fold increase of luminosity will again impose more technical
challenges for the experiment.
The expected number of interactions per crossing is around 40, twenty times
of current situation (or a hundred times of the LHCb initial design).
Fast timing resolution will be required in most subsystems to fight against
the combinatorial backgrounds caused by the pile-up.
Finer granularity in all tracking detectors is compulsory
under much higher multiplicity.
Radiation hardness will be more of concern especially for areas close
to the beampipe.
A daunting amount of 200 Tb of data will be produced every second, and has to
be reduced by four orders of magnitude before stored permanently.
New subsystems are being proposed in order to extend geometrical acceptance
for low-momentum tracks, and to improve $K/\pi$ separation at lower momentum.
A lot of development activities have been launched driven by these requirements,
while exploiting new technologies in detector and computing.
A framework Technical Design Report summarising these activities
was released recently~\cite{LHCb-TDR-023}.

\subsection{Physics prospects}
\label{sec:upgrade-prospect}
\begin{figure}
\begin{center}
 \includegraphics[width=0.8\textwidth]{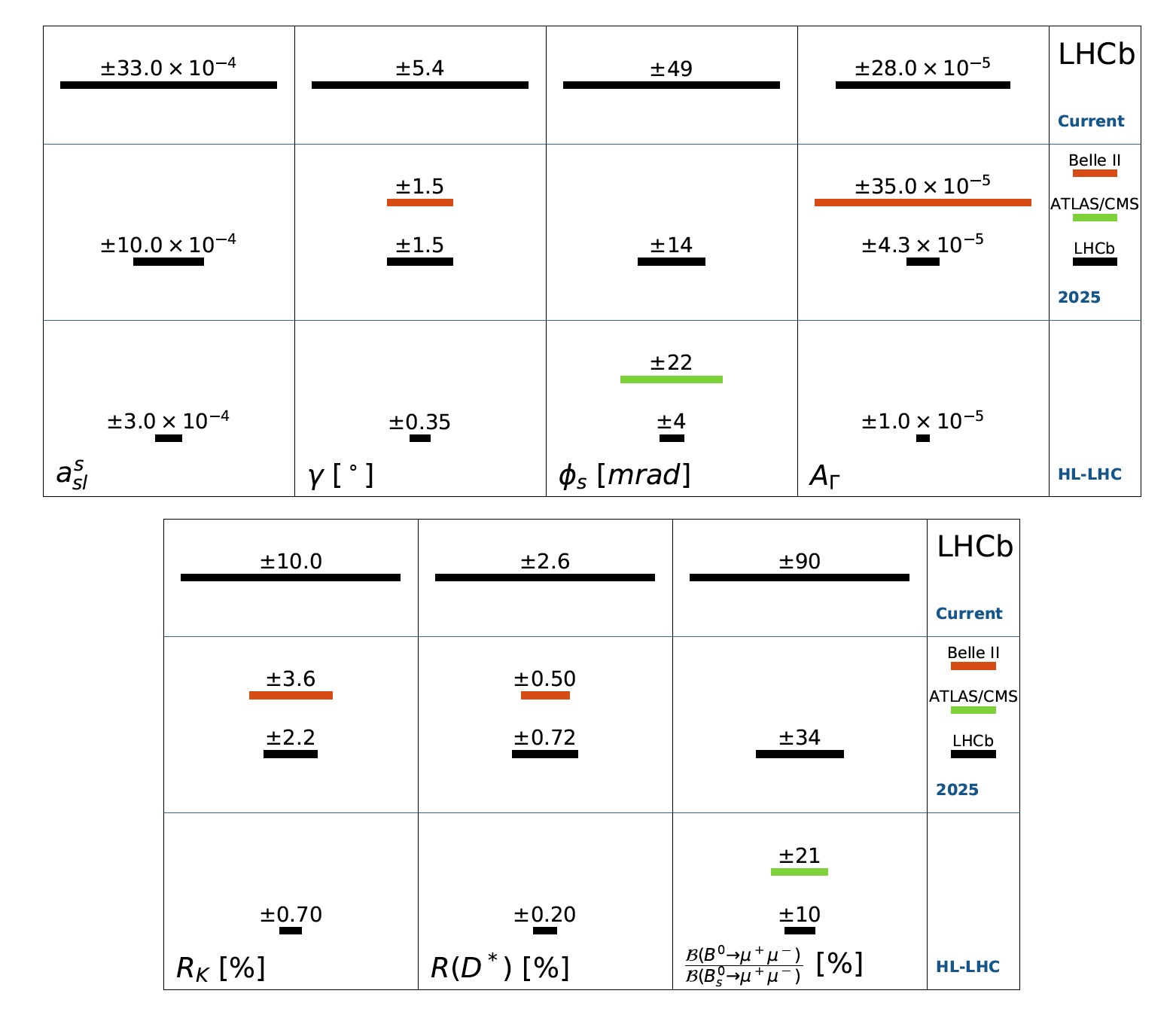}
\caption{Sensitivity to probe key CP violating variables, rare decay and lepton flavour universality tests expected from LHCb upgrades. Anticipated results from Belle II, ATLAS or CMS are listed when available. The figure is taken from Ref.~\cite{LHCb-PII-Physics}.}
\label{fig:upgrade-key-sensitivity}
\end{center}
\end{figure}

Before the HL-LHC or by the end of Run 3, LHCb will have taken $23 \invfb$ data,
drastically reducing the statistical uncertainties for most of channels
compared with current measurements. The expected projections are studied
in detail~\cite{LHCb-TDR-012} and updated with inputs from experiences in
Run I~\cite{LHCb-PAPER-2012-031,Gerson:1604468}.
The physics opportunities in Upgrade II with $300 \invfb$ 
have also been studied~\cite{LHCb-PII-Physics}, which concludes
that the energy scale probed by flavour observables will be doubled
with respect to pre-HL-LHC era. 
The sensitivity of a selection of key flavour observables after LHCb upgrades are listed in
Table~\ref{tab:upgrade-key-observables} and illustrated in
Fig.~\ref{fig:upgrade-key-sensitivity}, mostly from Ref.~\cite{LHCb-PII-Physics}
with minor updates when available. Note that Belle II will have completed
data taking when LHCb collects $23 \invfb$ data. Expected projection from Belle II,
ATLAS and CMS are listed for comparison when applicable.
A few highlights will be briefly mentioned here.

\begin{table}
  \begin{center}
    \caption{Sensitivity of selected key flavour observables for LHCb, ATLAS and CMS, taken from Ref.~\cite{LHCb-PII-Physics} with updates when available, and  Belle II sensitivities   from Ref.~\cite{Belle-II:2018jsg}. }
    \label{tab:upgrade-key-observables}
    \resizebox{\textwidth}{!}{%
    \begin{tabular}{lccccc}
    \hline
               & LHCb    &  LHCb       & LHCb         & Belle II & ATLAS \\
    Observable & current & ($23\invfb$) & ($300\invfb$) & ($50\invab$) & \& CMS \\
    \hline
    \textbf{CKM tests} & & & & & \\
    $\gamma$ (all modes) & $4^{\circ}$~\cite{LHCb-CONF-2020-003,LHCb-PAPER-2021-033}
    & $1.5^{\circ}$ & $0.35^{\circ}$ & $1.5^{\circ}$ & $-$ \\
    $\gamma$ ($B_s^0 \to D_s^+ K^-$) & $(_{-22}^{+17})^{\circ}$ & $4^{\circ}$ & $1^{\circ}$
    & $-$ & $-$ \\
    $\sin 2\beta$ & 0.04~\cite{LHCb-PAPER-2017-029} & 0.011
    & 0.003 & 0.005 & $-$ \\
    \multirow{2}{*}{$\phi_s$ ($B_s^0 \to \jpsi \phi$)} &
    \multirow{2}{*}{49 mrad~\cite{LHCb-PAPER-2014-059}} & \multirow{2}{*}{14 mrad} &
    \multirow{2}{*}{4 mrad} & \multirow{2}{*}{-} & 22 mrad~\cite{TheATLAScollaboration:2013ppf}\\
    & & & & & 5-6 mrad~\cite{CMS:2018wpq} \\
    $\phi_s$ ($B_s^0 \to D_s^+D_s^-$) & 170 mrad~\cite{LHCb-PAPER-2014-051} & 35 mrad
    & 9 mrad & $-$ & $-$ \\
    $\phi_s^{\ssbar{}s}$ ($B_s^0 \to \phi\phi$) & 154 mrad~\cite{LHCb-PAPER-2014-026} & 39 mrad
    & 11 mrad & $-$ & feasible~\cite{CMS:2017gvo} \\
    $a_{sl}^{s}$ & $33 \times 10^{-4}$~\cite{LHCb-PAPER-2016-013} & $10 \times 10^{-4}$
    & $3 \times 10^{-4}$ & $-$ & $-$ \\
    $|V_{ub}|/|V_{cb}|$ & 6\%~\cite{LHCb-PAPER-2015-013} & 3\% & 1\% & 1\% & $-$ \\
    \hline
    \textbf{Charm} & & & & & \\
    $\Delta \ACP$ & $2.9 \times 10^{-4}$~\cite{LHCb-PAPER-2019-006} & $1.7 \times 10^{-4}$
    & $3.0 \times 10^{-5}$ & $5.4 \times 10^{-4}$ & $-$ \\
    $\agamma$ & $1.3 \times 10^{-4}$~\cite{LHCb-PAPER-2020-045} & $4.2 \times 10^{-5}$
    & $1.0 \times 10^{-5}$ & $3.5 \times 10^{-4}$ & $-$ \\
    \hline
    \textbf{$\Bds \to \mu^+ \mu^-$} & & & & & \\
    $\frac{\BF(B^0 \to \mu^+ \mu^-)}{\BF(B_s^0 \to \mu^+ \mu^-)}$
    & 71\%~\cite{LHCb-PAPER-2021-007,LHCb-PAPER-2021-008} & 34\% & 10\%
    & $-$ & 21\%~\cite{CMS:2015iha,CMS:2018uxm} \\
    $\tau_{B_s^0 \to \mu^+ \mu^-}$ & 14\%~\cite{LHCb-PAPER-2021-007,LHCb-PAPER-2021-008}
    & 8\% & 2\% & $-$ & $-$ \\
    \hline
    \textbf{EW penguins} & & & & & \\
    $R_K$ ($B^+ \to K^+ \ellp\ellm$) & 0.044~\cite{LHCb-PAPER-2021-004} & 0.025 & 0.007
    & 0.036 & $-$ \\
    $R_{K^{*}}$ ($B^0 \to \Kstarz\ellp\ellm$) & 0.10~\cite{LHCb-PAPER-2017-013} & 0.031 & 0.008
    & 0.032 & $-$ \\
    \textbf{LFU tests} & & & & & \\
    $R_{D^*}$ ($\Bz \to \Dstarm \ellp \nu$) & 0.026~\cite{LHCb-PAPER-2015-025,LHCb-PAPER-2017-027}
    & 0.007 & 0.002 & 0.005 & $-$ \\
    $R_{\jpsi}$ ($B_c^+ \to \jpsi\ellp \nu$) & 0.24~\cite{LHCb-PAPER-2017-035}
    & 0.07 & 0.02 & $-$ & $-$ \\
    \hline
  \end{tabular}
  }
\end{center}
\end{table}

\textbf{Rare decays:} The decay of $B^0 \to \mu^+\mu^-$ is not very far from being observed
with imminent Upgrade I data, and its branching fraction with respect to $B_s^0 \to \mu^+ \mu^-$
will be measured with 10\% uncertainty giving a powerful test of minimal flavour violation.
A wide range of studies will be performed in $b \to s \ellp\ellm$ or $b \to d \ellp\ellm$ decays with
improved precision, so the current hint of discrepancy in $R_{K^{(*)}}$ with SM predictions
will be confirmed or excluded with confidence. A series of tests on lepton flavour
universality can be carried out in $b \to c \ellm \nu$ decays. The precision of $R_{D^*}$
will reach per mille level.

\begin{figure}
\begin{center}
  \includegraphics[width=0.48\textwidth]{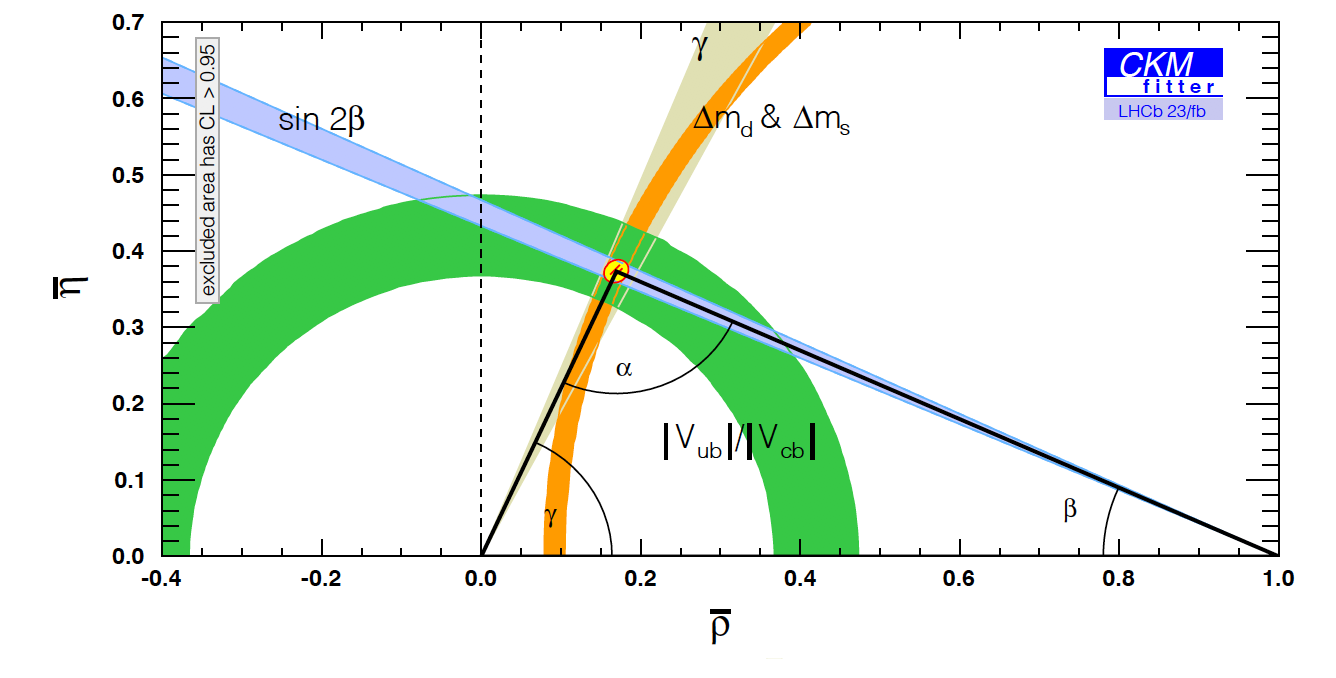}
  \includegraphics[width=0.48\textwidth]{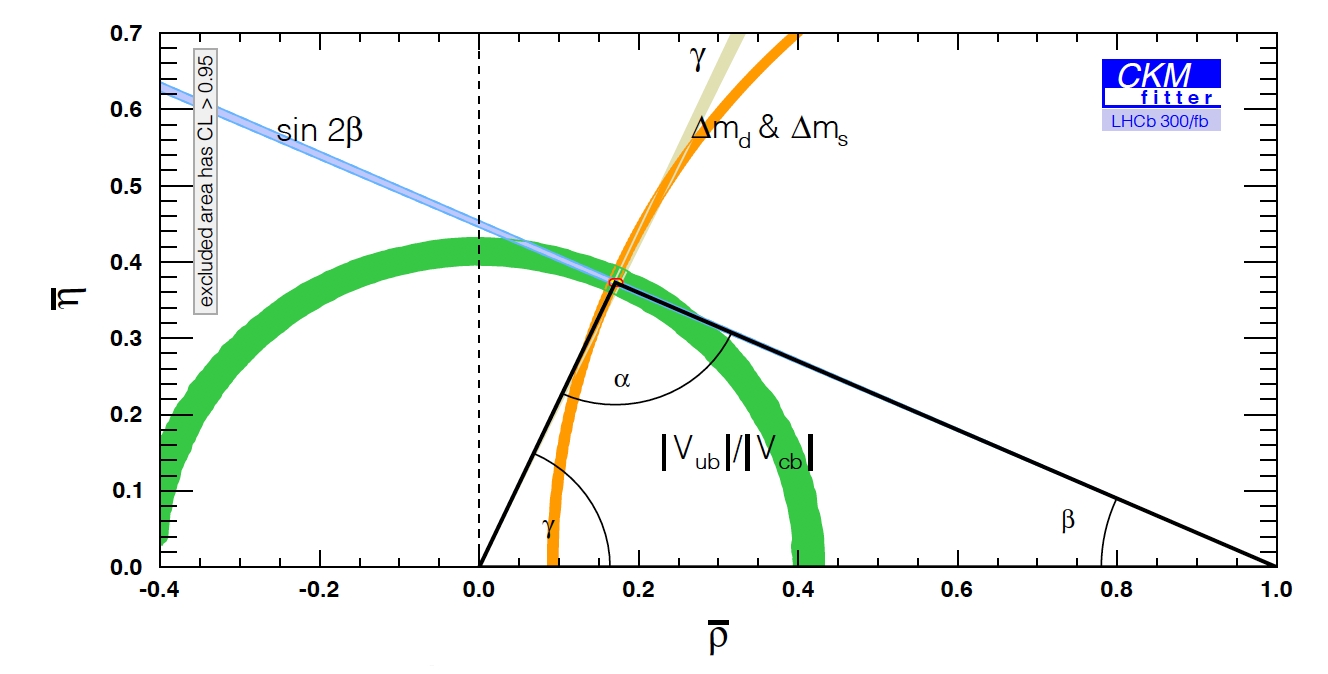}
\caption{LHCb constraints to the unitarity triangle with anticipated improvement from (left) Upgrade I and (right) Upgrade II. The figure is taken from Ref.~\cite{LHCb-PII-Physics}.}
\label{fig:upgrade-CKM-angles}
\end{center}
\end{figure}

\textbf{CKM tests:}  The CKM unitarity triangle will be so precisely determined in the future
that discrepancies between various measurement caused by physics beyond Standard Model will
be extremely difficult to hide, as shown in Fig.~\ref{fig:upgrade-CKM-angles}.
The angle $\gamma$, currently still the least well-known,
will be determined with an uncertainty of $1.5^{\circ}$ after Run 3, similar
as the precision expected from Belle II; The uncertainty will be further reduced to
$0.35^{\circ}$ after Upgrade II.
The expected precision on $B_s$ weak mixing angle $\phi_s$ will be pushed to a few mrad,
the same level as the current precision determined indirectly from CKM fit using
tree-level measurements.

\textbf{CP violation in charm:} After Upgrade II LHCb will be able to probe CP violation
in charm with a sensitivity of $\mathcal{O}(10^{-5})$, the only future facility promising
in observing indirect CP violation in charm which is predicted to be $\mathcal{O}(10^{-4})$
or less in the SM.

\textbf{Heavy-flavour spectroscopy:}
LHCb has demonstrated its capability as a general purpose
detector in the forward region, and it will be better equipped in this respect after Upgrades.
With data sample over an order of magnitude large than currently available, LHCb will
systematically update our knowledge on heavy hadron spectroscopy, including less-studied conventional hadrons, such as doubly heavy baryons, and those labelled
as exotic nowadays, such as pentaquark and tetraquark states.

\textbf{Beyond flavour physics:} 
In the high $\pt$ range LHCb could also make contribution complementary to ATLAS
and CMS, for instance to the precision determination of the effective weak mixing angle
$\sin\theta_W^2$ and the $W$ mass. Measurements of top pair and gauge boson production
at LHCb are also crucial to study the poorly known gluon parton distribution functions
at high-$x$ range. This is an important study in the QCD, which help to understand the
ubiquitous background for any new high-mass states in ATLAS or CMS.
LHCb will also push sensitivity in search for dark-photon and long-lived particles
predicted in several NP scenarios.

\subsection{Summary}
\label{sec:upgrade-summary}
This manuscript briefly reviews the recent experimental highlights using data collected
with the LHCb detector in its first 10-year operation. They are not only from the flavour physics benchmarks that the experiment
was designed for, but also include unexpected discoveries revealing LHCb's
capability as a general purpose detector in the forward region:
\begin{itemize}
    \item A large variety of new particles are discovered, either filling gaps in conventional heavy hadron spectroscopy or establishing new types of their own, like pentaquarks or tetraquarks;
    \item Some processes predicted to be extremely rare in the SM are observed, such as the $B_s^0 \to \mu^+ \mu^-$ decay. Precise measurements are performed on semileptonic and radiative FCNC beauty decays, where NP at high energy scale can be probed with promising sensitivity, and tensions with the SM are found in some cases;
    \item The CKM parameters are determined precisely using multiple approaches in a wide range of final states. The angle $\gamma$, which was the least known in the unitarity triangle, has been determined with an unprecedented precision of about $4^{\circ}$;
    \item Heavy flavour study in the charm sector witnesses a couple of milestones, such as the observation of $\Dz$-$\Dzb$ mixing in a single measurement and observation of non-zero mass difference between the two mass eigenstates in the $\Dz-\Dzb$ system. Precision in probing the CP violation in charm keeps pushing forward.
\end{itemize}
This is by far not a complete list~\cite{LHCbpapers}, and many  interesting topics studied at LHCb
are not covered  due to limited space. With the  Upgrade I detector in place, the LHCb experiment has resumed operation in 2022 and will continue to take data at a higher luminosity  while preparing for the future Upgrade II.
New exciting physics results are expected, which will continue to shape the landscape of heavy flavour physics and beyond.

\clearpage

\section*{Acknowledgements}

This work is partially supported by 
the National Key Research and Development Program of China under grant Nos. 2017YFA0402100, 2022YFA1601900,
National Natural Science Foundation of China (NSFC) under grant Nos. 11435003, 11575091, 11575094, 11925504, 11975015, 12175245,  12175005, 11705209, 12205312,
12275100, 11961141015, 12061141007,  Chinese Academy of Sciences, Fundamental Research Funds for the Central Universities, 
 Peking University Funds for the New Faculty Startup program.
We thank Franz Muheim and Niels Tuning for suggestions in improving the draft.

\bibliographystyle{LHCb}
\bibliography{main,standard,LHCb-PAPER,LHCb-CONF,LHCb-DP,LHCb-TDR,production}

\end{document}